\documentclass[11pt,oneside]{report} 

\usepackage[utf8]{inputenc}
\usepackage[a4paper,width=150mm,top=35mm,bottom=35mm]{geometry}

\usepackage{fancyhdr}
\usepackage{graphicx}
\graphicspath{ {figures/} }

\usepackage[
backend=bibtex,
style=phys, 
sorting=none,
maxbibnames=99,
maxcitenames=99,
giveninits=true,
url=false,
isbn=false,
doi=false,
eprint=true]
{biblatex}
\DeclareFieldFormat{labelnumberwidth}{\mkbibbrackets{#1}\space}

\usepackage[colorlinks=true,
			linkcolor=black,
			citecolor=blue,
			urlcolor=black]{hyperref}
\usepackage{xcolor}
\usepackage{multirow}
\usepackage{amsmath,amssymb,amsfonts,amsthm}
\usepackage{ytableau}
\usepackage{subcaption}
\usepackage{dsfont}
\usepackage{tocloft}
\usepackage{multirow}
\usepackage{pdfpages}

\numberwithin{equation}{section}

\makeatletter
\g@addto@macro\bfseries{\boldmath}
\makeatother

\theoremstyle{thmstyleone}
\newtheorem{theorem}{Theorem}
\newtheorem*{proposition}{Proposition}

\theoremstyle{definition}%
\newtheorem*{example}{Example}%
\newtheorem*{remark}{Remark}%
\newtheorem*{definition}{Definition}%

\newcommand{\ntr}{\underline{\mathrm{tr}}}
\newcommand{\tr}{\mathrm{tr}}
\newcommand{\Tr}{\mathrm{Tr}}

\newcommand{\E}{\mathbb{E}}
\newcommand{\I}{\mathbb{I}}

\newcommand{\Lin}{\mathcal{L}}
\newcommand{\Ord}{\mathcal O}
\newcommand{\Hil}{\mathcal H}
\newcommand{\nn}{\\ \nonumber}
\def\<{\langle}
\def\>{\rangle}
\def\cvec{\boldsymbol{\mathfrak{c}}}

\begin{document}
\pagenumbering{roman}



# Fluctuation of coherences in noisy mesoscopic quantum systems with diffusive transport

Soutenue par
**Ludwig Hruza**
Le 16 février 2024

Ecole doctorale n° 564
**Physique en Île-de-France**

Spécialité
**Physique Théorique**

arXiv:2408.03917v1 [math-ph] 7 Aug 2024

## RÉSUMÉ

La motivation de cette thèse est de trouver une description hydrodynamique fluctuante des effets cohérents quantiques pour des systèmes quantiques mésoscopiques ayant des propriétés de transport diffusif. Ces effets cohérents sont codés dans les cohérences (la fonction de Green à deux points) sur lesquelles nous nous concentrons en tant qu'éléments constitutifs d'une telle théorie. Notre approche est plutôt mathématique et n'est pas liée à des expériences concrètes. Nous étudions le « quantum symmetric simple exclusion process » (QSSEP), un modèle potentiellement iconique de fermions libres bruitées sur réseau dans 1D, qui possède la structure minimale pour saisir ce qui nous intéresse : les cohérences à longue portée et le transport diffusif. En moyenne, QSSEP se réduit au SSEP (symmetric simple exclusion process), un modèle jouet classique qui a été important dans le développement de la théorie des fluctuations macroscopiques, une description hydrodynamique fluctuante du transport diffusif dans des systèmes classiques. En étudiant QSSEP, nous espérons avancer vers une extension quantique cohérente de la théorie des fluctuations macroscopiques.

La thèse résume de nombreux résultats que nous avons obtenus sur QSSEP au cours des trois dernières années, tels que l'équation dynamique et la solution stationnaire exacte pour les fonctions de corrélation des cohérences à des échelles hydrodynamiques, la distribution de l'intrication quantique dans QSSEP et un argument pourquoi QSSEP pourrait être une description stochastique effective pour des systèmes quantiques mésoscopiques plus génériques. Beaucoup de ces résultats sont obtenu grâce à une relation entre la statistique des cohérences dans QSSEP et la théorie des probabilités libres. Nous consacrons un chapitre entier à cette relation et présentons, comme sous-produit, une méthode pour caractériser le spectre des sous-blocs d'une grande classe de matrices aléatoires structurées.

## MOTS CLÉS

Physique statistique hors d'équilibre, systèmes quantiques étendus, transport mésoscopique, coherences à longue portée, processus stochastique, chaînes de spin bruitées.

## ABSTRACT

The motivation for this thesis is to find a fluctuating hydrodynamic description of quantum coherent effects in mesoscopic quantum systems with diffusive transport properties. Coherent effects are inscribed into the coherences (two-point Green's function) on which we focus as the building blocks of such a theory. Our approach is rather mathematical and not related to concrete experiments. We study the quantum symmetric simple exclusion process (QSSEP), a potentially iconic model describing transport of noisy free fermions on a 1D lattice, which has the minimal structure to capture what we are interested in: Long-ranged coherences and diffusive transport. In mean, QSSEP reduces to the symmetric simple exclusion process (SSEP), a classical toy model that was important in the development of the macroscopic fluctuation theory, a fluctuating hydrodynamic description of diffusive transport in classical systems. Studying QSSEP we hope to make progress towards a quantum coherent extension of the macroscopic fluctuation theory.

The thesis summarizes many results we have obtained about QSSEP in the last three years, such as the dynamical equation and an exact stationary solution for correlation functions of coherences at hydrodynamic scales, the distribution of entanglement in QSSEP and an argument why QSSEP might be an effective noisy description for more generic mesoscopic quantum systems. Many of these results are due to a relation between the statistical properties of coherences in QSSEP and free probability theory. We devote a whole chapter to this relation and present, as a by-product, a method to characterize the spectrum of subblocks of a large class of structured random matrices.

## KEYWORDS

Out-of-equilibrium statistical physics, many-body quantum systems, mesoscopic transport, long-ranged coherences, stochastic processes, noisy spin chains.



\newpage

\setcounter{page}{1}
\chapter*{Acknowledgements}
Not without reasons most theses begin by thanking the supervisors and I won’t make an exception. Cher Denis, tu as toute ma gratitude pour la manière dont tu as encadré ma thèse, le temps que tu as consacré pour moi et la confiance que tu avais en moi. Je me rappelle bien comment tu m’as donné mes premières leçons sur le calcul d’Itō et sur l’équation de Lindblad dans la cuisine chez toi parce que le début ma thèse en automne 2020 était aussi le retour de la deuxième vague de Covid – et comme par hasard, on était voisins à 200m près. Je te remercie pour la patience que tu avais au début de ma thèse où j’avais l’impression d’avancer nulle part, mais aussi pour la liberté que tu m’as laissée et pour le plaisir de travailler ensemble quand ça commençait à avancer quand même un peu. C’est peut-être bizarre de te remercier pour toutes les idées que tu avais et qui nous ont fait progresser, mais c’était vraiment un aspect clé pour ma thèse. Ce que je vais garder pour la vie, c’est que même aujourd’hui il peut être plus rapide de résoudre une intégrale ou une équation différentielle à la main si on est prêt à faire marcher son cerveau, et la méthode magique et simple de s’approcher d’une question difficile : D'abord les cas $n=1,2,3$ et (très important) le cas $n=4$ explicitement à la main, puis il ne reste plus grand chose à faire.

Ensuite je voudrais remercier Tony Jin, avec qui j’ai eu la chance de pouvoir échanger au cours de toute ma thèse, et qui m’a invité pas seulement à venir pendant quelques jours à Genève mais aussi à venir à sa super école ``Les Gustins''. Quelques idées dans cette thèse sont nées grâce aux discussions avec Tony, notamment les moyennes unitaires locales pour reproduire la description effective stochastique de QSSEP, et j’espère profondément que notre projet en cours sur le model d’Anderson donnera bientôt lieu à notre premier papier commun.

D’un grand soutien m’était aussi Gabriel, qui était au même bureau que moi durant toute ma thèse, à qui je pouvais demander mes questions bêtes sans gêne et qui était là pour partager les joies et les inquiétudes quotidiennes.  Même si nos quelques essais pour commencer un projet scientifique n’ont finalement pas abouti, j’ai eu le plaisir de donner mes premiers TDs avec toi et tout particulièrement le cours ``physique pour tous''. Merci Gabriel.

I would also like to thank my other collaborators I had the pleasure to work with: Fabian Essler, who gave me hope in the difficult beginning of my thesis that I would nonetheless manage to do a numerical level-spacing analysis of our model. Michel Bauer and Philippe Biane, for the collaboration on SSEP and free probability to which -- I should remark in fairness -- my own contribution was rather minimal. And finally, Marko Medenjak. I still see him with his worn out white Covid mask and his beautiful blond hair in the orange ENS corridor. Being one of the first people I met in Paris, he introduced me to his climbing buddies, invited us a few times during Covid for excellent and substantial Slovenian lunches to his place. And his passing away stays as a painful memory in my mind.

I thank especially Silvia Pappalardi for our nice discussions about free probability, but also Laura Foini and Jorge Kurchan. Thanks to Ion Nechita for inviting me to a workshop to Toulouse and for his remarks about operator-valued free probability which finally encouraged me that it would be worth to dig deeper into the matter. Tanks to Lorenzo Pirolli for our discussion about QSSEP and entanglement (and for his unexpectedly cunning jokes), to Adam Nahum for discussions about random unitary circuits and his generous interest in our project, to Jacopo De Nardis for convincing me that numerics without analytical predictions is difficult, and to Valentina Ros for inviting me to speak at a very nice conference in Trieste with the slightly funny name “Youth in high dimensions”. 

I am very grateful to Tomaž Prosen and Ion Nechita for accepting the laborious task to review this thesis, and I also thank the other jury members Sarang Gopalakrishnan, Anna Minguzzi, Laura Foini and Adam Nahum.

Je dois aussi un grand merci à Christine Chambon, qui m’a sauvé la vie deux fois quand j’avais demandé trop tard mon ordre de mission ou quand faute d’un malentendu du mot ``nom d’usage'' j’avais réservé tous mes billets d’avions au nom de ``Ludwig Ludwig''. Merci aussi à tous les autres membres administratives du labo. Je pense à Yann, Laura, Olga ou Asma. 

Finally, I am very grateful to my office mates Paul, Mikhail and Leo, who were an excellent company in my last months of writing. To my former office mates Augustin, Vasilis, David, Manuel and Gautier who graduated a year before me. And to Christiana, Olalla, Paul and Elie for involved discussions about politics and cinema at lunch. I owe a big musical Meric to Ayako, Pierre, Marie, Cindy, Damien and Cécile for continuing the Lhomond Lunch Concerts in the most dedicated and caring manner I could have imagined.

I thank my climbing buddies Josh and Lorenzo (so that you have no excuse not to come to my defence). Je veux aussi dire merci à mon merveilleux group de théâtre, à Loman, Éve, Siham et tous les autres, qui m’ont aidé quelquefois à sortir de ma tête de physicien et à retrouver la relation avec mon corps. Et bien sûr à mon équipe de pianoforte -- Baptiste, Théophile et Natalia -- simplement pour la musique que nous avons faite sonner ensemble.

Merci Joséphine pour m’avoir soutenu jusqu’ici, pour ta patience et ta bienveillence -- même si c’est pas toujours facile -- et pour avoir corrigé mes fautes de français.

Und am Ende steht die Familie: Auch wenn ihr in der letzten Phase meiner Doktorarbeit nicht so viel von mir gehört habt, seid ihr mir eine große Stütze gewesen und ich habe oft an euch gedacht. Zu wissen, dass es dort am Bodensee, in Freiburg und in Kopenhagen Menschen gibt, die einen voll und ganz unterstützen, was immer man auch macht, das tut so gut, das könnt ihr euch gar nicht vorstellen.

\begin{flushright}
	Paris, le 22 décember 2023
\end{flushright}
\tableofcontents
\clearpage
\pagenumbering{arabic}
\chapter*{Index of notations}
\addcontentsline{toc}{section}{Index of notations}
\begin{tabular}{p{1.4cm} p{12cm}}
	$n(x,t)$ &Local particle density\\[5pt]
	$j(x,t)$ &Local particle current\\[5pt]
	$\rho_t$ &Density matrix at time $t$ \\[5pt]
	$Q$, $J$ & Quantum operators for a conserved charge and current\\[5pt]
	$N$ &Number of sites in a lattice system (sometimes used as the total number of particles in a continuous system) \\[5pt]
	$L$ & Physical length of the system \\[5pt]
	$L_\phi$ &Coherence length\\[5pt]
	$\ell$ &Mean free path \\[5pt]
	$D$ & Diffusion constant \\[5pt]
	$\sigma$ & Mobility or conductivity \\[5pt]

	$\tr$ & Trace on the one-particles Hilbert space (on $N\times N$ matrices) \\[5pt]
	$\ntr$ & Normalized trace $\ntr=\tr/N$ \\[5pt]
	$\Tr$ & Trace on fermionic or bosonic Fock space \\[5pt]
	
	$\eta$, $\xi$ & Normalized Gaussian white noise (in time and space) \\[5pt]
	$W_t$ & Complex Brownian motion with increments $dW_t=W_{t+dt}-W_t$\\[5pt]
	$B_t$ & Real Brownian motion\\[5pt]
	$(\cdots)^*$ & Complex conjugation (sometimes also $\overline{\cdots}$)  \\[5pt]
	$\overset{d}{=}$ & Equal in distribution (for random variables)\\[5pt]
	
	$\E[\cdots]^c$ & Connected expectation value or cumulant, same for $\<\cdots\>^c$ \\[5pt]
	$\varphi$ & Expectation value on non-commuting variables. For random matrices we often take $\varphi=\E[\ntr(\cdots)]$ \\[5pt]
	$\kappa_n$ & $n$-th free cumulant\\[5pt]
	$\pi$ & Partition of a set into subsets or ``parts'' $p\in\pi$ \\[5pt]

	$\sigma^{x,y,z}$ & Pauli matrices $\sigma^x = \begin{pmatrix} 0 & 1 \\ 1 & 0 \end{pmatrix}$, $\sigma^y = \begin{pmatrix} 0 & -i \\ i & 0 \end{pmatrix}$, $\sigma^z = \begin{pmatrix} 1 & 0 \\ 0 & -1 \end{pmatrix}$ \\[5pt]

	QSSEP & Quantum symmetric simple exclusion principle \\[5pt]
	SSEP & Symmetric simple exclusion principle \\[5pt]
	MFT & Macroscopic fluctuation theory
\end{tabular}
\chapter*{Publications}
\addcontentsline{toc}{section}{Publications}
Publications by the author discussed in this thesis
\begin{itemize}
	\item[\cite{Bernard2022DynamicsClosed}] \fullcite{Bernard2022DynamicsClosed}
	\item[\cite{Hruza2023Coherent}] \fullcite{Hruza2023Coherent}
	\item[\cite{Bernard2023Exact}] \fullcite{Bernard2023Exact}
	\item[\cite{Bernard2024Structured}] \fullcite{Bernard2024Structured}
\end{itemize}
Publications not discussed here
\begin{itemize}
	\item[\cite{Bauer2023Bernoulli}] \fullcite{Bauer2023Bernoulli}
\end{itemize}
Publications in preparation
\begin{itemize}
	\item[\cite{HruzaJin2024Fluctuating}] \fullcite{HruzaJin2024Fluctuating}
\end{itemize}
\chapter*{Introduction for amateurs\footnote{Not without a wink to Denis' wonderful habit of adding the comment "for (and by) amateurs" to the titles of his lecture notes this introduction is for amateurs in the proper meaning of the word.}}
\addcontentsline{toc}{chapter}{Introduction for amateurs}

Roughly speaking, this thesis deals with genuine quantum effects that might be observed in very small quantum systems consisting nonetheless of many degrees of freedom (or constituents such as atoms, electrons, molecules, vibration modes etc.) and being in a state that is \textit{not in equilibrium} (for example, think of a very small metallic conductor with electrons flowing through it). Situated within the vast field of statistical physics, our goal is not only to describe the mean behaviour of such systems, but also to characterize the fluctuations around the mean. In the following I will try to explain and motivate all important words that have appeared so far: "quantum effects", "small quantum system", "fluctuations" and "not in equilibrium".

A frequently used motivation for studying non-equilibrium systems is that ``nature'' itself is always out of equilibrium. When I recently mentioned this perspective to a friend, her valid objection was that to her nature seemed much more in equilibrium than out of equilibrium, and that equilibrium to her was an indispensable condition for life on earth. Without starting now a philosophical debate on whether the world is in or out of equilibrium, the anecdote is an interesting example for situations where the common understanding of a concept does not necessarily correspond to how a physicist uses it. In physics, a system with a large number of degrees of freedom  is out of equilibrium, if one can observe ``movement'' or ``change'' at macroscopic scales, that is, on scales where the individual particles are no longer resolved. One could also say, something is ``flowing'' or being ``transferred'' and that there is a \textit{current}, such as for example in a water tube where water molecules enter on one side and leave on the other side. On the contrary, the absence of ``movement'' or ``change'' is what we would call equilibrium, a situation that indeed seems much more dead than alive.

The dream of a physicist would be a universal framework to deal with all non-equilibrium systems -- in the same way as this is possible for systems in equilibrium (Boltzmann distribution etc.). However such a framework does not yet exist and there is not much hope to ever find such a single framework, because non-equilibrium systems seem to be so diverse in their behaviour. Indeed, their study has given rise to a whole plethora of interesting phenomena, both from a fundamental and from phenomenological perspective, which are absent in equilibrium. To just mention one of these, the particle density measured in a non-equilibrium system at positions $A$ and $B$ far away from each other will be correlated \cite{Dorfman1994Generic}, while in equilibrium this is not the case. This means that the density at $A$ is somehow affected by the density at $B$ (for example, because at an earlier time, a particle at $A$ was scattered in such a way that now it is at $B$). 

But to the best of our knowledge, the real world is described by quantum mechanics. And this implies that in addition to these classical correlations, there are correlations that cannot be explained by a classical picture of particles colliding with each other. These quantum correlations are called \textit{entanglement}. And their existence naturally leads to the question about how entanglement behaves out of equilibrium. Is it long-ranged, such as the classical correlation between the particle density at different positions? Is it maybe even stronger than classical correlations? Does this lead to a behaviour on macroscopic scales that differs from its behaviour based only on classical correlations?

Surprisingly, in most cases quantum correlations do not play a role on macroscopic scales. Or, to put it more precisely, which is also a bit more confusing at the same time: Quantum correlations quickly spread beyond the system and correlate the system with its surroundings (its environment). When afterwards one considers only the system, some quantum information that is now shared between system and environment is lost, and this makes the presence of quantum correlations inside the system to become practically invisible.

However, there are situations where quantum effects do become apparent, even systems with many degrees of freedom, notably if the system is small enough and the temperature is low. Such systems are called \textit{mesoscopic} and the actual size below which quantum effects become important, the so-called \textit{coherence length} $L_\phi$, depends on the specific system under consideration. For example, for electrons in a metallic conductor this is of the order of micrometers $L_\phi\approx 1\mathrm{\mu m}$ at a few hundred milli Kelvin $T\approx 100mK$ \cite{Steinbach1996Observation}. Well known effects of quantum mechanical origin at these scales are \textit{shot noise}, a fluctuation of the current due to the discreteness of charge and the Pauli exclusion principle, or the \textit{weak localization} correction to the mean conductivity.

Depending on the size of the system, one can also observe different forms of transport, i.e.\ different forms of ``movement'' within the system. For example, considering a conductor of size $L<\ell$ that is shorter than the mean free path of electrons  $\ell$, i.e.\ the length over which electrons propagate ``freely'' without interacting or ``colliding'' with any other particle, one speaks about \textit{ballistic} transport. Here the displacement of an electron is proportional to time, $\Delta x\sim t$. However, if the conductor is larger than the mean free path $L>\ell$, electrons on their way through the conductor will scatter with many other particles and this resembles the erratic movement of a random walk. As a result, the average displacement of an electron is now proportional to the square root of time $\Delta x\sim \sqrt{t}$ and one speaks about \textit{diffusive} transport. Note that the terms ballistic and diffusive have a meaning independently of whether the system is in the mesoscopic regime $L<L_\phi$ where quantum effects become important, or not.

In this thesis we are in particular interested in non-equilibrium systems with diffusive transport. For such diffusive systems a classical unifying mathematical framework has been recently developed that captures the complete statistics of particle density and current fluctuations in these systems, the so-called \textit{macroscopic fluctuation theory} (MFT) \cite{Bertini2015MFT}. In indeed, from a macroscopic perspective which does not resolve individual particles, one never has the complete information about the current microscopic state of the system, i.e. the position and velocity of all particles in the system, and as a result any macroscopic quantity will fluctuate. These fluctuations are usually very small. While proportional to one over the number of degrees of freedom in the system $1/N$, they can become important for quantities such as the total number of particles in a subsystem, which is a sum over an extensive number of degrees of freedom (proportional to $N$).

Coming back to mesoscopic quantum system with diffusive transport, i.e.\ systems where quantum correlations can have visible effects, one is tempted to ask if the macroscopic fluctuation theory can be extended to include quantum effects \cite{Bernard2021CanMFT}? This is the question which is at the 	heart of this thesis. In order to incorporate such quantum effects, we study, as building blocks of such a theory, the fluctuations of so-called \textit{coherence} between different positions in the system. Intuitively, the coherence between $A$ and $B$ characterizes the quantum correlation of finding a particle at position $A$ each time there is no particle at $B$, and vice versa. For the same reason that macroscopic quantities fluctuate, i.e.\ incomplete knowledge of the underlying microscopic state, one expects coherences to fluctuate when described at mesoscopic scales. Where the noise comes from is explained in more detail in Chapter \ref{chpt:goal_of_the_thesis}.

The approach to this question we will take in this thesis is to study a specific microscopic toy model, the so-called \textit{quantum symmetric simple exclusion process} (QSSEP). The reason we study this model, and not another maybe more realistic model, is because one can see QSSEP as a minimal model which has just enough structure for the phenomena we are interested in to appear. These are fluctuating coherences and diffusive transport in mean. Another reason is that QSSEP allows, to a large extend, to find analytical solutions with pen and paper. And indeed most of this thesis is concerned with analytical results, while numerical results only play a minor and supportive role.

\paragraph{Outline.} The thesis is structured as follows. Chapter \ref{chpt:concepts_from_non-equilibrium} introduces relevant concepts from the physics of non-equilibrium systems, both classical and quantum. It starts with an introduction to non-equilibrium and a review of paradigmatic results in Section \ref{sec:intro_to_non-eq}. We continue with a review of the macroscopic fluctuation theory for classical diffusive systems and related microscopic models such as the symmetric simple exclusion process in Section \ref{sec:classical_diffusive_transport}. Then, we switch to the quantum regime, and discuss the physics of open quantum systems with a particular focus on stochastic descriptions in Section \ref{sec:stochastic_quantum_dynamics}. Section \ref{sec:microscopic_models} is a summary of frequently used microscopic models of many-body quantum dynamics. Section \ref{sec:transport_quantum} discusses transport properties in integrable and chaotic quantum systems, giving a few examples for specific models, and notably recalling generalized hydrodynamics. Section \ref{sec:entanglement} on entanglement provides formal definitions of entanglement measures and intuitive explanations why entanglement is important for thermalization. We also make reference to the eigenstate thermalization hypothesis (ETH) here. Section \ref{sec:random_quantum_circuits} discusses results from random quantum circuits about entanglement spreading and diffusive transport in generic chaotic quantum systems. Finally, Section \ref{sec:mesoscopic_diffusive_conductors} tries to establish a link with the more applied and experimentally accessible physics of mesoscopic conductors, reviewing the Landauer-Büttiker formalism, and coherent effects such as weak localization and the universal conductance fluctuations.

We continue in Chapter \ref{chpt:goal_of_the_thesis} with a more technical outline of the goal of this thesis, than what we have provided here in this introduction. Chapter \ref{chpt:intro_to_QSSEP} represents the state of knowledge about QSSEP which we had when I started my thesis. And finally Chapter \ref{chpt:new_results_QSSEP} provides the new results about QSSEP which we have found during my thesis. An interesting insight was that the mathematical structure of QSSEP is tightly related to so-called \textit{free probability theory}, to which we devote a separate Chapter \ref{chpt:free_proba} which also includes a mathematical result on its own, that allows to find the spectrum of structured random matrices via a variational principle.

\paragraph{Unpublished results in this thesis.} A few results about QSSEP are here presented for the first time and do not yet appear in other publications: (1) The stochastic process in Eq.~\eqref{eq:dI} which reproduces the dynamics of cumulants of coherences in the scaling limit. (2) The generating function for loop-cumulants of coherences of the open QSSEP in Eq.~\eqref{eq:QSSEP_genfunc}. (3) The dynamics of entropy in Section \ref{sec:entanglement} and the time evolution of the generating function $F_0$ of the spectrum of QSSEP in Eq.~\eqref{eq:F_0_dynamics}.
\chapter{Relevant concepts from non-equilibrium physics}\label{chpt:concepts_from_non-equilibrium}
\section{Introduction to non-equilibrium}\label{sec:intro_to_non-eq}
The aim of statistical mechanics is to understand the macroscopic behaviour of systems based on the often intractable microscopic dynamics of the ``particles'', the constituents of the system. Macroscopic properties such as volume, particle density or temperature can take stationary values (e.g.\ if the system is in equilibrium) or be described by hydrodynamic equations (if it is not). But in addition to this, there are always \textit{fluctuations} around these mean values. They originate from the discreteness of atomic matter, from quantum and from thermal fluctuations\footnote{There is an argument within the eigenstate thermalization hypothesis that suggests that both are the same. That is, thermal fluctuations might be understood as quantum fluctuations \cite{Srednicki1996Thermal}}. A concept that appears often in this context is \textit{universality}. A macroscopic property is universal, if it does not depend on the precise microscopic laws, but rather applies to a whole class of microscopic models. In this sense, statistical mechanics is concerned with identifying universal behaviour of macroscopic properties and of their fluctuations. And indeed, all of this thesis is devoted to the understanding of fluctuation. A reference that we sometimes follow in this introduction is \cite{Derrida2016Cours}.

\paragraph{Equilibrium vs.\ non-equilibrium scenarios.}
A system is in equilibrium, if it is either in contact with a single heat or particle reservoir, or if it is completely isolated from any environment (a situation that never really exists in reality), and if it has evolved for a very long time such that all information about the initial state is lost and all currents have relaxed. In this case, the mean value of local observables, as well as their fluctuations, can be described by the well-known ensembles of statistical mechanics\footnote{That these ensembles are still a good description for local observables in an isolated quantum system is actually less trivial, because the unitary evolution always remembers the spectrum of the initial state. The problem finds an answer within the eigenstate thermalization hypothesis, see Section \ref{sec:entanglement}.}. For example, the probability to find the system at inverse temperature $\beta=1/kT$ in a configuration $\mathcal C$ with energy $E(\mathcal{C})$, is given by the Boltzmann distribution (canonical ensemble)
\begin{equation}
	P_{\text{eq},\beta}(\mathcal C)\sim e^{-\beta E(\mathcal C)}.
\end{equation}
And in the microcanonical ensemble at constant energy $E$ the probability is
\begin{equation}
	P_{\text{eq},E}(\mathcal C)\sim \delta(E(\mathcal C)-E).
\end{equation}

Non-equilibrium scenarios are everything else. Most importantly, they are distinguished by the presence of a current. The possibility of charge transfer causes the physics and the mathematical description to be very different from the equilibrium case (see e.g.\ the paragraph on Fokker-Planck below). One often considers the following scenarios:

In the \textit{quench} scenario, one studies the relaxation of an isolated system towards equilibrium starting in a state that is macroscopically different from its equilibrium state. In quantum mechanics this is often called a \textit{quantum quench}. Here the initial state is the eigenstate of some Hamiltonian that is abruptly changed into a new Hamiltonian at time zero. Thereafter, the initial state is no longer an eigenstate and observables undergo a non-trivial dynamics before they relax.

\begin{figure}
	\begin{center}
		\includegraphics[width=0.6\textwidth]{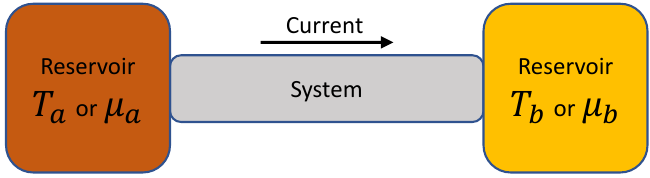}
		\caption{\label{fig:NESS} A system in contact with two heat reservoirs at temperatures $T_a$ and $T_b$, or two particles reservoirs at chemical potentials $\mu_a$ and $\mu_b$. The difference between the reservoirs induces a (heat or particle) current through the system}
	\end{center}
\end{figure}

Another scenario is that of a \textit{non-equilibrium steady state} (NESS). In this case, one couples a system to two or more reservoirs. This induces a steady current through the system that persists even at long times, see Figure \ref{fig:NESS}. At late times quantities become time-independent -- which is also the case in equilibrium systems. However, contrary to equilibrium, there is a current and the situation generally depends on the nature of the coupling to the reservoirs. For example, one can imagine a house as the ``system'' between a hot and cold reservoir, the outside and the heating inside the house. The temperature inside the house will depend how well its walls are isolated, that is, it depends on the coupling between system and cold reservoir. As a consequence, there is no general formula for a phase-space distribution $P_\text{NESS}(\mathcal C)$ of non-equilibrium steady states \cite{Ruelle1999}, such as there is in equilibrium. However, if one is only interested in macroscopic quantities, such as the current or the density profile, and one assumes the reservoirs to be Markovian (see below), then it is possible to extract universal properties of non-equilibrium steady states that do not depend on the nature of the coupling (e.g.\ the macroscopic fluctuation theory in Section \ref{sec:MFT}).

A clever method to study non-equilibrium steady states without caring about the precise coupling to reservoirs is the \textit{partition protocol}\footnote{The partition protocol has for example appeared in \cite{Rubin1971Abnormal,Lebowitz1977Stationary} (for classical systems) and \cite{Bernard2012Energy} (for quantum systems). But it is likely that the idea has already appeared earlier.}. Here an infinite, but isolated system is initialized in a \textit{domain-wall state}: Its left half is in equilibrium with a higher chemical potential (or temperature) than the right half, such that the particle (or energy) density profile resembles a step function. Applying Hamiltonian or unitary time evolution to this initial state, a current emerges around the discontinuity which tries to counterbalance the difference in densities. Scaling time and space appropriately (e.g.\ ballistically or diffusively, depending on the transport properties of the system), the current-carrying region looks like a non-equilibrium steady state. In this way, one can investigate non-equilibrium properties that are independent of how the system was put out of equilibrium. 

\paragraph{Hydrodynamics and Diffusion.} 
A big motivation to study non-equilibrium scenarios from the microscopic perspective is that actually there are a lot of phenomenological equations that describe universal non-equilibrium phenomena at hydrodynamic scales for a large variety of systems. However, most of these equations are extremely difficult to justify from first principles, that is, starting from Hamiltonian or quantum mechanics. For example Ref.\ \cite{Bonetto2000Fouriers} summarises theoretical attempts that have been made in the past 100 years to derive Fourier's law (Joseph Fourier, 1822), which describes the diffusive transport of heat through a medium. Other examples of hydrodynamics are the Euler equations for the flow of a fluid with zero viscosity, and the more general Navier-Stokes equation for non-zero viscosity.

We will be particularly interested in Fick's law, an analogue version of Fourier's law for the transport of particles instead of heat (Adolf Fick, 1855). It states that a system in contact with two particle reservoirs at densities $n_a>n_b$, such as in Fig.~\ref{fig:NESS}, carries a current that is inversely proportional to its length $L$
\begin{equation}\label{eq:ficks_law}
	j=D\frac{n_a-n_b}{L}.
\end{equation}
In particular, the current only depends on a single constant that is specific to the system, the diffusion constant $D$. In this sense, it is a universal law, applying to a large class of systems\footnote{Note that systems violating this law are for example integrable systems with ballistic transport or systems with a so-called anomalous Fourier law.}. For a small difference in the particle density, the right hand side becomes a derivative and one can formulate a local version of Fick's law: Any gradient in the system's density profile $n(x)$ induces a local current $j(x)$ that tries to counterbalance the gradient\footnote{Here we state everything in one dimension, but the generalization to three dimensions is straightforward.},
\begin{equation}
	j(x)=-D\big(n(x)\big)\partial_x n(x).
\end{equation}
Importantly, the diffusion constant now depends on the local particle density $n(x)$. Inserting this equation into the continuity equation $\partial_t n+\partial_x j=0$, one obtains a diffusion equation for the time evolution of the density profile
\begin{equation}
	\partial_t n(x,t)=\partial_x \big[ D\big(n(x,t)\big) \partial_x n(x,t) \big]
\end{equation}
If $D$ is a constant, a solution with initial condition $n(x,0)=\delta(x)$ is the well-known heat kernel $n(x,t)=\frac 1{\sqrt{4\pi D}}e^{-x^2/(4Dt)}$, a Gaussian distribution with variance
\begin{equation}\label{eq:diffusion_displacement}
	\<x^2\>= 2 D t.
\end{equation}
This is the iconic relation between the average displacement of a diffusing particle and the time it took. However, Fick's law tells us nothing about fluctuations around the mean density profile. We come back to this problem in Section \ref{sec:classical_diffusive_transport} when discussing the macroscopic fluctuation theory.

\paragraph{Stochastic descriptions.}
While it is often not possible to obtain hydrodynamic equations from deterministic Hamiltonian dynamics (or unitary quantum mechanics), this becomes much easier if one adds noise to the microscopic dynamics. The idea is that the noise is an effective description of very fast degrees of freedom that are only slightly affected by how the system evolves on long time scales. An assumption one usually makes in this context is the \textit{Markov assumption}: System and reservoir do not build up a memory of what happened in the past. The probability that during an infinitesimal time step the system and the reservoirs jump into a new configuration only depends on the probability of the present configuration -- and not on the probability of former configurations. 

The introduction of noise can also be seen as a way to address \textit{generic} systems. While a particular deterministic system might be difficult to solve and its behaviour can be very specific, the properties of an stochastic ensemble of systems might have universal character and in addition to this they are often easier to solve. 

A third reason to introduce noise comes from quantum mechanics where it originates from the measurement process.

\subparagraph{Langevin equation.}
A classic example how noise simplifies calculations is the Langevin equation.  It was introduced by Paul Langevin in 1908 in order to describe the Brownian motion used by Einstein in this classical article from 1905 to measure the size of molecules in a liquid from their mean displacement. Langevin's equation is Newton's law for a particle  under the influence of an external force $F$, a viscous force $-\gamma \dot x$ with friction coefficient $\gamma$, and a random force $\eta$ from collisions with other particles due to their thermal agitation at temperature $T$,
\begin{equation}\label{eq:langevin}
	m\ddot{x}=F(x)-\gamma \dot x + \sqrt{2\gamma k T} \,\eta
\end{equation}
The random force behaves like the sum of many independent random variables (if one neglects correlations between collisions) and is therefore modelled by a Gaussian white noise, $\<\eta(t)\eta(t')\>=\delta(t-t')$. Its prefactor $\sqrt{2\gamma k T}$ is chosen such that the equipartition theorem is satisfied, i.e.\ $\< v^2 \>=kT/m$. If the external force is $F=0$, one can integrate the Langevin equation and relate the mean square displacement of the particle to the friction coefficient $\gamma$,
\begin{equation}\label{eq:mean_square_displacement}
	\<(x(t)- x(0))^2\>\approx \frac{2kT}{\gamma} t.
\end{equation}
which holds for $t\gg \gamma/m$. In particular, a comparison to Eq.~\eqref{eq:diffusion_displacement} shows that the Langevin equation describes diffusive transport with diffusion constant $D=kT/\gamma$. The latter is called the Einstein-Smoluchowki relation. By Stokes' law $\gamma=6\pi\mu a$ the friction coefficient is related to $a$, the diameter of the particle, which Einstein suggested to determine via a measurement of the diffusion constant (the viscosity $\mu$ is assumed to be a known quantity).

Note that the Langevin equation can formally be derived as the effective description of a particle that is coupled linearly to an infinity of harmonic oscillators, see section 7.3 in \cite{Derrida2016Cours}.

\subparagraph{It\={o} and Fokker-Planck.}
The Langevin equation often appears in its over-damped limit where $m\ll\gamma$. Setting $\gamma=1$ and $\sqrt{2kT}=b$, we write it in the more general form,
\begin{equation}\label{eq:langevin_overdamped}
	\frac {dx} {dt} =F(x,t)+b(x,t) \,\eta(t).
\end{equation}
From the mathematical point of view, this equation is not yet well defined, since it is not clear if in a discretized version of this equation $b(x,t)$ is independent from $\eta(t)$ or not\footnote{Here we will only discuss the It\={o} convention and not the Stratonovich convention.}. If one makes the choice that it is independent, the resulting stochastic differential equation for $X_t\equiv x(t)$ is called an \textit{It\={o} process},
\begin{equation}\label{eq:Ito_process}
	dX_t=F(X_t,t)+b(X_t,t)\, dB_t
\end{equation}
where $dB_t=B_{t+dt}-B_t\approx\eta(t)dt$ is the increment of a Brownian motion. For calculations one uses the \textit{It\={o} rules}
\begin{align}\label{eq:Ito_rules}
	dB_t^2&=dt, & dB_t dt&=dt^2=0.
\end{align}

An easy consequence of Eqs.~\eqref{eq:Ito_process} and \eqref{eq:Ito_rules} is the Fokker-Planck equation, a partial differential equation for the probability $P(x,t)$ that the particle is at position $X_t=x$ at time $t$,
\begin{equation}
	\partial_{t}P(x,t)=-\partial_{x}\big[F(x,t)P(x,t)\big]+\frac{1}{2}\partial_{x}^{2}\big[b(x,t)^{2}P(x,t)\big].
\end{equation}
Setting back $b^2= 2kT$ we can draw a few conclusions from this equation:

Firstly, if $F=0$, the probability distribution satisfies a diffusion equation with diffusion constant $D=kT$ in accordance with Eq.~\eqref{eq:mean_square_displacement}. This shows that we recovered Fick's law for the trajectory of a typical particle.

Secondly, one can associate a conserved probability current to this equation,  $J(x,t)=[F(x,t)- kT \partial_x] P(x,t)$ such that $\partial_t P+\partial_x J=0$. Let us assume that the force $F=-V'$ derives from a potential (which is always true in 1D). In the steady state, the equation reduces to $\partial_x J=0$ and there are two possibilities: If the integral of $e^{-V(x)/kT}$ does not diverge (such that the probability distribution is normalizable), then the equation permits a solution with zero current and one obtains the Boltzmann distribution for a particle with energy $V(x)$,
\begin{equation}
	P_{\text{eq.}}(x)=\frac{1}{Z}e^{-V(x)/kT}, \qquad \text{with }Z=\int dx \, e^{-V(x)/kT}.
\end{equation}
Otherwise, one has to choose a solution with non-zero current $J$. In 1D, the solution can be found by variation of the constant. In higher dimension, however, there is no general solution. This is one of the reasons why the presence of a current complicates the study of non-equilibrium systems considerably.

\paragraph{Linear response theory.}
A regime that is meanwhile quite well understood is when the system is close to equilibrium. Then one can approximate observables linearly expanding around their equilibrium values. Important result in this context are the Onsager relations \cite{Onsager1931Reciprocal} and Kubo's fluctuation-dissipation theorem \cite{Kubo1966Fluctuation}, which groups together a few relations of similar nature such as the Einstein-Smoluchowki relation encountered above.
We briefly state the theorem and then derive a relation for charge fluctuations which we will reuse in Section \ref{sec:classical_diffusive_transport}.

Assume that the equilibrium energy $E_0(\mathcal C)$ of a system in a configuration $\mathcal C$ is perturbed by an observable $B(\mathcal C)$ at $t=0$ such that the new energy is $E(\mathcal C)=E_0(C)-U(t)B(\mathcal C)$, where $U$ is small. For example, $U(t)$ could be a potential and $B(\mathcal C)$ could be the total number of particles under the influence of the potential. Another observable $A(\mathcal C)$ will respond to this perturbation at linear order in $U$ as
\begin{equation}
	\<A_t\>_U=\<A\>_0+\int_0^t \chi_{AB}(t-t')\, U(t')\, dt'
\end{equation}
which defines the response coefficient $\chi_{AB}$. Kubo's fluctuation-dissipation theorem states then that the connected correlations of $A$ and $B$ at equilibrium are related to the response to the perturbation by
\begin{equation}\label{eq:dissipation-fluctuation}
	\<A_tB_0\>_0^c:=\<A_tB_0\>_0-\<A\>_0 \<B\>_0 \overset{!}{=} kT \int_t^\infty  \chi_{AB}(t') dt'.
\end{equation}
Note that equilibrium is time-translation invariant, so we write $\<A\>_0$ without time subscript. The derivation of Eq.~\eqref{eq:dissipation-fluctuation} is not very difficult and can be found for example in section 9.3 of \cite{Derrida2016Cours}. The idea is to write down the equilibrium distribution including the perturbation
\begin{equation}
	P_{\text{eq},U(t)}(\mathcal C) = \frac{e^{-\beta E_0(\mathcal C)+\beta U(t) B(\mathcal C)}}{\sum_{\mathcal{C}'} e^{-\beta E_0(\mathcal C')+\beta U(t) B(\mathcal C')}}
\end{equation}
and to expand up to first order in $U(t)$.

\subparagraph{Charge fluctuations.}
We apply this result to the situation in Figure \ref{fig:NESS}: Consider a system of length $L$ between two particle reservoirs, initially all in equilibrium with chemical potential $\mu_a$. At time zero, the right reservoir's chemical potential is lowered to $\mu_b$. The total energy is now $E(\mathcal C)=E_0(\mathcal C)-U R(\mathcal C)$ with $U=\mu_a-\mu_b$ and $R(\mathcal C)$ the number of particles in the right reservoir. This induces a flow of particles through the system towards the right reservoir, their total number at time $t$ being $Q_t=R_t-R_0$. The linear response coefficient in this case is defined as
\begin{equation}
	\<Q_t\>_U=\<R_t\>_U-\<R\>_0 = U\int_0^t \chi(t') dt'.
\end{equation}
According to the dissipation-fluctuation theorem we have \eqref{eq:dissipation-fluctuation},
\begin{equation}
	kT \int_0^t \chi(t') dt'= \<R_0R_0\>_0^c - \<R_t R_0\>_0^c = \frac 1 2 \<(R_t-R_0)^2\>_0,
\end{equation}
from which one finds a relation for charge fluctuations.
\begin{equation}\label{eq:charge_fluct}
	\<Q_t\>_U =\frac{U}{2kT} \<Q_t^2\>_0.
\end{equation}

The transported charge can also be understood as the integrated current between system and right reservoir
\begin{align}
	Q_t&=\int_0^t \bar\jmath(s) ds, &\bar \jmath(s)&=\frac{1}{L}\int_0^L j(x,s)dx
\end{align}
Here $\bar \jmath(s)$ is the spacially averaged current and $j(x,s)$ is the local current. At large times, the current becomes time-independent in mean, which implies $\<Q_t\>\sim t$, and it does not depend any more on the position $x$ where it is measured, $j(x,s)\to j$. In a diffusive system, the current also satisfies Fick's law $j=D(n_a-n_b)/L$, from which we can obtain the diffusion constant. If the reservoir densities $n_a$ and $n_b$ are close to a system's mean density $n$, they can be related to the difference in chemical potential $U=\mu_a-\mu_b$ with the help of the system's free energy density $f(n)$. One finds\footnote{The free energy $F$ of the system at volume $V$ and number of particles $N$ is related to the free energy density $f$ by $F(N,V)=:Vf(n)$ with $n=N/V$. Then $\mu:=dF(N,V)/dN = f'(n)$ and $\mu_a-\mu_b=f'(n_a)-f'(n_b)\approx (n_a-n_b)f''(n)$} $U=(n_a-n_b)f''(n)$. Therefore, we can restate Eq.~\eqref{eq:charge_fluct} as a relation between the diffusion constant at $n_a>n_b$ with $n_a\approx n\approx n_b$ and charge fluctuations at equilibrium $n_a=n=n_b$,
\begin{equation}\label{eq:diffusion_fluctuation}
	D(n)= \frac{f''(n)}{2kT}\,\frac{L \<Q_t^2\>_0}t
\end{equation}
Alternatively, one can express the charge fluctuations as current-current correlations
\begin{equation}
	\lim_{t\to\infty}\frac{\<Q_t^2\>_0}{t}=\lim_{t\to\infty}\frac{1}{t}\int_0^t ds\int_0^t ds'\< \bar\jmath(s)\bar\jmath(s')\>_0 \approx \lim_{t\to\infty} 2\int_{0}^t d\tau \,\< \bar\jmath(\tau)\bar\jmath(0)\>_0,
\end{equation}
which leads to the Green-Kubo formula \cite{Green1954Markoff,Kubo1957Statistica-Mechanical_II} for the diffusion constant\footnote{The factor $L$ comes from our somewhat unusual convention to consider the averaged current $\bar\jmath(x)$ and instead of the total current $j_L(s)=L \bar\jmath(s)$.}
\begin{equation}\label{eq:Green-Kubo}
	D = \lim_{t\to\infty}\lim_{L\to\infty} \frac{f''(n)}{kT} L\int_{0}^t d\tau\, \< \bar\jmath(\tau)\bar\jmath(0)\>_0.
\end{equation}

To conclude, note that if instead of two reservoirs we had considered an infinite system, initialized in a domain-wall state with particle density $n_a$ on the left and $n_b$ on the right, then the current would not be constant at large times, but rather decays as $1/\sqrt{t}$. As a result, $\<Q_t\>\sim\sqrt{t}$. Eq.~\eqref{eq:diffusion_fluctuation} still holds in this case, if we replace $1/t$ by $1/\sqrt{t}$, noting that $L\<Q^2\>_0/\sqrt{t}$ will be of order one for large $L$ and $t$.

\paragraph{Large deviations.}
Linear response theory characterizes the mean behaviour of systems close to equilibrium in terms of small fluctuations around equilibrium. But it tells us nothing about the probability of very rare occasions where an observable deviates a lot from its mean value, a so-called \textit{large deviation}.

The question of large deviations can be posed both for systems in equilibrium and out-of-equilibrium. The former usually has an easy answer, because the probability of each microscopic configuration, e.g.\ the Boltzmann distribution, is known. As we show in Appendix \ref{app:large_deviations}, the probability that a closed volume $V=L^d$ with $N$ particles at temperature $T$ has a density profile $n(x)$, that deviates from the average homogeneous density $\bar{n}=N/V$, is essentially given by the free energy density $f$ of the system,
\begin{align}\label{eq:density_large_deviation}
	P[n]&\sim e^{-V  I[n]}, & \text{with }\, I[n]&=\frac{1}{kT}  \frac{1}{V}\int_V dx\big[f(n(x))-f(\bar{n})\big].
\end{align}
The exponential scaling of the probability with the size of the system is called a \textit{large deviation principle} and $I[n]$ is called the \textit{rate function} or simply \textit{large deviation function}. It is an intensive $\Ord(1)$ quantity, since we divided it by $V$.

Quite remarkably, it turns out that the large deviation principle also holds for observables in non-equilibrium systems. The difficulty in this case is to find the explicit form of the rate function $I$ which has only been achieved in few situations such as the symmetric simple exclusion process (SSEP) in the steady state, see Eq.~\eqref{eq:SSEP_density_large_deviation}. The rate function in Eq.~\eqref{eq:density_large_deviation} is \textit{local}, meaning that it is an integral over local properties of the system. However, for an out-of-equilibrium systems the rate function usually involves non-local terms that depend on two or more positions. As a consequence, out-of-equilibrium systems can show long-ranged correlations, which is not possible for equilibrium systems. For example, the connected density correlations in SSEP in Eq.~\eqref{eq:SSEP_density_correlations} are long-ranged.

One can also formulate a large deviation principle for the current flowing through a system between two reservoirs. Eq.~\eqref{eq:charge_fluct} tell us that if $\<Q_t\>\sim t$ then also the fluctuations $\<Q_t^2\>^c\approx\<Q_t^2\>_0\sim t$. This is because the 2nd cumulant $\<Q_t^2\>^c$ close to equilibrium ($n_a\approx n_b$) is approximatively given by the 2nd cumulant $\<Q^2_t\>_0^c$ at equilibrium, and we also have $\<Q^2_t\>_0^c=\<Q^2_t\>_0$. The fluctuation of the instantaneous current $j=\frac{Q_t}{t}$ is then $\<j^2\>^c\sim 1/t$ and we could write
\begin{equation}
	j=\<j\>+\frac{\eta}{\sqrt{t}}
\end{equation}
with $\eta$ a centred random variable with $t$-independent variance\footnote{The precise expression is $\<\eta^2\>=\frac{2kT}{U}$, but this is not important.}. If also the higher cumulants of the charge scale as $\<Q_t^n\>^c\sim t$, then the current $j$ satisfies a large deviation principle of the form
\begin{equation}
	P\left(\frac{Q_t}{t}=j\right)\sim e^{-tF(j)}
\end{equation}
with rate function $F(j)$. Note that here we assumed that the probability of the current does not depend on the position where it is measured. This is expected after long times if the system has a finite relaxation time and the number of particles in the system is bounded.

To deal with cumulants of charge, it useful to define the cumulant generating function $\mu(\lambda)$
\begin{equation}\label{eq:charge_cumulants_intro}
	\<e^{\lambda Q_t}\>= e^{\mu(\lambda)t}
\end{equation}
which can be obtained from the rate function as a Legendre transformation
\begin{equation}
	\mu(\lambda)=\max_j(\lambda j-F(j))
\end{equation}
because
\begin{equation}
	\<e^{\lambda Q_t}\>=\int dj \, e^{\lambda t j}P(j)\approx e^{t\max_j(\lambda j - F(j))}.
\end{equation}

Note that in the case of an infinite system with a domain wall initial state, the transported charge scales as $Q_t\sim\sqrt{t}$ and the probability for a large current deviation is $P(\frac{Q_t}{\sqrt{t}}=j)\sim e^{-\sqrt{t}F(j)}$ with a new rate function $F(j)$.

Remarkably, current fluctuations in out-of equilibrium systems satisfy a general relation, the Gallavotti–Cohen fluctuation theorem \cite{Gallavotti1995Dynamical}, which we state here for the case of a system between particle reservoirs,
\begin{equation}
	\mu(\lambda)=\mu(-\lambda-\log\frac {n_a}{1-n_a}+\log\frac {n_b}{1-n_b}).
\end{equation}
In particular the relation holds for arbitrary large differences $n_a-n_b$ of the reservoir densities. For small differences $n_a-n_b\approx0$, one can expand the relation and find the fluctuation-dissipation relation Eq.~\eqref{eq:charge_fluct} which we derived in the regime of linear response.

\section{Classical diffusive systems and MFT}\label{sec:classical_diffusive_transport}
Classical diffusive systems satisfying Fick's law \eqref{eq:ficks_law} do not only share a common behaviour in mean, but also on the level of fluctuations. This is the insight of the macroscopic fluctuation theory (MFT) which we discuss below. Quite remarkably, the only information necessary to characterize these fluctuations, in addition to the diffusion constant $D(n)$, is the so-called \textit{mobility} $\sigma(n)$, the variance of the transported charge $Q_t$ at equilibrium and particle density $n$. For a system of physical length $L$ between two reservoirs at density $n_a$ and $n_b$, the two transport constants can be obtained at large times $t$ as
\begin{align}
	\frac{\<Q_t\>_{n_a\approx n\approx n_b}}{t}& =:D(n) \frac{n_a-n_a}{L} & \frac{\<Q_t^2\>_{n_a=n=n_b}}{t}=:\frac{\sigma(n)}{L}.
\end{align}
From the dissipation-fluctuation of charge in Eq.~\eqref{eq:diffusion_fluctuation}, the two constants are related by the free energy density of the system at equilibrium,
\begin{equation}\label{eq:D-sigma-f}
	D(n)=\frac{f''(n)}{2kT} \sigma(n).
\end{equation}

The MFT was very much inspired by the study of stochastic models of interacting particles on a discrete lattice, so-called \textit{stochastic lattice-gas models}, which can be solved exactly and any claim about fluctuations can be verified. The stochastic description enters into these models either as noise due to an uncontrolled environment, or as an effective description of the slow degrees of freedom under the random influence of the fast degrees freedom. Stochastic models can also be seen as representing a whole ensemble of deterministic systems, a specific realization of which would be very difficult to solve, but ensemble properties are analytically tractable. Below, we discuss two such models, \textit{independent random walkers} and the \textit{symmetric simple exclusion process} (SSEP). Then we outline the framework of MFT. We mostly follow a review \cite{Derrida2007Non-equilibrium} and a set of lecture notes \cite{Derrida2017Cours} by Bernard Derrida.

\subsection*{Independent random walkers}
As one of the simplest examples we consider -- mostly for pedagogic purposes -- independent random walkers on a 1D chain with $N$ sites and two boundary reservoirs. Since the model is non-interacting, the maths is simple and allows to nicely illustrate why the description in the continuum limit corresponds to the macroscopic fluctuation theory. For the SSEP (next section) this is possible, too, but already much harder.
 
\begin{figure}
	\centering\includegraphics[width=0.6\textwidth]{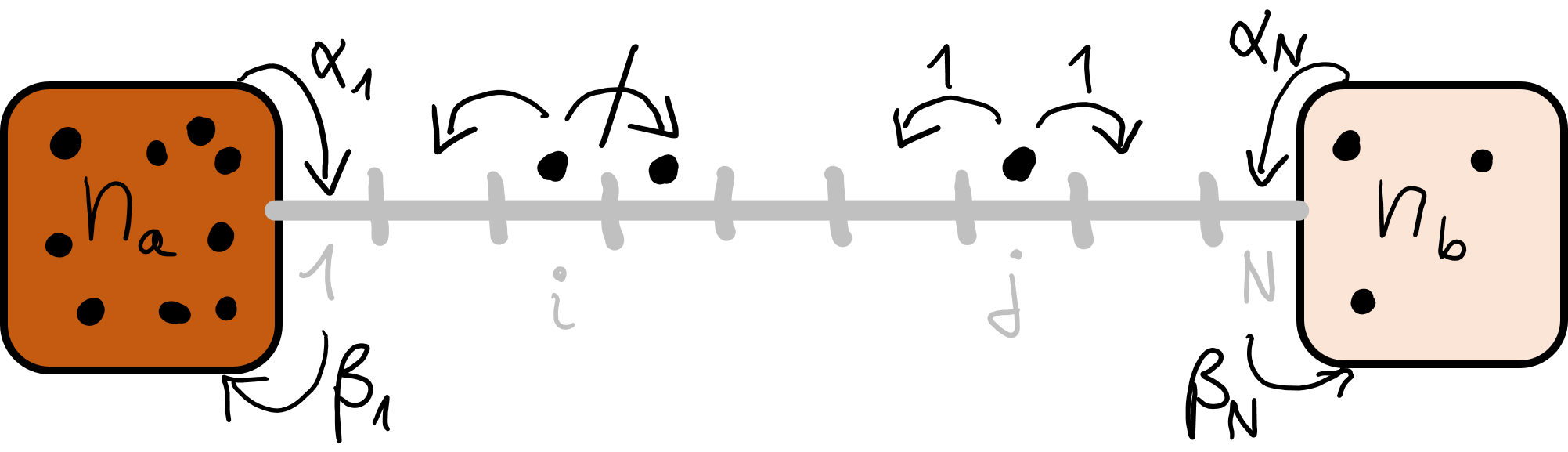}
	\caption{\label{fig:ssep} \textit{Independent random walkers}: Within a time step $dt$, particles hop to neighbouring sites with probability $dt$, while on the boundaries particles are injected and extracted with probability $\alpha dt$ and $\beta dt$. The reservoir densities are $n_a=\frac{\alpha_1}{\beta_1}$ and $n_b=\frac{\alpha_N}{\beta_N}$.\\
	\textit{Symmetric simple exclusion process}: In addition to random hopping there is the exclusion principle (crossed arrow), which prevents particles to hop to a neighbouring site that is already occupied. In this case the reservoir densities are related to the injection and extraction rates by $n_a=\frac{\alpha_1}{\alpha_1+\beta_1}$ and $n_b=\frac {\alpha_N}{\alpha_N+\beta_N}$.}
\end{figure}

\subparagraph{Definition.}
During a time interval $dt$ particles jump to neighbouring sites with probability $dt$, such as indicated in Figure \ref{fig:ssep}, but  without the exclusion constraint: The number of particles $n_i(t)$ on site $i$ at any time $t$ is not constraint. In the bulk,
\begin{equation}\label{eq:indep_walkers_bulk}
	n_i(t+dt)=\begin{cases}
		n_i(t)+1 &\text{w.p.  } (n_{i-1}(t)+n_{i+1}(t))dt \\
		n_i(t)-1 &\text{w.p.  } 2n_i(t)dt \\
		n_i(t)	 &\text{w.p.  } 1-(\text{sum of proba.})
	\end{cases}
\end{equation}
where w.p.\ means ``with probability''. And on the boundaries $i=1,N$, particles are injected and extracted with rates $\alpha_i$ and $\beta_i$, respectively. Here for the left boundary,
\begin{equation}\label{eq:indep_walkers_boundary}
	n_1(t+dt)=\begin{cases}
		n_1(t)+1 &\text{w.p.  } (\alpha_1+n_{2}(t))dt \\
		n_1(t)-1 &\text{w.p.  } (\beta_1+1)n_1(t)dt \\
		n_1(t)	 &\text{w.p.  } 1-(\text{sum of proba.})
	\end{cases}
\end{equation}
The average density obeys
\begin{equation}\label{eq:indep_walker_mean_evol}
	\partial_t\langle n_i\rangle =
	\begin{cases}
		\alpha_1-(1+\beta_1)\langle n_1\rangle+\langle n_2\rangle & i=1 \\
		\Delta^\text{dis}\langle n_i \rangle & i\in\text{bulk}\\
		\alpha_N-(1+\beta_N)\langle n_N \rangle + \langle n_{N-1} \rangle & i=N.
	\end{cases}
\end{equation}
where $\Delta^\text{dis}n_i=n_{i+1}+n_{i-1}-2n_i$ is the discrete Laplacian. The stationary solution of the mean density is
\begin{equation}
	\langle n_i \rangle = \frac{(n_b-n_a)(i-1+1/\beta_1)}{N-1+(1/\beta_1-1/\beta_N)}+n_a
\end{equation}
with $n_a=\alpha_1/\beta_1$ and $n_b=\alpha_N/\beta_N$. Since $\langle n_i \rangle \to \bar{n}(x):=(n_b-n_a)x+n_a$ in the limit of large $N$, with $x=i/N$, one sees that $n_a$ and $n_b$ can be interpreted as the particle densities of the left and right reservoir.

The independent walkers are one of the rare models in which the stationary measure has a product form -- which implies the absence of non-local correlations. One has
\begin{equation}\label{eq:indep_walker_measure}
	P_\infty(\{n_1,\cdots,n_N\}) = \prod_{i=1}^N \frac{\lambda_i^{n_i} e^{\lambda_i}}{n_i!},
\end{equation}
i.e.\ a product of Poisson distributions with expectation value $\lambda_i=\langle n_i \rangle$ corresponding to the mean particle density. In the limit of large $N$, one can use the Stirling's formula $n_i!\approx e^{n_i\log(n_i)-n_i}$ to express this measure as a large deviation principle,
\begin{align}\label{eq:indep_walker_rate_function}
	P_\infty[n]&\sim e^{-NI[n]}, & \text{with } I[n]&=\int_0^1 dx\, \left(\log \frac{n(x)}{\bar{n}(x)}-1\right)n(x)-\bar{n}(x).
\end{align}

Furthermore, the independent walkers have transport coefficients
\begin{align}\label{eq:indep_walker_coef}
	D(n)&=1 & \sigma(n)=2 n
\end{align}
The diffusion constant follows directly from the discrete Laplacian in Eq.~\eqref{eq:indep_walker_mean_evol}, while for the mobility one has to calculate the partition function of $P$ particles on $N$ sites $Z(N,P)=N^P/P!$, then the free energy density 
\begin{equation}\label{eq:free_energy_density}
	f(n):=-kT\lim_{N\to\infty} \frac 1 N \log Z(N,N n)=kT n (\log n-1)
\end{equation}
and finally use the dissipation-fluctuation relation \eqref{eq:D-sigma-f}. These coefficients only depend on the bulk properties and this allowed us to consider the partition function of $P$ particles -- even though, the number of particles is of course not fixed, once we have open boundary conditions.

\subparagraph{Dynamics of the generating function.} The correspondence with MFT is established via a dynamical equation for the moment generating function 
\begin{equation}\label{eq:indep_walker_Z}
	Z_t[a]=\langle e^{\sum_i n_i a_i} \rangle_t\overset{N\to\infty}{=}\langle e^{N\int  a(x) n(x) dx} \rangle_t
\end{equation}
where $n_i$ are random variables of a time dependent measure $\langle \cdots \rangle_t$. Using Eq.~\eqref{eq:indep_walkers_bulk}, we find
\begin{equation}
	Z_{t+dt}[a]=\Big\langle e^{\sum_i a_i n_i} \Big(1-\sum_i \big(e^{a_{i+1}-a_i}+ e^{a_{i-1}-a_i}-2\big)n_i dt\Big) \Big\rangle_t
\end{equation}
and therefore
\begin{align}
	\partial_t Z_t[a] =& \sum_i (e^{a_{i+1}-a_i}+ e^{a_{i-1}-a_i}-2) \frac{dZ_t[a]}{da_i}
\end{align}
Taking the continuum limit, $a(i/N):=a_i$ with $N\to\infty$, the sum turns into an integral, $\sum_i\approx N\int dx$; the exponentials can be expanded for small $a_{i+1}-a_i$ into a discrete Laplacian, which becomes $\Delta_i^\text{dis}\approx N^{-2}\Delta_x$; and $dZ/da_i\approx N^{-1}\delta Z/\delta a(x)$ becomes a functional derivative which is defined by integration against a test function and therefore has the additional factor of $N$. All this yields an extra factor of $N^{-2}$ on the right hand side which can be absorbed into a diffusive rescaling of time $t/N^2\to t$. Together we have
\begin{equation}\label{eq:indep_walker_dZdt}
\partial_t Z_t[a]= \int dx (a''(x)+a'(x)^2) \frac{\delta Z_t[a]}{\delta a(x)}
\end{equation}

\subparagraph{Dynamics starting from MFT.}
As a next step, we want to derive the time evolution of the moment generating function directly from the macroscopic fluctuation theory (MFT). That is, without knowledge of the microscopic laws of independent random walkers, and only on the basis of the transport coefficients $D=1$ and $\sigma(n)=2n$. We therefore anticipate the MFT equations \eqref{eq:MFT} which are stochastic differential equations for the density $n(x,t)$ and the current $j(x,t)$,
\begin{equation}\label{eq:indep_walker_MFT}
	\begin{cases}
	j=-D(n)\partial_x n+ \sqrt{\frac{1}{N}\sigma(n)}\;\xi\\
	\partial_t n+\partial_x j=0
	\end{cases}
\end{equation}
where $\langle \xi(x,t)\xi(x',t')\rangle=\delta(t-t')\delta(x-x')$ is a Gaussian noise independent in space and time. Given a density profile $\{n(t)\}:=\{n(x,t)|x\in[0,1]\}$ at time $t$ one finds the density profile at time $t+dt$ by rewriting the continuity equation as\footnote{Here we used the It\={o} convention, because the current is evaluated on the left of the interval $[t,t+dt]$, i.e.\ at time $t$.},
\begin{equation}\label{eq:indep_walker_dn}
	n(x,t+dt)=n(x,t)-\partial_x j(x,t) dt.
\end{equation}
The probability weight for this transition depends on both, the density and the current profile at time $t$, and is given by
\begin{equation}\label{eq:indep_walker_proba}
	\text{Pr}\big(\{n(t+dt)\}|\{n(t),j(t)\}\big)\sim e^{-N dt \int_0^1 \frac{[j+D(n)\partial_x n]^2}{2\sigma(n)}dx}.
\end{equation}
This follows from isolating $\xi$ in the MFT equations \eqref{eq:indep_walker_MFT} and writing its probability distribution as a path integral.

Now we evaluate $Z_{t+dt}$ by splitting the MFT average $\langle \cdots \rangle_{t+dt}$ into two parts, once taking the average up to time $t$ and then taking the average from $t$ to $t+dt$ conditioning on the density profile $\{n(t)\}$ at time $t$. Using Eq.~\eqref{eq:indep_walker_proba}, the average over $[t,t+dt]$ can then be evaluated. But since we only conditioned on $\{n(t)\}$, but not on $\{j(t)\}$, one has to sum over all current profiles which are compatible with the density profile via the continuum equation \eqref{eq:indep_walker_MFT},
\begin{align}
	Z_{t+dt}[a]&=\Big\langle \Big\langle e^{N\int_0^1 a(x)n(x,t+dt) dx}|\{n(t)\}\Big\rangle_{[t,t+dt]} \Big\rangle_{[0,t]}\\
	&= \Big\langle \sum_{\substack{\{j(t)\} \\ \text{ compatible}}} e^{-N dt \int_0^1 \frac{[j(x,t)+D(n)\partial_x n(x,t)]^2}{2\sigma(n)}dx} e^{N\int_0^1 a(x) [n(x,t)-\partial_x j(x,t) dt]dx}\Big\rangle_{[0,t]}.
\end{align}
In second exponential in the second line, we also replaced $n(x,t+dt)$ with the help of Eq.~\eqref{eq:indep_walker_dn}. The sum can be evaluated by a saddle point approximation and the optimal current which maximizes the sum is found to be $j=\sigma(n)a'-D(n)\partial_x n$. Substituting and performing a partial integration $\int a\partial_x j dx= -\int a'j dx$, we have
\begin{equation}
	Z_{t+dt}[a]=\Big\langle e^{-Ndt\int \big(a'D(n)\partial_x n-\frac{\sigma(n) (a')^2}{2}\big)dx}\,e^{N\int a n\, dx}\Big\rangle_{[0,t]}.
\end{equation}
Expanding the first exponential to first order in $dt$, and substituting $D=1$ and $\sigma(n)=2n$, we finally obtain
\begin{align}
	\partial_t Z_t[a]&= \Big\langle N \int dx \big(n(a')^2-a'\partial_x n\big) \; e^{N\int a n \,dx}\Big\rangle_{[0,t]}\nn
	&=\int dx\big(a'(x)^2+a''(x)\big)\frac{\delta Z}{\delta a(x)}
\end{align}
which is the same equation we obtained from the microscopic laws of the independent random walkers \eqref{eq:indep_walker_dZdt}. This shows that in the very easy example of independent random walkers, the MFT correctly reproduces the complete statistical behaviour of the underlying microscopic model in the continuum limit.

\subsection*{Symmetric simple exclusion process (SSEP)}\label{sec:SSEP}
If we add the exclusion principle to the random walkers, we obtain the \textit{symmetric simple exclusion process}, see Figure \ref{fig:ssep}: Particles cannot hop to a site that is already occupied. The SSEP can be modified to allow asymmetric hopping rates (ASEP, asymmetric simple exclusion process) or to forbid movement into one direction altogether (TASEP, totally asymmetric simple exclusion principle). These models have become a paradigm of classical out-of-equilibrium physics and have considerably contributed to the formulation of MFT. The first appearance in the literature, however, occurred in 1968 as a model of protein growth in ribosomes \cite{MacDonald1968Kinetics}. The name ``exclusion process'' was later coined by the mathematician Spreizer \cite{Spitzer1970Interaction}. Other important contributions are the exact solution of the ASEP by Derrida and co-workers \cite{Derrida1992Exact,Derrida1993Exact} and by Schütz \cite{Schutz1993Phase}, and the mapping of ASEP to the KPZ universality class \cite{Halpin-Healy1995Kinetic} (see also Eq.\eqref{eq:KPZ}). Today, exclusion processes appear in many different contexts, ranging from models of traffic flow to charge fluctuations in mesoscopic conductors \cite{Roche2005Mesoscopic}. We will come back to the latter in section \ref{sec:mesoscopic_diffusive_conductors}.

\subparagraph{Definition.} We could state the stochastic dynamics of SSEP similar to Eqs.~\eqref{eq:indep_walkers_bulk} and \eqref{eq:indep_walkers_boundary} with slight modifications. However, here we chose to present it as a Markov process for a $2^N$-dimensional probability vector $p_t$,
\begin{equation}\label{eq:ssep_markov_process}
	\partial_t p_t(\mathcal C)= \sum_{\mathcal{C}} M(\mathcal C,\mathcal C') p_t(\mathcal C').
\end{equation}
The entry $p_t(\mathcal C)$ corresponds to the probability of being in the configuration $\mathcal C=|\bullet \bullet \circ...\>$ at time $t$, where occupied and empty sites are represented as $\bullet$ and $\circ$. The transition matrix $M$ is given by its action on these configurations. In the bulk,
\begin{align}
	M|... \circ \circ ... \>&=M|... \bullet \bullet ...\>=0 \nn
	M|... \circ \bullet ...\>&= -|... \circ\bullet ...\>+|... \bullet\circ ...\>  \nn
	M|... \bullet\circ ...\>&= |... \circ\bullet ...\>-|... \bullet\circ ...\>.
\end{align}
These equations have to be understood locally, i.e.\ when projected onto the two specified sites, while the other sites $(...)$ do not change. At the boundaries, in addition to bulk dynamics between sites $(1,2)$ and $(N-1,N)$, there is particle injection and extraction,
\begin{align}
	M|\circ ...\>&=\alpha_1 M|\bullet...\> & M|...\circ\>&=\alpha_N M|...\bullet\>  \nn
	M|\bullet ...\> &= \beta_1 |\circ ...\> & 	M|...\bullet\> &= \beta_1 |...\circ\>.
\end{align}

The transport coefficients for SSEP are 
\begin{align}\label{eq:SSEP_coef}
	D(n)&=1 & \sigma(n)&=2n(1-n).
\end{align}
To obtain the diffusion constant, we consider the mean current $J$ in the steady state, which is everywhere the same, so for example equal to the current from $i$ to $i+1$,
\begin{equation}
	J=\< n_i (1-n_{i+1})\> - \< n_{i+1}(1-n_i)\> = \<n_i\> - \<n_{i+1}\>.
\end{equation}
By recursion, $J=\frac{\<n_1\>-\<n_N\>}{N-1}\approx \frac{n_a-n_b}{N}$ and therefore $D=1$. The mobility $\sigma$ is calculated in the same way as for the independent walkers around Eq.~\eqref{eq:free_energy_density}, using that the partition function of $P$ particles on $N$ sites with maximal occupation of one particle per site is $Z(N,P)=\frac{N!}{P!(N-P)!}$. The free energy density is then $f(n)= kT(n\log n + (1-n)\log(1-n))$ from which one gets $\sigma$ using the fluctuation-dissipation relation \eqref{eq:D-sigma-f}.

\subparagraph{Density fluctuations.}In contrast to the independent random walkers, the stationary measure of SSEP is no longer a product measure. In the continuum limit, the probability for a large deviation of the density profile $n(x)$ satisfies a large-deviation principle \cite{Derrida2001Free,Derrida2001Large}
\begin{equation}
	P_\infty[n|n_a,n_b]\sim e^{-NI[n|n_a,n_b]}
\end{equation}
with rate function\footnote{Here and in the following we suppress the arguments of $n=n(x)$ and $g=g(x)$}
\begin{equation}\label{eq:SSEP_density_large_deviation}
	I[n|n_a,n_b]=\int_0^1 dx\Big[(1-n)\log(\frac{1-n}{1-g})+n\log( \frac n g)+\log(\frac{g'}{n_a-n_b})\Big].
\end{equation}
The function $g$ is a solution of the differential equation
\begin{equation}
	n=g+\frac{g(1-g)g''}{g'^2}
\end{equation}
with boundary conditions $g(0)=n_a$ and $g(1)=n_b$. Note, that one can verify that the rate function of SSEP respects the MFT equations as will be explained around Eq.~(\ref{eq:SSEP_U}). Recently, a new derivation of the rate function was provided in Ref.~\cite{Bauer2023Bernoulli} which uses tool from free probability theory and the correspondence of SSEP with the quantum symmetric simple exclusion process (QSSEP) introduced in Chapter \ref{chpt:intro_to_QSSEP}.

Equivalently, one can express the moment generating function in large deviation form
\begin{equation}
	\<e^{N\int a(x) n(x) dx}\>_\infty\sim e^{N W[a]}
\end{equation}
where $W[a]$ is the cumulant generating function, which is related to $I$ by a Legendre transformation $W[a]=\max_{\{n(x)\}}\{\int a n - I[n]\}$. One finds,
\begin{equation}
	W[a]=\int_0^1 dx\Big[\log(1+g(e^a-1))-\log (\frac {g'}{n_a-n_b})\Big]
\end{equation}
and $g$ a solution of
\begin{equation}
	(1+g(e^a-1))g''=g'^2(e^a-1)
\end{equation}
and boundary conditions $g(0)=n_a$ and $g(1)=n_b$. Expanding $W[a]$ one finds the first few cumulants in the steady state for $x<y$,
\begin{align}\label{eq:SSEP_density_correlations}
	\<n(x)\>&=(n_b-n_a)x+n_a & \< n(x)n(y) \>^c&=-\frac{(n_a-n_b)^2}{N}x(1-y).
\end{align}
The mean density profile in the steady state is a slope that interpolates between the reservoir densities $n_a$ and $n_b$. The connected correlations of the density at positions $x$ and $y$ are non-local, and suppressed as $1/N$, according to the large deviation principle. Note that the connected correlations vanish in equilibrium, when $n_a=n_b$. In this sense, non-local correlations is a genuine out-of-equilibrium property.

\subparagraph{Current fluctuations.}
As shown in \cite{Derrida2004Current, Bodineau2004Current}, also current fluctuations of SSEP in the steady state satisfy a large deviation principle,
\begin{align}
	P\left(\frac{Q_t}{t}=j\right)&\sim e^{-tF(j)} & \<e^{\lambda Q_t}\>&=e^{\mu(\lambda)t}.
\end{align}
While there is no explicit expression for the rate function $F$, the cumulant generating function is
\begin{align}\label{eq:mu_SSEP}
	&\mu(\lambda)=\frac{1}{N}[\log(\sqrt{1+\omega}+\sqrt{\omega})]^2 \nn
	&\text{with }\omega(\lambda)=(e^\lambda-1)n_a+(e^{-\lambda}-1)n_b+(e^\lambda-1)(e^{-\lambda}-1)n_an_b.
\end{align}
Here we write down the first four cumulants of $Q_t$ for $n_a=1,n_b=0$,
\begin{align}\label{eq:SSEP_cumulants_explicit}
	\frac{\<Q_t\>}{t}&=\frac{1}{N}+\Ord(L^{-2}) \nn
	\frac{\<Q_t^2\>^c}{t}&=\frac{1}{3N}+\Ord(L^{-2}) \nn
	\frac{\<Q_t^3\>^c}{t}&=\frac{1}{15N}+\Ord(L^{-2}) \nn
	\frac{\<Q_t^4\>^c}{t}&=-\frac{1}{105N}+\Ord(L^{-2})
\end{align}
Interestingly, they are the same as the cumulants of charge in a mesoscopic disordered wire found in \cite{Lee1995Universal}. The reason is that the main property responsible for the fluctuation of electrons in disordered wires is the Pauli exclusion principle. The SSEP, being a classical model with noisy diffusive transport, has this property already built-in by hand.

We also note, that in the case of SSEP on an infinite line without boundary reservoirs and with an initial domain wall state at densities $n_a$ and $n_b$, one has $\<e^{\lambda Q_t}\>\sim e^{\sqrt{t}\mu(\lambda)}$ and the cumulant generating function has been shown to be \cite{Derrida2009Current},
\begin{equation}\label{eq:mu_infinite_SSEP}
	\mu(\lambda)=\frac{1}{\pi}\int_{\mathbb{R}} dk \log[1+\omega e^{-k^2}]
\end{equation}
with $\omega(\lambda)$ from Eq.~\eqref{eq:mu_SSEP}.

\subsection*{Macroscopic fluctuation theory (MFT)}\label{sec:MFT}
Inspired by the study of stochastic lattice gases, Bertini et al.\ \cite{Bertini2015MFT} have developed the \textit{macroscopic fluctuation theory} (MFT), a general approach to characterize density and current fluctuations in the non-equilibrium steady states of diffusive systems in contact with two or more reservoirs. In principle, this approach should be valid in any dimension, but it has mostly been verified with the help of microscopic models in one dimension. The only required information specific to the system (in addition to the boundary conditions imposed by the reservoirs) are two transport coefficients, the diffusion constant $D(n)$ and the mobility $\sigma(n)$. We follow the reviews \cite{Derrida2007Non-equilibrium,Bertini2015MFT}.

The basic equations on which MFT is based is fluctuating hydrodynamics,
\begin{equation}\label{eq:MFT}
	\begin{cases}
		j(x,t)=-D(n(x,t))\partial_x n(x,t)+ \sqrt{\frac{1}{N}\sigma(n(x,t))}\;\xi(x,t)\\
		\partial_t n(x,t)+\partial_x j(x,t)=0.
	\end{cases}
\end{equation}
The first equation is a fluctuating version of Fick's law with a Gaussian white noise of variance $\langle \xi(x,t)\xi(x',t')\rangle=\delta(t-t')\delta(x-x')$, multiplied by the mobility and suppressed by the number of degrees of freedom $N\gg1$. The small noise is responsible for the large deviation form of the density or the current as we will see below. The second equation is a simple continuity equation. Even though, current and density can fluctuate around the mean values allowed by Fick's law, the continuity equation has to be satisfied exactly at each moment. Combining the two equations one has
\begin{equation}\label{eq:MFT_combined}
	\partial_t n(x,t)=\partial_x\Big(D(n(x,t))\partial_x n(x,t)\Big) - \partial_x \Big(\sqrt{\frac{\sigma(n(x,t))}{N}}\;\xi(x,t)\Big)
\end{equation}

Integrating the first equation in \eqref{eq:MFT} (isolating $\xi$ and writing it as path integral) one obtains the probability to observe a certain density and current profile over some time interval. The probability for such a space-time trajectory $\{n,j\}:= \{n(x,t),j(x,t)\,|\,x\in[0,1],t\in[t_1,t_2]\}$, where $j$ and $n$ are related by the continuity equation, is
\begin{equation}\label{eq:MFT_prob}
	P_{t_1,t_2}(\{n,j\})\sim \exp\Big({-N\int_{t_1}^{t_2}dt\int_0^1 dx \frac {(j+D(n)\partial_x n)^2}{2\sigma(n)}}\Big).
\end{equation}
Intuitively this equation can be understood as follows: Defining $J(n):=-D(n)\partial_x n$ to be the hydrodynamic current (in contrast to the actual, fluctuating current $j$), the fluctuating trajectory $\{n,j\}$ leads to an excess current $j-J(n)=j+D(n)\partial_x n$ which could have been produced through an external electric field $j-J(n)=\sigma(n) E_\text{ext}$, with $\sigma(n)$ the electrical circuit conductivity at charge density $n$. Substituting this into Eq.~\eqref{eq:MFT_prob}, the exponent becomes $\int dt \int dx\, \sigma(n) E_\text{ext}^2$. This is the work delivered by the electric field to produce the excess current. In this sense,
\begin{equation}
	P(\{n,j\})\sim \exp(-\frac{N}{2}\times\text{ Work to produce the fluctuation } \{n,j\}).
\end{equation}
This is very similar to equilibrium thermodynamics, where the probability for a fluctuation is proportional to the exponential of minus the minimal work cost for the system to produces the fluctuation. For an isolated system, $P\sim e^{(S_\text{fluct}-S_\text{eq})/k_B}$, as has been noted by Einstein \cite{Einstein1910Theorie} by reversing Boltzmann's definition of entropy $S=k_B\ln W$.

From Eq.~\eqref{eq:MFT_prob}, one obtains the probability for a density profile $n(x)$ in the steady state by summing over all past trajectories $\{\hat n,\hat j\}$ that could have produced the desired density profile, starting with the mean density $\hat n(-\infty,x)=\bar n(x)$ and ending in the desired profile $\hat n(t,x)=n(x)$. Due to the exponential character of the probability weight, only the maximum will be relevant,
\begin{equation}
	P[n]=\max_{\{\hat n,\hat j\}} P_{-\infty,t}(\{\hat n,\hat j\}).
\end{equation}
This means, that the rate function for the density is
\begin{equation}
	I[n]=\min_{\{\hat n,\hat j\}} \left[ \int_{-\infty}^t dt'\int_0^1 dx \frac {(\hat j+D(\hat n)\partial_x \hat n)^2}{2\sigma(\hat n)} \right]
\end{equation}
Differentiating $U(x):=\delta I[n]/\delta n(x)$, one can show (see section IX in \cite{Derrida2007Non-equilibrium} for details) that $U$ satisfies a Hamilton-Jacobi equation (first derived in \cite{Bertini2001Fluctuations})
\begin{equation}\label{eq:mft_ham_jacobi}
	\int_0^1 dx \left[\Big(\frac{D(n) n'}{\sigma(n)}-U'\Big)^2-\Big(\frac{D(n) n'}{\sigma(n)}\Big)^2\right]\frac{\sigma(n)}{2}=0
\end{equation}
from which one can in principle find $U'(x)$ and then $I$ -- though one only knows how to do this for certain choices of $D$ and $\sigma$. 

Much easier is to verify that a given rate function satisfies this equation. For the independent random walkers the rate function is given in Eq.~\eqref{eq:indep_walker_rate_function} and leads to 
\begin{align}
	U(x)&=\log(n(x)/\bar{n}(x)), & U'&=n'/n-\bar n'/\bar n.
\end{align}
Here one should remember that $\bar n(x)=(n_b-n_a)x+n_a$ and $\bar n'(x)=n_b-n_a$. Substituting this with $D(n)=1$ and $\sigma(n)=2n$ into Eq.~\eqref{eq:mft_ham_jacobi}, and performing a partial integration with boundary conditions $n(0)=\bar n(0)$ and $n(1)=\bar n(1)$, the integrand vanishes. 

For SSEP one can do the same, though the calculation is a bit trickier. From the rate function  in Eq.~\eqref{eq:SSEP_density_large_deviation} on finds,
\begin{equation}\label{eq:SSEP_U}
	U(x)=\log\left(\frac{n(x)(1-g(x))}{g(x)(1-n(x))}\right).
\end{equation}
Substituting this with $D(n)=1,\sigma(n)=2n(1-n)$, the Hamilton-Jacobi equation of MFT will be verified.

As far as the current or charge is concerned, it is also possible to obtain a large deviation function from the basic Eq.~\eqref{eq:MFT_prob} in this case, either using the additivity principle (see section XIII in \cite{Derrida2007Non-equilibrium}) or directly (see section IV.F \cite{Bertini2015MFT}), but this is beyond the scope of the present review.

We also note, that recently the time dependent MFT equations with the transport coefficients of SSEP have been solved exactly on an infinite line with initial domain wall state \cite{Mallick2022Exact}, and that the solution agrees with the charge cumulant generating function of the SSEP on an infinite line in Eq.~\eqref{eq:mu_infinite_SSEP}. This underlines the importance of MFT not only for the stationary, but also the time dependent regime.

\section{Stochastic description of open quantum systems}\label{sec:stochastic_open_quantum_systems}
While an ideal quantum system evolves unitarily and maintains coherence, in reality a system always interacts with its uncontrolled environment and slowly decoheres. In other words, the quantum system is \textit{open}. The unitary evolution of a closed quantum system must therefore be replaced by a more general description that allows decoherence. This is the Lindblad equation which we introduce below. Then we present another method, pioneered in quantum optics, where the environment is modelled as an effective noise that acts on the system. In mean this produces a Lindblad equation, but it allows to study how the density matrix of the system fluctuates as a result of the interaction with the environment. These developments will allow us to study non-equilibrium steady states of quantum systems between two reservoirs in the sense of Figure \ref{fig:NESS}. For its historical importance we will also discuss the Caldeira-Legett model, that tries to reproduce the classical Langevin equation from unitary quantum mechanics and has served as a huge inspiration in the study of decoherence. Finally, we outline the idea of quantum trajectories in the context of continuous weak measurements. This provides us with a complementary perspective on how stochastic quantum dynamics can arise. 

\subsection*{Lindblad equation}\label{sec:lindblad}
\begin{figure}
\begin{center}
		\includegraphics[width=.5\textwidth]{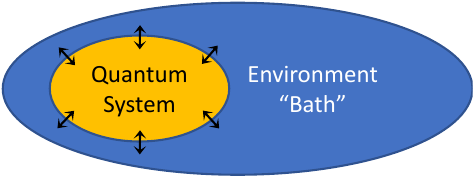}
		\caption{\label{fig:system_bath} A quantum system coupled to its environment, usually taken to be an infinitely large reservoir or ``bath''. They evolve together as a closed system and in the end the environment is traced out.}
\end{center}
\end{figure}
The usual approach to study open quantum systems is portrayed in Figure \ref{fig:system_bath}. One treats a system together with its environment (the ``reservoir'' or ``bath'') as a single closed quantum system and studies the time evolution of its density matrix $\rho(t)$ with a Hamiltonian consisting of a system, a bath and their interaction
\begin{equation}\label{eq:Ham_S_B_I}
	H=H_S+H_B+H_I.
\end{equation}
Note that this approach also applies to the case of more than one bath\footnote{Contrary to the question about the statistical ensemble of a system in contact with more than one reservoir, here we don't run into troubles, because we are concerned with a specific microscopic dynamics (that might a priori not be described by an ensemble).}. The dynamics of the system alone is recovered as a quantum average over the degrees of freedom of the bath,
\begin{equation}\label{eq:rho_B}
	\rho_S(t)=\Tr_B(\rho(t)).
\end{equation}
As a result, the dynamics of $\rho_S$ is no longer unitary, but it should still be trace-preserving and completely positive. 

We briefly outline the steps and the assumptions that lead to the Lindblad equation. The starting point is a self-consistent expansion of the dynamics of $\rho_S$ to second order in $H_I$. Then, under the assumption that the coupling between system and bath is weak, one approximates the complete state as a product state $\rho(t)\approx\rho_S(t)\otimes\rho_B$ which implies that the bath is almost unaffected by the system (the inverse is of course not true) and can be well approximated by its initial state (\textit{Born approximation}). Furthermore, assuming that excitations in the bath which have been induced by the system decay on a time scale $\tau_B\ll\tau_S$ that is much shorter than the relaxation time of the system $\tau_S$, there is no memory build-up and the dynamics of $\rho_S$ becomes local in time (\textit{Markov approximation}). In some situations, it is furthermore required to perform the so-called \textit{rotating wave approximation}, justified if the internal evolution of the bare system happens at time scales $\tau_0\ll\tau_B\ll\tau_S$ much shorter than the relaxation times of both, the bath and the system. These approximations lead to an equation, simultaneously discovered by Lindblad \cite{Lindblad1976Generators} and Gorini–Kossakowski–Sudarshan \cite{Gorini1976Completely} in 1976, which describes the most general permissible evolution of $\rho_S$,
\begin{equation}\label{eq:lindblad_equation}
	\partial_t \rho_S=-i[H_S,\rho_S]+\sum_i \left(L_i\rho_S L_i^\dagger -\frac 1 2 \{L_i^\dagger L_i,\rho_S\}\right).
\end{equation}
The so-called jump operators $L_i$ are specific to the system and the environment one considers. We will often refer to the non-hermitian part (the second term on the right hand side) as the \textit{Lindbladian} and denote it as $\mathcal{L}(\rho_S)$. For a derivation of this equation we refer to chapter 3 in Breuer and Petruccione \cite{Breuer2002Theory}. An alternative derivation of the Lindblad equation in quantum optics via a stochastic formulation will be outlined in the next paragraph.

\subsection*{Stochastic quantum dynamics}\label{sec:stochastic_quantum_dynamics}
A peculiarity of quantum mechanics is that the density matrix, the analogue of the classical phase space distribution, is a dynamical variable. As such, it can be itself a random variable and fluctuate. From this point of view, tracing out the bath in Eq.~\eqref{eq:rho_B} only corresponds to the mean density matrix of the system $\bar\rho_S$. If we want to keep track of how $\rho_S$ fluctuates due to its interaction with the bath, one needs to treat the bath as an effective noise and keep track of its statistics. This is the idea of \textit{quantum noise} which we discuss now on the basis of the standard model of quantum optics. We follow lecture notes by Peter Zoller \cite{Zoller1997Quantum}.

\subparagraph{Model from quantum optics.}
Resorting to the situation in Eq.~\eqref{eq:Ham_S_B_I}, consider a system $H_S$, e.g.\ a two-level transition in an atom, coupled to a bath of bosonic harmonic oscillators representing the modes $b(\omega)$ of an electromagnetic field in second quantization,
\begin{equation}
	H_B=\int_0^\infty d\omega \, \omega\, b(\omega)^\dagger b(\omega).
\end{equation}
They are weakly coupled in the linear response regime
\begin{equation}
	H_I=\frac{i}{\sqrt{2\pi}}\int_0^\infty d\omega \, \kappa(\omega) [c-c^\dagger] [b(\omega)^\dagger+b(\omega)].
\end{equation}
The bath operators satisfy the canonical commutation relations
\begin{equation}
	[b(\omega),b(\omega)^\dagger]=\delta(\omega-\omega').
\end{equation}
and the system operator $c$ can be though of as the dipole moment of the system which interacts with the electromagnetic field. For simplicity, assume that under the bare dynamics $H_S$ it evolves as $c(t)=e^{-i\omega_0 t}$, where $\omega_0$ is the resonance frequency of the system.

Within the \textit{rotating-wave approximation} one assumes that the system couples only to a small band of frequencies $[\omega_0-\theta,\omega_0+\theta]$  with $\omega_0\gg \theta$ around its resonance frequency, and that these frequencies are much larger than the relaxation frequency $\omega_S=\tau_S^{-1}$ of the system. If the coupling $\kappa(\omega)$ is a smooth function, it will be approximately constant in the relevant frequency range and we set it to one. 
Integrating out the fast dynamics due to $H_S+H_B$ by going to the interaction picture, one has $b(\omega)\to e^{iH_Bt}b(\omega)e^{-iH_Bt}=b(\omega)e^{-i\omega t}$ and $c\to c \,e^{-i\omega_0 t}$, and the interaction Hamiltonian becomes
\begin{equation}\label{eq:H_I}
	H_I(t)=i\left(b^{(\theta)}(t)^\dagger c-b^{(\theta)}(t)c^\dagger\right) \qquad \text{with } b^{(\theta)}(t) = \frac{1}{\sqrt{2\pi}}\int_{\omega_0-\theta}^{\omega_0+\theta} d\omega \,b(\omega)e^{-i(\omega-\omega_0)t}.
\end{equation}
Here we neglected $ c\, b(\omega) e^{-i(\omega+\omega_0)}+\text{h.c.}$, since these terms oscillate very fast and average out. This is the rotating-wave approximation. In case $H_S$ includes other frequencies than $\omega_0$, the rotating frame with frequency $\omega_0$ will not completely absorb the system's bare evolution, but leaves a residual part, which we will call $H_0$. Then a state in the interaction picture obeys
\begin{equation}\label{eq:evolution_interaction_pic}
	\partial_t |\psi(t)\>=-i\left(H_0+H_I(t)\right)|\psi(t)\>.
\end{equation}

Within the \textit{Markov approximation} one furthermore takes $\theta\to\infty$. As a consequence, excitations in the bath at different times become uncorrelated. From Eq.~\eqref{eq:H_I} one finds
\begin{equation}\label{eq:bath_commutator}
	[b(t),b(t')^\dagger]=\delta(t-t') \qquad \text{with } b(t) = \frac{1}{\sqrt{2\pi}}\int_{\mathbb R} d\omega \,b(\omega)e^{-i(\omega-\omega_0)t}.
\end{equation}

\subparagraph{Quantum stochastic calculus.}
To find an efficient method that allows us to extract the dynamics of the system alone, we will interpret the bath operators as an effective non-commuting noise that acts on the system and construct its statistical properties. This is the starting point of quantum stochastic calculus of It\={o} type that has been developed by Hudson and Parthasarathy in \cite{Hudson1984Quantum} (we will not discuss the Stratonovich convention here). It is based on the quantum stochastic process defined by
\begin{align}
	W_t:=\int_0^t ds\,b(s).
\end{align}
We also introduce an expectation value for such a process $\<\cdots\>:=\Tr_B(\cdots\rho_B)$, the quantum average over the bath. For simplicity, we consider the bath initially to be in the vacuum state $\rho_B=|0\>\<0|$ with $b(\omega)|0\>=0$. Since the coupling to the system is weak and the bath is infinitely large, one assumes that it stays close to its initial state also at later times, $\rho(t)\approx \rho_S(t)\rho_B$ (Born approximation). Then, for example $\<W_tW_t^\dagger\>=t$ and $\<W_t^2\>=\<(W_t^\dagger)^2\>=0$ in complete analogy to a complex Brownian motion, but also $\<W_t^\dagger W_t\>=0$. In It\={o} convention, one defines the increments of this process as
\begin{align}
	dW_t:=W_{t+dt}-W_t=\int_t^{t+dt}ds \, b(s),
\end{align}
which satisfies $\<dW_tdW_t^\dagger\>=dt$ and $[dW_t,dW_t^\dagger]=dt$. For a stochastic integral $\int_0^t ds f(s) dW_s$ this definition implies that the increment is independent of and commutes with the integrand. The integrand $f(s)$ can be any function of $W_s$ and $W_s^\dagger$, so it includes the operators $b(s')$ and $b(s')^\dagger$ up to time $s$. But the increment $dW_s$ includes only operators between $s$ and $s+ds$. Therefore $[f(s),dW_s]=0$ due to Eq.~\eqref{eq:bath_commutator}, and the vacuum expectation value can be taken independently, $\<f(s)dW_s\>=\<f(s)\>\<dW_s\>$.

In practise, one can simplify the stochastic equations according to the quantum It\={o} rules (here for the bath in the vacuum state), 
\begin{align}
	dW_tdW_t^\dagger&=dt\nn
	dW_tdt=dW_t^\dagger dt&=0	 \nn
	dW_t^\dagger dW_t=dW_t^2=(dW_t^\dagger)^2=dt^2&=0.
\end{align}
This means that a stochastic process, call it $A$, agrees with the process $B$ obtained after the application of these rules, in the sense that $\<(A-B)^2\>\to0$ when $dt\to0$.

\subparagraph{Quantum stochastic Schrödinger equation.}
Let us come back to the time evolution of states in Eq.\eqref{eq:evolution_interaction_pic}. Integrating over a small time step $dt$, one can express the state at time $t+dt$ in terms of the stochastic increments $dW_t$ as
\begin{equation}
	|\psi(t+dt)\>=e^{-idH_t}|\psi(t)\>
\end{equation}
with Hamiltonian increment
\begin{equation}\label{eq:ham_increment}
	dH_t= H_0 dt + i\,dW_t^\dagger \,c-i\,dW_t\,c^\dagger.
\end{equation}
To find the contribution up to order $dt$, the It\={o} rules require to expand the exponential $e^{-idH_t}$ up to second order, since $dH_t^2=c^\dagger c\, dt$. Therefore, states evolve according to the \textit{quantum stochastic Schrödinger equation}
\begin{equation}\label{eq:stochastic_state_evolution}
	d|\psi(t)\>:=|\psi(t+dt)\>-|\psi(t)\>= -i dH_t-\frac 1 2 c^\dagger c dt.
\end{equation}
Note that the It\={o} convention enforces the extra contribution $-\frac 1 2 c^\dagger c$ to the noise averaged Hamiltonian $H_0$. 

Similarly, one can study the time evolution of the density matrix of system and bath $\rho(t+dt)=e^{-idH_t}\rho(t)e^{iH(t)}$, with increments $d\rho(t) :=\rho(t+dt)-\rho(t)$. One obtains,
\begin{equation}
	d\rho(t) =-i[dH_t,\rho(t)]  + dW_t^\dagger\, c\rho(t)c^\dagger \,dW_t - \frac 1 2\{\rho(t),c^\dagger c\} dt
\end{equation}
Note that here some terms vanished, since the bath is in the vacuum. Alternatively, one applies the It\={o} rules in a cyclic fashion, as if one took the trace $\Tr_B$ over the bath. In mean, i.e.\ when taking the trace over the bath $\bar\rho_S(t)=\<\rho(t)\>$, this reduces to a Lindblad equation with a single jump operator $L=c$,
\begin{equation}\label{eq:quantum_optics_lindblad}
	\partial_t \bar\rho_S(t) = -i[H_0,\bar\rho_S(t)] +  c\bar\rho_S(t)c^\dagger - \frac 1 2\{\bar\rho_S(t),c^\dagger c\}
\end{equation}
The bar on $\bar\rho_S$ should emphasize that this only captures the mean behaviour of the system, and not the fluctuations.

\subparagraph{High bath occupation and classical noise.}
In the case, where the bath is not in the vacuum, but in a state with mean photon occupation number $n$, i.e.\ $\<b(t)^\dagger b(t')\>=n\delta(t-t')$, $\<b(t) b(t')^\dagger\>=(n+1)\delta(t-t')$ and $\<b(t)b(t')\>=\<b(t)^\dagger b(t')^\dagger\>=0$, and one has to adjust the It\={o} rules according to
\begin{align}
	dW_t^\dagger dW_t&=\kappa^2 ndt & dW_tdW_t^\dagger=\kappa^2(n+1)dt.
\end{align}
Here we reinstated the coupling constant as $\kappa(\omega)=\kappa$ (which was set to unity) into the definition of $dW_t$ and $b(t)$. In the limit of a high bath occupation $n\to\infty$, one has to reduce the coupling $\kappa\to0$ such that $\kappa^2 n$ stays constant. This situation corresponds to a classical limit, and indeed the quantum noise becomes a classical commuting noise
\begin{equation}
	[dW_t, dW_t^\dagger]= \kappa^2\to0.
\end{equation}

\subparagraph{Summary.}
This example showed that open quantum systems can be described in a stochastic fashion through a Hamiltonian increment in the form of Eq.~\eqref{eq:ham_increment}. The description involves the system's own evolution $H_0$ as well as the noise terms $dW_t$ that can be either of quantum (non-cummuting) or of classical (commuting) nature, depending on the properties of the bath. This is the way how the stochastic nature of the toy model QSSEP, which most of this thesis is devoted to, can be understood.

\subsection*{Caldeira-Leggett model} Based on the system-bath picture discussed above, Caldeira and Leggett were interested in understanding how classical and irreversible dynamics can emerge from the underlying quantum description. Indeed, many situations of system-reservoir coupling at finite temperature are well described by the classical Langevin equation \eqref{eq:langevin}, for example charge fluctuations on the capacitor of an LRC-circuit. But at low temperatures, quantum effects become visible. Therefore Caldeira and Leggett \cite{Caldeira1983Path} in 1983 asked the simple question: ``Can one reconcile damped equation of motion with the process of quantization?''

One should note that in contrast to the situation encountered in quantum optics, their approach is adapted to situations where the time scale of the system's internal evolution $\tau_0\equiv\omega_0^{-1}\sim \tau_S$ is of the same order as the system's relaxation time. Therefore the rotating-wave approximation is not valid here. Furthermore, their equation for the reduced density matrix $\rho_S$ has a memory kernal, in general, and becomes Markovian only at large temperatures $kT\gg \hbar \omega_0 $ (we resituate $\hbar$ in this paragraph). In this limit, they are able to show that their equation for $\rho_S$ reproduces the Fokker-Planck equation for the probability of a classical particle obeying the Langevin equation. We give a short overview of their approach below.

Consider a system $H_S$ (variables $x,p$) under the influence of a potential $V(x)$ with resonance frequency $\omega_0$, coupled to a bath of particles $H_B$ (variables $X_k,P_k$) behaving like independent harmonic oscillators with frequency $w_k$, initially at equilibrium with temperature $T$. The coupling $\kappa_k$ is weak in the sense that the system-bath interaction can be approximated by linear response,
\begin{align}
	H_S&=\frac{p^2}{2m}+V(x) \nn
	H_B&=\sum_k\left(\frac{P_k^2}{2M}+ \frac{1}{2}M\omega_k^2X_k^2\right) \nn
	H_I&=x\sum_k \kappa_k X_k.
\end{align}
To find an equation for the reduced density matrix $\rho_S$, Caldeira and Leggett use the influence-functional approach of Feynman and Vernon \cite{Feynman1963Theory}.
By requiring that for $kT\gg \hbar \omega_0$ this reproduces the Langevin equation with friction coefficient $\gamma$, they phenomenologically adjust the density of the harmonic oscillators to
\begin{equation}
	\rho(\omega)=
	\begin{cases}
		\frac{2 M \gamma \omega^2}{\kappa(\omega)^2 \pi}, &\omega<\Omega \\
		0, &\omega>\Omega,
	\end{cases}
\end{equation}
where $\kappa(\omega_k)=\kappa_k$ are the coupling constants and $\Omega$ is a high frequency cutoff. At arbitrary temperatures, this leads to a non-Markovian equation for $\rho_S$ (see eq.~(3.38) in \cite{Caldeira1983Path}), while for $2kT\gtrsim\hbar\Omega\gg \hbar \omega_0$ the equation becomes Markovian and reads
\begin{equation}
	\partial_t \rho_S(t)=-\frac{i}{\hbar} [H_S',\rho_S(t)]-\frac{i\gamma}{2m\hbar}[x,\{p,\rho_S(t)\}]-2\gamma k T\frac{1}{2\hbar^2} [x,[x,\rho_S(t)]].
\end{equation}
Here $H_S'=H_S-x^2\sum_k \frac{\kappa_k^2}{2M \omega_k^2}= H_S-x^2 \Omega \gamma/\pi$ is the renormalized Hamiltonian of the system. This is a consequence of the introduction of the high frequency cutoff $\Omega$. The equation is not yet of Lindblad form, but can be brought to Lindblad form by adding a term which is negligible in the high temperature limit (see eq.~(3.414) in Breuer \cite{Breuer2002Theory}).

\subsection*{Quantum trajectories}\label{sec:quantum_trajectories}
Another situation in which a quantum system undergoes a stochastic evolution is in the context of a continuous weak measurement. We follow \cite{Bernard2016Statistical}. Assume that a quantum system $\rho$ interacts for a short time with a probe $|\varphi\>$ through a unitary $U$, and afterwards the probe is measured and projected onto the state $|s\>$
\begin{equation}
	\rho\to U\Big(\rho\otimes|\varphi\>\<\varphi|\Big)U^\dagger \to \frac{F_s\,\rho\, F_s^\dagger}{\pi(s)}.
\end{equation}
$F_s:=\<s|U|\varphi\>$ is the back-action of the probe's measurement on the system for an outcome $s$, which occurs with probability $\pi(s)=\Tr(F_s\rho F_s^\dagger)$. Repeating this protocol, each time with a new probe, $\rho$ undergoes a stochastic dynamics that depends on the measurement outcomes $(s_1,s_2,\cdots)$. If the probe $|\varphi\>$ is a two-level system $s=\pm$ and has approximately equal overlap with both $|+\>$ and $|-\>$ such that the measurement outcomes form a (biased) random walk\footnote{If one of the measurement basis states has a much large overlap with $|\varphi\>$, then most of the time the probe will be measured in this state, and the occasions where it is measured in the other state are rare events described by a Poisson process. Therefore the resulting quantum trajectory would not be driven by a Brownian motion, but rather by a Poisson process.}, one can expand the operators $F_s$ in terms of the time step $dt$ over which they act on the system, $F_{\pm}=\frac{1}{\sqrt{2}}(\mathds{1}\pm\sqrt{dt}N+\cdots)$. In the limit of an infinitesimal time step $dt$, the stochastic dynamics becomes a so-called \textit{quantum trajectory}
\begin{equation}
	d\rho_t = -i[H,\rho_t]dt+\Lin_N(\rho_t)+D_N(\rho_t)dB_t
\end{equation}
where $dB_t$ is the increment of a real Brownian motion, $H$ is the system's Hamiltonian without monitoring, $\Lin_N$ is a Lindbladian for the (non-hermitian) measurement operator $N$ (determined through $F_{\pm}$ which itself depends on $U$) and $D_N$ is called the "stochastic innovation term",
\begin{align}
	\Lin_N(\rho)&=N\rho N^\dagger - \frac 1 2\{N^\dagger N,\rho\} \nn
	D_N(\rho)&=N\rho + \rho N^\dagger - \rho \Tr(N \rho+\rho N^\dagger).
\end{align}
At the same time, the measurement series $(s_1,s_2,\cdots)$ undergoes a random walk $X_n=\sqrt{dt}\sum_{i=k}^n s_k$. In the limit of continuous time ($t=ndt$) it becomes
\begin{equation}
	dX_t=\Tr(N\rho_t+\rho_t N^\dagger)dt + dB_t,
\end{equation}
i.e.\ it is a Brownian motion with a bias that depends on $\rho_t$.

\subparagraph{Computational technique.} Note that quantum trajectories, not for density matrices but for states, can be used as a computational technique: When solving a Lindblad equation \eqref{eq:lindblad_equation} for a density matrix $\rho_t$ with $N^2$ entries, it can be less costly to sample the stochastic evolution of the $N$ entries of a state $|\psi_t\>$ that in mean follows the same Lindblad equation and therefore $\rho_t=\overline{|\psi_t\>\<\psi_t|}$. The method builds on the fact that the Lindblad equation \eqref{eq:lindblad_equation} (for simplicity for a single jump operator) can be rewritten in terms of a non-hermitian Hamiltonian $H_\text{eff}=H-\frac{i}{2}L^\dagger L$ and a remaining part $L\rho L^\dagger$. Evolving a state with this Hamiltonian
\begin{equation}
	|\tilde \psi_{t+dt}\>=(1-iH_\text{eff}\,dt)|\psi_t\>
\end{equation}
the norm is no longer conserved since $\<\tilde \psi_{t+dt}|\tilde \psi_{t+dt}\>=1-r\,dt+\Ord(dt^2)$. One can understand $r:=\<\psi_{t}|L^\dagger L |\psi_{t}\>$ as the rate at which the state undergoes a ``jump''. Therefore one associates the following quantum trajectory to states
\begin{equation}
	|\psi_{t+dt}\> =
	\begin{cases}
	 	\frac{(1-iH_\text{eff}\,dt)|\psi_t\>}{\sqrt{1-r\,dt}} &\text{with prob.\ }1-rdt\\[10pt]
	 	\frac{L|\psi_t\>}{\sqrt{r}}  &\text{with prob.\ }rdt.
	\end{cases}
\end{equation}
The stochastic evolution agrees with the Lindblad equation in mean and provides a computational advantage, if the sampling can be done efficiently. For more details see \cite[sec.~III.G]{Landi2022Nonequilibrium}  and references therein.

\section{Microscopic models}\label{sec:microscopic_models}
From a theoretical point of view, the richness of many-body quantum physics rests very much upon the plethora of microscopic models that have been introduced within the last hundred years. One could almost say that each model possesses its own character, behaving similar in some circumstances and different in others. And it is only through the acquaintance with many of these models, that one finally gets the feeling to understand a little bit of what is actually happening. We find it therefore inevitable to list a few of these models and their properties, restricting mostly to one dimension.
\paragraph{Free Fermions.}
Undoubtedly the simplest model of many-body quantum dynamics are free fermions. On a one dimensional chain with $N$ sites they are defined by the Hamiltonian
\begin{equation}\label{eq:free_fermions}
	H = -J\sum_j^N \left(c_{j+1}^\dagger c_j+c_j^\dagger c_{j+1}\right) = \sum_k\epsilon(k) n(k),
\end{equation}
with the usual anti-commutation relations $\{c_j^\dagger,c_k\}=\delta_{ik}$ for the fermionic creation and annihilation operators, and 
\begin{align}\label{eq:free_fermions_energy}
	\epsilon(k)&=-2J\cos(k),& n(k)&=c(k)^\dagger c(k), & c(k)&=\frac{1}{\sqrt{N}}\sum_{j=1}^N c_j e^{-ikj}.
\end{align}
Imposing periodic boundary conditions the momenta are quantized, $k_j=2\pi j/N$ with $j=1,\cdots,N$ and a $P$-particle eigenstate can be characterized in terms of these momenta as
\begin{equation}
	|k_1,\cdots,k_P\>= \frac{1}{\sqrt{P}}c(k_1)^\dagger\cdots c(k_P)^\dagger|0\>
\end{equation}

Free fermions are integrable (because they can be exactly diagonalized) and as such they posses an infinite number of conserved  $[H,Q^{(n,\alpha)}]=0$ and mutually commuting $[Q^{(n,\alpha)},Q^{(m,\beta)}]=0$ charges with local densities (following \cite{Essler2022Short}),
\begin{equation}\label{eq:free_fermions_Q}
	Q^{(n,\alpha)}=\sum_j \epsilon^{(n,\alpha)}(k_j)\,n(k_j)=-2J \sum_j Q^{(n,\alpha)}_j, \qquad \text{with }\alpha=0,1,
\end{equation}
where $\epsilon^{(n,0)}_k:=-2t\cos(nk)$ and $\epsilon^{(n,1)}_k:=-2t\sin(nk)$. The associated locally conserved densities are
\begin{equation}
	Q^{(n,\alpha)}_j=\frac{1}{2}(i^\alpha c_{j+n}^\dagger c_j + \text{h.c.}).
\end{equation}
Their Heisenberg equations of motions are equivalent to local conservation laws
\begin{equation}
	\partial_t Q_j^{(n,\alpha)} = i[H,Q_j^{(n,\alpha)}]= -(J_{j}^{(n,\alpha)}-J_{j-1}^{(n,\alpha)})
\end{equation}
with current operator
\begin{equation}\label{eq:free_fermions_J}
	J_j^{(n,\alpha)} =-\frac{J}{2} \left(  i^{1-\alpha} (c_j^\dagger c_{j+n+1} - c_{j+1}^\dagger c_{j+n}) + \text{h.c.} \right)
\end{equation}
In particular, the local fermion number and associated current are
\begin{align}
	Q_j^{(0,0)}&= c_j^\dagger c_j & J_j^{(0,0)}&=-J\,(ic_j^\dagger c_{j+1}-ic_{j+1}^\dagger c_j)
\end{align} 

\paragraph{XXZ Heisenberg spin chain.}
This is one of the simplest and most studied interacting, but also integrable models in one dimension.
\begin{equation}\label{eq:XXZ}
	H= \sum_j \left(\sigma_j^x \sigma_{j+1}^x + \sigma_j^y \sigma_{j+1}^y + \Delta \sigma_j^z \sigma_{j+1}^z \right).
\end{equation}
Note that the XX part can be written as $2(\sigma_j^+ \sigma_{j+1}^- + \sigma_j^- \sigma_{j+1}^+)$ with $\sigma_j^\pm = \frac 1 2 (\sigma_j^x\pm i \sigma_j^y)$. Through a Jordan-Wigner transformation
\begin{align}\label{eq:jordan_wigner}
	c_j^\dagger&=e^{(-i\pi\sum_{k<j}\sigma_k^+\sigma_k^-)} \sigma_j^+ & 	c_j&=e^{(i\pi\sum_{k<j}\sigma_k^+\sigma_k^-)} \sigma_j^- & 2n_j-1=\sigma_j^z
\end{align}
with $n_j=c_j^\dagger c_j$, it can be mapped to spinless fermions,
\begin{equation}
	H= \sum_j \left(2(c_{j+1}^\dagger c_j+c_j^\dagger c_{j+1})+\Delta(2n_j-1)(2n_{j+1}-1)\right).
\end{equation}
If $\Delta=1$, the model possesses an obvious global $SU(2)$ symmetry generated by $\sum_j \sigma_j^\alpha$ with $\alpha=x,y,z$. Otherwise only the total spin-$z$ component $S^z=\frac 1 2 \sum_j\sigma_j^z$ is conserved. But being integrable, there are, apart from $H$ itself, an infinite number of other conserved charges and locally conserved currents. For example, the local spin current associated to $S^z$ is
\begin{equation}
	J^{(S)}_j=\sigma_j^x \sigma_{j+1}^y-\sigma_j^y\sigma_{j+1}^x = 2(ic_j^\dagger c_{j+1} - ic_{j+1}^\dagger c_j).
\end{equation}
For a review how to solve this model with the algebraic Bethe ansatz see for example \cite{Faddeev1996Algebraic}.

\paragraph{Lieb-Liniger model.}
This integrable model describes bosons interacting via a $\delta$-potential and was solved exactly by Lieb and Liniger in 1963 \cite{Lieb196Exact}. In can be approximately realized in experiments where it has been used to test generalized hydrodynamics \cite{Schemmer2019Generalized,Malvania2021Generalized}. Following \cite{Essler2022Short}, the Hamiltonian for $P$ particles in first quantization reads
\begin{equation}
	H=\sum_j^P -\frac{1}{2m}\partial_{x_j}^2 + 2c\sum_{j<k}^P \delta(x_j-x_k)
\end{equation}
and in second quantization
\begin{equation}\label{eq:LL}
	H=\int dx \left(\frac{1}{2m}\partial_x\Phi^\dagger(x)\partial_x \Phi(x) + c \,\Phi^\dagger(x)^2 \Phi(x)^2\right)
\end{equation}
with $\Phi(x)$ a complex bosonic field satisfying $[\Phi(x),\Phi(y)^\dagger]=\delta(x-y)$.

A $P$-quasiparticle eigenstate of the Hamiltonian can be expressed as
\begin{equation}
	|k_1,\cdots,k_P\>=\frac{1}{\sqrt{P!}}\int dx_1\cdots dx_P \psi_k(x_1,\cdots,x_P) \Phi^\dagger(x_1)\cdots\Phi^\dagger(x_P)
\end{equation}
where the wave function $\psi_k(x_1,\cdots,x_P)$ has an exact form in terms of the celebrated Bethe ansatz
\begin{equation}
	\psi_k(x_1,\cdots,x_P)=\frac{1}{\mathcal N} \sum_{\sigma\in S_P} \text{sgn}(\sigma) e^{i\sum_{j=1}^P k_{\sigma(j)} x_j \prod_{j>k}}[k_{\sigma(j)}-k_{\sigma(k)}-ic]
\end{equation}
and the set $k=\{k_j\}$ plays the role of momenta or rapidities. Imposing periodic boundary conditions, $\psi_k(\cdots,x_j+L,\cdots)=\psi_k(\cdots,x_j,\cdots)$ with $L$ the lenght of the system, the set $k$ is determined by the Bethe equations
\begin{equation}
	k_j L + \sum_{k=1}^P 2 \arctan\left(\frac{k_j-k_k}{c}\right) = 2\pi I_j, \qquad j=1,\cdots,P.
\end{equation}
Here $I_j$ are integers if $P$ is odd, and half-integers if $P$ is even. Each set $\{I_j\}$ is in one-to-one correspondence with $\{k_j\}$ and can be used equivalently to characterize the state.

Finally note that the states $|k_1,\cdots,k_P\>$ are also eigenstates of the conserved charges
\begin{equation}
	Q^{(n)}|k_1,\cdots,k_P\>=\sum_j^{P} k_j^n \,|k_1,\cdots,k_P\>.
\end{equation}
So a single quasiparticle carries the charge $Q^{(n)}(k)=k^n$. The hamiltonian corresponds to $H=Q^{(2)}$.

\paragraph{Bose-Hubbard model.}
The Bose-Hubbard in any dimension is defined by bosonic operators $[b_i,b_j^\dagger]=\delta_{ij}$ on a regular lattice. It consists of a nearest neighbour hopping $\<ij\>$ and an interacting term,
\begin{equation}\label{eq:Bose-Hubbard}
	H=-t\sum_{\<ij\>}(b_{i}^\dagger b_{j}+\text{h.c.}) + \frac U 2 \sum_{i} n_{i} (n_{i}-1)
\end{equation}
The model is attractive for $U<0$ and repulsive for $U>0$, the latter case being the more popular one. In particular, if $U/t\gg1$, the model is in the \textit{hard-core boson regime}. In 1D on an infinite lattice and in the hard-core boson regime it becomes the integrable Lieb-Liniger model. In 2D and higher, the model is non-integrable and has been used to verify ETH \cite{Rigol2008Thermalization}, which justifies to call it a quantum chaotic model.

\paragraph{Anderson model.}
This model, famous for its transition between a diffusive and a localized regime and extensively studied in the context of mesoscopic transport, can be defined as a nearest neighbour hopping of spinless fermions on a $3D$ lattice with sites $i=(i_x,i_y,i_z)$ under the influence of random on-site potentials $V_i$ of variance $W$ (also known as \textit{disorder strength}) ,
\begin{align}\label{eq:Anderson}
	H&=-t\sum_{\<ij\>} c_i^\dagger c_j + \sum_i V_i \,c_i^\dagger c_i, & V_i&\in[-W/2,W/2]
\end{align} 
The hopping can also be inhomogeneous with components $t_x,t_y,t_z$ in the corresponding directions. The potentials $V_i$ are independently and uniformly distributed in $[-W/2,W/2]$. The critical disorder strength is $W_c\approx 16.5t$ \cite{MacKinnon1981One-Parameter,Pichard1981Finite}. Below this value the model is diffusive, above its become \textit{Anderson localized}. In the diffusive regime the model can be used to explore coherent effects described within the scope of mesoscopic transport, such as weak localization or the universal conductance fluctuations \cite{Markos2006Numerical}.

\section{Transport in integrable and chaotic quantum systems}\label{sec:transport_quantum}
Many-body quantum systems can broadly be distinguished into two classes: \textit{Integrable} systems, in which the presence of extra conserved charges prevents the decay of currents and therefore there is no thermalization in the sense of the standard ensembles of statistical mechanics. And \textit{chaotic} systems, which are non-integrable and thermalize at late times. If there exists a third class of systems, that displays \textit{many-body localization} has recently been a source of controversial debate \cite{Abanin2021Distinguishing} and it seems now quite likely that many-body localization is a finite size effect that does not survive in the thermodynamic limit \cite{Suntajs2020Quantum,Sierant2022Challenges}.

Naively one expects integrable systems to behave as \textit{ballistic} conductors, i.e.\ there is no bulk resistivity and the displacement of charge is proportional to time, while chaotic systems should display \textit{diffusive} transport. However, there are exceptions to this general paradigm as we will see below. Also note, that at zero temperature, even diffusive metals behave ballistically, so the interesting regime to study is at finite temperature \cite{Bertini2021Finite}.

To speak about transport, a system should possess conserved charges $[H,Q]=0$, with locally conserved densities $Q(x,t)$ that can be ``transported'' \footnote{Our notation in this section is to denote operators by capital letters, expectation values by small letters and to state everything for a continuous space variable $x$.},
\begin{align}\label{eq:local_cons_law}
	Q&=\int Q(x,t) dx & \partial_t Q(x,t) + \partial_x J(x,t)&=0,
\end{align}
and $J(x,t)=e^{iHt}J(x)e^{-iHt}$ is the associated local current operator in the Heisenberg picture. Integrable systems possess an infinite number of such conserved charges, non-integrable system only a few, such as energy or total spin. 

Traditionally, transport has been studied in terms of the conductivity $\sigma$, which is the linear response of the current to a driving field,
\begin{equation}
	\<J_L(\omega)\>=\sigma(\omega) E(\omega).
\end{equation}
Here $J_L(\omega)=\int dx \int_0^L dt J(x,t)e^{-i\omega t}$ is the total current operator in the frequency domain built from the local current $J(x,t)$ which results from a time-dependent driving field $E(t)=(1/2\pi)\int dt E(\omega) e^{i\omega t}$ and $\<\bullet\>:=\Tr(e^{-\beta H}\bullet)/Z$ is the thermal quantum expectation value with $Z=\Tr(e^{-\beta H})$ and $\beta=1/kT$. Within Kubo's linear-response theory \cite{Kubo1957Statistical-Mechanical}, the conductivity can be determined from current-current correlations at equilibrium
\begin{equation}\label{eq:Kubo_conductivity}
	\sigma(\omega)=\lim_{t\to\infty}\lim_{L\to \infty}  \frac{1}{L}\int_0^\infty dt \int_0^{\beta} d\lambda\, e^{-i\omega t} \<J_L(0) J_L(t+i\lambda)\>_0,
\end{equation}
where $L$ is the physical length of the system. For a derivation in an early review see \cite{Zwanwig1965Time-Correlation}. 

To extract a hydrodynamic description, one should be interested in the long wave-length modes where $\omega\to 0$. Here, the imaginary part of $\sigma$ is almost zero since $\Im\sigma(-\omega)=-\Im\sigma(\omega)$, and one decomposes the real part generically into a singular and a regular part,
\begin{equation}
	\Re\sigma(\omega)=2\pi D_W \delta(\omega) + \sigma_{\text{reg}}(\omega).
\end{equation}
The prefactor $D_W$ (not to be confused with the diffusion constant) is called the \textit{Drude weight} and its meaning is that it characterizes the presence of ballistic transport. This can be also understood by analogy to the classical Drude model which consists of charged particles accelerated by an electric field and damped by a force proportional to the velocity \cite{AshcroftMermin},
\begin{equation}\label{eq:drude_conductivity}
	\sigma(\omega)=\frac{\sigma_0}{1-i\omega \tau}, \qquad \sigma_0=\frac{ne^2\tau}{m}.
\end{equation}
Here $\tau$ is the average time between collisions and $n$ is the density of particles (electrons) with charge $e$ and mass $m$. This implies $\Re\sigma(\omega)=\sigma_0/(1+\omega^2\tau^2)$. If collisions happen all the time, $\tau\to0$, such that the effective dynamics is diffusive, then $\Re(\sigma)$ has no singular part and $D_W=0$. However, if particles move ballistically and do not collide, $\tau\to\infty$, the conductivity has a singular part with $D_W=ne^2/(2m)$.

In case the integral in Eq.~\eqref{eq:Kubo_conductivity} does not converges, there is also the possibility for anomalous transport which is neither ballistic nor diffusive: The width of a perturbation growths as $\sigma \sim t^\alpha$ with $0<\alpha<1$. For example, the Heisenberg XXZ spin chain with $\Delta=1$ is superdiffusive with $\alpha=2/3$ \cite{Znidaric2011Spin}. For a review on superdiffusion in spin chains see \cite{Bulchandani2021Superdiffusion}. There the Drude weight is zero, but the current-current correlation function decays so slow that the diffusion constant in the Green-Kubo formula \eqref{eq:Green-Kubo} diverges. More details about finite-temperature transport can be found in the review \cite{Bertini2021Finite}. The study of anomalous transport also has a long history in classical system, in particular in anharmonic chains modelling the flow of heat, see \cite{Lepri2016Heat} for a recent review.

\paragraph{Integrable systems.}
In integrable systems, the interaction of particles can be represented in terms of the motion of non-decaying collective excitations, so-called quasi-particles, that scatter without particle production. In other words, the multi-particle scattering of quasi-particles can be described by subsequent 2-particle scattering events, mathematically expressed by the Yang-Baxter equation. Ballistic transport as a general feature of integrable models was first conjectured in \cite{Castella1995Integrability}. And generalized hydrodynamics, see below, describes very successfully the hydrodynamic evolution of conserved charges in ballistic integrable systems. However, there are exceptions where transport is diffusive or superdiffusive. Prominently, the XXZ spin chain from Eq.~\eqref{eq:XXZ} is ballistic for $\Delta<1$, but diffusive for $\Delta>1$ \cite{Znidaric2011Spin,Bertini2021Finite}. See also \cite{Bernard2016Conformal} for a review of transport in integrable systems described by conformal field theory.

\paragraph{Chaotic systems.}
The notion of chaotic quantum systems is not as well defined as the notion of integrability. One definition of chaotic quantum systems is that their classical counter part is chaotic in the sense that ergodicity can be proven rigorously. However, this comprises only a small number of systems, and therefore it is better to regard chaotic quantum systems as those systems that thermalize and that satisfy the eigenstate thermalization hypothesis (see Section \ref{sec:ETH}). Another characteristic of chaotic systems is that the level spacing statistics is described by random matrix theory (see Section \ref{sec:level-spacing}).

The prominent class of chaotic quantum system are random unitary circuits with a conserved charge. Transport in this case is generically diffusive, see Eq.\eqref{eq:diffusion_ruc}. Another example of a chaotic system is to add a staggered magnetic field $\sum_{j\text{ even}}^{N}h_{j}\sigma_{j}^{z}$ to the integrable XXZ model in Eq.~\eqref{eq:XXZ}. Driven out-of-equilibrium by a Lindbladian, transport in this model was studied for example in \cite{Prosen2009Matrix} (with $\Delta=0.5$ and $h_{2j}=-0.5$) where it was shown that the non-equilibrium steady state could be efficiently simulated by matrix product states since the density matrix had a low so-called operator space entanglement. Yet another chaotic model is the Bose-Hubbard model in 2D (or 3D) from Eq.~\eqref{eq:Bose-Hubbard} which was numerically shown to satisfy the eigenstate thermalization hypothesis.

\paragraph{Generalized Hydrodynamics.}
Built solely on the assumption of local entropy maximization (sometimes referred to as local thermodynamic equilibrium), a hydrodynamic theory for the evolution of conserved charges in integrable models has been put forward in Refs.\ \cite{Castro2016Emergent,Bertini2016Transport}, now referred to as generalized hydrodynamics (GHD). It has been experimentally verified for the Lieb-Liniger model \cite{Malvania2021Generalized,Schemmer2019Generalized}.

The starting point is the \textit{generalized Gibbs ensemble} (GGE) \cite{Rigol2007Relaxation} which describes stationary expectation values of observables in isolated quantum systems with more conserved charges $Q=(Q_1,Q_2,\cdots)$ than just the energy -- so in particular in interacting integrable systems \cite{Ilievski2015Complete},
\begin{equation}
	\rho_\beta = \frac{1}{Z} e^{-\sum_n \beta_n Q_n} \qquad \text{with } Z=\Tr(e^{-\sum_n \beta_n Q_n}).
\end{equation}
The GGE is a maximal-entropy ensemble which retains the minimal information of the initial state
\begin{equation}
	\< Q_n\>_\beta :=\Tr(Q_n\rho_\beta)= \<\psi_0|Q_n|\psi_0\>.
\end{equation}
This fixes the value of the potentials $\beta=(\beta_1,\beta_2,\cdots)$. By translation invariance of $\rho_\beta$, expectation values of the local densities $q=(q_1,q_2,\cdots)$ do not depend on space and are therefore in one-to-one correspondence with the potentials. As a consequence, the local currents $j=(j_1,j_2,\cdots)$ can be seen as a function of the local densities
\begin{align}
	q_n&:=\<Q_n(x)\>_\beta & j_n(q)&:=\<J_n(x)\>_\beta.
\end{align}
with $Q_n(x)$ and $J_n(x)$ now in the Schrödinger picture.

If the system is initialized in an inhomogeneous state and slowly relaxes towards a homogeneous state, physical properties usually vary only space-time scales much larger than microscopic scales. The assumption of GHD is that the space-time scales over which physical properties are approximately constant are thermodynamically large such that locally the system at $(x,t)$ is always well described by a generalized Gibbs ensemble with smoothly varying potentials $\beta(x,t)$. Then the local conservation laws \eqref{eq:local_cons_law} become hydrodynamical equations for the evolution of the local densities $q_n(x,t)=\<Q_n(x)\>_{\beta(x,t)}$,
\begin{equation}\label{eq:conservation_law_GHD}
	\partial_t q_n(x,t)+\partial_x j_n(q(x,t))=0.
\end{equation}
Diagonalizing the Jacobi matrix $\partial j(q)/\partial q=R\,\text{diag}(v_1,v_2,\cdots)R^{-1}$ we obtain an equation for the ``hydrodynamic normal modes'' $n:=R^{-1}q $,
\begin{equation}
	\partial_t n_n + v_n(x,t) \partial_x n_n(x,t)=0.
\end{equation}
The difficulty is to find the normal mode velocities $v_n(x,t)$ for a given model. For integrable models, this is possible. 

Following \cite{Essler2022Short}, we outline how to do this in two cases. For free fermions Eq.~\eqref{eq:free_fermions} one can directly evaluate densities and currents with the help of Eqs.~\eqref{eq:free_fermions_Q} and \eqref{eq:free_fermions_J},
\begin{align}
	q_{(n,\alpha)}(x,t)&=\int \frac{dk}{2\pi} \epsilon^{(n,\alpha)}(k) \rho_{x,t}(k) \\
	j_{(n,\alpha)}(x,t)&=\int \frac{dk}{2\pi} \epsilon'(k) \epsilon^{(n,\alpha)}(k) \rho_{x,t}(k).
\end{align}
We introduced $\rho_{x,t}(k)$ as the density of fermions with momentum $k$ in the generalized Gibbs ensemble, and $\epsilon(k)=-2J\cos(k)$ their energy from Eq.~\eqref{eq:free_fermions_energy}. But actually, there are many microscopic states with particle density $\rho_{x,t}$. Following \cite{Essler2022Short} we can view $\rho_{x,t}$ as a \textit{macrostate}, specifying the state of the system on hydrodynamic scales. In particular, substituting into Eq.~\eqref{eq:conservation_law_GHD} one finds,
\begin{equation}
	\partial_t\rho_{x,t}(k) + \epsilon'(k)\partial_x \rho_{x,t}(k).
\end{equation}
So $\rho_{x,t}(k)$ is indeed a hydrodynamical normal mode with velocity $\epsilon'(k)$. Here the velocity does not depend on the macrostate $\rho_{x,t}$. This changes when considering interacting models, such as the Lieb-Liniger model from Eq.~\eqref{eq:LL}. In this case one finds,
\begin{equation}
	\partial_t \rho_{x,t}(k) +\partial_x\left(v_{\rho_{x,t}}(k)\rho_{x,t}(k)\right),
\end{equation}
where $\rho_{x,t}(k)$ is the quasiparticle density. We refer to \cite{Essler2022Short} for explicit expressions.

\section{Entanglement}\label{sec:entanglement}
The property that really makes many-body quantum systems to be so different from classical systems is entanglement. As soon as one considers two or more particles, quantum mechanics allows this fascinating concept of entanglement to enter the game. Two or more particles are entangled if their state cannot be written as a product of the individual states. This seemingly uniquely quantum mechanical concept has many astonishing and, at first sight, sometimes contradicting consequences: Isolated quantum systems can thermalize and appear classical, precisely because they are very entangled  (see below). On the other hand, entanglement is the reason why quantum algorithms can be faster than classical ones. And the main challenge in building quantum computers is that they tend to interact with their environment and internally loose entanglement\footnote{If we assume a two-particle ``quantum computer'' to be in a maximally entangled Bell state, not entangled with a third particle in state zero $|\psi\rangle=\frac 1 {\sqrt{2}}(|00\rangle+|11\rangle)|0\rangle$, then the mutual information between the two particles of the ``quantum computer'' is $I(1:2)=S_1+S_2-S_{12}=2\ln(2)$. However, if we entangle it with the third particle, $|\psi\rangle=\frac 1 {\sqrt 2}(|000\rangle+|111\rangle)$, e.g.\ via a CNOT gate, then $I(1:2)=\ln(2)$. The information gained about particle 2 when learning the state of particle 1, is reduced by one bit, because their reduced state is now described by a mixed density matrix $\rho_{12}=\frac 1 2 |00\rangle\langle00|+\frac 1 2|11\rangle\langle11|$.}. 

Entanglement turns out to be a key concept in the understanding of equilibrium and non-equilibrium properties of many-body quantum systems from the theoretical point of view. As gets clear when discussing random quantum circuits in the next section, it acts as an organizing principle of many-body quantum phases. From the computational point of view, entanglement is an indicator of the computational hardness to simulate a quantum system on a classical computer. This is exploited in matrix product states and tensor networks (see \cite{Orus2014Practical} for an introduction).

\paragraph{Entanglement Entropy.}
We start, by giving the formal definition of the entanglement entropy. Many different measures of entanglement have been proposed, see e.g. \cite{Plenio2006Introduction}. Here we will focus on bipartite entanglement measures, that quantify the entanglement between two parts, $A$ and its complement $\bar A$, of a system in a pure state $\rho=|\psi\rangle\langle\psi|$. In particular we consider the \textit{von Neumann entropy}
\begin{equation}\label{eq:van_neumann}
	S_A:=-\Tr(\rho_A \log \rho_A)
\end{equation}
where $\rho_A=\Tr_{\bar A}(\rho)$ is the reduced density matrix on $A$. We also consider the family of \textit{Renyi entropies}
\begin{equation}
	S_A^{(q)} :=\frac{1}{1-q}\log\mathrm{Tr}(\rho_A^{q}),
\end{equation}\label{eq:renyi}
which reduce to the von Neumann entropy in the limit where $q\to1$. If not further specified, we use the term \textit{entanglement entropy} for any of these entropies and simply denote it by $S_A$. Let us summarize a few important properties:
\begin{itemize}
	\item The entanglement entropy only depends on the eigenvalues $\{\lambda_i\}_i$ of $\rho_A$ and is therefore invariant under unitary transformations that act separately on $A$ or $\bar A$.
	\item Writing $S_A=-\sum_i\lambda_i\ln \lambda_i$, the von Neumann entropy is equivalent to the Shannon entropy for a classical probability distribution $\{\lambda_i\}_i$.
	\item Via a Schmidt decomposition $|\psi\rangle=\sum_i \sqrt{\lambda_i}|\phi_i\rangle |\xi_i\rangle$ one sees that $\rho_A$ and $\rho_{\bar A}$ have the same eigenvalues and therefore the entanglement entropy is symmetric $S_A^{(q)}=S_{\bar{A}}^{(q)}$. 
	\item The Schimdt decomposition also shows that the $0$th Renyi entropy is the logarithm of the number of terms appearing in a decomposition into product states, $S_A^{(0)}=\log(\#\{\text{nonzero }\lambda_i\})$. This number  is also called the \textit{bond dimension} between $A$ and $\bar A$. Furthermore, $S_A^{(0)}\ge S_A^{(q)}$ is an upper bound for all higher Renyi entropies.
	\item The von Neumann entropy takes the maximal value $S_A=\log(\dim(\Hil_A))$ for a maximally mixed state $\rho_A=\frac 1 {d_A} \mathbb I_{d_A}$ with $d_A=\dim(\Hil_A)<\dim(\Hil_{\bar A})$ (also known as infinite temperature state). It is minimized for a pure state $\rho_A=|\psi\rangle\langle\psi|$ where $S_A=0$.
	\item The entanglement entropy is subadditive. For two (not necessarily complementary) parts $A$ and $B$ of the system, $S_{A\cup B}\le S_A+S_B$. This ensures that the mutual information, defined below, is always greater or equal than zero.
\end{itemize}

When dealing with mixed states, as they appear for example in the description of open quantum systems, the entanglement entropy is no longer a meaningful concept. Instead one can consider the \textit{mutual information} between two (not necessarily complementary) regions $A$ and $B$
\begin{equation}
	I(A: B):=S_{A}+S_{B}-S_{A\cup B}
\end{equation}
where the part, that is only due to the state being mixed is subtracted. This is a straightforward generalization of the mutual information between two correlated random variables $X$ and $Y$ in classical information theory, which quantifies the information we gain about $X$ when learning $Y$. Mutual information measures the total quantum and classical correlations, the latter being defined as the maximal correlations left after erasing any entanglement \cite{Groisman2005Quantum}.

We remind the reader that a mixed state $\rho$ is entanglement between $A$ and $\bar A$ if it is not \textit{separable}, meaning that we cannot write it as a classical mixture of product states $\rho_A^i\otimes\rho_{\bar A}^i$ on $A$ and $\bar A$,
\begin{equation}
	\rho=\sum_i p_i \,\rho_A^{i}\otimes \rho_{\bar{A}}^{i}
\end{equation}
While product states have mutual information zero, a generic separable state only satisfies $I(A:\bar A)<\min(S_A,S_{\bar{A}})$.

\paragraph{Many-body entanglement in equilibrium.}
We summarize a few properties of the entanglement entropy in different many-body systems in equilibrium and justify the statement that entanglement can be seen as an organizing principle. 
\begin{itemize}
	\item For ground states of 1D critical systems (gapless), the entanglement entropy has a universal form $S_A=c/3\,\ln(\ell_A)$ due to Calabrese and Cardy \cite{Calabrese2009Entanglement}, with $\ell_A$ the length of the subsystem $A$, from which one easily learns the central charge $c$ of the conformal symmetry algebra of the model. 
	\item For ground states of  gapped systems in 1D \cite{Hastings2007Area}, and more generally, for ground states of systems in any dimension with short ranged correlations, the entanglement entropy $S_A$ is proportional to the boundary area of the region $A$ and said to satisfy an \textit{area law} \cite{Eisert2010Colloquium}. This is in contrast to typical energy eigenstates of thermalizing systems, which have an extensive entanglement entropy (\textit{volume law}) as we will see in the paragraph on ETH. 
	\item To distinguish equilibrium from non-equilibrium, the mutual information is a useful concept. It has been shown that in equilibrium, the mutual information between two adjacent regions of a Gibbs state with local interactions scales like the boundary area between the regions \cite{Wolf2008Area} -- essentially because the two point correlations or coherences $G_{ij}$ between two lattice points $i$ and $j$ decay fast enough. As we will see later, the non-equilibrium steady state of QSSEP, and of non-interacting random unitary circuits, satisfies a volume law for the mutual information. 
	\item The entanglement entropy also plays an important role in organizing topological phases of matter \cite{Wen2017Colloquium}.
\end{itemize}

\paragraph{Measurement of entanglement entropy.}
In principle, quantum state tomography \cite{James2001Measurement} allows to fully reconstruct the density matrix in an experiment and then calculate the entanglement entropy. However the number of measurements required growths as the dimension of the Hilbert space and soon becomes out of reach for many-body systems. In addition to this, for each measurement, the system has to be reliably prepared in the exact same state, which can be challenging. An alternative method to measure the entanglement entropy is to evolve $n$ replica of the same system at once and to interfere them (e.g.\ via a beam splitter) at the desired time, from which one can obtain quantities such as $\Tr(\rho_A^n)$. In \cite{Islam2015Measuring}, the 2nd Renyi entropy was measured in this way for a Bose-Hubbard model realized in an optical lattice of four atoms (in each replica).

\paragraph{Thermalization and ETH.}\label{sec:ETH}
Entanglement is also an important concept to explain why isolated quantum systems can \textit{thermalize} at long times, i.e.\ that time average of (few-body) observables can be described through the ensembles of statistical mechanics \textit{as if} the system was in a microcanonical state
\begin{equation}\label{eq:rho_micro}
	\rho_E=\sum_{E_i\approx E}|E_i\rangle\langle E_i|
\end{equation}
or in a canonical state (Gibbs state)
\begin{equation}\label{eq:rho_canonical}
	\rho_\beta=e^{-\beta H}/Z, \qquad \text{with } Z=\Tr(e^{-\beta H}).
\end{equation}
The paradox that makes this question so hard to answer, is that the unitary evolution of an initial state $\rho_0$ will always preserve its spectrum and therefore some information about the initial state is retained. But this is in contradiction with the ensembles of statistical mechanics. The popular paradigm to explain thermalization is that ``the system acts as its own reservoir'' and the notion of entanglement can make this statement precise: Under unitary time evolution, the system becomes more and more entangled and any information that was initially accessible through a small number of degrees of freedom can now only be retrieved by probing an extensive number of degrees of freedom. But this is practically impossible, experiments to measure the particle or energy density are always local, and therefore the spreading of entanglement ensures that the information about the initial state is effectively lost.

This line of thought has been put on solid ground in context of the Eigenstate thermalization hypothesis (ETH) \cite{Deutsch1991Quantum,Srednicki1994Chaos,Srednicki1999Approach} -- also noting one of the first numerical verifications \cite{Rigol2008Thermalization} with the astonishing insight that five particles on 21 lattice sites can be enough for statistical mechanics to apply. ETH states that energy eigenstates in thermalizing systems are special. They play the role of ``typical'' configurations in a classical many-body system, which are very vaguely defined by the fact that macroscopic observables take the same value on almost all ``typical'' configurations\footnote{Think about a box with gas particles partitioned in two and take the macroscopic observable to be the number of particles in the left half. Then, on a large majority of configurations, this observable takes the same value, with the exception of very few ``atypical'' configurations, where for example all particles are in the left half and no particle is in the right half. Thermalization then means to reach a ``typical'' configuration and one does not need to make reference to ergodicity.}. This becomes easier in the quantum setting: Energy eigenstates are very well defined, and ETH states that any (few-body) observable, when expressed in the energy basis $A_{ij}:=\langle E_i|A|E_j\rangle$, is of the form (with $E_+=(E_i+E_j)/2$ and $\omega=E_i-E_j$)
\begin{equation}\label{eq:ETH}
	A_{ij}=A(E_+)\delta_{ij}+ e^{-S(E_+)/2} f_A(E_+,\omega)R_{ij},
\end{equation}
where crucially the diagonal element $A(E_+)$ is a smooth function of energy and is equal to the microcanonical or canonical expectation value at this energy. This has the striking consequence that a single energy eigenstate of the many-body system is enough to compute thermal averages: In Eq.~\eqref{eq:rho_micro}, we could replace the microcanonical energy shell $\rho_E$ by a single state with energy $E_i\approx E$. Intuitively, the state's entanglement structure is rich enough to act as its own reservoir when traced over the remaining part, not probed by the observable $A$. 

Coming back to Eq.~\eqref{eq:ETH}, the off-diagonal elements are exponentially suppressed by the thermodynamic entropy $S(E_+)$, multiplied by an observable-dependent, smooth envelope function $f_A(E_+,\omega)$ that falls of rapidly with $\omega$ and enforces a band structure. The off-diagonal terms are furthermore random due to the complex random variable $R_{ij}$ with zero mean and unit variance. Its precise distribution was initially assumed to be Gaussian. However it has been noticed in \cite{Foini2019Eigenstate} that this distribution is actually more complicated than just Gaussian (see Sec.~\ref{sec:general_ETH}).

Finally, let us note that a similar ansatz can also be used to show that the thermodynamic entropy of an isolated, homogeneous quantum system, defined through the common relation $dS_\text{td}=\delta Q/T$, is related to the von Neumann entropy of a single eigenstate of this system, by tracing out a larger half of the system and calculating the von Neumann entropy of the smaller half
\begin{equation}
	S_A= V_A/V_\text{tot}\; S_\text{td},
\end{equation}
where $V_A$ and $V_\text{tot}$ is volume (number of degrees of freedom) of part $A$ and of the whole system \cite{Deutsch2010Thermodynamic}. This result has been refined in \cite{Murthy2019Structure}, finding a correction to the entanglement entropy proportional to the system's heat capacity, if the two halfs are of equal size. 

\section{Random quantum circuits}\label{sec:random_quantum_circuits}
Simulating the time evolution of a particular non-integrable, thermalizing system is very hard and analytical calculations are usually impossible. Instead, one can content oneself with studying statistical properties of an ensemble of \textit{generic} systems that have in common only the minimal structure of any isolated quantum system: \textit{local interactions} and \textit{unitary evolution}. This line of thought is similar to random matrix theory for the spectrum of heavy nuclei \cite{French1981Random-matrix} or for the scattering matrix of mesoscopic systems \cite{Beenakker1997Random-matrix}: While individual instances are analytically intractable, the statistical ensemble obey simple universal “laws”. This is the idea of \textit{random quantum circuits}. They provide a tractable setting to understand universal phenomena that could occur in more structured models. In particular, they allow to understand the spreading of entanglement in terms of hydrodynamic equations that are independent of the microscopic details. Here we follow two reviews on this the subject, \cite{Fisher2023Random} and \cite{Potter2022Entanglement}.

\subsection*{Unstructured circuits}
In the simplest case, following \cite{Nahum2017Quantum}, one considers an infinite chain of spins with local Hilbert space dimension $d$ and starts in a pure product state $|\psi_0\rangle$. At each time step $t$, a unitary gate is applied to a randomly selected neighbouring pair of sites. The unitary gates are Haar random, i.e.\ they are uniformly chosen from the unitary group $U(d^2)$. 
\subparagraph{Entanglement as a surface growth.}The entanglement entropy $S_x:=S_{[x,\infty]}$ of the region to the right of site $x$ can increase during one time step only if a unitary gate acts on the sites $(x-1,x)$. If so, its new value is equal to the minimum of $S_{x-1}$ and $S_{x+1}$ plus the contribution due to entangling the sites $(x-1,x)$. For the 0th Renyi entropy, which is the logarithm of the bond dimension, this contribution is almost surely equal to one, if we take logarithms to the base $d$, because a typical unitary gate will fully entangle the sites $(x-1,x)$. That is,
\begin{equation}
	S_x(t+1)=\min\{S_{x-1}(t),S_{x+1}(t)\}+1.
\end{equation}
Here we did not write the superscript for the 0th Reny entropy, because it turns out that this update rule is also true in the limit $d\to\infty$ for all higher Renyi entropies, and it is believed to capture universal properties of entanglement growth also for finite $d$. In a coarse grained continuum limit, where one averages over blocks of spins and rescales time such that one unit of time corresponds to the application of one unitary per bond on average, the update rule turns into a Kardar-Parisi-Zang equation originally introduced to describe the growth of a surface's height profile $S_x$ of a surface with time $t$ \cite{Kardar1986Dynamic},
\begin{equation}\label{eq:KPZ}
	\partial_t S_x=\nu \partial_x^2 S_x-\frac \lambda 2 (\partial_x S_x)^2 + \eta_x(t)+c
\end{equation}
Here $c$ is the average growth rate, $\eta_x(t)$ is a white noise (uncorrelated in space and time), $\nu$ describes a diffusive smoothening of sharp features of the surface and $\lambda$ describes the nonlinear dependence on the slope of the surface.

\begin{figure}
	\includegraphics[width=\textwidth]{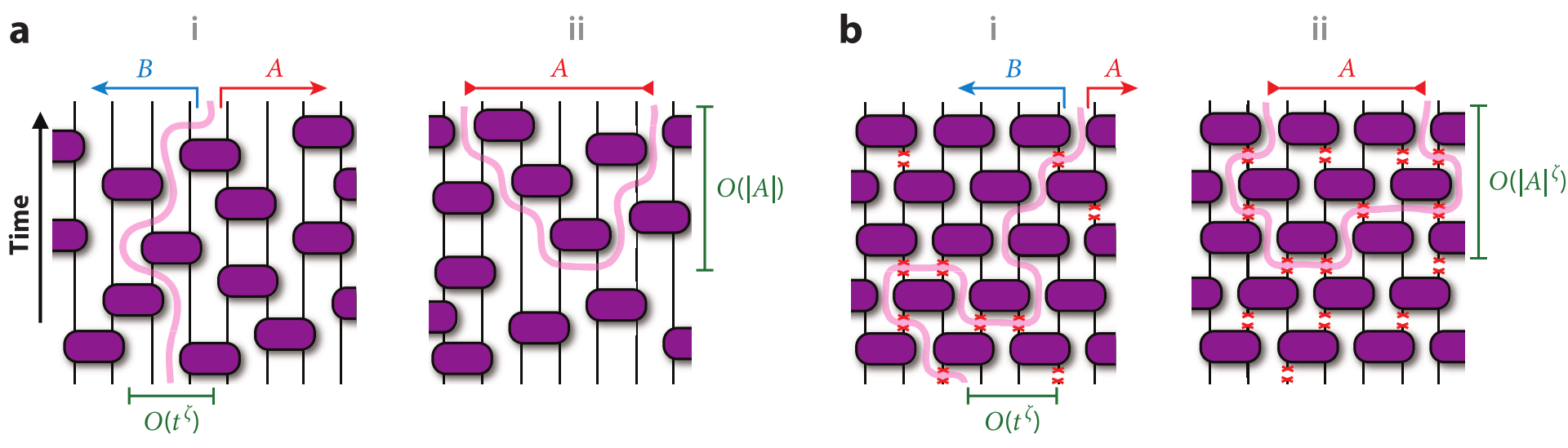}
	\caption{\label{fig:circuit} (a) The minimal cut through a random unitary circuit and (b) for a monitored dynamics with randomly selected single-site projective measurements. The wandering of the minimal cut is controlled by an exponent $\zeta=2/3$ of the directed polymer in a random environment. Figure reproduced from \cite{Fisher2023Random}}
\end{figure}

From this equation one learns that the entanglement of a region $[x,\infty)$, averaged over all circuit realization, growths linearly in time with some speed $v_E$ and a subleading correction $\sim t^\beta$ with KPZ exponent $\beta=1/3$, which also controls the entanglement broadening (the fluctuations around its mean value). That is,
\begin{align}
	 \langle S_x(t) \rangle&= v_E t + B t^{\beta} & \langle S_x(t) \rangle_c^{1/2}=C t^\beta,
\end{align}
where $C/B$ is universal, but the constants $v_E$ and $B$ are not. For $d\to\infty$ the update rule is exactly solvable and one finds $v_E=1/2$. Interestingly, entanglement spreads slower than the maximal speed of operator spreading $v_B=1$ allowed by the circuit geometry.
\subparagraph{Minimal cut.}
Alternatively, the same result can be obtained from the idea of the so-called \textit{minimal cut}. The entanglement entropy between a finite region $A$ and its complement $B$ is upper bounded by the 0th Renyi entropy, i.e.\ by the logarithm of the bond dimension. When starting from a product state, the bond dimension at time $t$ is equal to the minimal number of lines one has to cut in the tensor network representation of the circuit (see Figure \ref{fig:circuit}) to connect the two boundaries of $A$ at time $t$ -- either by going back to the initial product state (see panel a.i), or if this is too far, by traversing the circuit, respecting its causal order (see panel a.ii). For a region of length $|A|=\ell_A$, this implies that $S_A(t<t_\star)=2 s_\text{eq} v_E t$ growths on average linearly with time (the factor 2 is due to the two boundaries) and saturates to its equilibrium value $S_A(t>t_\star)=s_\text{eq}\ell_A$ at the moment $t_\star=\ell_A/(2v_E)$ where it becomes less costly to traverse the circuit directly than to go back to the initial state. Here $s_\text{eq}$ is the thermodynamic entropy density at equilibrium. The description via the minimal cut is valid for all Renyi entropies when $d\to\infty$, but many phenomenological aspects survive at finite $d$. This idea led to the more general ``entanglement membrane'' picture \cite{Zhou2020Entanglement}.

\subsection*{Circuits with a conserved charge}\label{sec:circuits_with_charge}
Unstructured random quantum circuits have no notion of transport, because there is no conserved charge that could be transported. Following \cite{Khemani2018Operator,Rakovszky2018Diffusive}, we add a locally conserved $U(1)$ charge $Q=\sum_x \sigma_x^z$ by splitting the Hilbert space of each spin into a conserved and a neutral part $\mathbb C^2 \otimes \mathbb C^d$ and constraining the two-site unitaries to be uniformly random of the form
\begin{equation}\label{eq:charge_conserving_unitary}
	U\in\begin{pmatrix} \uparrow\uparrow & & & \\ & \uparrow\downarrow, &\downarrow\uparrow & \\ & \uparrow\downarrow, &\downarrow\uparrow &  & \\ & & &\downarrow\downarrow	\end{pmatrix} \otimes U(d^2).
\end{equation}
If the unitary gates are applied in a \textit{brick wall structure}, meaning that at each odd time step $t$, unitaries are applied to all pairs $(x,x+1)$ with $x$ odd, and vice versa at even time steps, then the local charge $q(x,t)=\langle\psi_t|\sigma_x^z|\psi_t\rangle$ behaves on average like the average of a random walk,
\begin{equation}
	q(x,t+1)=q(x+1,t+1)=\frac{q(x,t)+q(x+1,t)}{2},
\end{equation}
because a Haar averaged unitary acting on $(x,x+1)$ will evenly distribute the charge between the two sites. In a coarse-grained continuum description, this naturally leads to a diffusion equation
\begin{equation}\label{eq:diffusion_ruc}
	\partial_t q=D\partial_x^2 q
\end{equation}
with diffusion constant $D=1/2$. In this sense, the random unitary circuit with a conserved $U(1)$ charge provides one of the simplest models for diffusion in a many-body quantum system. This also has an influence on the dynamics of entanglement entropy. Renyi entropies with $q>1$ grow diffusively, $S_x^{(q)}(t)\sim\sqrt{t}$, while the von Neumann entropy, $S^{(1)}_x(t)\sim t$, remains ballistically \cite{Rakovszky2019Sub-ballistic}.

\subparagraph{Charge fluctuations.}
In Ref. \cite{McCulloch2023Counting}, a closed random unitary circuit with conserved $U(1)$ has been used to study charge fluctuations. Their main finding is that these agree with charge fluctuations in the SSEP (see Section \ref{sec:SSEP}). In other words, charge fluctuations in random unitary circuits behave essentially classically and can be described by MFT -- even at the level of a single circuit realization. We outline their work below.

The quantity of interest are the cumulants (with respect to repeated quantum measurements $\<\cdots\>$) of the total charge $Q_t=R_t-R_0$ that flows during a time $t$ from the left half to the right half of the closed circuit. Here we denote the total charge operator in the right half by $R$ and in the left half by $L$. The study makes use of the partition protocol (see beginning of Section \ref{sec:intro_to_non-eq}), initializing the circuit in a domain wall state $\rho\sim e^{\mu L - \mu R}$ such that the difference in chemical potential between left and right half is $2\mu$. Since classically in the situation of an infinite domain wall, the total charge scales as $Q_t\sim\sqrt{t}$, one expects that
\begin{equation}
	\<e^{\lambda Q_t}\>=e^{\sqrt{t} \chi(\lambda)}.
\end{equation}
And one is interested in the cumulant generating function $\chi$ for a single realization of the circuit.

\begin{figure}
	\begin{center}
		\includegraphics[width=.7\textwidth]{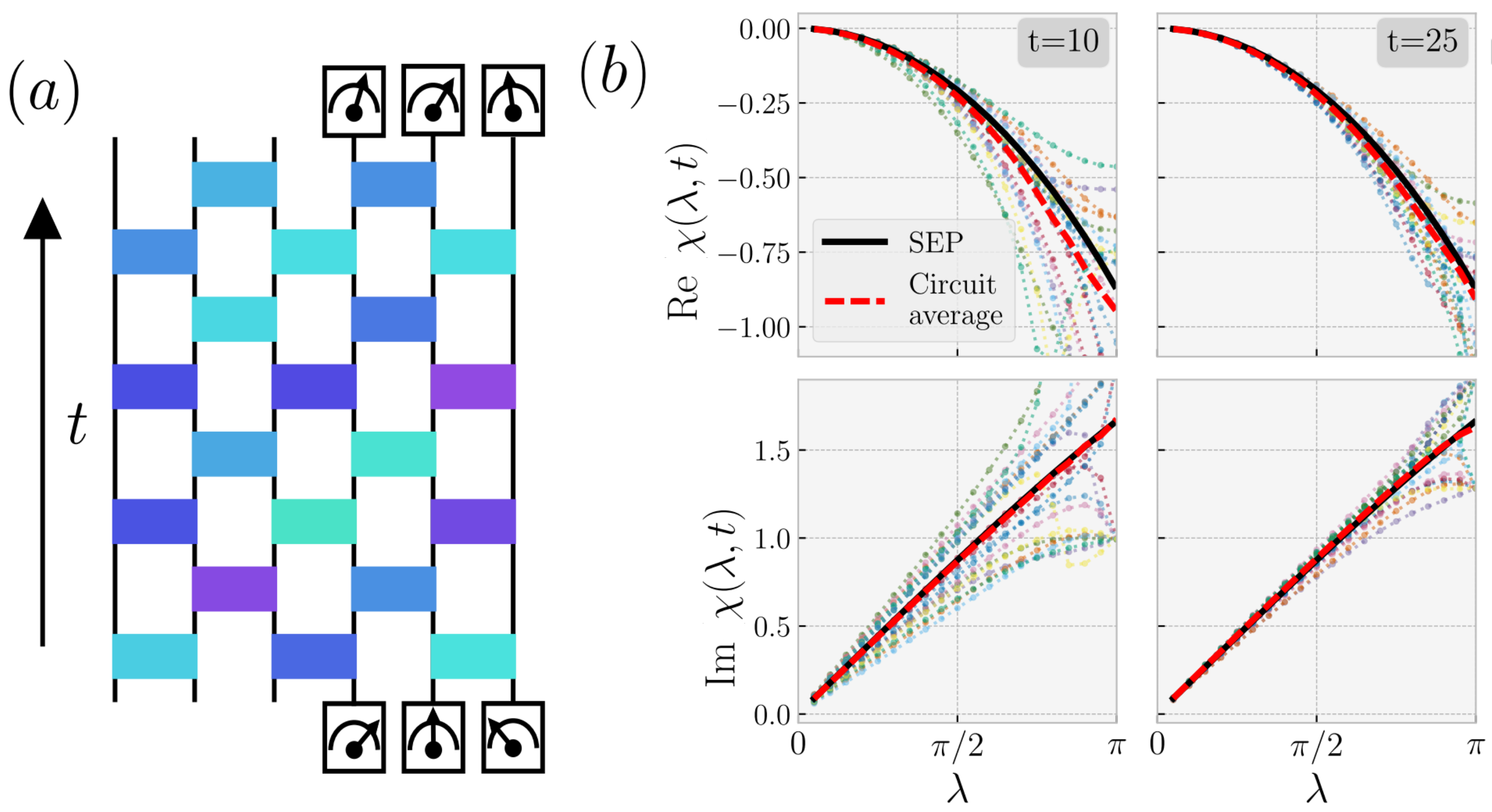}
		\caption{\label{fig:charge_fluctuations_circuit} Reproduced from \cite{McCulloch2023Counting}. (a) A two-time measurement of the total charge $R$ in the right half of the circuit in brickwork geometry. (b) Numerical TEBD simulations for the cumulant generating function in different circuit realizations at two different times. As time growths, individual realizations approach the circuit averaged value. This shows that the cumulant generating function is self-averaging at late times. Furthermore, one sees that the numerical data agrees with the theoretical prediction of SSEP from Eq.~\eqref{eq:mu_infinite_SSEP}.}
	\end{center}
\end{figure}

Measuring the charge $Q_t$ in a quantum system requires a two-time measurement protocol\footnote{Such a protocol appeared for example in \cite{Tang2014Full,Bernard2016Conformal}, though it is difficult to say when it was used the first time.} of the right half's charge operator $R$, once at time zero and once at time $t$, see Figure \ref{fig:charge_fluctuations_circuit} (a). The outcomes $q_0$ and $q_t$ of these measurements will occur with probability $Pr(q_0,q_t)=\Tr(\rho\, P_{q_0} U_t^\dagger P_{q_t} U P_{q_0})$.
Here $P_{q_0}$ and $P_{q_t}$ are projectors on sectors with charge $q_0$ and $q_t$ in the right half of the system, i.e. $R\, P_{q_t}|\psi\>=q_t\,P_{q_t}|\psi \>$. The total charge $Q_t$ that has been transferred is $q_t-q_0$. Then one has
\begin{equation}
	\<e^{\lambda Q_t}\>=\sum_{q_t, q_0} e^{\lambda(q_t-q_0)}P(q_0,q_t)=\sum_{q_t,q_0} \Tr[\rho \,P_{q_0} U_t^\dagger P_{q_t} e^{\lambda R} U_t e^{-\lambda R} P_{q_0}]
\end{equation}
Due to the choice of $\rho$, which is a function of the charge operator $R$ on the right half of the circuit, the sum over projectors on $q_0$ will not change $\rho$. Furthermore $\sum_q P=1$. Therefore,
\begin{equation}\label{eq:charge_fluctuations_protocol}
	\<e^{\lambda Q_t}\>=\Tr[\rho\, U_t^\dagger e^{\lambda R} U_t e^{-\lambda R}]
\end{equation}
With this the protocol, the authors show numerically (see Figure \ref{fig:charge_fluctuations_circuit} (b)) and analytically that the generating function for a single circuit realization is self averaging and equal to
\begin{equation}
	\chi(\lambda)=\overline{\chi(\lambda)}+\Ord(t^{-1/2}).
\end{equation} 
Here $\overline{\chi}$ is the circuit averaged generating function which turns out to be equal to that of SSEP in Eq.~\eqref{eq:mu_infinite_SSEP} with particle densities $n_a=\frac{e^{\mu}}{1+e^{\mu}}$ and $n_a=\frac{e^{-\mu}}{1+e^{-\mu}}$.

\subparagraph{Experimental evidence of MFT in quantum systems.} 
Motivated by the result from random unitary circuits, charge fluctuations were recently studied experimentally in an isolated chaotic quantum system without classical randomness, implemented with cold atoms in the group of Immanuel Bloch \cite{Wienand2023Emergence}. More precisely, the authors realize a two-ladder ($\alpha=1,2$) Bose-Hubbard model
\begin{equation}
	H=-J\sum_{i, \alpha=1,2}(a_{\alpha,i}^\dagger a_{\alpha,i+1}+\text{h.c.})-J_\perp \sum_i (a_{1,i}^\dagger a_{2,i}+\text{h.c.}) + \frac U 2 \sum_{\alpha,i} n_{\alpha,i} (n_{\alpha,i}+1)
\end{equation}
with bosonic operators $a_{\alpha,i}$ and $n_{\alpha,i}=a_{\alpha,i}^\dagger a_{\alpha,i}$ on sites $i$. The experiments were performed on $2\times 50$ sites in the hard-core regime $U/J>6.5$, where each site is occupied at maximum by a single atom (with $\gtrsim 97\%$ probability), and in the chaotic (non-integrable) regime $J_\perp/J=1$, where the ladders are strongly coupled and the mean dynamics is diffusive.

\begin{figure}
	\centering
	\begin{subfigure}[b]{0.33\textwidth}
		\centering
		\includegraphics[width=\textwidth]{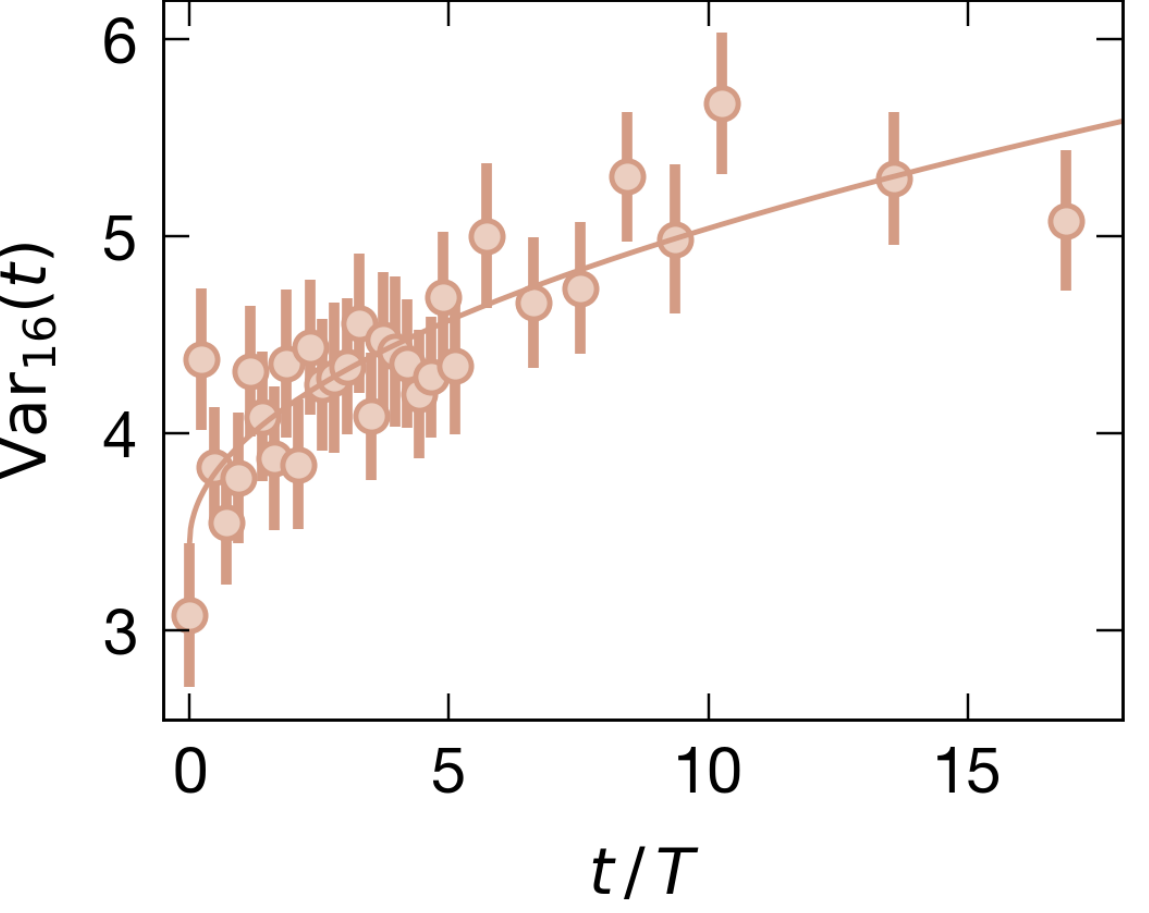}
		\caption{}
	\end{subfigure}
	\hfill
	\begin{subfigure}[b]{0.66\textwidth}
		\centering
		\includegraphics[width=\textwidth]{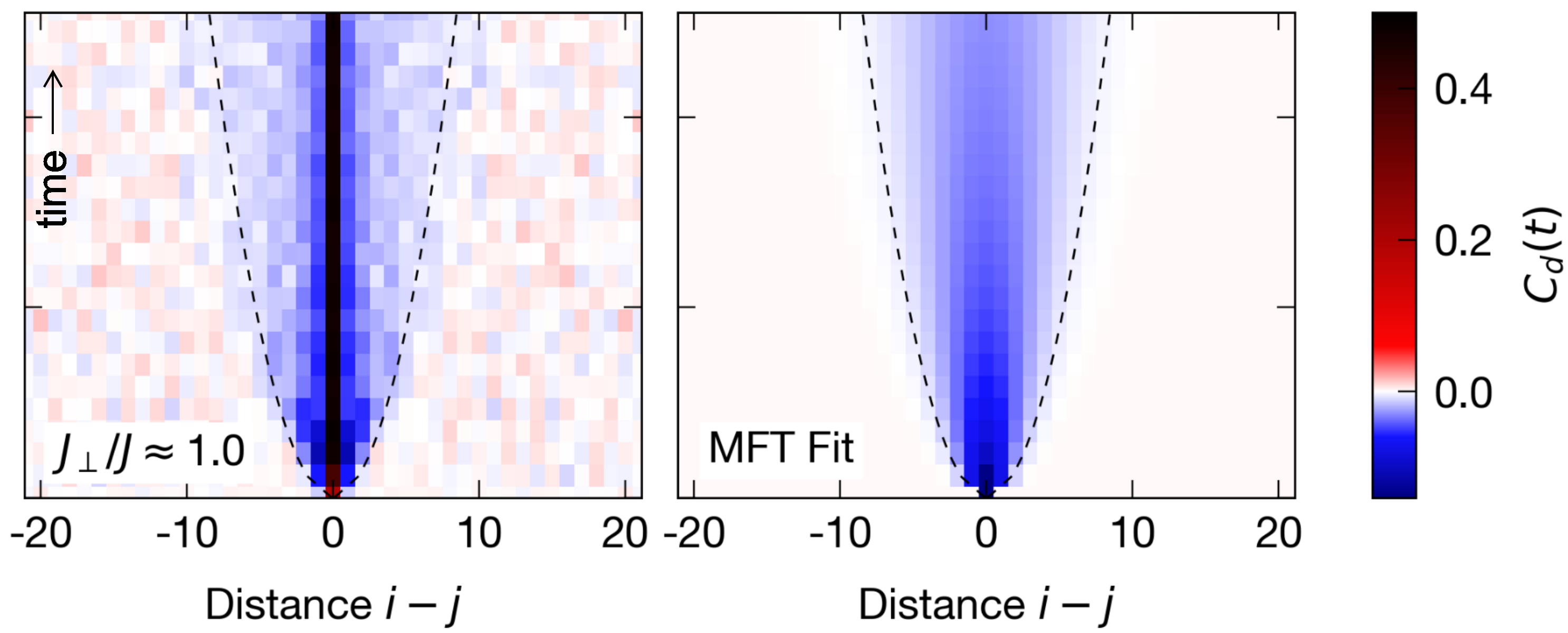}
		\caption{}
	\end{subfigure}
	\caption{\label{fig:bloch_paper} Reproduced from \cite{Wienand2023Emergence}. (a) Growth of the variance $\text{Var}_L(t):=\<Q_L(t)^2\>^c$ of the number of atoms that moved into or out of a subsystem of size $L$. The solid line is the MFT prediction \eqref{eq:MFT_prediction}. (b) Growth of connected density-density correlations $C_d(t)=\sum_{i} \<(n_{1,i}+n_{2,i})(n_{1,i+d}+n_{2,i+d})\>^c$ between all possible sites with distance $d=i-j$.}
\end{figure}

With initial condition a charge density wave (even sites empty and odd site occupied) in both ladders, they study the charge fluctuations of the total number of atoms $Q_L(t)=N_L(t)-N_L(0)$ that moves into or out of a subsystem of length $L$. Here $N_L(t)$ denotes the number of atoms in this subsystem at time $t$. Under the assumption that the diffusion constant $D(n)=D$ is independent of the local density (which is justified in the situation of an initial charge density wave since the density soon becomes homogeneous), and furthermore that the mobility is $\sigma(n)=2n(n-1)$ as in SSEP (which the authors assume tacitly), the growth of the variance of charge fluctuations $\text{Var}_L(t):=\<Q_L(t)^2\>^c$ (with $\<\cdots\>$ the quantum expectation value) can be predicted by MFT as
\begin{equation}\label{eq:MFT_prediction}
	\text{Var}_L(t)\approx \sqrt{\frac{2Dt}{\pi a^2}}
\end{equation}
where $a$ is the lattice spacing. The result is shown in Figure \ref{fig:bloch_paper} (a). With the help of the MFT prediction, the a priori unknown diffusion constant can be fitted to the experimental data. Then it can be compared to the diffusion constant obtained from a measurement of the density-density correlations whose form is also be predicted from MFT, see Figure \ref{fig:bloch_paper} (b). The two values $D=1.11(25)Ja^2/\hbar$ and $D=0.88(5)Ja^2/\hbar$ obtained from charge fluctuations and from density-density correlations, respectively, seem to agree.

To sum up, the experiment shows, that charge and density fluctuations in chaotic many-body quantum systems are well described by MFT. This means that the classical noise that appears in the MFT equations \eqref{eq:MFT} is reproduced from the stochastic nature of quantum measurements. In this sense, $\<\cdots\>_\text{MFT}=\<\cdots\>_\text{QM}$.

\subsection*{Current-driven circuits}\label{sec:current-driven_circuit}
\begin{figure}[t]
	\centering
	\begin{subfigure}{0.37\linewidth}
		\includegraphics[width=\textwidth]{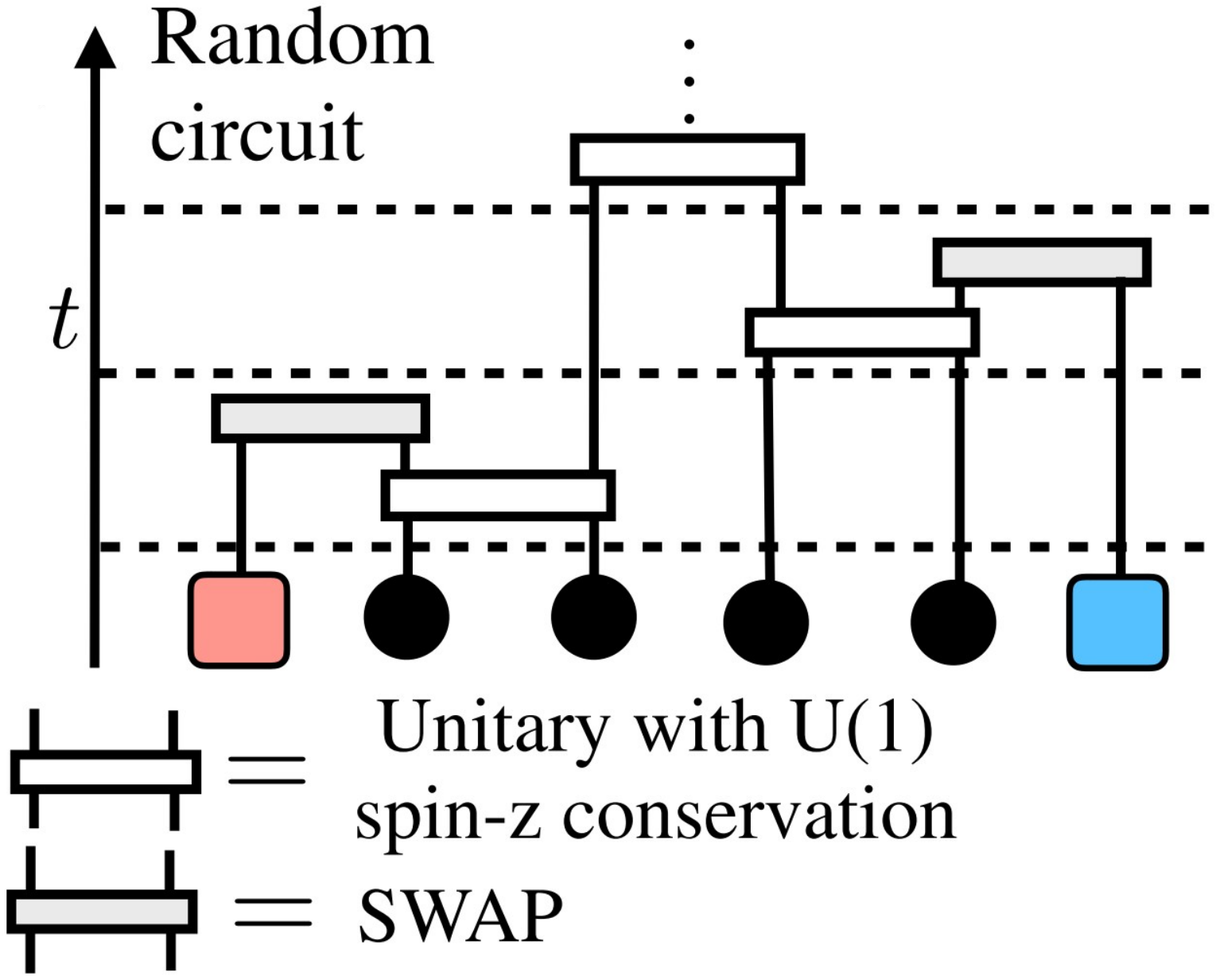}
	\end{subfigure}
	\hfill
	\begin{subfigure}{0.58\linewidth}
		\includegraphics[width=\textwidth]{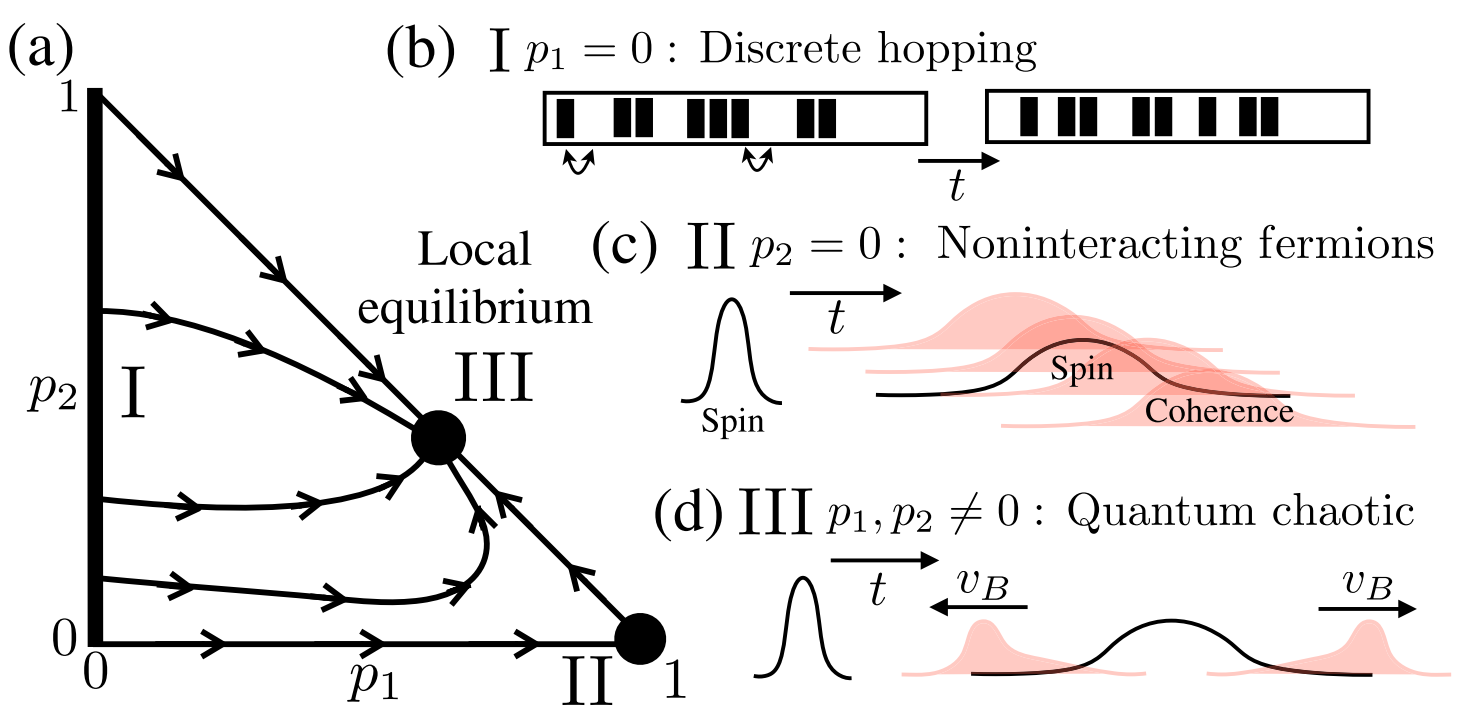}
	\end{subfigure}
	\caption{\label{fig:current-driven_circuit} (A) The random unitary circuit with random geometry interacts on its boundaries with the reservoirs via a SWAP gate. Figures reproduced from \cite{Gullans2019Entanglement}}
\end{figure}

Up to now we have considered isolated random circuits that come to an equilibrium at late times. But we can change this and couple a circuit with a locally conserved charge to two boundary reservoirs which maintain a current through the system at late times. In \cite{Gullans2019Entanglement}, this is done for a spin-$1/2$ chain of length $N$ (no neutral spins, $d=1$). Unitary U(1) conserving gates in the form of Eq.~\eqref{eq:charge_conserving_unitary} act on a randomly chosen pair of neighbouring sites in each time step, see Figure \ref{fig:current-driven_circuit} (left). Whenever such a pair is a boundary and a reservoir site, the unitary is a deterministic SWAP gate that exchanges the spin on the boundary site with a ``fresh'' spin on the reservoir site. The reservoirs (L: left, R: right) are prepared in a state with fixed magnetization in the range $m_{L/R}=\pm \delta/2\in[-\frac 1 2,\frac 1 2]$,
\begin{equation}
	\rho_{m_{L/R}}=\frac 1 2 (\mathds{1}\pm \delta \sigma^z).
\end{equation}
Note that due to the interaction with the reservoirs the system is no longer pure. It is described by a density matrix $\rho_t$ that also is a random variable, since it depends on the circuit realization. At late times, the authors show that the measure $\E$ of $\rho_t$ converges to an ensemble of (what they call) nonequilibrium attracting states. Here, the mean density matrix $\bar \rho=\E[\rho_\infty]$ is diagonal and equal to the steady-state probability measure of the SSEP, which we discussed in Section \ref{sec:SSEP}. Most of the work is then devoted to the mean entanglement properties of this ensemble, for the following refined classes of bulk unitaries: With probability $p_1$ they are ``non-interacting fermion'' gates that can perform ``partial swaps''; with probability $p_2$ they are interacting gates; and with remaining probability $1-p_1-p_2$ they are gates that are neither interactions nor partial swaps. This leads to a phase diagram for the late time entanglement structure shown in Figure \ref{fig:current-driven_circuit} (right). Note that the different classes of unitary gates are such that the mean density matrix $\bar\rho$ is the same in all of these phases.

Phase III ($p_1\neq0, p_2\neq1$) reflects the generic behaviour of random circuits, the ``quantum chaotic'' regime. For small $\delta$, a perturbative expansion shows that the average (von Neumann) mutual information between two halves of the system is suppressed by $1/N^2$ at leading order in $\delta$,
\begin{equation}
	\E[I(A:\bar A)]=\Ord(\delta^2/N^2)+\Ord(\delta^4).
\end{equation}
This is due to the fact that coherences spread ballistically and rapidly disappear into the reservoirs, so that no extensive entanglement can build up in the system. In particular, transport in this phase is not in the \textit{mesoscopic regime} in the sense of Section \ref{sec:mesoscopic_diffusive_conductors}

Phase II ($p_2=0$) is the regime of ``noisy noninteracting fermions'' and shows a mutual information that is extensive in the system size. Again for small $\delta$ the authors find
\begin{equation}
	\E[I(A:\bar A)]=\delta^2 z^2(1-z^2)N+\Ord(N^0+\delta^4),
\end{equation}
where $z\in(0,1)$ parametrizes the position of the cut between the halves $A$ and $\bar A$. This phase is (by definition) in the mesoscopic regime and the toy model QSSEP which we will introduce in the next chapter allows us to study this regime analytically in greater detail -- in particular characterizing the complete (large deviation) probability distribution in a non-perturbative way. Interestingly, the mutual information in a current-driven 3D Anderson model, with similar size in every direction $N_x\approx N_y\approx N_z$, is also extensive (proportional to the volume) \cite{Gullans2019Entanglement}, even though the Anderson model has two conserved quantities: In addition to spin, there is also energy conservation. This is one of the reasons why it would be interesting to compare properties of QSSEP to the 3D Anderson model.

Finally, phase I ($p_1=0$) corresponds to an ensemble of product states that can be mapped to the dynamics of SSEP and has zero mutual information.

\subsection*{Monitored circuits}
A realistic quantum computer undergoes not only unitary evolution, but also non-unitary operations resulting from measurements and
noisy couplings to the environment, that tend to irreversibly destroy quantum information by revealing it. This is the motivation to include an additional element into the study of isolated random quantum circuits: Measurements.

The circuit $K_m$ now consists of unitaries interleaved with single-site projective measurements that occur with probability $p$ per time step and produce outcomes $m=(m_1,m_2,\cdots)$,
\begin{equation}
	|\psi_m(t)\rangle=K_m|\psi(0)\rangle /\sqrt{p_m} \qquad p_m=\langle \psi(0)|K_m^\dagger K_m|\psi(0)\rangle.
\end{equation}
The monitored circuit dynamics is both nonlinear (due to $p_m$) and nonunitary. Formally, each realization with outcome $m$ can bee seen as a quantum trajectory $\rho_m=|\psi_m\rangle\langle\psi_m|$ resulting from the unravelling of an open system dynamics. 

Importantly, the measurements induce an entanglement phase transition \cite{Skinner2019measurement,Li2019Measurement}. In a weak monitored regime $p<p_c$ the entanglement entropy still satisfies a volume law as before. One can therefore identity decoherence-free subspaces, in which the dynamics is effectively unitary, and use them as the code space for quantum error correction. In this sense, the volume law phase is the one where a quantum computer can operate. Strong monitoring with $p>p_c$ will destroy this phase and leads to an area law entanglement. The critical value $p_c$ is not universal and depends on the local spin dimension $d$ and on the circuit geometry. For a square lattice (brick wall geometry) and $d\to\infty$ it is $p_c=1/2$.

\begin{figure}
	\centering
	\begin{subfigure}{0.48\linewidth}
		\includegraphics[width=\textwidth]{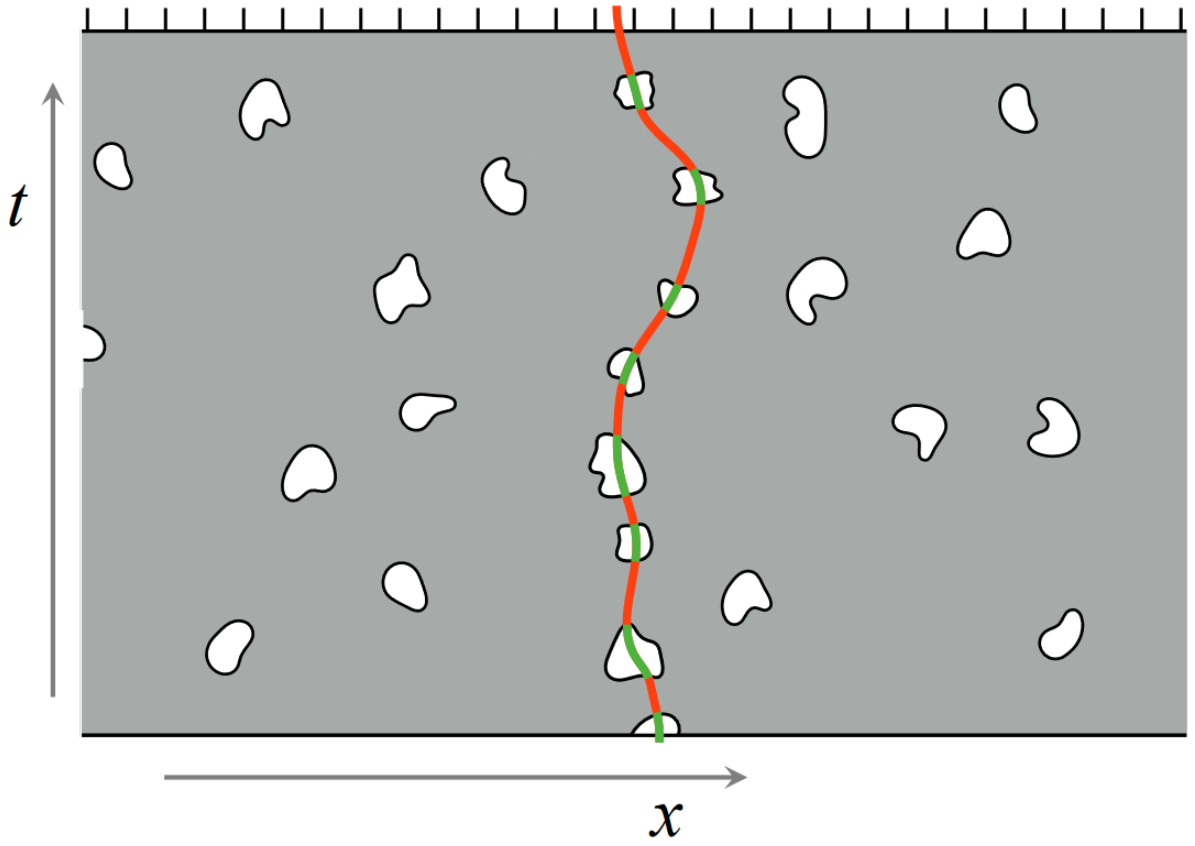}
	\end{subfigure}
	\hfill
	\begin{subfigure}{0.49\linewidth}
		\includegraphics[width=\textwidth]{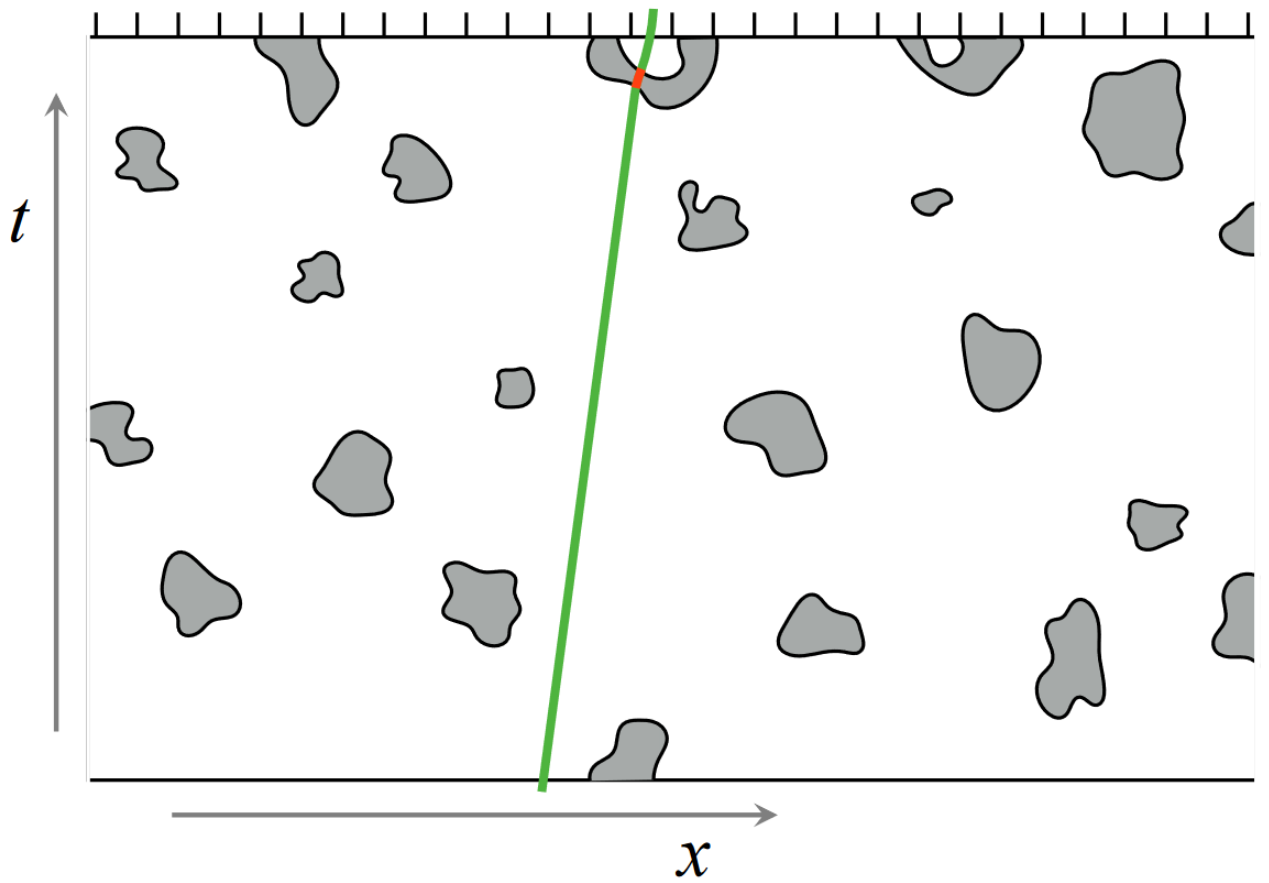}
	\end{subfigure}
	\caption{\label{fig:islands} (Left) A low measurement rate $p<p_c$ leads to white islands that are completely disconnected from the rest of the circuits. (Right) These islands become very large if the measuring rate is high $p>p_c$. The line of the minimal cut is coloured in green on segments where it is free of cost and in red where it has to break bonds. Figures reproduced from \cite{Skinner2019measurement}.}
\end{figure}

The phase transition can again be understood intuitively via the minimal cut, which provides an estimate for the 0th Renyi entropy and an upper bound for all higher Renyi entropies, see Figure \ref{fig:circuit} (b). The difference to before is that bonds on which a measurement has been applied do not need to be cut any more. For a low measuring rate $p<p_c$, this produces small islands which are completely disconnected from the rest of the circuit and along which the minimal cut passes without any cost of cutting a bond, see Figure \ref{fig:islands} (left). But the total cost is nevertheless proportional to $t$, and $S_A^{(0)}$ saturates to a value proportional to $|A|$. However, if the measuring rate is high $p>p_c$, see Figure \ref{fig:islands} (right), the islands (still in white) become very large and the minimal cut is almost free of cost. Only a fixed number of bonds near the starting point of the cut must be broken and $S_A\sim t^0$ becomes independent of $t$.

Note that experimental verifications of the predictions of entanglement in monitored quantum circuits are rather hard to implement due to the \textit{post selection} barrier: To measure the entanglement entropy of a typical state $\rho_m$ via quantum state tomography, one has to repeated the experiment exponentially many times, before a trajectory $\rho_m$ with the same measurement outcome $m$ is reproduced. 

\section{Mesoscopic diffusive conductors}\label{sec:mesoscopic_diffusive_conductors}
While the preceding sections have discussed many-body quantum systems in non-equilibrium states from a rather abstract and theoretical perspective, the study of mesoscopic conductors is directly related to the very practical question about the behaviour of very small electric conductors as they appeared early on in semiconductor electronics with the miniaturization of transistors. Theoretically, \textit{mesoscopic} conductors are distinguished by the fact that the electron coherence length $L_\phi$ is comparable or greater than the lenght of the conductor, $L\leq L_\phi$. This allows effects of quantum mechanical origin to become important. Here we focus in particular on diffusive conductors in which frequent elastic scattering events due to disorder cause the mean free path $\ell\ll L$ to be much shorter than the conductor. Conversely, a perfect conductor with $\ell>L$ would show ballistic transport. Note that for strong disorder, conductors tend to become localized (Anderson localization). In this section we will only focus on the \textit{diffusive regime} where the localization length $\ell\ll L \ll \lambda$ is much greater than the size of the conductor. From the microscopic perspective, diffusive conductors are well described by the 3D Anderson model in Eq.~\eqref{eq:Anderson} if the disorder strength is below the critical value $W_c\approx16.5$.

\paragraph{Thermal noise and shot noise.} 
Historically, a lot of attention has been put on characterizing current noise, because experimentally the current is a very accessible quantity. An excellent review on this subject has been provided by Blanter and Büttiker \cite{Buttiker2000Shot}.

A current of charge carries flowing through a conductor might fluctuation due to two reasons. Firstly, at finite temperature, there is \textit{thermal noise}. The mean occupation number of a state is given by the Fermi-Dirac or the Bose-Einstein distribution $\<n\>=f$. And since $n^2=n$ for fermions, the mean squared fluctuations of electrons is given by $\<(\Delta n)^2\>=f(1-f)\>$ where $\Delta n = n-\<n\>$. 

Secondly, even at zero temperature, there is \textit{shot noise}, which is a consequence of the quantization of charge and it depends on the exchange statistics of the charge carriers, i.e.\ if they are fermions or bosons. Contrary to thermal noise which also leads to fluctuations in equilibrium, shot noise can only be observed if the system is in a current-carrying non-equilibrium state. Shot noise can be understood as a special form of \textit{partition noise} which is caused by a half-transparent mirror (or beamsplitter) on which particles are reflected and transmitted with probability $R$ and $T$. For an incoming beam with thermal occupation $\<n\>=f$ the fluctuations of the transmitted and reflected beam can be shown to be $\<(\Delta n_R)^2\>=Rf(1-Rf)$ and $\<(\Delta n_T)^2\>=Tf(1-Tf)$. In particular, these fluctuations do not vanish at zero temperature. In a conductor, this situation occurs when electrons scatter elastically on impurities. To understand the effect of the exchange statistics one can now consider two identical particles, incident on the same beamsplitter from different sides. Treating this problem by quantum mechanics, one can show that the probability for both particles to end up on different sides of the beamsplitter is $T^2+R^2\pm 2TR |J|^2$ where $J$ is the overlap of the wave functions of the two particles and $+$ is for bosons and $-$ for fermions. As soon as the particles ``see'' each other, i.e.\ they are likely to arrive at the beamsplitter at the same time, there is an enhancement or a suppression of the classical probability $T^2+R^2$ which depends on the exchange statistics of the particles. In a mesoscopic conductor the wave functions of all charge carries overlap, and therefore, in a very idealized way, one can imagine shot noise to arise from a collective scattering on a single beamsplitter.

In a conductor which is $m=L/L_\phi$ times longer than its coherence length, this intuition would need to be replaced by the independent scattering on $m$ beamsplitters. As a result, the current is averaged over all $m$ segments and the shot noise power gets suppressed by $1/m$. In a macroscopic conductor shot noise is absent.

\paragraph{Classical vs.\ quantum coherent effects.}
We should stress, that even though shot noise has its origin in the quantized nature of charge carries, it can be described classically by Boltzmann-Langevin methods, if one adds the exchange statistic of particles by hand. In fact, up to small corrections (weak localization) this is the case for all ensemble- or disorder-averaged quantities. To distinguish the ensemble- or disorder-average from the quantum expectation value $\<\bullet\>$, we denote it by $\E[\bullet]$. Blanter and Büttiker \cite{Buttiker2000Shot} write: ``The picture which emerges is, therefore, that like the conductance, the ensemble-averaged shot noise is a classical quantity. Quantum effects in the shot noise manifest themselves only if we include weak localization effects or if we ask about fluctuations away from the [ensemble]-average''. Indeed, as we shall see below (in the paragraph on charge fluctuations), shot noise in diffusive mesoscopic conductors is actually described by the classical SSEP. 

Still, the scatting approach of Landauer and Büttiker introduced below allows to treat shot noise in a completely coherent way and therefore allows to address genuine quantum coherent effects such as the fluctuations between different samples. Such fluctuations can also be studied in the same sample, if one varies external parameters, such as the magnetic field, the chemical potential or the Fermi energy. All of this has the same effect on the fluctuation of the conductance, which is referred to as the \textit{ergodic hypothesis}. To see quantum coherent effects, for example in the variations between samples, one needs to consider quantities like $\E[\<\bullet\> \<\bullet\>]= \E[\Tr(\bullet \rho)\Tr(\bullet\rho)]$. They are quadratic in the density matrix $\rho$ and can therefore probe the fluctuation of off-diagonal elements of the density matrix. These off-diagonal elements are sometimes called \textit{coherences} and we try to characterize them with our toy model QSSEP in the next chapters.

\begin{figure}
	\begin{center}
		\includegraphics[width=0.5\textwidth]{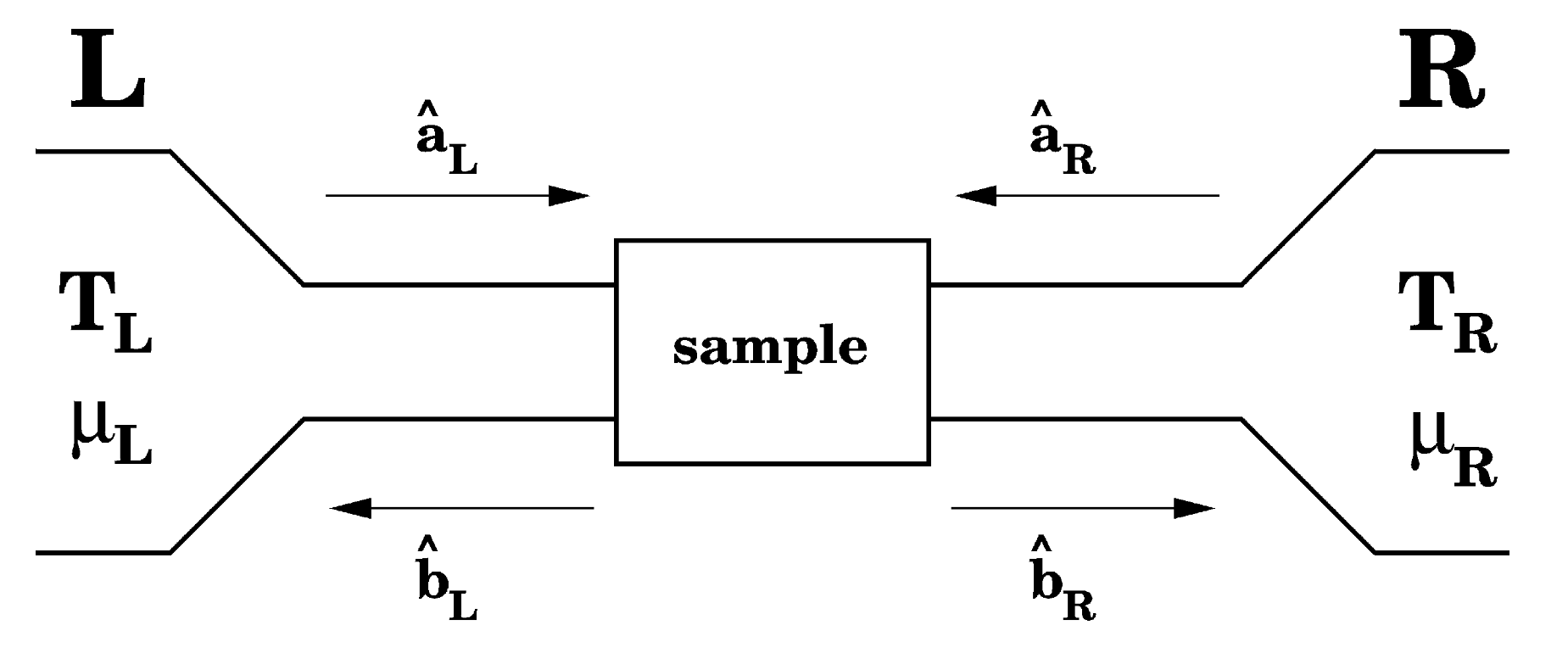}
		\caption{\label{fig:landau_buttiker} In- and outgoing scattering states for a sample with a single transverse mode. Reproduced from \cite{Buttiker2000Shot}}
	\end{center}
\end{figure}

\paragraph{Scattering approach (Landauer-Büttiker formalism).} The scattering approach tries to relate transport properties of a mesoscopic conductor to its quantum mechanical scattering properties. It applies to non-interacting systems in a non-equilibrium state. Following \cite{Buttiker2000Shot}, one considers a sample in contact with two reservoir or leads ($\alpha=L,R$) which are themselves assumed to be perfect ballistic conductors, but infinitely large, such that their temperature and chemical potential is not changed by absorbing or emitting electrons, see Fig.~\ref{fig:landau_buttiker}. Energy levels in the leads are occupied according to the Fermi-Dirac distribution
\begin{equation}
	f_\alpha(E)=\frac{1}{e^{(E-\mu_\alpha)/kT_\alpha}+1}, \qquad \alpha=L,R.
\end{equation}
Even though the dynamics of the scattering problem is determined by a Hamiltonian and a unitary scattering matrix, one should note that we are dealing with an irreversible open-system dynamics since the reservoirs are a perfect source and sink of uncorrelated Fermi-distributed electrons. 

The scattering approach relates so-called \textit{scattering states} on the left and the right side of the sample, see Fig.~\ref{fig:landau_buttiker}. The operators $a_\alpha^\dagger(E)$ and $a_\alpha(E)$ create and annihilate incoming electrons with energy $E$ on side $\alpha=L,R$, while $b_\alpha^\dagger(E)$ and $b_\alpha(E)$ create and annihilate outgoing electrons. The number operator in the leads is $n_\alpha(E)=a_\alpha^\dagger(E) a_\alpha(E)$ and its expectation value in the bath is just the Fermi-Dirac distribution
\begin{equation}\label{eq:scattering_expectation}
	\<a_\alpha^\dagger(E) a_\beta(E')\>=\delta_{\alpha,\beta}\delta(E-E')f_\alpha(E),  \qquad \alpha=L,R.
\end{equation}
For example, an incoming scattering state on the left $a_L(E)^\dagger|0\>$ can be understood as an eigenstate of sample plus leads with energy $E$ that is composed of: An incoming ($\rightarrow$) wave $\psi^\text{in}_L\sim e^{i k z}$ and a reflected outgoing ($\leftarrow$) wave $\psi^\text{out}_L\sim r_L\, e^{-i k z}$ in the left lead, a transmitted outgoing ($\rightarrow$) wave $\psi^\text{out}_R \sim t_R \,e^{i kz}$ in the right lead, and a more complex wave function inside the sample. Note that $r_L, t_R$ and the momentum $k$ depend on $E$. A common choice is to normalize the incoming plane waves to unit current.

A 3D conductor also has transverse directions $x,y$ in which the momenta $k_x,k_y$ are quantized (assuming periodic boundary conditions). Denoting transverse modes by $n$ and their energy in a left or right scattering state by $E_{\alpha,n}$, the momenta $k_{\alpha,n}$ in $z$-direction (the direction of propagation) are constraint by the total energy $E$ as $E=\hbar^2 k_{\alpha,n}^2/2m + E_{\alpha,n}$. This means that there is only a finite number of so-called ``transport channels'' $N_{\text{ch},\alpha}$ in each lead, one for each $k_{\alpha,n}$. We should now add the transverse mode index $n$ to the operators, for example $a_{L,n}$, and we denote $\vec a_{L}=(a_{L,1},\cdots,a_{L,N_L})$. The outgoing scattering states are then related to the incoming ones in terms of reflection and transmission matrices $r_L,r_R$ and $t_L,t_R$,
\begin{equation}
	\begin{pmatrix}
		\vec b_{L} \\ \vec b_{R}
	\end{pmatrix}
	= s
		\begin{pmatrix}
		\vec a_{L} \\ \vec a_{R}
	\end{pmatrix}
	\qquad
	\text{with }
	s=	\begin{pmatrix}
		r_L&t_R\\ t_L&r_R
	\end{pmatrix}.
\end{equation}
The scattering matrix $s=s(E)$ is assumed to be known from the Hamiltonian of the sample. Due to flux conservation, $s$ is unitary. If furthermore the Hamiltonian is real, which we will assume now, it is also symmetric, so $t_L=t_R=:t$.

To find formulas for the current and for its fluctuations, we can write the current density operator in lead $\alpha=L,R$ (far away from the sample) as
\begin{equation}
	J_\alpha(r,t) = \frac{\hbar e}{2 m i}[\psi_\alpha(r,t)^\dagger \nabla \psi_\alpha(r,t) - \nabla \psi_\alpha(r,t) \psi_\alpha^\dagger(r,t)],
\end{equation}
where the field operators create and annihilate electrons at $r=(x,y,z)$ in the leads
\begin{equation}
	\psi_\alpha(r,t)=\sum_{n=1}^{N_{\text{ch},\alpha}} \int d E \, \frac{e^{-iEt/\hbar}\,\chi_{\alpha,n}(x,y)}{\sqrt{2\pi \hbar^2 k_{\alpha,n}/m}}\left(a_{\alpha,n}(E)e^{ikx}+b_{\alpha,n}(E)e^{-ikx}\right).
\end{equation}
Here $\hbar^2 k_{\alpha,n}^2/2m=E-E_{\alpha,n}$ and $\chi_{\alpha,n}$ is the transverse component of the wave function. After some calculation, using Eq.~\eqref{eq:scattering_expectation} and integrating $I(z,t)=\int dx dy \,J(r,t)$, one finds the average current to be
\begin{equation}
	\<I\>=\frac{e}{h}\int dE\, \Tr[t(E)^\dagger t(E)](f_L(E)-f_R(E)).
\end{equation}
Since we are in the stationary regime, the average current does not depend on where it is measured. Denoting $0\leq T_n(E)\leq 1$ the eigenvalues of the matrix $t^\dagger(E)t(E)$, this gives the Landauer formula for the conductance $I=GV$ in the zero temperature limit when evaluated at Fermi energy $E_F$,
\begin{equation}\label{eq:Landauer_conductance}
	G=\frac{e^2}{h}\sum_n^{N_\text{ch}} T_n(E_F),
\end{equation}
where $N_\text{ch}=\min(N_{\text{ch},L},N_{\text{ch},R})$. Furthermore, one can show that the quadratic current fluctuation (its disorder average is referred to as the \textit{noise power}) is
\begin{equation}\label{eq:I_fluctuation}
	\<(I-\<I\>)^2\>=\frac{2 e^2}{h}\sum_n \int \left[T_n\left(f_L(1\mp f_L)+f_R(1\mp f_R)\right)\pm T_n(1-T_n)(f_L-f_R)^2\right].
\end{equation}
Here ($-$) is for charge carries with Fermi-Dirac statistics and ($+$) for Bose-Einstein statistics, and $f_\alpha$ has to be adopted accordingly. The first two terms come from equilibrium or thermal noise, and only the third term which is non-linear in $T_n(E)$ is the non-equilibrium or shot noise contribution.

\paragraph{Mean conductance and universal fluctuations.}
The conductance in a disordered conductor is usually not a self-averaging quantity. Following \cite{Akkermans_Montambaux_2007}, the mean conductance should correspond to the classical Drude conductivity $\sigma_0$ given by
\begin{align}
	\E[G]&=\sigma_0 \frac{S}{L}, & \sigma_0&=e^2\rho(E_F)D,
\end{align}
where $S$ is the surface perpendicular to the current in $d$ dimensions and $L$ is the length of the conductor\footnote{The conductance $G=\sigma S/L$ is specific to the size of the sample, the conductivity is independent of it.}. With the density of states $\rho(E)=n(E)d/(2E)$ of free electrons and the diffusion constant $D=v_F^2 \tau/d$ one finds the usual form of the Drude conductivity which we already encountered in Eq.~\eqref{eq:drude_conductivity}.
Alternatively, substituting the density of states $\rho(E_F)=k_F^2/(2\pi^2 \hbar v_F)$ for a quadratic dispersion relation ($E=k^2/2m$) in $d=3$ dimensions, one finds
\begin{equation}
	\E[G]=\frac{e^2}{h} \frac{k_F^2 S}{3\pi} \frac{\ell}{L}\approx \frac{e^2}{h}g, \qquad g=N_\text{ch} \,\frac{\ell}{L},
\end{equation}
where we identified the mean free path with $\ell=v_F \tau$. Since $N_\text{ch}=\frac{k_F^2 S}{4\pi}$ represents the number of transverse channels in the conductor, comparing to Eq.~\eqref{eq:Landauer_conductance}, one finds that the mean transmission probability is approximately $\E[T]=\ell/L$. However, in a metallic diffusive wire, the distribution of $T$ turns out to be bimodal in the sense that open channels( $T\sim 1$) and closed channel ($T\ll 1$) coexist. This is a consequence of the Fermi statistics and leads to the famous $1/3$ suppression of the noise power compared to the value one would obtain from independent Poisson events, see \cite{Buttiker2000Shot}.

Remarkably, the quadratic fluctuation of the conductance
\begin{equation}
	\E[G^2]^c\sim (e^2/h)^2/\beta,
\end{equation}
is a universal number that is independent of the disorder strength (therefore independent of the mean free path $\ell$), and only depends on the geometry of the sample, as well as the symmetry of the Hamiltonian $\beta$ \cite{Lee1987Universal} (also see Chpt.\ 11 in \cite{Akkermans_Montambaux_2007}). In this sense, mesoscopic diffusive conductors show \textit{universal conductance fluctuations}. From a classical perspective one would imagine that each segment of the order of $\ell$ is an independent subsystem and the conductance would be just the average of $N=(L/\ell)^d$ independent contributions, whose relative fluctuations would vanish as $1/\sqrt{N}$. This shows that one has to take into account quantum coherent effects to explain the universal conductance fluctuations.

\begin{figure}
	\centering
	\begin{subfigure}[b]{0.45\textwidth}
		\centering
		\includegraphics[width=\textwidth]{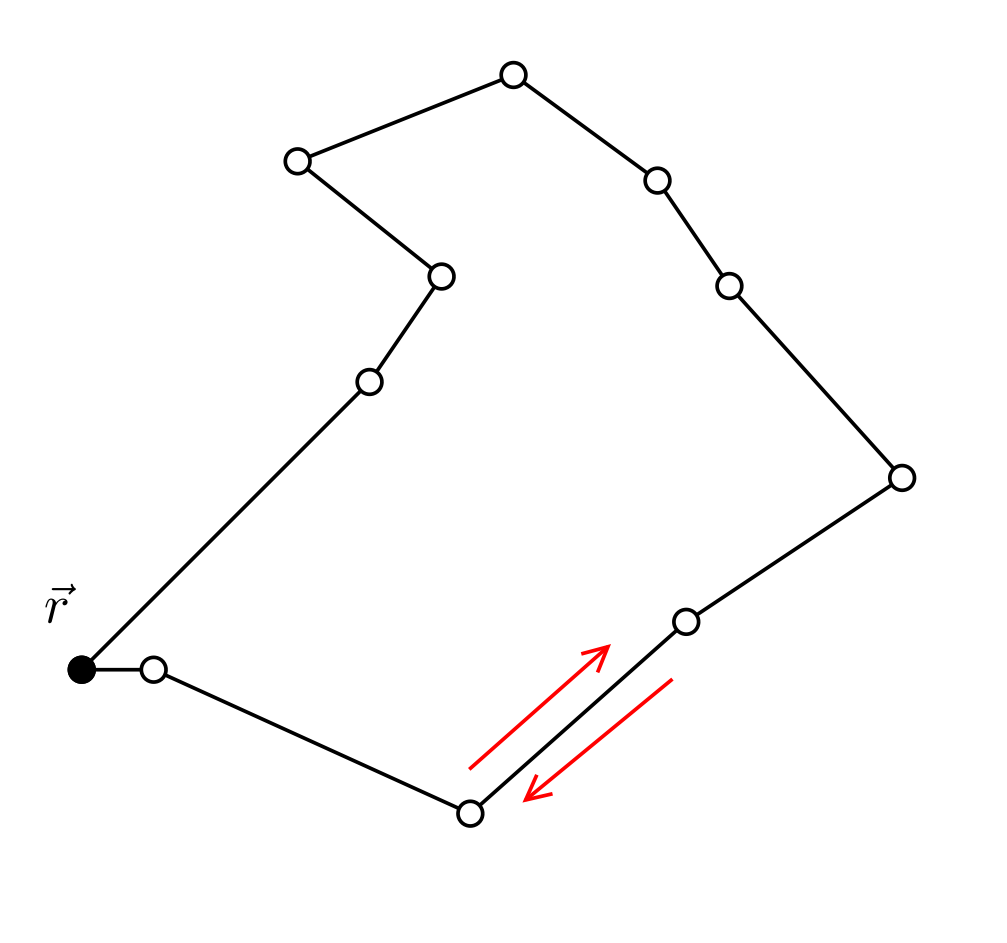}
		\caption{}
	\end{subfigure}
	\hfill
	\begin{subfigure}[b]{0.47\textwidth}
		\centering
		\includegraphics[width=\textwidth]{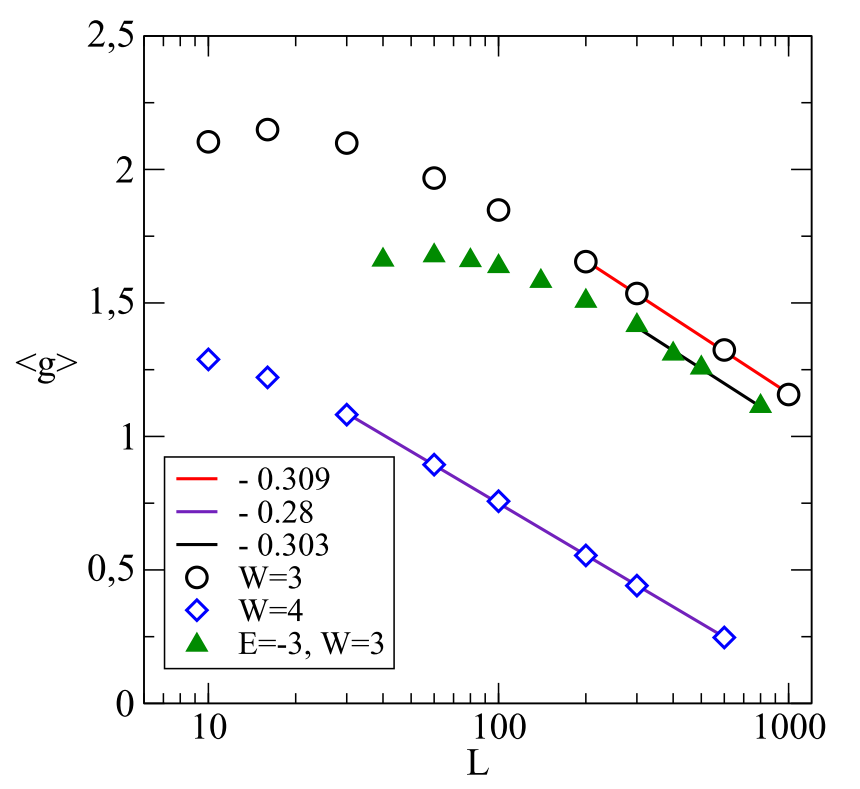}
		\caption{}
	\end{subfigure}
	\caption{\label{fig:weak_localization} Reproduced from \cite{Markos2006Numerical}. (a) Coherent backscattering of an electron in a disordered conductor. (b) The average conductance (in unit of $e^2/h^2$) as a function of $\log(L)$ in the 2D Anderson model with different disorder strength $W$. As predicted (solid lines) by the effect of weak localization, the numerical data points show a linear decrease of the mean conductance with $\log(L)$, with slope $\pi^{-1}\approx 0.318$.}
\end{figure}

\paragraph{Weak localization.}
Experimentally it is observed that a diffusive conductor shows a negative correction $\delta g$ to its average conductance $g$, when the temperature is lowered which increases the coherence length $L_\phi$ (due to the absence of the inelastic electron-phonon scattering). This so-called \textit{weak localization} correction comes from the coherent back scattering of electrons on random impurities in the sample, see Fig.\ \ref{fig:weak_localization}. If the system is time reversal invariant, forward and backward trajectories can be added up coherently and lead to constructive interference. This enhances the probability for electrons to stay localized and slightly lowers the conductance of the sample. 

The effect of coherent back scattering on the conductance can be schematically expressed as the probability that the electron takes any of the available trajectories $n$ with amplitudes $A_n$ that returns to the same point, $\delta g_\text{q}\sim\E[|\sum_n A_n|^2]$. However, the classical Drude conductance already includes the classical contribution $\delta g_\text{cl}\sim\E[\sum_n |A_n|^2]$ for backscattering. The weak localization correction is therefore $\delta g=\delta g_\text{q}-\delta g_\text{cl}$ and one has (in units of $e^2/h$) \cite{Akkermans_Montambaux_2007}
\begin{equation}
	\delta g=
	\begin{cases}
		-1/3, & \text{quasi-}1D\\
		-\frac{1}{\pi}\log(L/\ell),& d=2 \\
		-\frac{1}{2\pi}L/\ell, & d=3
	\end{cases}
\end{equation}
Since there is no dissuive regime in strictly one dimension, one considers the quasi-1D case, where a 3D conductor is much longer than wide. 

The probability to return to the same point $r$ within a time $t$ can also be written as,
\begin{equation}
	P(r,t)=\E\big[|\sum_n A_n|^2\big]=\E\big[|\<r|e^{-iHt/\hbar}|r\>|^2\big]\sim\E\big[|\<\Omega|c_r(t)^\dagger c_r(0)|\Omega\>|^2\big]
\end{equation} 
passing from first to sequent quantization with $|\Omega\>$ the ground state at Fermi energy. In terms of a density matrix, this can be expressed as
\begin{equation}\label{eq:weak_loc_quadratic_rho}
	P(r,t)\sim\E\big[\Tr(c_r^\dagger(t) c_r(0)\rho)\Tr(c_r^\dagger(0)c_r(t)\rho)\big].
\end{equation}
Again this shows that we are dealing with a quantity quadratic in the density matrix.

\paragraph{Charge fluctuations in disordered conductors.}
The statistics of charge fluctuations (also known as \textit{full counting statistic}) due to shot noise in mesoscopic disordered conductors at low temperatures has been developed by Levitov et al.\ in \cite{Levitov1993Charge,Lee1995Universal}. It provides a way to characterize shot noise beyond the noise power from Eq.~\eqref{eq:I_fluctuation}, which is just the second cumulant of charge. Similarly to the large deviation principle for charge fluctuations in classical systems in Eq.~\eqref{eq:charge_cumulants_intro}, one is interested in
\begin{equation}
	\<e^{\lambda Q_t}\>=e^{t\mu(\lambda)}
\end{equation}
where $Q_t$ is the total transported charge up to time $t$. But here samples with macroscopically equivalent parameters can differ by the realisation of disorder, so one is usually interested in the disorder-averaged generating function $\E[\mu(\lambda)]$. The approach taken by Levitov et al.\ is based on the Landauer-Büttiker formalism which describes transport though $N_\text{ch}$ independent channels with transmission probabilities $T_n(E)$. To find the disorder average, they assume that the variables $\nu_n$ defined by $T_n=1/\cosh^2(\nu_n)$ are uniformly distributed with probability density $P(\nu_n)=g$ up to some cutoff $\nu_n<\nu_c$, where $g=N_\text{ch}\ell/L$ is the average conductance (in units of $e^2/h$), $\ell$ is the mean free path and $N_\text{ch}$ the number of open channels. This leads to
\begin{equation}
	\E[\mu(\lambda)]=\frac{GV}{e}\sinh^{-1}(\sqrt{e^\lambda-1}),
\end{equation}
with average conductance $G=\frac{e^2}{h} g$ and $V$ the applied voltage. The expression is equivalent to the cumulant generating function for SSEP in Eq.~\eqref{eq:mu_SSEP} with $n_a=1$ and $n_b=0$ and the replacement of $1/N$ (for SSEP) by $\E[\<Q_t\>/t]=\frac{eV}{h}g$ (here). With this identification the first few cumulants of charge in mesoscopic disordered conductors are precisely given by Eq.~\eqref{eq:SSEP_cumulants_explicit}.

As explained in the introductory paragraphs, genuine quantum effects that cannot be described by a classical theory only appear in the fluctuations due to disorder (between samples). Based on a result in \cite{Beenakker1993Nonlogarithmic,Macedo1994Exact}, Ref.~\cite{Lee1995Universal} provides the following expressions for the variance of the (quantum) cumulants of the transferred charge
\begin{align}\label{eq:disorder_fluctuations}
	\E\Big[\big(\<Q_t\>/t\big)^2\Big]^c &= \frac{2}{15\beta}(eV/h)^2 & 
	\E\Big[\big(\<Q_t^2\>^c/t\big)^2\Big]^c &= \frac{46}{2835\beta}(eV/h)^2 \nn
	\E\Big[\big(\<Q_t^3\>^c/t\big)^2\Big]^c&= \frac{1136}{1447875\beta}(eV/h)^2, & &
\end{align}
where $\beta=1$ for real symmetric Hamiltonians and $\beta=2$ for complex hermitian ones (without time-reversal symmetry). We provide these expressions here, because it might be interesting to compute the same quantity within the toy model QSSEP\footnote{I was made aware of this fact thanks to Tony Jin} discussed in the following chapters.
\chapter{Goal of the thesis}\label{chpt:goal_of_the_thesis}
The preceding chapter has introduced several aspects of out-of-equilibrium statistical physics, both in classical and in quantum systems. A particular emphasis was laid on situations that are far away from equilibrium where linear response theory is no longer applicable and observables are described by a large deviation principle. In such a situation, we saw that the macroscopic behaviour of classical systems with diffusive transport could be completely characterized by the macroscopic fluctuation theory (MFT). Albeit the MFT equations are in general difficult to solve, they provide a universal framework to describe the complete statistics of density and current fluctuations in such systems. 

\paragraph{A quantum MFT.}
A natural question is whether the MFT can be extended to the quantum regime. In other words, how to describe the statistics of large fluctuations in a non-equilibrium quantum system? This question has first been raised in the open questions section of Ref.~\cite{Bernard2016Conformal} and it was further refined in Refs.~\cite{Bauer2017Stochastic, Bernard2021CanMFT}. The question can be split up in two questions. 

Firstly, in diffusive non-equilibrium quantum systems, one can wonder if fluctuations that arise from the quantum measurement of the local number operator $\hat n(x)$ or current operator $\hat j(x)$ are described by MFT. As we saw in Section \ref{sec:circuits_with_charge} when discussing random unitary circuits with a conserved charge, this question has recently been answered positively in \cite{McCulloch2023Counting,Wienand2023Emergence}. Though, one should keep in mind that these results are only first examples with a simple diffusion constant $D(n)=1$ and to completely answer the question, one should understand the effect of more complicated dependences of the diffusion constant $D(n)$ on the mean density $n$.

Secondly, one can ask how to incorporate quantum coherent effects into MFT. This is the question we are concerned with in this thesis. In analogy, such a theory could be called a \textit{quantum mesoscopic fluctuation theory}. As we saw when discussion mesoscopic conductors in Section \ref{sec:mesoscopic_diffusive_conductors}, such effects appeared in the presence of disorder $\E$ if one considers quantities which are quadratic (or of higher power) in the density matrix, such as for example the weak localization correction in Eq.~\eqref{eq:weak_loc_quadratic_rho}. Generally, coherent effects are encoded into the off-diagonal elements of a density matrix $\rho_t$. As building blocks of our theory we therefore focus on spatial \textit{coherences}, defined as
\begin{equation}
	G_{ij}(t):=\Tr(\rho_t \,c_i^\dagger c_j),
\end{equation}
with $c_i^\dagger$ and $c_i$ (spinless) fermionic operators on a discrete lattice. Within condensed matter physics these objects would be called equal-time Green's function. Intuitively, $|G_{ij}|^2$ characterizes the probability that each time there is a fermion at site $i$ there is none at site $j$ and vice versa. Alone from this definition, it is clear, that fluctuations of such objects cannot be described by MFT because there is no classical variable corresponding to it. Contrary to the approach within the physics of mesoscopic transport, which explains specific coherent effects of experimental importance, here we would like to build a hydrodynamic theory of coherences from which more complicated coherent effects could then be derived.

\paragraph{Origin of noise.} What is the origin of the noise $\E$ that causes coherences to fluctuate? The easiest answer is that the noise corresponds to different realizations of disorder in a sample. Then fluctuations such as $\E[|G_{ij}|^2]$ would tell us how coherences fluctuate for varying disorder, which can be achieved by changing the sample, or by varying a magnetic field or the Fermi energy in the sample. 

However, coherent effects can also show up for a single sample (i.e. for a single realization of disorder). For example, the weak localization correction can be of same order or large than the variance of the conductance \cite{Markos2006Numerical}, such that its effect should be important already for a single realization of disorder. Furthermore, one could consider the entanglement entropy $S_A=\Tr(\rho_A\log \rho_A)$, which crucially depends on coherences, and which is also a meaningful quantity for a single realization of disorder\footnote{Of course, one should keep in mind that in order to measure the entanglement entropy, one either needs to prepare the same sample multiple times in exactly the same state (state tomography), or one needs to interfere identical copies of the sample.}.

Therefore we would like to propose a second perspective on the origin of noise. Originally, we consider a weakly interacting quantum system without noise and with Hamiltonian $H_0+H_I$, such that on hydrodynamical space-time scales the interacting part $H_I$ leads to diffusion, while phase coherence is maintained throughout the system and coherent effects can show up. The interacting part $H_I$ is then replaced by an effective classical stochastic process such that, at hydrodynamical scales, all properties of the original system are reproduced by the stochastic description. In this sense,
\begin{equation}
	H_0+H_I \approx H_0' + \text{Noise}.
\end{equation}
The density matrix undergoes an almost free and noisy evolution with a possibly renormalized hamiltonian $H_0'$. While the evolution of the mean density matrix is rather simple in this description, copies of the same system $\rho_t\otimes \rho_t$ are now correlated by the noise and undergo a non-trivial evolution. This leads to a non-trivial fluctuation of coherences. For example, at second order, their fluctuations are defined as
\begin{equation}
	\E[G_{ij} G_{ji}]=\E[\Tr(\rho_t c_i^\dagger c_j) \Tr(\rho_t c_j^\dagger c_i)].
\end{equation}
Note that in a similar way one could view the Anderson model as an effective stochastic model. Instead of treating the flow of electrons through a disordered conductor in a completely quantum mechanical way, where impurities, electrons, phonons etc.\ and all interactions are contained in the quantum description, one approximates the system on appropriate scales by free electrons with a random on-site potential. Important properties, such as the diffusive transport of electrons, are retrieved only by taking the (classical) disorder average. 

We should admit a difference to the Anderson model for which the static noise (the disorder) is reproducible -- simply by reusing the same sample. In contrast, the approach we have in mind does not necessary allow to reproduce the noise since it is of dynamical nature. As a consequence, it is actually not possible to measure the fluctuations such as $\E[|G_{ij}|^2]$ directly, because this would require to preform many quantum measurements with the \text{same} noise $\omega$, to obtain the quantum average $G_{ij}(\omega)$, and then vary the noise to obtain $\E[|G_{ij}|^2]=\sum_\omega P(\omega)|G_{ij}(\omega)|^2$ where $P(\omega)$ is the probability for a given noise realization. But this shall not bother us too much, since here our goal is only to build an effective theory from which one would ideally be able to retrieve a prediction for actually measurable coherent effects in the original system.

To summarize, our aim is to develop a theory for the fluctuation of coherences in mesoscopic diffusive quantum systems. The noise can represent either static disorder in the sample, in which case coherent effects show up as fluctuations between samples. Or it can be understood as an effective stochastic description of a weakly interacting system without noise. Here the ultimate goal is to make prediction for coherent effects in a single sample. In both perspectives, the description through noise helps to find universal features shared by \textit{generic} systems, while each realization of the noise corresponds to a complicated \textit{particular} system. This is the same argument we gave in the context of random unitary circuits.

\paragraph{Possible approaches.} To build such a theory, there are in principle two ways. Firstly, one could try to directly quantize the MFT equations \eqref{eq:MFT}, replacing the density and the current by operators $\hat n$, $\hat j$ which has been explored in Ref.~\cite{Bauer2017Stochastic}. However, this leads to a number of formal and conceptual problems, such as the question how to deal with the equation for the current ($j=-D(n)n'+\sqrt{\sigma/N}\,\xi$), which in quantum mechanics becomes a constraint between operators, or the question whether one should quantize only the fluctuations $\hat n=n_\text{cl}+\delta \hat n$ on top of a classical background, which allows to use use a scalar diffusion constant $D(n_\text{cl})$, or whether the diffusion constant $D(\hat n)$ becomes itself an operator. Conceptually, however, such an approach would neither guarantee that the resulting quantum theory is actually a relevant description of diffusive quantum systems, nor would it detect the potential need to introduce other system specific coefficients besides diffusion constant $D$ and mobility $\sigma$ to describe of coherent effects. 

For these reasons, we have chosen to follow a different approach, which parallels the development of MFT from SSEP and other stochastic microscopic models. Studying an appropriate stochastic toy model, we hope to gain insight into the mathematical structure inherent to the fluctuation of coherences. In the spirit of random quantum circuits, such a toy model should only consist of a \textit{minimal structure} necessary to describe coherent effects. In random quantum circuits this minimal structure is locality, unitarity (and the conservation of charge). However, when discussing entanglement in the current-driven circuit from Ref.~\cite{Gullans2019Entanglement} in Section \ref{sec:current-driven_circuit}, we saw that generic unitary gates drive the system out of the mesoscopic regime (phase III). Coherences spread ballistically and quickly get lost into the reservoir. For such systems we do not expect the need to go beyond classical MFT.

The situation changes, when the unitary gates are non-interacting. Coherences spread slower (diffusively) and as a result phase coherence was maintained. The toy model QSSEP which we consider in this thesis to can be though of as a continuous-in-time version of this non-interacting random unitary circuit -- though historically QSSEP was formulated before and independently of the model from Ref.~\cite{Gullans2019Entanglement}. QSSEP has the advantage over the unitary circuit approach that it is analytically tractable and many quantities can be computed exactly. Having made clear our motivation, we now turn to the definition of QSSEP and the investigation of its properties. 
\chapter{Introduction to QSSEP}\label{chpt:intro_to_QSSEP}
This chapter presents the state of knowledge about the \textit{quantum symmetric simple exclusion process} (QSSEP) when I started my thesis and mainly builds on two articles by Tony Jin, Denis Bernard (and Michele Bauer) that introduced QSSEP: \cite{Bauer2019Equilibrium} in the closed case and \cite{Bernard2019Open} in the open case. A good summary of these results is also provided in \cite{Bernard2021CanMFT}. A small exception is the last section of this chapter, that tries to explain how coherences (defined below) could be measured in an experiment.
\section{Definition and basic properties}
We directly start off with the definition of QSSEP, leaving the motivation and the different approaches that led to this definition to the next section. The model describes spinless fermions on a one-dimensional chain coupled to reservoirs on its boundaries, and it consists of two parts: In the bulk, fermions can hop to neighbouring sites with noisy amplitudes -- the classical noise can be though as originating from reservoirs coupled to each link, in the sense of Section \ref{sec:stochastic_quantum_dynamics}. On the two boundaries, the chain can exchange fermions with the reservoirs, which induces a current and keeps the system out of equilibrium.

Formally, we describe the state of the chain through its density matrix $\rho_t$. In the bulk it evolves as
\begin{equation}\label{eq:rho_qssep}
	\rho_{t}\rightarrow \rho_{t+dt} = e^{-idH_{t}}\rho_{t}e^{idH_{t}},
\end{equation}
with a stochastic Hamiltonian increment representing noisy free fermions,
\begin{equation}\label{eq:dH_qssep}
	dH_{t}=\sum_{j=1}^N \left(c_{j+1}^{\dagger}c_{j}dW_{t}^{j}+c_{j}^{\dagger}c_{j+1}d\overline{W}_{t}^{j}\right).
\end{equation}
Here $N$ denotes the number of sites of the chain and the spinless fermions satisfy the usual commutation relations $\{c_j^\dagger,c_k\}=\delta_{jk}$ and $\{c_j,c_k\}=\{c_j^\dagger,c_k^\dagger\}=0$. Furthermore, $dW_t^j:=W_{t+dt}^j-W_{t}^j$ are the increments of complex Brownian motions which are independent on each link $(j,j+1)$. Their expectation value will be denoted by $\E$. The Brownian motions are written in It\={o} convention such that in any stochastic differential equation we can apply the It\={o} rules 
\begin{align}\label{eq:ito_rules_complex}
	dW_{t}^{i}d\overline{W}_{t'}^{j}&=\begin{cases} \delta^{ij}dt, &t=t'\\ 0, & t\neq t'\end{cases} \quad
	& dW_{t}^{i}d{W}_{t}^{j}=dW_t^idt=d\overline W_t^idt=dt^2&=0.
\end{align}

On the two boundaries $j\in\{1,N\}$ fermions get injected into the system with rate $\alpha_j$ and extracted from it with rate $\beta_j$. This is described by the Lindblad equation (cf.\ Eq.~\eqref{eq:lindblad_equation})
\begin{equation}\label{eq:qssep_boundary}
	\partial_t \rho_t=\Lin_\mathrm{bdry}(\rho_t) \qquad \text{with } \mathcal{L}_\mathrm{bdry}=\alpha_{1}\Lin_{1}^{+}+\beta_{1}\Lin_{1}^{-}+\alpha_{N}\Lin_{N}^{+}+\beta_{N}\Lin_{N}^{-}.
\end{equation}
Here $\mathcal{L}_{j}^{+}(\bullet)=c_{j}^{\dagger}\bullet c_{j}-\frac{1}{2}\{c_{j}c_{j}^{\dagger},\bullet\}$
and $\mathcal{L}_{j}^{-}(\bullet)=c_{j}\bullet c_{j}^{\dagger}-\frac{1}{2}\{c_{j}^{\dagger}c_{j},\bullet\}$
model particle injection and extraction, respectively\footnote{For example, the density matrix of an isolated empty site $\tau{}_{t}=|0\rangle\langle0|$ that evolves according to $\partial_{t}\tau_{t}=\alpha\mathcal{L}^{+}(\tau_{t})$ will be occupied after a time interval $dt$ with probability $\alpha dt$. That is,\[|0\rangle\langle0|\to\alpha dt|1\rangle\langle1|+(1-\alpha dt)|0\rangle\langle0|\]}. As we will see in Eq.~\eqref{eq:steady_mean_density}, the injection and extraction rates can be related to the particle densities of the right and left reservoir according to
\begin{align}\label{eq:na_nb}
	n_a&=\frac{\alpha_1}{\alpha_1+\beta_1} & n_b&=\frac{\alpha_N}{\alpha_N+\beta_N}.
\end{align}

We already encountered the Lindbladian for particle extraction $\Lin^-$  in Eq.~\eqref{eq:quantum_optics_lindblad} in the context of a quantum optics model. There, the reservoir (the electromagnetic field in a cavity) was empty (in the vacuum state), such that the injection of particles into the system (excitation of modes of the atom) was not possible. So $\Lin^+$ was absent there.

The full evolution of the system, taking into account the bulk and the boundary, can now be written as a stochastic differential equation for $d\rho_t:=\rho_{t+dt}-\rho_t$,
\begin{equation}\label{eq:drho_open_qssep}
	d\rho_{t}=-i[dH_{t},\rho_{t}]-\frac{1}{2}[dH_{t},[dH_{t},\rho_{t}]]+\mathcal{L}_\mathrm{bdry}(\rho_{t})dt.
\end{equation}
Here we expanded the exponentials in Eq.~\eqref{eq:rho_qssep} up to second order since these terms can yield an $\Ord(dt)$ contribution if we apply the It\={o} rules. Note that the density matrix $\rho_t$ has become itself a random variable.

The model presented so far is also referred to as the \textit{open QSSEP} to distinguish it from the \textit{closed QSSEP}. In the latter case, one takes away the boundary reservoirs and closes the chain periodically. As a consequence, there is no current in the stationary state, which leads to a behaviour that resembles an equilibrium situation. If not explicitly referred to it, we will always consider the open QSSEP in this thesis. 

We stress that as long as there is a current, the property that makes QSSEP unique is its bulk evolution, and not the mechanism (here a Lindblad driving) that maintains the system out of equilibrium. Though we did not explore other mechanism, e.g.\ a partition protocol, we expect that the situations will qualitatively be very similar.

\paragraph{Conserved $U(1)$ charge and current.} As a first property, we show that QSSEP has a locally conserved charge, the number operator $\hat n_j=c_j^\dagger c_j$. It generates a $U(1)$ symmetry on each site by conjugation with the unitary
\begin{equation}
	U_\theta = e^{i\sum_j \theta_j \hat n_j}.
\end{equation}
With the help of the relations $U_\theta c_j U_\theta^\dagger = e^{-i\theta_j} c_j$ and $U_\theta c_j^\dagger U_\theta^\dagger = e^{i\theta_j} c_j^\dagger$ one can verify that for any angle $\theta_j$,
\begin{align}
	U_\theta dH_t U_\theta^\dagger &\overset{d}{=} dH_t & \Lin_\mathrm{bdry}(U_\theta \rho_t U_\theta^\dagger)=\Lin_\mathrm{bdry}(\rho_t)
\end{align}
The first equality holds in distribution, since the Brownian increments get multiplied by a phase, the second holds exactly. This shows that if $\rho_t$ is a solution of Eq.~\eqref{eq:drho_open_qssep}, then so is $U_\theta \rho_t U_\theta^\dagger$.

Since the number operator undergoes a stochastic evolution, the associated conserved current $J_j$ is also of stochastic nature. From Eq.~\eqref{eq:drho_open_qssep} one can obtain an equation of motion for operators in the Heisenberg picture\footnote{The Heisenberg equation for an operator $O_t$ is obtained from duality of operators and density matrices with respect to the trace, $\Tr(\rho_t O_0)=\Tr(\rho_0 O_t)$.}. In the bulk, this leads to $d\hat n_j=d \hat J_{j-1} - d \hat J_{j}$ with conserved current
\begin{equation}
	d\hat J_j= (n_j - n_{j+1}) dt +i(c_j^\dagger c_{j+1}d\overline W_t^j-c_{j+1}^\dagger c_jdW_t^j).
\end{equation}

\paragraph{Diffusive transport.} As a second property, we show that transport in QSSEP is diffusive in mean. We denote $\bar{n}_j=\E[\Tr(\rho_t \,\hat {n}_j)]$ the mean particle density at site $j$. From Eq.~\eqref{eq:drho_open_qssep} one finds that
\begin{align}\label{eq:density_discrete}
	\partial_{t}\bar{n}_{j}(t) & =\Delta_j \bar{n}_{j}(t)+\sum_{p\in\{1,N\}}\delta_{jp}(\alpha_{p}-(\alpha_{p}+\beta_{p}) \bar{n}_{p}(t)),
\end{align}
Here $\Delta_j n_j=n_{j+1}+n_{j-1}-2n_j$ is the discrete Laplacian which gets truncated on the boundaries, i.e.\ $\Delta_1 n_1=n_2-n_1$ and $\Delta_N n_N=n_{N-1}-n_N$. The equation suggests that the mean density diffuses with diffusion constant $D(\bar{n})=1$. Furthermore, the mobility is $\sigma(\bar{n})=2\bar{n}(1-\bar{n})$. This is, because the mean dynamics of QSSEP can be mapped onto the classical SSEP, as we will see in the next paragraph.

At late times, $\bar n_j$ attains a stationary solution which interpolates linearly between the two reservoirs\footnote{The expression for the mean density in the discrete case, with $a=\frac{1}{\alpha_1+\beta_1}$ and $b=\frac{1}{\alpha_N+\beta_N}$, is
\begin{equation*}
	\bar n_j= \frac{n_a \left(N-j + b\right) + n_b\left(j-1 + a\right)}{ N-1+ b + a}.
\end{equation*}
}. For $N\to\infty$ and $x=i/N$ one finds
\begin{equation}\label{eq:steady_mean_density}
	\bar{n}(x)=(n_b-n_a)x+n_a.
\end{equation}
This shows why $n_a$ and $n_b$, related to the injection and extraction rates by Eq.~\eqref{eq:na_nb}, can be interpreted as the particle densities of the left and right reservoir.

\paragraph{Correspondence with the classical SSEP.}
If one takes the expectation value of Eq.~\eqref{eq:drho_open_qssep}, and one neglects the boundary contribution for a moment, then the only term left is the second term, $\Lin_\text{SSEP}(\rho_t):=-\frac{1}{2}[dH_{t},[dH_{t},\rho_{t}]]$. It reads,
\begin{align}\label{eq:lindblad_ssep}
	\Lin_\text{SSEP}(\rho_t)= \sum_j \left(\ell^-_{j}\, \rho_t\, \ell^+_{j}+\ell^+_{j}\, \rho_t\, \ell^-_{j} -\frac{1}{2}\{\ell^+_{j}\ell^-_{j}+\ell^-_{j}\ell^+_{j}, \rho_t\}\right).
\end{align}
The reason we call it $\Lin_\text{SSEP}$ is firstly that it is of Lindblad form. But secondly and more importantly, it corresponds to the dynamics of the classical SSEP. Indeed, writing $\rho_t$ in the number eigenbasis, its diagonal elements represent the probability $p_t(\mathcal C)$ to be in one of the $2^N$ classical configurations $\mathcal C$ with a well defined particle number on each site. The action of $\Lin_\text{SSEP}$ is then equivalent to the action of the SSEP transition matrix $M$ on the probability vector $p_t$, see Eq.~\eqref{eq:ssep_markov_process}. This is also true if we include the boundary terms. Therefore, transport properties of QSSEP in mean are completely equivalent to SSEP. Note that spinless fermions are a very appropriate way to obtain the dynamics of SSEP in a quantum system: They naturally satisfy the Pauli exclusion principle.

More generally, the generating function of density correlations in SSEP can be obtained from the mean density matrix in QSSEP $\bar \rho:=\E[\rho]$ as
\begin{equation}
	Z[a]=\langle e^{\sum_{j}a_{j}n_{j}}\rangle_\mathrm{ssep}=\mathrm{Tr}\big(\bar{\rho}\,  e^{\sum_{j}a_{j}\hat{n}_{j}}\big),\label{eq:seep-Q-SSEP}
\end{equation}
where $n_j$ is the local particle density in SSEP and $\hat n_j=c_j^\dagger c_j$ is the number operator.

\paragraph{Reduction to the one-particle sector.} 
An important feature is that all information about QSSEP is encoded into the two-point function $G_{ij}(t):=\Tr(\rho_t c_i^\dagger c_j)$ (which we will call \textit{coherences} in the following). As such, for each realization of the noise the dynamics reduces from $2^N$-dimensional Fock space to the $N$-dimensional one-particles sector, which considerably simplifies calculations. This is because the noisy dynamics of QSSEP preserves fermionic Gaussian states of the form
\begin{equation}
	\rho=\frac{1}{Z}\exp(\vec c^{\:\dagger} M \vec c)
\end{equation}
with $Z=\Tr\,[\exp(\vec c^{\:\dagger} M \vec c)]=\det(1+e^M)$, $\vec c=(c_1,\cdots,c_N)$, and $M$ a Hermitian matrix of size $N$ which now encodes all information about the state of the system. The Hamiltonian part $dH_t$ of QSSEP preserves these states, since it is quadratic in the fermion operators. One can also check that $\Lin_{bdry}$ preserves these states. Finally, as shown in Appendix \ref{app:fermionic_gaussian}, one can express the matrix $G=(G_{ij})$ through $M$ as
\begin{equation}
	G=\frac{e^M}{1+e^M}.
\end{equation}
So indeed, all information about QSSEP is contained in the the two-point function.

\section{Related models}
We put QSSEP into the context of other quantum models with a relation to SSEP and explain how it can be ``derived'' from a noisy Heisenberg spin chain in the strong noise limit.

\paragraph{A quantum version of SSEP.}
The idea to study the classical stochastic dynamics of SSEP in a quantum systems is not new. However, all previous work has only focused on deterministic evolutions in terms of a Lindbladian where the density matrix itself is not a stochastic variable. For example, adding the SSEP-Lindbladian \eqref{eq:lindblad_ssep} and the boundary driving \eqref{eq:qssep_boundary} to the evolution under a free fermion Hamiltonian, Temme, Wolf and Verstraete \cite{Temme2012Stochastic} investigated the interplay between classical stochastic and quantum coherent transport processes. This was further explored by Eisler \cite{Eisler2011Crossover} (in the closed case) who analytically characterized the crossover between ballistic and diffusive transport, as the coherent or decoherent hopping rates are modified.

The new idea in QSSEP is that the density matrix itself becomes a stochastic variable and that the SSEP-Linbladian characterises only the mean evolution. It can be thought of a stochastic unravelling of the SSEP-Lindbladian and the density matrix undergoes a quantum trajectory. To make this explicit, we rewrite Eq.~\eqref{eq:drho_open_qssep} as
\begin{equation}\label{eq:drho_qssep}
	d\rho_{t}=-i[dH_{t},\rho_{t}]+\Lin_\text{SSEP}(\rho_t)+\Lin_\text{bdry}(\rho_t).
\end{equation}
On a formal level, the origin of the noise (hidden in $dH_t$) can be explained in two ways (referring to Sec.~\ref{sec:stochastic_open_quantum_systems}): Either, within the the system-bath picture, one can imagine each link (or each site, see the paragraph on noisy spin chains below) to be coupled to a classical Markovian bath. Or, within the weak-measurement picture, one imagines the system to be continuously monitored by the interaction with spin-1/2 probes. For the physical motivation of the noise, we refer to the discussion in the preceding Chapter \ref{chpt:goal_of_the_thesis}.

\paragraph{A noisy spin chain.}
Another way to introduce QSSEP is to consider a noisy spin chain in the strong coupling limit. This is actually how it was originally derived \cite{Bauer2017Stochastic}. More precisely, we consider the Heisenberg XX spin chain
\begin{equation}\label{eq:h_xx}
	h^{xx}=\epsilon \sum_j \left(\sigma_j^x \sigma_{j+1}^x + \sigma_j^y \sigma_{j+1}^y\right)=2\epsilon \sum_j \left(\sigma_j^+ \sigma_{j+1}^- + \sigma_j^- \sigma_{j+1}^+\right)
\end{equation}
with $\sigma^\pm=\frac{1}{2}(\sigma^x \pm i\sigma^y)$, and couple each site to a Markovian bath. In Section \ref{sec:stochastic_quantum_dynamics} we saw that such a coupling introduces an effective noise on each site of the system. We will choose the noise to be a locally random $z$-rotation $\sigma_j^z dB_t^j$ where $B_t^j$ is a real normalized Brownian motion, independent on each site. In particular, this choice preserves the local $U(1)$ charge, i.e. the local spin density $n_j=\frac{1}{2}(1+\sigma_j^z)$ on each site. The full dynamics can be formulated via the stochastic Hamiltonian
\begin{equation}
	dH_t = h^{xx}dt + \sqrt{2\eta}\sum_j \sigma_j^z dB_t^j
\end{equation}
where $\eta$ parametrised the strength of the noise and the density matrix evolves according to $\rho_{t+dt}=e^{-idH_t}\rho_t e^{idH_t}$.

If the noise becomes very strong $\eta\to \infty$, one can identify an emergent slow mode dynamics in terms of a slow variable $s=t/\eta$. Without the term $h^{xx}$ in the stochastic Hamiltonian, the noise slowly causes the initial state $\rho_0$ to be projected onto the submanifold of states that are invariant under local $z$-rotations -- with small oscillations around their constant mean value, but without any interesting dynamics. The presence of $h^{xx}$, however, causes a late time dynamics to emerge on this submanifold. One can extract this dynamics by going to an interaction picture with ``free part'' $H_0=\sqrt{2\eta}\sum_j \sigma_j^z B_t^j$ (where we integrated $\int_0^t dB_{t'}^j=B_{t}^j$) and an ``interacting part'' $H_I=h^{xx}$,
\begin{align}
	\hat \rho_t&=e^{i\sqrt{2\eta}\sum_j \sigma_j^z B_t^j}\, \rho_t\, e^{-i\sqrt{2\eta}\sum_j \sigma_j^z B_t^j} \\
	\hat H_t&= e^{i\sqrt{2\eta}\sum_j \sigma_j^z B_t^j} h^{xx} e^{-i\sqrt{2\eta}\sum_j \sigma_j^z B_t^j} 
\end{align}
Variables $\hat \rho,\hat H_t$ with a hat are in the interaction picture. This transformation filters out the fast oscillation of the components of $\rho_t$ which are not invariant under local $z$-rotations, and leaves the remaining invariant components unchanged such that late time observables do not get modified.

The conjugation of $h^{xx}$ can be evaluated with the help of $[\sigma^z,\sigma^\pm]=\pm \sigma^\pm$. One finds
\begin{equation}
	\hat H_t=2\epsilon \sum_j \sigma_j^+ \sigma_{j+1}^- W_t^j(\eta) + \sigma_j^- \sigma_{j+1}^+ \overline{W}^j(\eta)
\end{equation}
where $W_t^j(\eta)=e^{i\sqrt{2\eta}(B_t^j-B_t^{j+1})}$ and the bar denotes complex conjugation. Since in distribution $B_{t}=\sqrt{\eta}B_{t/\eta}$, we can introduce the slow variable $s=t/\eta$ as 
\begin{equation}
	W_t^j(\eta) dt=e^{i\eta\sqrt{2}(B_s^j-B_s^{j+1})}\,\eta\, ds=:dW_s^j(\eta).
\end{equation}
In the limit $\eta\to\infty$ with $s=t/\eta$ fixed, it can be shown that $dW_s^j(\eta)\to dW_s^j$ converges to a normalised complex Brownian motion (see \cite{Bauer2017Stochastic} appendix A for details). The It\={o} rules are those of Eq.~\eqref{eq:ito_rules_complex}. Therefore, we can write the Hamiltonian increment $d\hat H_s:=\hat H_t dt$ in terms of the slow variable $s$ as
\begin{equation}\label{eq:dH_qssep_spin}
	d\hat H_s=2\epsilon \sum_j \sigma_j^+ \sigma_{j+1}^- dW_s^j + \sigma_j^- \sigma_{j+1}^+ d\overline{W_s}^j
\end{equation}
Once we transform this via the Jordan-Wigner transformation \eqref{eq:jordan_wigner} from spins to spinless fermions, it becomes the bulk Hamiltonian \eqref{eq:dH_qssep} of QSSEP.

\section{Fluctuation of coherences}
With the aim in mind to find a hydrodynamic theory for the fluctuations of coherences
\begin{equation}
	G_{ij}(t)=\mathrm{Tr}(\rho_{t}\,c_{i}^{\dagger}c_{j}),
\end{equation}
we should now investigate how coherences behave in QSSEP. Note that interchanging $i$ and $j$ leads to complex conjugation, $G_{ji}=\overline G_{ij}$. From the construction of the model it is not at all clear that they will survive at large times, or that they are long ranged. This was investigated in \cite{Bernard2019Open}: While zero in mean, the fluctuations of coherences are indeed long-ranged and they survive in the steady state. Furthermore, they satisfy a large deviation principle, since the leading order of the $n$-th cumulant is $N^{1-n}$. Adopting the notation $\mathbb E_t$, where the time dependence of $G_{ij}(t)$ is transferred to the measure $\E$, the authors find that
\begin{align}
	\E_\infty[G_{ij}]&=0 \nn
	\mathbb{E_{\infty}}[G_{ij}G_{ji}]^{c}&=\frac{1}{N}(\Delta n)^{2}x(1-y) \nn
	\E_\infty[G_{i_1i_2}\cdots G_{i_ni_1}]^c&=\frac{1}{N^{n-1}}(\Delta n)^n g_n(x_1,\cdots,x_n).
\end{align}
Here $\Delta n=n_{a}-n_{b}$ is the difference in the particle density and $x=i/N\in[0,1]$ are continuous coordinates in the large $N$ limit. We usually set $n_a=0$ and $n_b=1$ (without loss of generality) if not stated otherwise. A closed solution for the leading order contributions $g_n$ will be given in Section \ref{sec:steady_state_solution}. Note that already at second order we get an interesting insight: Long-ranged fluctuations are non-zero only if the system is out-of-equilibrium, i.e.\ $\Delta n\neq 0$. This has also been recognised as a generic feature of classical systems such as SSEP, compare to Eq.~\eqref{eq:SSEP_density_correlations}. Here, long-ranged correlations of the density can only exist out-of-equilibrium. In equilibrium they would vanish. The discussion shows that the emergence of coherent effects is tightly bound to the fact that we consider non-equilibrium states.

In principle, all these results can be obtained from the stochastic evolution of coherences. A longer calculation, based on Eq.~\eqref{eq:drho_open_qssep} and the proper use of the It\={o} rules, yields\footnote{Note that the $-1$ in the third line cancels partly with $-2G_{ij}$ in the first line once an index is on the boundary. This is because we don't have periodic boundary conditions.}
\begin{align} \label{eq:dG_ij}
	dG_{ij} =
	&\delta_{ij}(G_{i+1,i+1}+G_{i-1,i-1})dt -2G_{ij}dt \nonumber
	\\
	&-i(G_{i,j-1}dW_{t}^{j-1}+G_{i,j+1}d\overline{W}_{t}^{j}-G_{i-1,j}d\overline{W}_{t}^{i-1}-G_{i+1,j}dW^{i}) 
	\\  &+\sum_{p\in\{1,N\}}\left(\delta_{pi}\delta_{pj}\alpha_{p}-\frac{1}{2}(\delta_{ip}+\delta_{jp})(\alpha_{p}+\beta_{p}-1)G_{ij} \right)dt.\nonumber
\end{align} 
In a more compact form, the stochastic evolution of the $N\times N$ matrix $G$ is\footnote{In some of the older papers \cite{Bauer2019Equilibrium,Bernard2019Open} the coherences have been defined as $G_{ij}=\Tr(\rho_t c_j^\dagger c_i)$. The reason for this choice is that then the evolution of $G$ reads $G(t+dt)=e^{-idh_t}G(t)e^{idh_t}$ which resembles the evolution of the density matrix, $\rho_{t+dt}=e^{-idH_t}\rho_t e^{idH_t}$.} This is the reason why in some of the older papers
\begin{equation}\label{eq:G_evolution}
	G(t+dt)=e^{i\,dh_t}G(t)e^{-i\,dh_t}+\mathcal L(G)dt
\end{equation}
with 
\begin{align*}
	dh_t=
	\begin{pmatrix}
		0 				&dW_t^1 	& 					&0 			\\
		d\overline W_t^1&\ddots 	&\ddots 			& 			\\
						&\ddots		&\ddots				& dW_t^{N-1}\\
		0				&			&d\overline W_t^{N-1}& 0		\\
	\end{pmatrix}
\end{align*}
and
\begin{equation*}
	\mathcal{L}(G)_{ij}=\sum_{p\in{1,N}}(\delta_{pi}\delta_{pj}\alpha_p - \frac 1 2 (\delta_{ip}+\delta_{jp})(\alpha_p+\beta_p)G_{ij}).
\end{equation*}
Note that this form is very suitable for numeric simulations of QSSEP. One only needs to choose a finite time step $dt$ and approximates $dW_t^i$ as independent complex Gaussian variables with variance $dt$.

\paragraph{$U(1)$ invariant measure.}
The local charge conservation in QSSEP, leads to an $U(1)$ invariant expectation value of coherences. This means that 
\begin{align}
	&\E[G_{i_1j_1}G_{i_2j_2}\cdots G_{i_nj_n}]\neq 0& &\Leftrightarrow & \{i_1,\cdots,i_n&\} \text{ is a permutation}\nn &&&&\text{of }&\{j_1,\cdots,j_n\}.
\end{align}
In other words, only the expectation value of loops and of products of loops is non-zero.

Another way to see this is to multiply the coherences with local phases 
\begin{equation}\label{eq:QSSEP_U1}
	\tilde{G}_{ij}=e^{-i\theta_{i}}G_{ij}e^{i\theta_{j}}.
\end{equation}
Then one can check, that the transformed coherences $\tilde G$ still satisfy Eq.~\eqref{eq:dG_ij} if the Brownian motions get multiplied by a phase, $d\tilde{W}_{t}^{j}=e^{i(\theta_{j+1}-\theta_{j})}dW_{t}^{j}$. But $d\tilde{W}_{t}^{j}$ and $dW_{t}^{j}$ have the same distributions and therefore $\tilde G_{ij}$ and $G_{ij}$ have the same distribution, if initially their off-diagonal elements where all zero. If not, Eq.~\eqref{eq:dG_ij} will cause any contribution that is not $U(1)$ invariant to vanish exponentially with time.

\begin{figure}
	\centering
	\includegraphics[width=.4\textwidth]{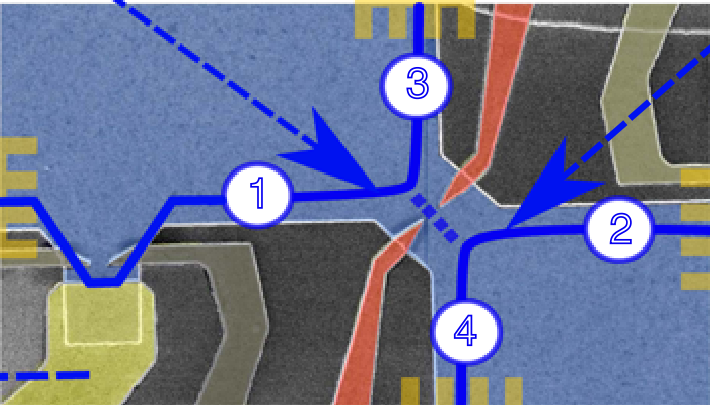}
	\caption{\label{fig:interference_feve} Reproduced from \cite{Feve2019Quantum}. A sketch of the experimental setup. The two-dimensional electron-gas is represented in blue and the edge channels as blue lines. They can interfere via a quantum point contact (in red). Channels 1 and 2 function as inputs for time dependent voltages and the coherences are measured through the current noise in output 3.}
\end{figure} 

\section{Measuring coherences}
We have already insinuated in Chapter \ref{chpt:goal_of_the_thesis} that a direct measurement of the fluctuation of coherences is only possible if the noise is reproducible. For example, the noise could represent static disorder in a given sample. In this case, the protocol is: (i) Obtain the quantum average $G_{ij}(\omega)$ by repeated measurements with the same noise realization $\omega$. (ii) Repeat step (i) for different noise realizations $\omega$. (iii) Calculate the fluctuation, for example $\E[G_{ij}G_{ji}]=\sum_\omega P(\omega)|G_{ij}(\omega)|^2$ where $P(\omega)$ is the probability for a given noise realization.

In contrast to this, if the noise is understood as an effective description of the fast degrees of freedom of an interacting Hamiltonian, then the same realization of the noise cannot be reproduced in an experiment, and the fluctuation of coherences cannot be measured directly. The construction presented below has therefore to be understood in the first sense, i.e.\ noise represents a static disorder.

\begin{figure}
	\centering
	\includegraphics[width=0.4\textwidth]{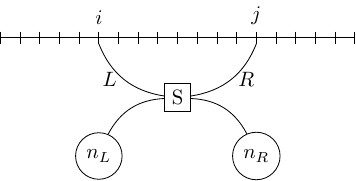}
	\hspace{0.5cm}(a)
	\includegraphics[width=0.2\textwidth]{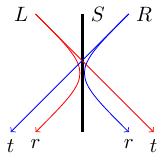}
	\hspace{0.5cm}(b)
	\includegraphics[width=0.2\textwidth]{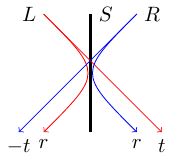}
	\caption{Two wires are attached to the system at sites $i$ and $j$ such that
		only one fermion can enter at a time. First the fermions in the wire
		are allowed to interact via the beam splitter $S$. Then their occupation
		number $n_{L}$ and $n_{R}$ is measured on each side. In the first
		measurement (a) one uses a symmetric beam splitter, which allows to
		measure the imaginary part of $G_{ij}$. In the second measurement
		(b), one needs to use a beam splitter where the fermion that is transmitted
		from $R$ to $L$ accumulates a phase $\pi$, while it does not accumulate
		this phase when being transmitted in the other direction. In this
		way one can measure the real part of $G_{ij}$. \label{fig:experiment}}
\end{figure}

The idea to measure coherences, is pretty standard: One probes the system at two sites in a completely coherent way, interferes the two signals via a beamsplitter, and measures each output separately, see Fig.\ \ref{fig:experiment}. Unfortunately, in real experiments, the coherent coupling and the interference of signals from two sites of a diffusive conductor, such as a dirty metal, seems currently out of reach since there are too many uncontrolled degrees of freedom in the system\footnote{Private discussions with Gwendal Fève and Meydi Ferrier}. However, for well controlled, usually ballistic conductors with a very low electron density, such interference experiments have been possible. In Ref.\ \cite{Feve2013Coherence} Bocquillon et al.\ realized an interference experiment for single electron wave-packets from independent synchronized sources, where the role of the beam splitter was played by a quantum point contact. In Ref.\ \cite{Feve2019Quantum} the same group extended on this result and was able measure coherences between ballistic spin-polarized one-dimensional conductors, realized as the outer-edge-channels of a two-dimensional electron gas (GaAs/AlGaAs), see Fig.~\ref{fig:interference_feve}. But to our knowledge, such a setup has not been yet realized for diffusive conductors.

To conclude, we give a theoretical description how coherences could be measured in an ideal interference experiment. Inspired by \cite{Gullans2019Entanglement} we make the protocol outlined there a bit more precise\footnote{Taking a paragraph from our article \cite[app. A]{Hruza2023Coherent}.}:

\begin{itemize}
	\item The total state of system, left and right wire is described by a state
	in the Hilbert space $\mathcal{H}_{S}\otimes\mathcal{H}_{L}\otimes\mathcal{H}_{R}$.
	Let us assume that the system is in a pure state and that initially
	the wires are empty and not coupled to the system,
	\[
	|\psi^{(0)}\rangle=|\psi_{S}\rangle|0,0\rangle.
	\]
	\item Now we couple the two wires to the system. A very simple description
	of this coupling could be given by the unitary evolution with $U_{int}=e^{-i\lambda(c_{L}^{\dagger}c_{i}+c_{R}^{\dagger}c_{j}+h.c.)}$,
	where $\lambda$ is the product of coupling strength and the time
	during which we allow the wires to couple to the system, and $c_{L}$
	($c_{R}$) are fermionic operators on the left (right) wire. If we
	tune the coupling strength and duration such that $\lambda\ll1$ is
	small, we can neglect $\mathcal{O}(\lambda^{2})$ terms and find,
	\[
	|\psi^{(1)}\rangle:=U_{int}|\psi^{(0)}\rangle\approx|\psi_{S}\rangle|0,0\rangle-i\lambda\left(c_{i}|\psi_{S}\rangle|1,0\rangle+c_{j}|\psi_{S}\rangle|0,1\rangle\right)
	\]
	\item Next, the fermions in the left and right wire interfere in a beam
	splitter. Written in the basis $\{|00\rangle,|01\rangle,|10\rangle,|11\rangle\}$
	the beam splitter can in general be described by the scattering matrix
	\[
	S=\left(\begin{array}{cccc}
		1\\
		& r' & t\\
		& t' & r\\
		&  &  & rr'-tt'
	\end{array}\right),
	\]
	where $r$ and $t$ ($r'$ and $t'$) are the reflection and transmission
	amplitudes for a fermion incident from the left (right) 
	side\footnote{To obtain the last entry $rr'-tt'$, one has take into account that the wave function is antisymmetric,
	$|1,1\rangle=|\phi_{L}\rangle_1|\phi_{R}\rangle_1-|\phi_{R}\rangle_2|\phi_{L}\rangle_2$.
	Here $1$ and $2$ label the fermion, whereas $L$ and $R$ label states in the left and right wire. After the beamsplitter this state becomes
	\begin{align*}
		|1,1\rangle  \to  \big(r|\phi_{L}\rangle_1+t|\phi_{R}\rangle_1\big)\big(r'|\phi_{R}\rangle_2+t'|\phi_{L}\rangle_2\big)
		-\big(r'|\phi_{R}\rangle_1+t'|\phi_{L}\rangle_1\big)\big(r|\phi_{L}\rangle_2+t|\phi_{R}\rangle_2\big),
	\end{align*}
	which leads to the entry $(rr'-tt')$.
	}.
	Unitarity demands $|r|^{2}+|t|^{2}=1$,  $|r'|^{2}+|t'|^{2}=1$ and
	${r}^*t'+{t}^*r'=0$ (the condition $|rr'-tt'|^{2}=1$ is then
	automatically fulfilled). The following choices allow to measure the (a) imaginary and (b) real part of $G_{ij}$:
	\begin{itemize}
		\item[(a)] $r=r'$ and $t=t'$. Expressing $r=\sin\theta$ and
		$t=i\cos\theta$ to fulfil the unitary constrains one obtains
		\begin{align*}
			|\psi^{(2,a)}\rangle&=S^{(a)}|\psi^{(1)}\rangle\nn
			&\begin{aligned}
				=|\psi_{S}\rangle|0,0\rangle
				&-i\lambda c_{i}|\psi_{S}\rangle\left(\sin\theta|1,0\rangle +i\cos\theta|0,1\rangle\right)  \\
				&-i\lambda c_{j}|\psi_{S}\rangle\left(\sin\theta|0,1\rangle+i\cos\theta|1,0\rangle\right).
			\end{aligned}
		\end{align*}
		\item[(b)] $r=r'$ and $t=-t'$. This can be expressed as $r=\sin\theta$ and $t=\cos\theta$,
		\begin{align*}
			|\psi^{(2,b)}\rangle&=S^{(b)}|\psi^{(1)}\rangle\nn
			&\begin{aligned}
				=|\psi_{S}\rangle|0,0\rangle
				&-i\lambda c_{i}|\psi_{S}\rangle\left(\sin\theta|1,0\rangle+\cos\theta|0,1\rangle\right)\\
				&-i\lambda c_{j}|\psi_{S}\rangle\left(\sin\theta|0,1\rangle-\cos\theta|1,0\rangle\right).
			\end{aligned}
		\end{align*}
	\end{itemize}
	\item Finally, one measures the particle number $n_{L}=c_{L}^{\dagger}c_{L}$
	and $n_{R}=c_{R}^{\dagger}c_{R}$ of the left and right outgoing beams. Denoting
	averages$\<\psi_S|\cdots|\psi_{S}\rangle=\langle...\rangle_{S}$
	and $G_{ij}=\langle c_{j}^{\dagger}c_{i}\rangle_{S}$ one finds for
	case (a)
	\begin{align*}
		\langle n_{L}\rangle^{(a)} & =\lambda^{2}\left(\sin^{2}\theta\langle n_{i}\rangle_{S}+\cos^{2}\theta\langle n_{j}\rangle_{S}-2\sin\theta\cos\theta\mathrm{\Im}(G_{ij})\right)\\
		\langle n_{R}\rangle^{(a)} & =\lambda^{2}\left(\cos^{2}\theta\langle n_{i}\rangle_{S}+\sin^2\theta\langle n_{j}\rangle_{S}+2\sin\theta\cos\theta\Im(G_{ij})\right).
	\end{align*}
	Choosing an angle $\theta=\pi/4$ gives the imaginary part of $G_{ij}$,
	\[
	2\lambda^{2}\Im(G_{ij})=\langle n_{R}\rangle^{(a)}-\langle n_{L}\rangle^{(a)}.
	\]
	For the case (b), one gets
	\begin{align*}
		\langle n_{L}\rangle^{(b)} & =\lambda^{2}\left(\sin^{2}\theta\langle n_{i}\rangle_{S}+\cos^{2}\theta\langle n_{j}\rangle_{S}-2\sin\theta\cos\theta\mathrm{\Re}(G_{ij})\right)\\
		\langle n_{R}\rangle^{(b)} & =\lambda^{2}\left(\cos^{2}\theta\langle n_{i}\rangle_{S}+\sin^2\theta\langle n_{j}\rangle_{S}+2\sin\theta\cos\theta\Re(G_{ij})\right).
	\end{align*}
	Choosing the same angle $\theta=\pi/4$ gives the real part of $G_{ij}$,
	\[
	2\lambda^{2}\Re(G_{ij})=\langle n_{R}\rangle^{(b)}-\langle n_{L}\rangle^{(b)}.
	\]
\end{itemize}

\chapter{New results about QSSEP}\label{chpt:new_results_QSSEP}
This and the next chapter present results we developed during my thesis. While the present chapter only focuses on results about QSSEP, the next chapter gives further insights into the free probability structure of QSSEP and contains proofs about statement made here.
\section{Scaling limit and dynamics of coherences}\label{sec:scaling_limit_and_dynamics}
Ultimately we would be interested in a hydrodynamical description of coherences  -- similar to MFT. On these scales, we can hope that universal features of coherences become apparent and that details of the underlying microscopic model get washed out. Therefore we should take a scaling limit where the number of sites $N\to\infty$, keeping the physical length $L=1$ of the system fixed to one. At the same time, we have to rescale time, to avoid the equations of motions to become trivial. Diffusion of the mean density suggests that we should take a diffusive scaling, that is
\begin{equation} \label{eq:scaling-limit}
	i \to x=i/N\in[0,1],\quad t/N^{2}\to t.
\end{equation}

In this limit, the $n$-th order cumulants of coherences, whose indices are arranged in a loop, will scale as $N^{1-n}$ at leading order. We denote their leading contribution by
\begin{equation}\label{eq:def_g}
	g_n(x_{1},x_2\cdots,x_{n};t):=\lim_{N\to\infty}N^{n-1}\mathbb{E}_{N^{2}t}[G_{i_{1}i_{2}}G_{i_{2}i_{3}}\cdots G_{i_{n}i_{1}}]^c.
\end{equation}
We will see that these loops $g_n$ are the building blocks from which all other correlations functions can be obtained, at leading order. We will show below towards the end of this section that $g_n$ satisfies the equation
\begin{align}\label{eq:g_n_evolution}
	&(\partial_{t}-\Delta)g_n(x_{1},\cdots,x_{n})
	\\ \nonumber
	&=\sum_{i,j=1;\,i<j}^{n}2\,\delta(x_{i}-x_{j})\partial_{i}g_{|j-i|}(x_{i},\cdots,x_{j-1})\partial_{j}g_{|n-j+i|}(x_{j},\cdots,x_{i-1}),
\end{align}
with $\Delta\equiv\sum_{i=1}^{n}\Delta_{x_{i}}$, and with boundary conditions
\begin{equation}
	g_n(x_{1},\cdots,x_{n})=
	\begin{cases}
		n_a \text{ or } n_b & \text{ for } n=1 \text{ and } x=0,1\\
		0  &\text{ for } n\ge 2 \text{ and some } x_i\in\{0,1\}
	\end{cases}.
\end{equation}
That is, only the mean density $g_1=\bar n$ depends explicitly on the reservoirs' densities $n_a$ and $n_b$, and higher order loops take lower order loops as source terms. This \textit{triangular structure} can be visualised as
\begin{equation}
	\raisebox{-0.5\height}{\includegraphics{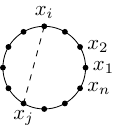}}
	\longrightarrow
	\raisebox{-0.5\height}{\includegraphics{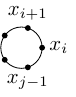}}
	\,
	\raisebox{-0.5\height}{\includegraphics{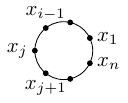}}.
\end{equation}

\begin{figure}
	\vskip 0.5 truecm
	\centering
	\includegraphics[width=0.45\textwidth]{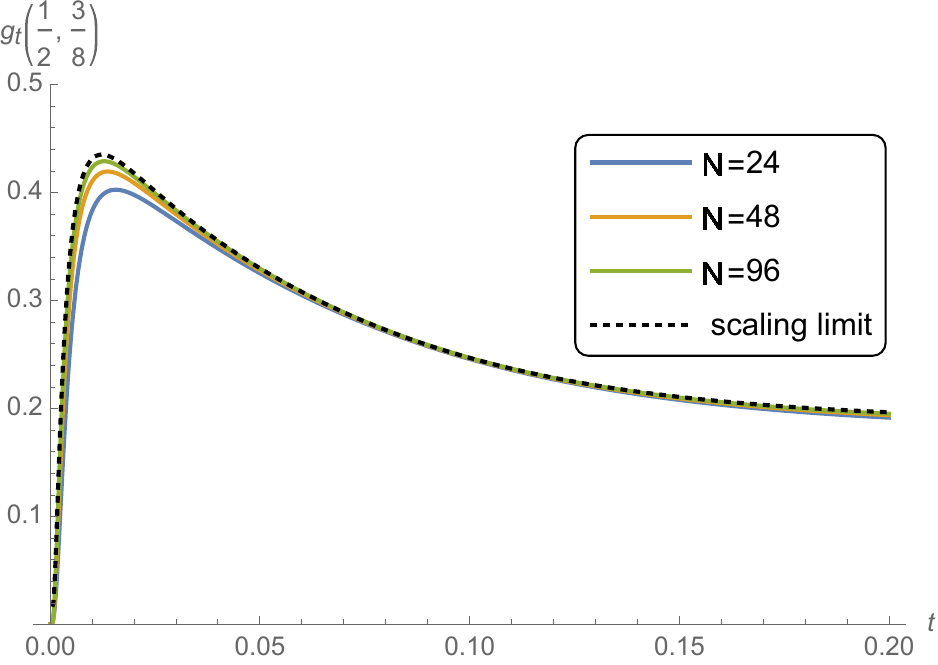}
	\hspace{1cm}
	\includegraphics[width=0.45\textwidth]{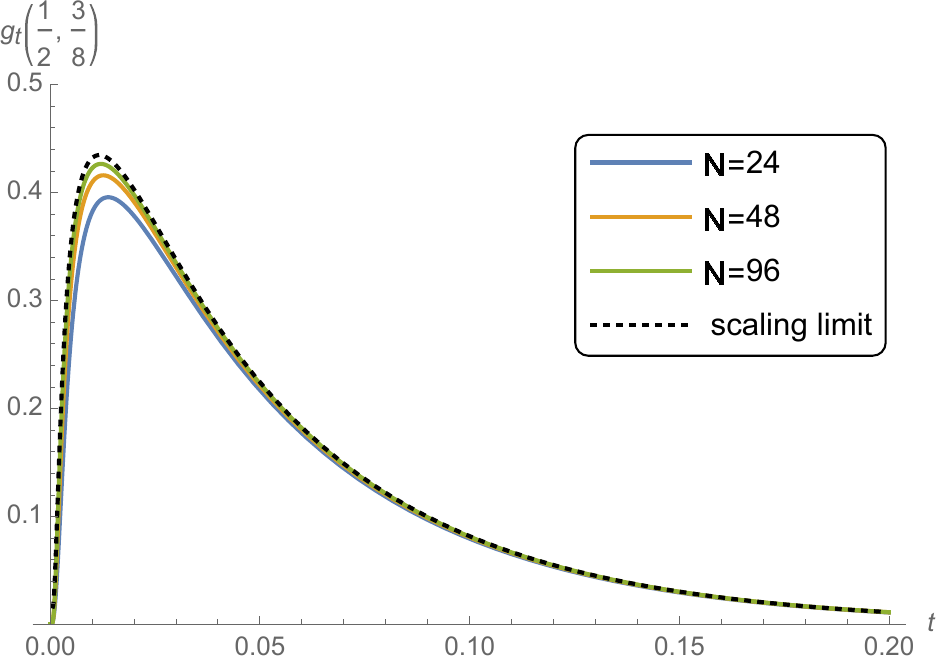}
	\caption{Comparison of the time evolution between the discrete 2nd cumulant $g_{ij}^\text{dis}(t)$ and its analytical solution in the scaling limit $g_2(x,y;t)$. The boundary conditions are $n_{a}=1$, $n_{b}=0$ (left) and $n_{a}=1/2$, $n_{b}=1/2$ (right). The initial state was chosen to be a Heaviside step function $g_2(x,y,0)=\Theta(1/2-x)$.}
	\label{fig:g_2}
\end{figure}

In order to check that, that the hydrodynamical Equations~\eqref{eq:g_n_evolution} agree with the microscopic equation \eqref{eq:G_evolution}, in \cite{Hruza2023Coherent} we performed a numerical test at order $n=2$, inspired by a similar test in our earlier work on the closed QSSEP \cite{Bernard2022DynamicsClosed}: The analytic solution for $g_2(x,y;t)$ was compared to a numerical solution
of the discrete and coupled differential equations for $g_{ij}^\text{dis}(t):=N(\mathbb E_{N^{2}t}[G_{ij}G_{ji}]-\delta_{ij}\mathbb E_{N^{2}t}[G_{ii}]^2)$
with different values of $N$. Figure \ref{fig:g_2} shows that the agreement is excellent.

\paragraph{Stochastic dynamics of coherences in the scaling limit.}
In view of our aim to find a quantum coherent extension of MFT which could be called a \textit{mesoscopic fluctuation theory}, we should try to promote Eq.~\eqref{eq:g_n_evolution}, the scaling limit for the time evolution of loop-cumulants, to a stochastic equation for individual coherences. In other words, we try to find the scaling limit of Eq.~\eqref{eq:dG_ij}. Unfortunately, this question is to a large extend still unanswered -- except for the following somewhat artificial answer. 

Let $I_{x,t}$ be a matrix of size $N$ whose elements are functions of $x\in[0,1]$ and $t$, undergoing the stochastic process (in It\={o} convention)

\begin{align}\label{eq:dI}
	dI_{x,t} &= \Delta_x I_{x,t} dt + \sqrt{\frac{2}{N}}\,\partial_x (I_{x,t} dW_{x,t}), &
	dW_{x,t}^{ij}d{W}_{y,t}^{kl}=\delta(x-y)\delta^{il}\delta^{jk}\, dt,
\end{align}
where $dW_x$ is a Hermitian matrix consisting of complex Brownian increments $dW_{x,t}^{ij}$ independent in space and time. Then the cumulants of coherences can be identified with
\begin{equation}\label{eq:def_gn_Ix}
	g_n(x_1,x_2\cdots,x_n;t)=\E[\ntr(I_{x_1,t}I_{x_2,t}\cdots I_{x_n,t})]^c.
\end{equation}
One can check that this indeed reproduces Eq.~\eqref{eq:g_n_evolution}. The stochastic equation we have proposed here for $I_{x,t}$ is actually similar to the MFT equation \eqref{eq:MFT_combined}, just that here the equation is matrix-valued. However, it does not yet provide us with a hydrodynamic description of coherences, because we do not know how the matrix $I_{x,t}$, which is defined at a single point, can be related to coherences $G(x,y):=G_{ij}$ in the scaling limit $x=i/N,y=j/N$, which are defined at two points. 

\subparagraph{Verification at 2\textsuperscript{nd} order.} For $n=2$ we check that Eq.~\eqref{eq:dI} leads indeed to the correct dynamics for the cumulants of coherence. One needs to evaluate
\begin{align}
	d(I_{x,t}I_{y,t})
	&=d I_{x,t} I_{y,t} + I_{x,t} dI_{y,t} + dI_{x,t}dI_{y,t}\nn
	&=(\Delta_x+\Delta_y) I_{x,t} I_{y,t} \,dt + \frac{2}{N} \partial_x\partial_y \big(\delta(x-y)\,I_{x,t}\,\tr(I_{y,t})\big) dt,
\end{align}
where we neglected noisy terms in the last line, since they vanish under $\E$. Note that using the It\={o} rules for the last term, the effect of $dW_{x,t}^{ij}dW_{y,t}^{kl}$ is to decouple the matrices $I_{x,t}$ and $I_{y,t}$ with respect to the trace. Furthermore, one has to assumes that the expectation value (not the cumulant) of terms evaluated at the same position factorizes at leading order in $N$. Here this implies 
\begin{equation}
	\E[\delta(x-y)I_{x,t}\,\tr(I_{y,t})]\approx\delta(x-y)\E[I_{x,t}]\E[\tr(I_{y,t})].
\end{equation}
For coherences in QSSEP, this factorization is actually true, see Eq.~\eqref{eq:QSSEP_factorization}. Taking the normalized trace one obtains
\begin{align}
	&\partial_t \,\E[\ntr(I_{x,t}I_{y,t})]\nn &=(\Delta_x+\Delta_y) \E[\ntr(I_{x,t} I_{y,t})] + 2\partial_x\partial_y \left(\delta(x-y)\E[\ntr(I_{x,t})]\E[\ntr(I_{y,t})]\right).
\end{align}
This equation has now to be transformed into an equation for cumulants $\E[\cdots]^c$, which is exactly the same problem as transforming Eq.~\eqref{eq:loop_moment_evolution} (see below) to Eq.~\eqref{eq:g_n_evolution} outlined at the end of this section. With the definition in Eq.~\eqref{eq:def_gn_Ix}, one then obtains
\begin{equation}
	\partial_t g_2(x,y)=(\Delta_x+\Delta_y)g_2(x,y) + 2\delta(x-y)\partial_x g_1(x)\partial_y g_1(y),
\end{equation}
which is in accordance with Eq.~\eqref{eq:g_n_evolution}.

\paragraph{Derivation of Equation \eqref{eq:g_n_evolution}.} This derivation is contained in \cite{Bernard2022DynamicsClosed, Hruza2023Coherent}. Here we present an overview of the quite lengthy calculations.

\subparagraph{Step 1.} The first step is to arrive at an equation for the moments $\E[G_{i_{1}i_{2}}\cdots G_{i_{n}i_{1}}]$ (non-connected part) with indices arranged in a loop. This is done in two stages. First, we consider Eq.~\eqref{eq:dG_ij} without the boundary terms and find an equations for these moments on the discrete lattice, from which we extract the scaling and the equation in continuous space at leading order in $N$. In the second stage we consider the boundary terms on the lattice and figure out what are the correct boundary conditions which the continuous equations need to be supplemented by.

We denote loops by $A_{i_1,...,i_n}:=G_{i_{1}i_{2}}\cdots G_{i_{n}i_{1}}$ and abbreviate $[\cdots]=\E[\cdots]$. Using Eq.~\eqref{eq:dG_ij} without boundary terms (third line), we evaluate $d[A_{i_1,...,i_n}]$ in It\={o} convention. That is, we have to keep track not only of terms $[dGG\cdots]$, but also of $[dGdG\cdots]$. Denoting the discrete Laplacian $\Delta_i A_i:=A_{i+1}+A_{i-1}-2 A_i$ and skipping the details, one find a diffusion equation with source term
\begin{align}\label{eq:discrete_evolution}
	\partial_t[A_{i_1,...,i_n}]=\sum_{k=1}^n \Delta_{i_k}[A_{i_1,...,i_n}] +\sum_{k<l}^n \left( \delta_{i_k,i_l} \Delta_{i_k}-(\Delta_{i_k} \delta_{i_k,i_l})\right) [B_{i_k,i_l}].
\end{align} 
with $[B_{i_k,i_l}]=[A_{i_k,i_{k+1}...,i_{l-1}} A_{i_l,i_{l+1}...,i_{k-1}}]+[A_{i_l,i_{k+1}...,i_{l-1}} A_{i_k,i_{l+1}...,i_{k-1}}]$. Note that the first term of $B_{i_k,i_l}$ is the expectation value of the product of two loops that form when pinching the original loop at $i_k$ and $i_l$. The second term is equivalent with $i_k$ and $i_l$ interchanged.

In the scaling limit \eqref{eq:scaling-limit} we denote $A_n(x_1,...,x_n):=A_{i_1,...,i_n}$. Then the discrete Laplacian becomes $\Delta_i A_i\to N^{-2}\Delta_x A(x)+\Ord(N^{-4} A(x))$ and $\delta_{ij}\to N^{-1}\delta(x-y)+\Ord(N^{-2})$. We see that the second term in Eq.~\eqref{eq:discrete_evolution} will scale one order of $N$ lower than the first one. Before stating the continuous version of \eqref{eq:discrete_evolution}, we rewrite the second term by partial integration (against a test function)
\begin{equation}
	\big( \delta(x-y) \Delta_{x}-(\delta''(x-y)\big) [B_n(x,y)]=\partial_{x}\partial_{y}\big(\delta(x-y)[B_n(x,y)]\big).
\end{equation}
Rescaling time $t/N^2\to t$ and replacing $B(x,y)=2[A(x,\cdots)A(y,\cdots)]$ which is allowed due to the presence of $\delta(x-y)$ , we have
\begin{align}
	\partial_t [A_n(\vec x)]
	&=\sum_{k=1}^n \Delta_{x_k}[A_n(\vec x)] \\
	&+\frac 2{N} \sum_{k<l}^n \partial_k\partial_l\big( \delta(x_k-x_l)[A_{l-k}(x_k,x_{k+1},...,x_{l-1}) A_{n-l+k}(x_l,x_{l+1},...,x_{k-1})]\big) \nonumber
\end{align}
This is a diffusion equation with a source term suppressed as $1/N$.

From this equation we can extract the scaling of the leading order of loops. We expand $[A_n]=[A_n]^{(0)}+N^{-1}[A_n]^{(1)}+N^{-2}[A_n]^{(2)}+\cdots$.
For simplicity, we assume that the initial state is a product state, i.e. in the beginning $[A_n]=0\; \forall n\ge 2$, so except for the density at $n=1$. At a later time, $[A_n]$ can only be non-zero, if the source term, proportional to $ N^{-1}[A_{l-k}A_{n-l+k}]$, is non-zero. One can verify that the source term (when all indices are different) also satisfies a diffusion equation with a new source term of order $N^{-1}$ lower, which consists of the product of three loops. If we iterate this reasoning $n-1$ times, we end up with a diffusion equation whose source term is a product of $n$ loops of size one, i.e.\ $[A_{i_1}\cdots A_{i_n}]$. This final source term satisfies a pure diffusion equation without source term. Since the initial condition is factorized, it stays factorizes at all times, i.e. it is equal to the product of densities $[A_{i_1}]\cdots[A_{i_n}]$, which is non-zero\footnote{
	Explicitly: If $g_0(x,y)=f_0(x)f_0(y)$ and $(\partial_t-\Delta)g_t(x,y)=0$, then $g_t(x,y)=f_t(x)f_t(y)$ is a solution at all times if $f$ satisfies $(\partial_t-\Delta)f_t(x)=0$.
}.
Counting all powers of $1/N$ encountered in the $n-1$ iterations, we learn that at leading order $[A_n]= 1/N^{n-1}[A_n]^{(n-1)}$. Iterating backwards from here to the first source term we encountered, one can convince oneself that also this term factorizes, i.e. $[A_{l-k}A_{n-l+k}]=[A_{l-k}][A_{n-l+k}]$ at leading order, which is what we wanted to show\footnote{We can relax the assumption on the initial product state: As shown in \cite{Bauer2019Equilibrium}, the steady state distribution of $G$ (for the closed QSSEP) is unique. By the argument given above, in particular the steady state moments of loops have the desired properties (scaling with $N$ and factorization). But the steady state can also be reached from a state that is not a product state. So there must be some finite time at which moments of loops start to satisfy the desired properties and from this time onwards the equations preserve these properties.}.

To summarize, we found (with $\Delta=\sum_{k=1}^n \Delta_{x_k}$)
\begin{align}\label{eq:loop_moment_evolution}
	&(\partial_t-\Delta)[A_n(\vec x)]^\#\\
	&=2\sum_{k<l}^{n}\partial_{k}\partial_{l}
	\left(\delta(x_{k}-x_{l}) [A_{l-k}(x_{k},\cdots,x_{l-1})]^\#[A_{n-l+k}(x_{l},\cdots,x_{k-1})]^\#\right) \nonumber
\end{align}
where we denoted the leading order as $[A_n]= 1/N^{n-1}[A_n]^\#$.

Now we turn to the second stage, that is to find appropriate boundary conditions for this equation. Considering the boundary term (third line) in Eq.~\eqref{eq:dG_ij}, one has to add the following contribution to the right hand side of the Eq.~\eqref{eq:discrete_evolution}
\begin{equation}
	+\sum_{k=1}^n \sum_{p\in\{1,N\}}\left(\delta_{p,i_k}\delta_{p,i_{k+1}}\alpha_{p}[A_{i_1,...,\hat i_k,...,i_n}]-\frac{1}{2}(\delta_{p,i_{k}}+\delta_{p,i_{k+1}})(\alpha_{p}+\beta_{p})[A_{i_1,...,i_n}] \right),
\end{equation}
where $\hat i_k$ means that $i_k$ is missing from the list\footnote{One also has to truncate any discrete Laplacian on the boundaries as explained below Eq.~\eqref{eq:density_discrete}} and $n+1\equiv 1$. We see that the boundary contribution for the loop of order $n$ couples to a loop of order $n-1$. This makes it very difficult to extract the correct boundary conditions in the scaling limit analytically. Fortunately, it turns out that bulk and boundary distinguish themselves by a separation of time scales: The moments of loops approach their steady state values on the boundaries almost immediately, while in the bulk they continue to evolve. As a consequence, we can use the steady state values as boundary conditions. 

This is explained in greater detail in \cite[sec.~IV.C]{Hruza2023Coherent}. There one constructs an analytic solution for the only feasible case $n=1$, which describes the mean density already encountered in Eq.~\eqref{eq:density_discrete}. From there one learns that in the scaling limit the mean density approaches its steady state value in the bulk in a time $t\sim\Ord(1)$ while on the boundary $t\sim\Ord(1/N^2)$. The boundary ``thermalizes'' almost immediately. This justifies to use the steady state values as boundary conditions in the scaling limit, i.e. $[A_1(0)]=n_{a}$ and $[A_1(1)]=n_{b}$. To justify this claim for $n>1$ we did the numerics shown in Fig.~\ref{fig:g_2}.

\subparagraph{Step 2.} The second and final step consists of transforming Eq.~\eqref{eq:loop_moment_evolution} for the evolution of moments of loops into Eq.~\eqref{eq:g_n_evolution} for the evolution of their cumulants. The complete derivation can be found in \cite[appendix E.2]{Hruza2023Coherent}. The derivation of Eq.~\eqref{eq:g_n_evolution} crucially depends on the fact that moments and cumulants of loops are related by a sum over non-crossing partitions, such as we will see in Eq.~\eqref{eq:QSSEP_moment_cumulants} in the next section. This is a property reminiscent of free probability theory.

\section{Signs of free probability}
Motivated by the need to find the time evolution of the cumulants of coherences from the time evolution of the moments, in Ref.~\cite{Hruza2023Coherent} we expanded moments into cumulants. As we review in Section \ref{sec:intro_free_proba}, this can be done as
\begin{equation}\label{eq:moments_cumulants}
	\mathbb{E}[X_{i_{1}}\cdots X_{i_{n}}]=\sum_{\pi\in P(n)}\prod_{p\in\pi}\mathbb{E}[X_{i_{p(1)}}X_{i_{p(2)}}\cdots]^{c}.
\end{equation}
where $\pi$ is a partition of the set $\{1,\cdots,n\}$ into subsets, called parts $p=\{p(1),p(2),\cdots\}$. Applying this formula to coherences in QSSEP, we made the surprising observation that terms corresponding to \textit{crossing} partitions, vanished. For example $\pi=\{\{1,3\},\{2,4\}\}$ is a crossing partitions for $n=4$, while $\pi=\{\{1,2\},\{3,4\}\}$ and $\pi=\{\{1,4\},\{2,3\}\}$ are \textit{non-crossing}. Denoting the set of non-crossing partitions by $NC(n)$ and $|p|$ the number of elements in part $p$, we found that in the limit $N\to\infty$,
\begin{equation}\label{eq:QSSEP_moment_cumulants}
	\mathbb E[G_{i_1i_2}...G_{i_ni_1}]
	=\sum_{\pi\in NC(n)}\mathbb \delta_{\pi^*}(i_1,...,i_n)\prod_{p\in \pi}\mathbb E[G_{i_{p(1)}i_{p(2)}}...G_{i_{p(|p|)}i_{p(1)}}]^c.
\end{equation}
Here $\pi^*$ is the Kreweras complement (a kind of dual non-crossing partition of $\pi$ defined around Eq.~\eqref{eq:G_moments_etappe}) and $\delta_{\pi^*}(i_1,\cdots,i_n)=\prod_{p\in\pi^*} \delta_{i_{p(1)},\cdots,i_{p(|p|)}}$ is a product of Kronecker deltas that sets all indices belonging to the same part $p\in\pi^*$ to be equal.

Summing over all indices, replacing the sums by integrals over continuous variables and using the definition of $g_n$ from Eq.~\eqref{eq:def_g}, one finds the equivalent expression
\begin{align}\label{eq:QSSEP_moments_G_cumulants}
	\E[\ntr(G^n)]&=\sum_{\pi\in NC(n)} \int  \delta_{\pi^*}(\vec x) \,g_\pi(\vec x)\,d\vec x, &
	g_\pi(\vec x)=\prod_{p\in\pi} g_n(\vec x_p)
\end{align}
with $\ntr=\tr/N$ the normalized trace, $\vec x=(x_1,\cdots,x_n)$ and $\vec x_p=(x_{p(1)},x_{p(2)},\cdots)$. Furthermore, $\delta_{\pi^*}(\vec x)=\prod_{p\in\pi^*} \delta(\vec x_p)$ and $\delta(\vec x_p)=\delta(x_{p(1)}-x_{p(2)})\delta(x_{p(2)}-x_{p(3)})\cdots$.

These formulas are reminiscent of the expansion of moments into free cumulants within the framework of so-called \textit{free probability theory}. We leave their proof to Section \ref{sec:moment-cumulant-expansion}. But it is important to note that the proof of these formulas requires only three minimal assumption about the measure $\E$ of the matrix elements of $G$. In this sense, the formulas apply to a much large class of random matrices.
 
\paragraph{Three properties.}
The class of matrices for which Eqs.~(\ref{eq:QSSEP_moment_cumulants}, \ref{eq:QSSEP_moments_G_cumulants}) hold is characterized by the following three properties:
\begin{enumerate}
	\item[(i)] Local $U(1)$-invariance, meaning that in distribution, $G_{ij}\stackrel{d}{=}e^{-i\theta_i}G_{ij}e^{i\theta_j}$ for any angles $\theta_i$ and $\theta_j$;	
	\item[(ii)] Expectation values of loops of order $n$ without repeated indices scale as $N^{1-n}$, meaning that $\mathbb E [G_{i_1 i_2}G_{i_2i_3}\cdots G_{i_n i_1}]=\mathcal O(N^{1-n})$ for all indices $i_k$ distinct;
	\item[(iii)] Factorization of the expectations value of products of loops at leading order, \\ $\mathbb E[G_{i_1 i_2}\cdots G_{i_{m} i_1}\,G_{j_{1} j_{2}}\cdots G_{j_n j_{1}}]=\mathbb E [G_{i_1 i_2}\cdots G_{i_{m} i_1}]\,\mathbb E [G_{j_{1} j_{2}}\cdots G_{j_{n} j_{1}}](1+\Ord(\frac{1}{N}))$, even if $i_1=j_1$.
\end{enumerate}
For QSSEP, the first property is just Eq.~\eqref{eq:QSSEP_U1}. The second property is Eq.~\eqref{eq:def_g}. And the third property follows from the fact that thanks to Eq.~\eqref{eq:g_n_evolution} $g_n(x_1,\cdots,x_n)$ stays finite if two arguments $x_1=x_k$ become equal. In other words, if $i_1=i_k$ for some $k\in\{1,\cdots,n\}$, then
\begin{equation}\label{eq:QSSEP_factorization}
	\underbrace{\E[G_{i_{1}i_{2}}\cdots G_{i_{n}i_{1}}]^c}_{\Ord(N^{1-n})}=\E[G_{i_{1}i_{2}}\cdots G_{i_{n}i_{1}}]-\underbrace{\E[G_{i_1 i_2}\dots G_{i_{k-1} i_1}] \E[G_{i_1 i_{k+1}}\dots G_{i_{n} i_1}]}_{\Ord(N^{2-n})}.
\end{equation}
The product of expectation values on the right hand side scales as $\Ord(N^{2-n})$, i.e.\ more dominant than the left hand side. Therefore the first term on the right hand side must factorize and cancel the second term, this is property (iii).
\section{Steady state solution}\label{sec:steady_state_solution}
Initially, the connection between QSSEP and free probability was made in a slightly different manner by Philippe Biane \cite{Biane2021Combinatorics}. He showed that the connected fluctuations of coherences $g_n$ in the steady state could be understood as the free cumulants (defined in Eq.~\eqref{eq:def_free_cumulants}) of the indicated function $\I_x(y)=1_{y<x}$, where $y$ is distributed according to the Lebesgue measure $dy$ on $[0,1]$. In other words, for $t=\infty$ one has
\begin{equation}\label{eq:g_steady}
	\sum_{\pi\in NC(n)} g_\pi (\vec x)=\E[\I_{x_1}(y)\cdots \I_{x_n}(y)] =\min(\vec x), \qquad \vec x\equiv(x_1,\cdots,x_n)
\end{equation}
with notation as in Eq.~\eqref{eq:QSSEP_moments_G_cumulants}. The corresponding moments are given by a simple integration $\phi_n(\vec x)=\E[\I_{x_1}(y)\cdots \I_{x_n}(y)]=\int \I_{x_1}(y)\cdots \I_{x_n}(y)\,dy$. The approach by Biane is actually quite surprising, because here, the ``free cumulants'' $g_n$ belong to a family of commuting variables $\mathbb I_x(y)$. But free probability usually appearers in the context of non-commuting random variables, such as large random matrices. In this sense, our approach to free probability outlined in the last section is perhaps less surprising.

Biane's result on the steady state of QSSEP can also be derived from the time evolution of $g_n$ in Eq.~\eqref{eq:g_n_evolution}. Let us show this, following our article \cite{Hruza2023Coherent}. Defining
\begin{equation}\label{eq:phi_g}
	\phi_n(\vec x)=\sum_{\pi\in NC(n)} g_\pi(\vec x),
\end{equation}
one finds that $\phi_n$ satisfies exactly the same equation as $g_n$,
\begin{align}\label{eq:phi_equation}
	&(\partial_{t}-\Delta)\phi_n(\vec x )
	\\ \nonumber
	&=\sum_{i,j=1;\,i<j}^{n}2\,\delta(x_{i}-x_{j})\partial_{i}\phi_{|j-i|}(x_{i},\cdots,x_{j-1})\partial_{j}\phi_{|n-j+i|}(x_{j},\cdots,x_{i-1}),
\end{align}
However the boundary conditions are different. For simplicity we consider the case $n_a=0,n_b=1$ in the following. If some $x_i\in\{0,1\}$ lies on the boundary, then 
\begin{align}\label{eq:phi_boundary}
	\phi_n(x_{1},\cdots,x_i,\cdots,x_{n})= x_i\, \phi_n(x_{1},\cdots,\hat{x}_{i},\cdots,x_{n}) 
\end{align} 
where the hat on $\hat{x}_{i}$ indicates that $x_{i}$ is missing from the
set $\{x_{1},\cdots,x_{n}\}$. Then it remains to check that $\phi_n(\vec x)=\min(x_1,\cdots,x_n)$ is a stationary solution of this equation and that the solution is unique.

To conclude, we expand Eq.~\eqref{eq:g_steady} up to $n=4$ and provide explicit expressions for the connected correlations of coherences in QSSEP,
\begin{align}
	g_{1}(x_{1})&= x_{1}\nn
	g_{2}(x_{1},x_{2})&= \min(x_{1},x_{2})-x_{1}x_{2}\nn
	g_{3}(x_{1},x_{2},x_{3})&= \min(x_{1},x_{2},x_{3})-x_{1}\min(x_{2},x_{3})_{\circlearrowleft3}+2x_{1}x_{2}x_{3}\nn
	g_{4}(x_{1},x_{2},x_{3},x_{4})&=\min(x_{1}, x_{2}, x_{3}, x_{4})-x_{1}\min(x_{2}, x_{3}, x_{4})_{\circlearrowleft4} \nn
	&-\min(x_{1}, x_{2}) \min(x_{3}, x_{4})_{\circlearrowleft2} + 2x_{1}x_{2}\min(x_{3}, x_{4})_{\circlearrowleft4} \nn
	&+ x_{1}x_{3}\min(x_{2}, x_{4})_{\circlearrowleft2}-5x_{1}x_{2}x_{3}x_{4}.
\end{align}
Here ${\circlearrowleft q}$ denotes the distinct $q$ cyclic permutation of the indices of the term.

\section{QSSEP as a noisy mesoscopic systems}\label{sec:noisy_mesoscopic_systems}
\begin{figure}
	\centering
	\includegraphics[width=0.5\textwidth]{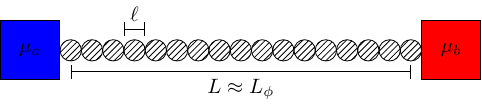}
	\caption{\label{fig:ballistic_cell}Schematic representation of a diffusive mesoscopic system of length $L$ between two reservoirs which is decomposed into small cells comparable to the mean free path $\ell$ in which transport is ballistic. Above these scales transport is diffusive and still phase coherence up to the coherence length $L_\phi$}
\end{figure}
This section is inspired from our article \cite[sec.~II.A]{Hruza2023Coherent}, though some of the arguments made in the article have been re-examined and improved\footnote{I am grateful to Adam Nahum for the constructive discussion on this section}. The aim is to develop a picture in which QSSEP is the effective stochastic description of a generic diffusive mesoscopic system. Such a picture could emerge on hydrodynamic scales, when averaging over time and length scales corresponding to the mean free path $\ell$, see Fig.~\ref{fig:ballistic_cell}. Instead of showing this rigorously for a specific model, here we will start with a generic weakly-interacting fermionic system which satisfies classical MFT and introduce a noise average $\mathbb E$ that allows to reproduce the three important properties of QSSEP (i)-(iii). Our approach is rather heuristic than rigorous.

\begin{figure}[h]
	\centering
	\includegraphics{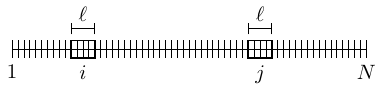}
	\caption{Ballistic cells $I$ and $J$ centred around sites $i$ and $j$. Noise emerges by averaging over all quadratic unitary transformations that only act on the individual cells and conserve the number of particles inside each cell.}\label{fig:coarse-graining}
\end{figure}

The main assumption is a separation of time scales: The original system can be decomposed into \textit{ballistic cells} of size $\ell$ that evolve as isolated free systems when observed during times shorter than a typical time scale $t_\ell$, see Figure \ref{fig:coarse-graining}. Only when the observation time is larger, ballistic cells can become correlated over lager distances and particles diffuse. In a first step we define the noise average of coherences $G_{ij}$ as a time average over $t_\ell$ plus a residual long-ranged noise, denoted by $\overline{\cdots}$. This promotes $G_{ij}$ to a random variable that is only sensitive to the long time behaviour (with respect to $t_\ell$) of the system. In a second step, we exploit the assumption of a separation of time scales to replace the time average by a local unitary average, that explores all particle conserving quadratic unitaries within a ballistic cell, but doesn't mix particles between cells,
\begin{equation}\label{eq:fluctuation_G}
	\mathbb E_t [G_{ij}]:=\frac 1{t_\ell} \int_t^{t+t_\ell} \overline{G_{ij}(t')} dt'
	=\overline{\Tr( \rho_t [c_i^\dagger c_j ]_U)}
\end{equation}
Here we denote the Haar average by $[O]_U:=\int d\mu(U) U^\dagger O U$ and $U=U^{(i)}U^{(j)}$ is a product of two quadratic unitaries acting respectively only on cells  $I:=\{i-\ell/2,\cdots,i+\ell/2\}$ and $J:=\{j-\ell/2,\cdots,j+\ell/2\}$, see Fig.~\ref{fig:coarse-graining}. If two indices are closer than a distance of $\ell$, i.e.\ $|i-j|<\ell$, then we take the two unitaries to be the same, $ U^{(i)}= U^{(j)}$. Otherwise, if $|i=j|>\ell$, they are independent.

Averages of several variables will evolve replicas. For example, 
\begin{equation}
	\mathbb{E}[G_{ij}G_{kl}]=\overline{\Tr(\rho\otimes\rho \cdot [c_i^\dagger c_j \otimes c_k^\dagger c_l]_{U^{(i)}U^{(j)}\otimes U^{(k)}U^{(l)}})},
\end{equation}
where the unitaries have been written out explicitly. This shows that the local unitary average can introduce correlations between $G_{ij}$ and $G_{kl}$ only if some of the indices are the same, for example if  $j=k$ and $i=l$. In the following we will justify the three properties (i)-(iii) on the basis of local unitary invariance alone.
\paragraph{Remark on the literature.}The idea that fluctuations of a chaotic quantum-many-body system can be characterised through a definition of ergodicity by unitary invariance has already been put forward in \cite{Bauer2020Universal}. The consequences of restricting such a global unitary invariance to local sectors of fixed energy has been explored in \cite{Jin2020Equilibration} and \cite{Foini2019Eigenstate} (the latter in the context of ETH, the former independently of ETH and unfortunately without getting much attention) though both talk about very similar things). Here we use similar ideas, but instead of local in energy, we restrict the unitary invariance to be local in space. 

\paragraph{(i) $U(1)$-invariance.}
By construction, the coarse-grained description has a local unitary invariance. In particular it is invariant under $U(1)$ transformations of the form $V=\exp(\sum_i \theta_i \hat n_i)$. These also conserve the particle number $\hat n_i=c_i^\dagger c_i$ on each site. For any $n$ replica operator, $O=O_1\otimes\cdots\otimes O_n$, the $U(1)$ invariance means that
\begin{equation}
	[(V^{\otimes n})^\dagger O V^{\otimes n}]_{U^{\otimes n}}=[O]_{U^{\otimes n}}
\end{equation}
This is only possible, if the total charge of $O$ under $U(1)$ is zero on each site. Since $c_i^\dagger$ and $c_i$ carry the charges $+1$ and $-1$, $O$ must be composed of an equal number of operators $c_i^\dagger$'s and $c_i$'s in order to have a non-zero expectation value\footnote{Note that $[\hat n_i,c_i^\dagger]=c_i^\dagger$ and  $[\hat n_i,c_i]=-c_i$}. As a consequence, only those expectation values $\E[G_{i_1,j_n}\cdots G_{i_n,j_n}]$ are non-zero, where the set $\{i_1,\cdots,i_n\}$ is a permutation of $\{j_1,\cdots,j_n\}$. For example in Eq.~\eqref{eq:fluctuation_G}, one needs $i=l,j=k$ or $i=j,k=l$.

\paragraph{(ii) Scaling of loops with system-size.}
The correct scaling when indices are not repeated follows from the fact that the original system satisfies the classical macroscopic fluctuation theory (MFT) \cite{Bertini2015MFT},
\begin{equation}
	 \langle n_{i_1} \cdots n_{i_n} \rangle_t^c \sim N^{1-n},
\end{equation}
where $N$ is the number of sites in the system.
Here $n_i$ denotes the classical particles density at site $i$ and $\langle\cdots\rangle_t$ is the MFT average.
The correspondence to the quantum description is
\begin{equation}\label{eq:density_classical_quantum}
	\langle n_{i_1} \cdots n_{i_n} \rangle_t = \mathbb E[\Tr(\rho_t \hat n_{i_1}\cdots \hat n_{i_n})]
\end{equation}
with $\mathbb E$ the average introduced above.
Since in the effective description $\rho_t$ evolves only through quadratic unitaries, we use Wick's theorem to expand this, see Appendix in \cite{Bernard2019Open}. Assuming by induction that loops of size $n-1$ scale as $N^{2-n}$, one can show that the only terms that survive for large $N$ in this expansion are loops of size $n$. For all indices distinct, that is
\begin{equation}
	\langle n_{i_1} \cdots n_{i_n} \rangle_t^c =(-)^{n-1}\frac{1}{n}\sum_{\sigma\in S_n} \mathbb E[G_{\sigma(i_1)\sigma(i_2)}\cdots G_{\sigma(i_n)\sigma(i_1)}],
\end{equation}
where the sum is over all permutations of $n$ elements. Therefore loops of order $n$ must also scale as $\mathbb E [G_{i_1 i_2}G_{i_2i_3}\cdots G_{i_n i_1}]\sim N^{1-n}$.

\paragraph{(iii) Factorisation of products of loops.}
If indices are repeated, we would like to show that the expectation value of loops factorizes. Here we will show that $\mathbb E[G_{ij}G_{ji}G_{il}G_{li}]=\mathbb E[G_{ij}G_{ji}]\mathbb E[G_{il}G_{li}]$ at leading order in $N$ and $\ell$. To do so, we will actually evaluate the resulting Haar averages. 

Note that for some quadratic unitary $U=e^{c^\dagger M c}$ with $c=(c_1,\cdots,c_\ell)$ and $M$ anti-hermitian, we have 
\begin{align}
	U^\dagger c_i U &= \sum_j u_{ij} c_j & U^\dagger c_i^\dagger U &= \sum_j u_{ij}^* c_j^\dagger & u:&=e^M\in U(\ell).
\end{align}
Furthermore, the non-zero average of up to four Haar unitaries $u$ are given by
\begin{align}
	[u_{aa'}u_{bb'}^*]&=\frac{1}{\ell}\delta_{ab}\delta_{a'b'} \label{eq:haar_averages_2}\\
	[u_{aa'} u_{bb'}^* u_{cc'} u_{dd'}^*] \label{eq:haar_averages_4}
	&= \frac{1}{\ell^2-1} (\delta_{ab}\delta_{a'b'}\delta_{cd}\delta_{c'd'}+\delta_{ad}\delta_{a'd'}\delta_{bc}\delta_{b'c'})+\Ord(\ell^{-3})
\end{align}
Let us denote by $[G_{ij}]$ the local unitary Haar average alone, such that $\mathbb E[G_{ij}]=\overline{[G_{ij}]}$. For example, using  Eq.~\eqref{eq:haar_averages_2} we have
\begin{equation}
	[G_{ij}]= \Tr(\rho_t [U^\dagger c_i^\dagger c_j U]_U)=\sum_{a\in I, \,b\in J} [{u_{ia}^{(i)}}^*u_{jb}^{(j)}] G_{ab} =\delta_{ij} \frac {1}{\ell}\sum_{a\in I} G_{aa}
\end{equation}
where $u^{(i)}$ and $u^{(j)}$ are independent Haar unitaries corresponding to cells $I$ and $J$.
Similarly, but only for all indices distinct, one evaluates
\begin{align}
	[G_{ij}G_{ji}] &=\frac{1}{\ell^2} \sum_{a\in I, \,b\in J} G_{ab} G_{ba} + \mathcal O(\ell^{-3})\\
	[G_{ij}G_{jk}G_{kl}G_{li}] &= \frac{1}{\ell^4} \sum_{a\in I, b\in J, c\in K, d\in L} G_{ab}G_{bc}G_{cd}G_{da} + \mathcal O(\ell^{-5})
\end{align}
Finally evaluating the case where two indices coincide $i=k$, one finds with the help of Eq.~\eqref{eq:haar_averages_4}
\begin{equation}
	[G_{ij}G_{ji}G_{il}G_{li}] = \frac{1}{\ell^2(\ell^2-1)}\sum_{a\in I, b\in J, c\in I, d\in L} (G_{ab}G_{ba} G_{cd}G_{cd}+G_{ab}G_{bc}G_{cd}G_{da}).
\end{equation}
The first term is equal to $[G_{ij}G_{ji}][G_{il}G_{li}]$, while the second term has rather the structure of $[G_{ij}G_{ji'}G_{i'l}G_{li}]$ but with $i$ and $i'$ treated as if they were in different cells. The only thing left to do is to take the long-ranged residual noise average. Here we need to assume $\overline{[G_{ij}G_{ji}][G_{il}G_{li}]}\approx \overline{[G_{ij}G_{ji}]}\;\,\overline{[G_{il}G_{li}]}$, since each factor is a sum over many terms which might decorrelate the long-ranged residual noise. Therefore, at leading order in $\ell$
\begin{equation}
	\mathbb E[G_{ij}G_{ji}G_{il}G_{li}] \approx \underbrace{\mathbb E[G_{ij}G_{ji}]}_{\sim 1/N} \underbrace{\mathbb E[G_{il}G_{li}]}_{\sim 1/N} + \underbrace{\mathbb E[G_{ij}G_{ji'}G_{i'l}G{li}]}_{\sim 1/N^3}.
\end{equation}
Using the scaling with $N$ from the last paragraph, one finds, that the second term is sub-leading\footnote{One needs to be a bit careful in taking the limits $1\ll \ell\ll N$}. This proves the claim about factorization of the expectation value for this specific example. Any other case can be done in an analogous manner.

\paragraph{Analogy with ETH.}\label{sec:general_ETH}
Simultaneously to our work it has been observed in \cite{Pappalardi2022ETH} that there is a very similar link to free probability in the context of the eigenstate thermalization hypothesis (ETH) which also builds on the idea of a local unitary invariance. Indeed, on the mathematical level, fluctuations of spatial coherences $G_{ij}$ in 1D mesoscopic systems seem to behave in complete analogy to matrix elements $A_{ij}=\langle E_i|A|E_j\rangle$ of observables in the energy basis of a closed system that obeys ETH: Both satisfy properties (i)-(iii). In the context of ETH, $A_{ij}$ is to be understood a random variable with respect to an fictitious ETH-random-matrix-ensemble that captures its typical behaviour. 

Comparing to \cite{Foini2019Eigenstate}, in which the authors introduce the three properties (i)-(iii) in the context of ETH, which they call ``general ETH'', one sees that the reasoning we have presented here is very similar: In general ETH, local unitary invariance follows from an average over small energy windows, whereas we considered averages over small windows in space (and time). Furthermore, for us the scaling parameter is the system size $N$, whereas for ETH it is the density of states $e^{S(E_+)}$ at the mean energy.

\section{Entanglement entropy}\label{sec:qssep_entanglement}
For the closed QSSEP, the stationary entanglement entropy has been calculated in \cite{Bernard2021Entanglement}. Here we show that the Renyi mutual information of the open QSSEP in the steady state satisfies a volume law, stressing that coherences are long-ranged and extensive. We also explore to what extend the dynamics of entanglement in QSSEP can be characterised analytically. Though, it turns out that the dynamical equations for the spectrum of coherences $G$, a crucial ingredient, do not close. The only information about entanglement growth we could obtain so far is through numerical simulations, which suggests a diffusive growth for all Renyi and van Neumann entropies.

Though our analytical calculations hold for all Renyi entropies, we usually consider the 2nd Renyi entropy. This is firstly in order to compare with the numerical estimates in Ref. \cite{Gullans2019Entanglement} and secondly since the phenomenology of the Renyi entropies in the steady state is usually the same\cite{Rakovszky2019Sub-ballistic}. The following paragraphs are taken from \cite[sec.\ 2]{Bernard2023Exact}, except for the last one on the entropy dynamics, which we present here the first time.

\paragraph{Definition of mutual information.}
Recalling the definition in Eq.~\eqref{eq:renyi}, consider the $q$th Renyi entropy of a segment $I\subset[0,1]$ of QSSEP of length $\ell_I$,
\begin{equation}
	S_I^{(q)} :=(1-q)^{-1}\log\mathrm{Tr}(\rho_I^{q}),
\end{equation}
where $\rho_I$ is the system's density matrix reduced to the segment $I$. Here we switched to a continuous description $x=i/N$ and view the system to be defined on $[0,1]$. The Reny entropies can be expressed in terms $d\sigma_I(\lambda)$, the spectrum or eigenvalue density of the matrix of coherences reduced to this segment $G_I:=(G_{i_1,i_2})_{x_1,x_2\in I}$. The intensive part of the Renyi entropies is then\footnote{Note that the integration limits are $\lambda\in[0,1]$ because for Gaussian fermionic states the relation $G=\frac{e^M}{1+e^M}$ bounds the eigenvalues $\lambda$ of $G$ to be in this range.}
\begin{equation}
	s_I^{(q)}:= \frac{S_I^{(q)}}{N}=\frac{\ell_I}{1-q} \int_0^1\! d\sigma_I(\lambda) \log[ \lambda^q +(1-\lambda)^q].
\end{equation}
In the limit $q\to1$ one obtains the (intensive part of the) van Neumann entropy
\begin{equation}
	s_I^{(1)}:=- \ell_I \int\! d\sigma_I(\lambda) \,[\lambda \log(\lambda) + (1-\lambda)\log(1-\lambda)].
\end{equation}
Since the system is in a mixed state, the entanglement entropy of a subsystem is not a meaningful quantity. Instead, we consider the (intensive part) of the mutual information between two adjacent intervals $I_1=[0,c]$ and $I_2=[c,1]$. 
\begin{equation} \label{eq:mutual-extensive}
	i^{(q)}(I_1:I_2) := s^{(q)}_{I_1} + s^{(q)}_{I_2} - s^{(q)}_{I_1\cup I_2}.
\end{equation}
where the contribution that is solely due to the state being mixed is subtracted.

\begin{figure}[t]
	\centering
	\begin{subfigure}{0.49\linewidth}
		\includegraphics[width=\textwidth]{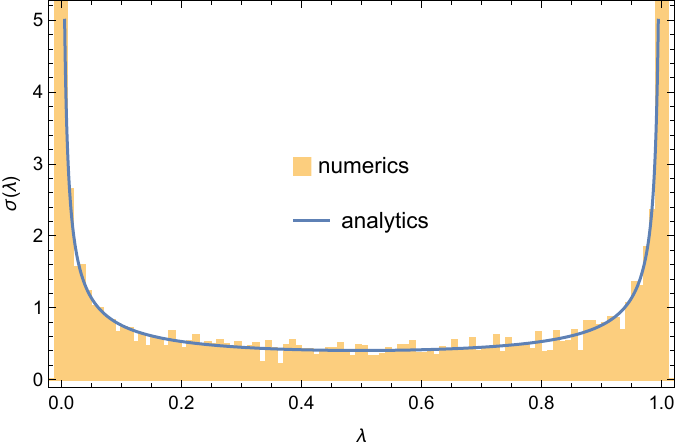}.
	\end{subfigure}
	\hfill
	\begin{subfigure}{0.49\linewidth}
		\includegraphics[width=\textwidth]{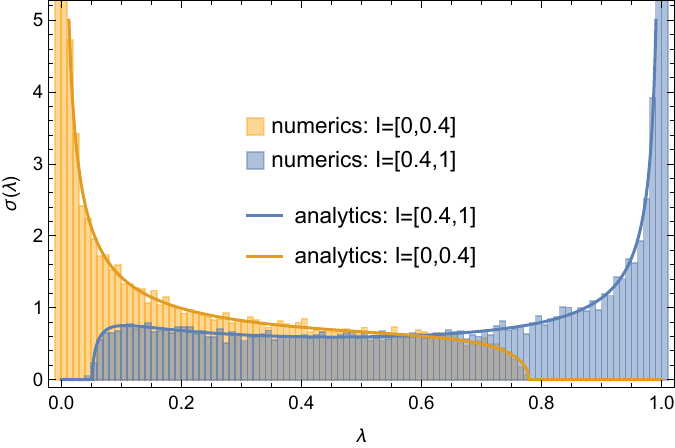}.
	\end{subfigure}
	
	\caption{\label{fig:spectrum}(Left) The spectral density $\sigma_I(\lambda)$ on the complete interval $I=[0,1]$. A comparison between the analytical prediction in Eq.~\eqref{eq:spectrum_c1} and a numerical simulation of $G$. The histogram of eigenvalues of $G$ corresponds to a single realization of the stochastic evolution of $G$.
		(Right) Spectral density $\sigma_I(\lambda)$ for the intervals $I=[0,0.4]$ and $I=[0.4,1]$. The support of the spectra is larger than the intervals $I$ and therefore the mutual information scales as the volume.}
\end{figure}

\subsection*{Steady state entanglement}
The spectrum $d\sigma_I$ on which the mutual information depends can be found using the variational principle \eqref{eq:action} for the spectrum of subblocks of structured random matrices. One also needs the exact solution \eqref{eq:g_steady} for the loop-cumulants $g_n$ of QSSEP in the steady state. In \cite[app.\ B and C]{Bernard2023Exact}, we show that for the interval $I_2=[c,1]$ this leads to the spectrum (again for $n_a=0,n_b=1$)
\begin{equation}\label{eq:spectrum_c1}
	d\sigma_{[c,1]}(\lambda) = \frac{d\lambda}{\pi \lambda(1-\lambda)}\,
	\frac{\theta}{\theta^2 + \log^2(re^{1/c})} \,  1_{\lambda\in[z_l(c),1]}.
\end{equation}
Here, $\theta$ and $r$ are functions of $\lambda$ implicitly defined through the (transcendental) equations
\begin{align}\label{eq:r_theta}
	1+ \log r &= r \xi \cos \theta,\quad \theta = r\xi \sin\theta ,
\end{align}
with $\xi=e^{1/c}(\frac{1-c}{c})(\frac{\lambda}{1-\lambda})$. The left boundary of the spectrum (the other being $\lambda=1$) is 
\begin{equation}\label{eq:left_boundary}
	z_l(c)=\frac{c}{c+(1-c)e^{1/c}}.
\end{equation}
From this, the spectrum of the reflected interval $I_1=[0,c]$ is obtained as
\begin{equation}
	d\sigma_{[0,c]}(\lambda)= d\sigma_{[1-c,1]}(1-\lambda)
\end{equation}
because $[0,c]$ is equivalent to $[1-c,1]$ by the exchange of the left and right reservoirs $n_a\leftrightarrow n_b$, which is equivalent to $\lambda\to1-\lambda$. 

For the spectral density in the generic case $I=[c,d]$, see \cite[sec.\ 3.3]{Bernard2024Structured}. Also note that in case of generic reservoir densities $n_a,n_b$, the eigenvalues of $G_I$ are 
\begin{equation}\label{eq:lambda_na_nb}
	\lambda_{n_a,n_b}=n_a + (n_b-n_a)\lambda,
\end{equation}
where $\lambda$ are the eigenvalues for $n_a=0,n_b=1$ distributed as above.

\paragraph{Volume law.}
The volume scaling of the mutual information can be inferred from the fact that the support of the spectra $d\sigma_{[0,c]}$ and $d\sigma_{[c,1]}$ are larger than the intervals $[0,c]$ and $[c,1]$ themselves. This is illustrated in Fig.~\ref{fig:spectrum}. As a consequence, the effective spectral density $d\sigma_c^{\text{eff}}=c\,d\sigma_{[0,c]}+(1-c)d\sigma_{[c,1]}-d\sigma_{[0,1]}$, over which one integrates in Eq.~\eqref{eq:mutual-extensive}, cannot be zero. 

\begin{figure}[t]
	\centering\includegraphics[width=0.6\linewidth]{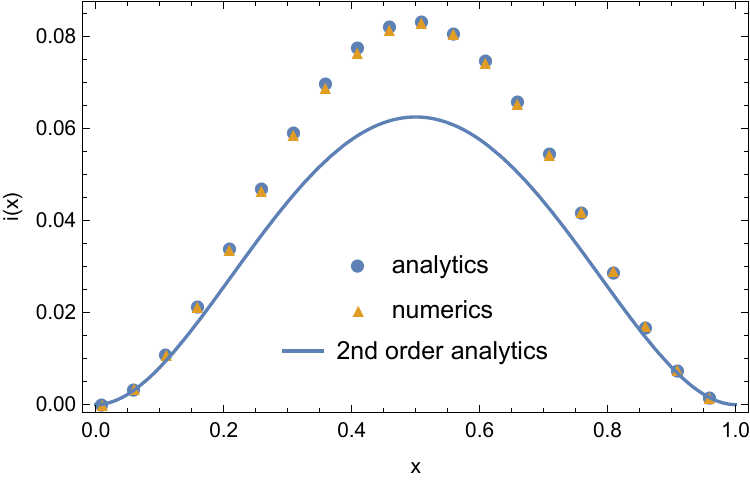}
	\caption{\label{fig:mutual-info}The intensive part $i(c):=i^{(2)}([0,c]:[c,1])$ of the $2$nd Renyi mutual information as a function of the cut at $c$. The ``analytical'' data points are obtained from Eq.~\eqref{eq:spectrum_c1} via a numerical solution of Eq.~\eqref{eq:r_theta}. They differ quite substantially from the second order contribution based solely on $g_2(x,y)$, which shows that the higher order local free cumulants $g_{n\ge3}$ are important. This is compared to ``numerical'' data points from a numerical simulation of the QSSEP dynamics on $N=100$ sites with discretization time step $dt=0.1$. Instead of averaging over many noisy realizations, we exploit the ergodicity of QSSEP and perform a time average of a single realization between $t=0.15$ and $t=0.4$. The QSSEP dynamics reaches its steady state at approximately $t=0.1$.}
\end{figure}

Fig.~\ref{fig:mutual-info} shows that the analytic result for the mutual information agrees perfectly with a numerical simulation. We included the result for the non-interacting random unitary circuit from Ref.~\cite{Gullans2019Entanglement} that takes into account only the second order fluctuations of coherences, but not their higher moments. The message is that the higher order fluctuations $g_n$, despite being very small in system size, are important for quantities such as the mutual information that involves an extensive sum over coherences.

To conclude, we point out that the volume law breaks down in the equilibrium case of equal reservoirs $n_a=n_b$. Due to Eq.~\eqref{eq:lambda_na_nb}, all eigenvalues are then equal to $n_a$ and the effective spectral density $d\sigma^\text{eff}_c$ is zero at leading order in $N$. To confirm an area law one would need to study the sub-leading terms.

\subsection*{Dynamics of entanglement}
\begin{figure}[t]
	\centering
	\includegraphics[width=0.475\textwidth]{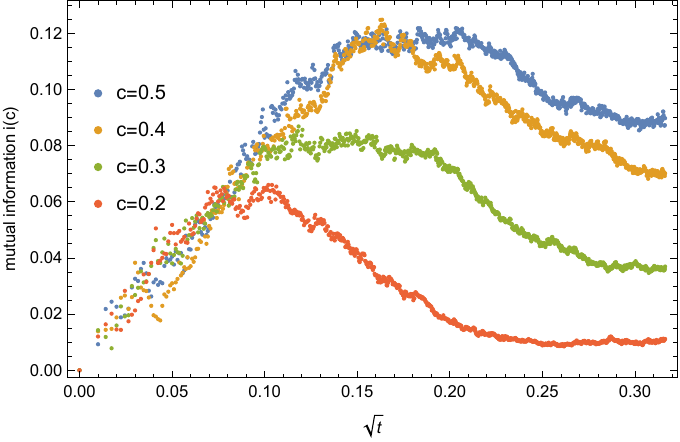}
	\hspace{0.1cm}
	\includegraphics[width=0.49\textwidth]{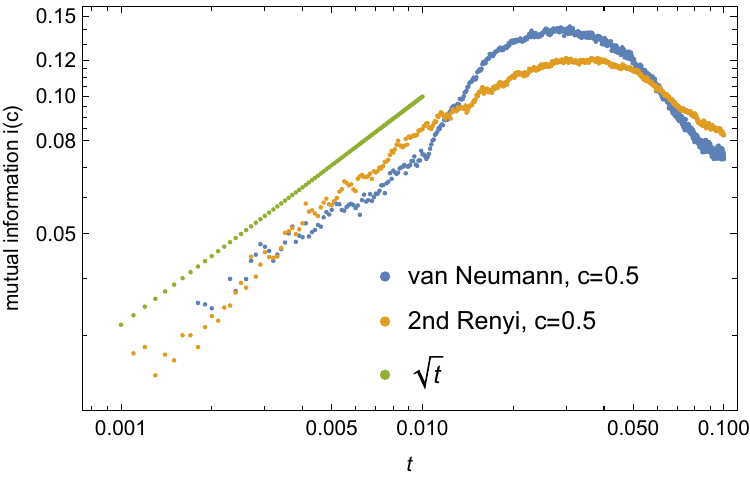}
	\caption{\label{fig:mutual_info_growth} (Left) The growth of the 2nd Renyi mutual information $i(c):=i^{(2)}([0,c]:[c,1])$ as a function of the square root of time for different values of the cut $c$. For each choice of $c$, the initial condition was a domain wall at $c$. One sees that initially $i(c)\sim \sqrt{t}$ with a slope independent of $c$ and that $i(c)$ approaches the steady state value at larger times. The numerical simulation was done on $N=100$ sites with discretization time step $dt=0.1$. Each curve is a single realization of the QSSEP dynamics. (Right) A log-log-plot to compare growth of the mutual information $i(c)$ between the 2nd Renyi and the van Neumann entropy, averaged over 10 realization of the noise. In both cases the initial growth is as $\sqrt{t}$, but at later times, the van Neumann entropy abruptly changes its slope, which suggest that superdiffusive contributions to the entanglement take over at this time.}
\end{figure}

For the analytical part, this section assumes knowledge of the variational principle Eq.~\eqref{eq:action} for the spectrum of subblocks of $G$ which is discussed in Section \ref{sec:spectrum_of_subblocks}. The less mathematically interested reader may content himself with the numerical simulation in Fig.~\ref{fig:mutual_info_growth}.  It shows that the mutual information growths initially with the square root of time, $i(c)=D_I\sqrt{t}$ where $D_I\approx 0.8$ independently of the cut $c$. Interestingly the value is different from the diffusion constant $D=1$, but this might be a finite size effect.

The dynamics of entanglement entropy and mutual information in QSSEP could in principle be inferred from the time evolution of the eigenvalue density $d\sigma_I$. This means, we are looking for the time evolution of the resolvent \eqref{eq:resolvent_I}, which can be rewritten as $G_I=\int_I a_z(x) \,dx$, with $a_z$ a solution of Eq.~\eqref{eq:a-b}. As we show now, the resulting equations do not close.

From the dynamics of the cumulants of coherences $g_n$ in Eq.~\eqref{eq:g_n_evolution}, we can derive a dynamical equation for the generating function $F_0$ from Eq.~\eqref{eq:F_0}. Under the assumption that $p(0)=p(1)=0$ one finds,
\begin{equation}\label{eq:F_0_dynamics}
	\partial_t F_0[p]=\int \left(p(x)\,\partial_x^2\frac{\delta F_0}{\delta p(x)}[p] +\left(p(x)\,\partial_x\frac{\delta F_0}{\delta p(x)}[p]\right)^2\right) dx.
\end{equation}
Consider the generating function $F[h](z)=\text{ext}_{a,b}\, F[a,b,h](z)$ in Eq.~\eqref{eq:action}. If we plug in $a_z$ and $b_z$ as solutions of the extremization conditions \eqref{eq:a-b}, we have
\begin{equation}
	\partial_t F[h](z)= \int_0^1 \Big(\underbrace{\frac{\partial F[a_z,b_z,h]}{\delta a_z(x)}}_0 \partial_t a_z(x)+\underbrace{\frac{\partial F[a_z,b_z,h]}{\delta b_z(x)}}_0 \partial_t b_z(x)\Big)dx-\partial_t F_0[a_z].
\end{equation}
Here the variation of $F[a,b,h]$ with respect to $a$ vanishes at the saddle point $a=a_z$. With $b_z(x)=\frac{\delta F_0}{\delta p(x)}[p=a_z]$, this leads to 
\begin{equation}
	\partial_t F[h](z) =-\int_0^1 \left(a_z(x) b_z''(x)+(a_z(x)b_z'(x))^2\right)dx.
\end{equation}
Since $\partial_t\partial_z F=\partial_z\partial_t F$ and $\partial_z F=(z-hb)^{-1}$ from Eq.~\eqref{eq:resolvent}, we can further simplify $\partial_t\int  (z-hb)^{-1}\, dx=-\partial_z\int(a_z b_z''+(a_zb_z')^2)dx$. For $h_I=1_I$ and $x\in I=[x_1,x_2]$, we can replace $b_z$ using the relation $a_z=(z-b_z)^{-1}$. Then,
\begin{equation}
	\partial_t \int_I  a_z \,dx=-\partial_z\int_I \Big(\frac{a_z'}{a_z}\Big)'dx=-\partial_z \left( \frac{a'_z(x_2)}{a_z(x_2)}-\frac{a_z'(x_1)}{a_z(x_1)}\right).
\end{equation}
This shows that in order to obtain the time evolution of the resolvent $G_I=\int_Ia_z\, dx$, we have to know the derivative of $a_z$ with respect to $x$ on the boundary of the interval. As a consequence, the equation does not close and we cannot proceed further. However, it might be possible to derive the initial growth of the mutual information with $\sqrt{t}$ by expanding in small $t$. But we have not tried this yet.
\section{Large deviation principle}\label{sec:large_deviation_principle}
Ultimately it would be nice to be able to express the probability of coherences $G$ as a large deviation principle,
\begin{equation}
	\mathbb P[G]\stackrel{N\to\infty}{\asymp} e^{-N \,I[G]}
\end{equation}
where $I[G]$ is the so-called \textit{rate function}. This function can be obtained as the Legendre transform of the cumulant generating function $w[Q]$ defined as
\begin{equation}\label{eq:closed_QSSEP_genfunc}
	\E[e^{N\, \tr(GQ)}]\stackrel{N\to\infty}{\asymp} e^{N w[Q]}
\end{equation}
where $Q$ has finite rank.

\paragraph{Closed QSSEP.} For the close QSSEP in the steady state, $w[Q]$ can be directly obtained from the connection between Haar randomly rotated matrices and free cumulants in Eq.~\eqref{eq:large-HCIZ}. Indeed, in this case, the matrix of coherences $G$ is distributed as $UG_0U^\dag$ where $U$ is a Haar distributed unitary of size $N$ and $G_0$ is the initial matrix of coherences. Denoting $\kappa_n$ the free cumulants of the spectral measure of $G_0$, we have
\begin{equation}
	w[Q]=\sum_n \frac 1 n \kappa_n \,\tr(Q^n).
\end{equation}

\paragraph{Open QSSEP.} A direct generalization of this formula to the open case is
\begin{equation}\label{eq:QSSEP_genfunc}
	w[Q]=\sum_n \frac 1 n\int g_n(\vec x) q(x_1,x_n)\cdots q(x_2,x_1) d\vec x
\end{equation}
where $Q_{ij}=\frac{1}{N} q(x,y)$ with $x=\frac i N, y=\frac j N$ and $\vec x=(x_1,\cdots,x_n)$. The proof goes as follows:

Consider $X=\tr(GQ)$ as a scalar random variable, then the relation between moments and classical cumulants in Eq.~\eqref{eq:classical_cumulant_genfunc} allows us to write
\begin{equation}
	Nw[Q]=\sum_{n\ge1} \frac{N^n}{n!} \E[(\tr(GQ))^n]^c=\sum_{n\ge1} \frac{N^n}{n!} \sum_{ij}\E[\tilde G_{i_1j_1}\cdots \tilde G_{i_nj_n}]^c,
\end{equation}
where we abbreviate $\tilde G_{i_1j_1}= G_{i_1j_1} Q_{j_1i_1}$ and also $i=(i_1,\cdots,i_n)$ and $j=(j_1,\cdots,j_n)$. In the sum over $i$ one can keep only distinct indices, taking away all the terms where two or more indices are equal, which introduces an error that is only sub-leading in $N$. Then, due to $U(1)$ invariance of the measure, one has
\begin{equation}
	\sum_{i \text{ distinct}} \sum_{j}\E[\tilde G_{i_1j_1}\cdots \tilde G_{i_nj_n}]^c= \sum_{i \text{ distinct}} \sum_{\sigma\in S_n}\E[\tilde G_{i_1i_{\sigma(1)}}\cdots \tilde G_{i_ni_{\sigma(n)}}]^c
\end{equation}
where $\sigma\in S_n$ is a permutation of $n$ elements. Permutations consisting of a complete cycle such as $\sigma=(1\cdots n)$ produce terms of the form $\E[G_{i_1i_2}\cdots G_{i_ni_1}]^c\sim\Ord(N^{1-n})$, while all other permutations, consisting of more than one cycle produce sub-leading terms. For example $\sigma=(1)(2\cdots n)$ leads to $\E[G_{i_1i_1} G_{i_2 i_3}\cdots G_{i_ni_2}]^c\sim\Ord(N^{-n})$. Therefore, one keeps the $(n-1)!$ complete cycles which all give the same contribution
\begin{equation}
	Nw[Q] = \sum_{n\ge1} \frac{N^n}{n} \sum_{i \text{ distinct}} \E[G_{i_1i_2}\cdots G_{i_ni_1}]^c Q_{i_1i_n}\cdots Q_{i_2i_1}.
\end{equation}
Adding terms where indices $i$ are equal to the sum will again only make a sub-leading error. Therefore, one can replace the sum by an integral, using the scaling of $Q$ and $G$, which leads to Eq.~\eqref{eq:QSSEP_genfunc}.

\section{Test for integrability}
One the level of the mean $\bar \rho_t=\E_t[\rho]$, QSSEP corresponds to SSEP by construction, and is therefore integrable. But what about the fluctuations in QSSEP, are they also integrable? For example, the quadratic fluctuations of an operator $O$ are given by $\E_t[\Tr(\rho O)\Tr(\rho O)]$. In other words, they are encoded into the two-replica mean 
\begin{equation}
	\bar \rho^{(2)}_t=\E_t[\rho\otimes \rho].
\end{equation}
One can show that it satisfies $\partial_t \bar \rho^{(2)}=\Lin^{(2)}(\bar \rho_t^{(2)})$ with a Lindbladian that is obtained from the one-replica Lindbladian in Eq.~\eqref{eq:lindblad_ssep} via the replacement
\begin{equation}
	\ell^\pm_{j} \to \ell^\pm_{j}\otimes \mathbb{I} + \mathbb{I}\otimes \ell^\pm_{j} = \ell^\pm_{1,j} + \ell^\pm_{2,j},
\end{equation}
where fermions in different replicas commute. Now we can ask, is the two-replica Lindbladian integrable? More generally, is the $R$-replica Lindbladian integrable? 

This is the question we explored in \cite{Bernard2022DynamicsClosed} in the context of the closed QSSEP and its asymmetric version QASEP\footnote{Actually, we worked in the Heisenberg picture $\Tr(\rho_t O)=\Tr(\rho_0 O_t)$ where operators are time dependent and evolve in mean with a dual Lindbladian, $\partial_t\bar O_t=\Lin^*(\bar O_t)$. For QSSEP, however, $\Lin=\Lin^*$, so we continue working with $\Lin$.}. Here we will only comment on the QSSEP. First, we explain the algebraic structure of the two-replica Lindbladian which turns out to possess a global $\mathfrak{gl}(4)$ symmetry and can be represented through its generators. Then we compare the two-replica Lindbladian to known integrable spin chains in order to see if they can be identified, which would prove integrability for the two-replica Lindbladian. Finally we present a numerical investigation on the level spacing statistic of the two-replica Lindbladian, which is an indicator of integrability or integrability-breaking.

\paragraph{Algebraic structure.}
As derived in our paper, the two-replica Lindbladian has a global $\mathfrak{gl}(4)$ symmetry and can be expressed in terms of $\mathfrak{gl}(4)$-generators on each site $G^{AB}$ ($A,B=1,\cdots,4$). They satisfy the commutation relations 
\begin{equation}\label{eq:gl_commutator}
	\big[G^{AB},G^{CD}\big]=\big(\delta^{BC}G^{AD}-\delta^{DA}G^{CB}\big)
\end{equation}
and they act on the local $16$-dimensional fermionic Hilbert space of a site in the two-replica system. For concreteness, in Appendix \ref{app:gl4_generators}, we have them written out as matrices acting on an explicit local basis. The two-replica Lindbladian takes the form
\begin{equation}\label{eq:lindblad_gl4}
	\Lin^{(2)} = \sum_j \Big( \sum_{A,B} G_{j+1}^{AB}G_j^{BA} - \frac{1}{2}(C_{j+1}+C_j) - 2 \Big),
\end{equation} 
where

counts the total $\mathfrak{u}(1)$ charge on site $j$. In this form, the global $\mathfrak{gl}(4)$ symmetry is evident, since $\Lin^{(2)}$ commutes with $\sum_j G_j^{AB}$. In addition to this, $\Lin^{(2)}$ commutes with $C_j$ on each site. Therefore the action of $\Lin^{(2)}$ on each site splits up into five invariant sectors corresponding to $C_j=0,\pm1,\pm2$. The splitting of the dynamics into sectors has also be been termed ``fragmentation'' in Ref. \cite{Essler2020Integrability}.

Since $C$ also commutes with all generators $G^{AB}$, the $16$-dimensional representation on each site is reducible into the five sectors with $C=0,\pm1,\pm 2$. Splitting $\mathfrak{gl}(4)=\mathfrak{sl}(4)\oplus\mathfrak{u}(1)$, the irreducible $\mathfrak{sl}(4)$ representations in each sector are: A $6$-dimensional representation (the antisymmetric rank-2 tensor representation, with $\mathfrak{sl}(4)\equiv\mathfrak{so}(6)$ this is equivalent to the vector representation of $\mathfrak{so}(6)$), a $4$-dimensional one (the vector representation of $\mathfrak{sl}(4)$) and its conjugate, a 1-dimensional one (the scalar representation) and its conjugate. That is,
\begin{equation}
	\ytableausetup{smalltableaux,centertableaux}
	[16]=[1]\oplus [4] \oplus [6] \oplus [\bar 4] \oplus [\bar 1] \equiv 
	\bullet \oplus \ydiagram{1} \oplus \ydiagram{1,1} \oplus \overline{\ydiagram{1}} \oplus \overline \bullet ,
\end{equation}
In terms of $\mathfrak{sl}(4)$ generators $J^{AB}:=G^{AB} - \frac{1}{4}\delta^{AB}(C+2)$ such that $\sum_A J_j^{AA}=0$, the two-replica Lindbladian is
\begin{equation}
	\Lin^{(2)}  = \sum_j\Big( \sum_{A,B} J_{j+1}^{AB}J_j^{BA}+ \frac{1}{4}\,C_{j+1}C_j - 1\Big).
\end{equation}

Also note, that the same algebraic construction as in \eqref{eq:lindblad_gl4} can be applied to the case of $R$ replica if instead of $\mathfrak{gl}(4)$ we use $\mathfrak{gl}(2R)$ as the symmetry algebra. In this case one has to modify the definition of $C=\sum_A G^{AA}-\frac{R}{2}$ and there are $R+1$ invariant sectors with charge $0,\pm1,\cdots,\pm R$.

\paragraph{Integrable spin chains.}
The algebraic construction above provides us with a two-replica Lindbladian in the form of a $\mathfrak{gl}(4)$ spin chain with five invariant sectors on each site. Here we investigate if any of these sectors can be identified with known integrable spin chains. We will focus on particular global sectors that have the same $C$-charge on every site,
\begin{equation}
	\cvec_0\equiv (0,\dots,0)\ ,\quad
	\cvec_{\pm1}\equiv (\pm1,\dots,\pm 1)\ ,\quad
	\cvec_{\pm2}\equiv (\pm2,\dots,\pm 2)\ .
\end{equation}
These five sectors are the only gapless ones and they contain the zero modes (steady states) of the Lindbladian, see \cite[section 5.1 and 5.2]{Bernard2022DynamicsClosed} for details. In all other non-homogeneous sectors, the Lindbladian exhibits a finite spectral gap. Hence states belonging to these sectors decay exponentially fast in time, even in the large system size limit.

\subparagraph{The $\cvec_{\pm 1}$ sector.}
Here the local two-replica Lindbladian $\sum_{AB} J^{AB}\otimes J^{BA}$ (up to a constant) acts on the tensor product of two $\mathfrak{sl}(4)$ vector representations, which is reducible into symmetric and antisymmetric rank-2 tensors, $\ydiagram{1} \otimes \ydiagram{1}= \ydiagram{2}\oplus \ydiagram{1,1}$.
The associated projectors of $\sum_{AB} J^{AB}\otimes J^{AB}$ onto these subspaces are the identity $1$ and the permutation operator $P$ and one finds
\begin{equation}
	\Lin^{(2)}=\sum_j (P_{j,j+1}+\frac{1}{4}).
\end{equation}

This is the $\mathfrak{sl}(4)$ version of the isotropic Heisenberg spin chain, and it is known to be integrable \cite{Sutherland1975model}. In other words, the $\cvec_{\pm 1}$ sector is integrable.

\subparagraph{The $\cvec_0$ sector.}
This sector corresponds to the vector representation $\square$ of  $\mathfrak{so}(6)$ and one needs to evaluate how the local Lindbladian acts on $\ydiagram{1} \otimes \ydiagram{1} = [20] \oplus [15] \oplus [1]$. It decomposes into traceless symmetric rank-2 tensors, anti-symmetric rank-2 tensors and the trace. The projectors on these subspace are (with $d=6$)
\begin{equation}
	P_S=\frac{1}{2}(1+P)-\frac{1}{d}Q,\ P_A=\frac{1}{2}(1-P),\ P_\bullet= \frac{1}{d}Q
\end{equation}
with $P$ the permutation operator and $Q$ the so-called trace operator. Computing explicitly the tensor Casimir $\sum_{AB} J^{AB}\otimes J^{AB}$, see \cite[appendix B]{Bernard2022DynamicsClosed}, one finds
\begin{equation}
	\Lin^{(2)}= \sum_j (P_{j;j+1}-Q_{j;j+1}-1)\ .
\end{equation}
The Hamiltonian of the $\mathfrak{so}(6)$ integrable spin chain \cite{reshetikhin1985integrable} is $H= \sum_j (P_{j;j+1} - \frac{1}{2} Q_{j;j+1})$. Due to the different factor in front of $Q$, the two-replica Lindbladian cannot be identified with this known integrable model and the question, if it is integrable, stays open.

A similar analysis applies to higher number of replicas. In particular, because they are associated to the $\mathfrak{sl}(2R)$ vector representations, the dynamics in the sectors with all $C$ equal to $\pm(R-1)$ are always integrable in the usual sense (they are mapped to the $\mathfrak{sl}(2R)$ analogues of the isotropic Heisenberg spin chain). Whether the other sectors are integrable is an open question.

\paragraph{Level-spacing statistics.}\label{sec:level-spacing} In the following we mainly cite from \cite[sec.\ 4.2]{Bernard2022DynamicsClosed}: Most studies that have been conducted on the level-spacing statistics of integrable spin chains deal with the case where the global symmetry algebra is $\mathfrak{u}(1)$ or $\mathfrak{sl}(2)$. But the symmetry algebra $\mathfrak{gl}(4)$ encountered here is of higher rank. As explained below, this leads to a new source of degeneracies of eigenvalues, which makes the level statistics potentially differ from the usual $\mathfrak{sl}(2)$ case. To our knowledge, this is the first time this problem has been addressed in the literature. The result of our analysis is that the $\cvec_{\pm 1}$ sectors are indeed integrable, while for the $\cvec_0$ sector is probably not. Effects of integrability breaking perturbations become visible only for sufficiently large system sizes \cite{szaszschagrin2021weak, Modak2014Universal} and we suspect that the maximal system size we could use here was too small to get a completely consistent picture in this sector.

\subparagraph{Eigenvalues in RMT.} Let us start by recalling some known results on the eigenvalue and level-spacing statistics of integrable and non-integrable models in the context of random matrix theory. Integrable Hamiltonians possess the very particular property that their eigenvalues are i.i.d. random variables, as if the Hamiltonian was just a random diagonal matrix. This was first conjectured by Berry and Tabor \cite{berry1977level} and has been confirmed in many explicit examples such as the XXX Heisenberg chain \cite{Poilblanc1993Poisson,Kudo2005Level}. Importantly, the spacing $s_n=e_n-e_{n-1}$ between adjacent eigenvalues follows an exponential distribution
\begin{equation}\label{eq:poisson}
	p(s)=e^{-s}.
\end{equation} 
To be precise, this holds only for the so-called ``unfolded spectrum'' of the Hamiltonian, where one performs a local change of variable on the eigenvalues $e_n$ such that the density of the new variables $\hat e_n$ is uniform (see Appendix \ref{app:unfolding}). Instead, as showed in \cite{atas2013distribution}, one can also consider the ratio of consecutive spacings $r_n=s_{n}/s_{n-1}$ whose distribution is independent of the local density of eigenvalues and is given by
\begin{equation}\label{eq:ratio}
	p(r)=\frac{1}{1+r^2}\ .
\end{equation}

In contrast to integrable Hamiltonians, the eigenvalues of a generic Hamiltonian -- a random Matrix -- tend to repel each other. The spacing between eigenvalues of a $2\times2$ random matrix in the GOE (Gaussian Orthogonal Ensemble) -- which would be the appropriate ensemble to deal with since the Q-SSEP Linbladian is symmetric and real -- has a probability distribution know as \textit{Wigners surmise}
\begin{equation}\label{wigner}
	p(s)=\frac{\pi s}{2}e^{-\pi s^2/4}\ .
\end{equation}
This turns out to be a good approximation also for the level-spacing of large GOE random matrices. In particular, there is a zero probability to find consecutive eigenvalues with spacing zero. The same is true if instead of the spacing one again considers the ratio of adjacent spacings $r$, which behaves as $p(r)\sim r^\beta$
for small $r$, where $\beta=1,2,4$ is the Dyson index of the matrix ensemble \cite{atas2013distribution}.

The ratio of adjacent spacings has the nice property, that the average of
\begin{equation}
	\tilde r=\mathrm{min}(r,1/r)\in[0,1]
\end{equation}
over the given ensemble is a constant, and can therefore be used to classify the ensemble which a numerical distribution might belong to. One finds\footnote{For Poisson statistics the derivation is easy: Since the probability of consecutive spacings $s_1$ followed by $s_2$ is equivalent the inverse order, i.e. $\text{Prob}(s_1,s_2)=\text{Prob}(s_2,s_1)$, it follows that $r=s1/s2$ and $1/r=s_2/s_1$ have the same distributions. Therefore $\text{Prob}(\tilde r)=2p(r)\Theta(1-r)$, from which the average value $\langle \tilde r \rangle=2\ln2-1$ can be computed.}, \begin{align}
	\langle \tilde r \rangle_\text{Poisson}&=2\ln2-1\approx0.3863, & \langle \tilde r \rangle_\text{GOE}&\approx0.5359.
\end{align}

\subparagraph{Reduction to symmetry sectors.} Before discussing the results, let us also comment on the reduction of the Lindbladian to its remaining symmetry sectors. The eigenvalues in each symmetry sector are statistically independent and therefore one should treat each sector independently. In practice, we bring the Lindbladian to block diagonal form with respect to all its mutually commuting symmetries $I_i$, $[\mathcal{L},I_i]=0$, $[I_i,I_j]=0$. After fixing to $\cvec_{\pm1}$ or $\cvec_0$, the maximally set of commuting symmetries are translation $T$, the three Cartan elements of $\mathfrak{sl}(4)$ (which are the analogues of the magnetization for $\mathfrak{sl}(2)$) and depending on the choice of the three Cartan elements, a permutation $F$ of the states (which is the generalization of a spin flip in the $m=0$ sector for $\mathfrak{sl}(2)$). The three Cartan elements $(J_1^z,J_2^z,\frac{C_1-C_2}{2})$ are built from $\mathfrak{gl}(2)$ one-replica operators that are embedded into the two-replica operators as (with $A=\sum_j A_j$).
\begin{align}
	J_{1,j}^z&=G_j^{11}-G_j^{22} &
	C_{1,j}&=G_j^{11}+G_j^{22}-1 \nn
	J_{2,j}^z&=G_j^{44}-G_j^{33} &	
	C_{2,j}&=G_j^{33}+G_j^{44}-1
\end{align}
\begin{figure}[t]
	\centering
	\begin{subfigure}[b]{0.32\textwidth}
		\centering
		\includegraphics[width=\textwidth]{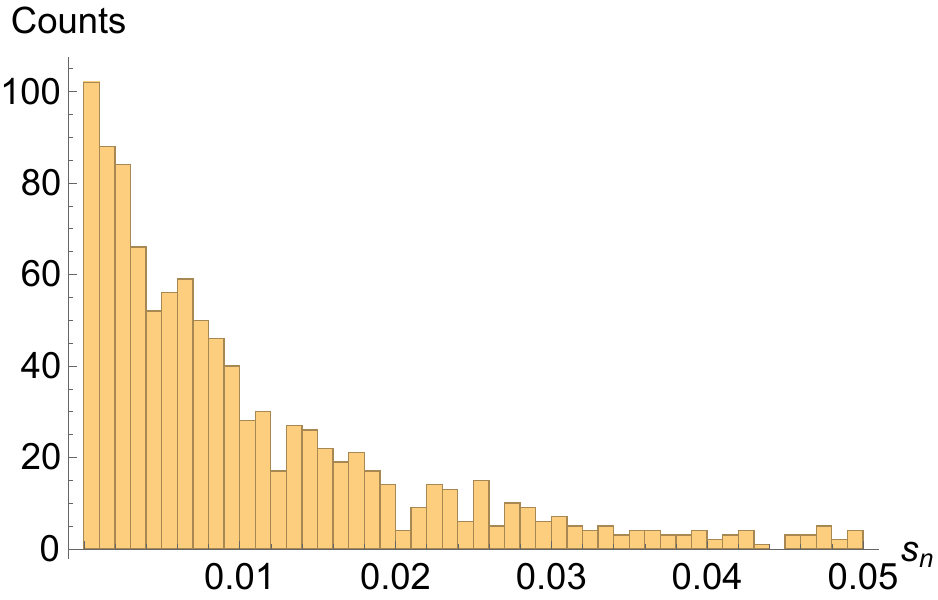}
		\caption{Spacings $s_n=|e_n-e_{n-1}|$ of adjacent eigenvalues}
	\end{subfigure}
	\hfill
	\begin{subfigure}[b]{0.32\textwidth}
		\centering
		\includegraphics[width=\textwidth]{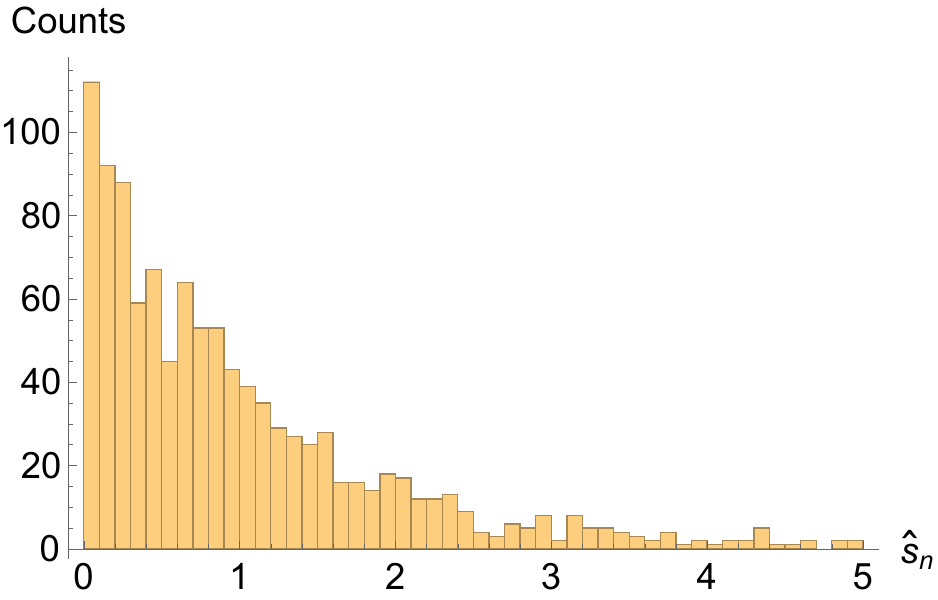}
		\caption{Spacings $\hat s_n=|\hat e_n-\hat e_{n-1}|$ of unfolded eigenvalues $\hat e_n$}
		\label{fig:c=1_unfolded}
	\end{subfigure}
	\hfill
	\begin{subfigure}[b]{0.32\textwidth}
		\centering
		\includegraphics[width=\textwidth]{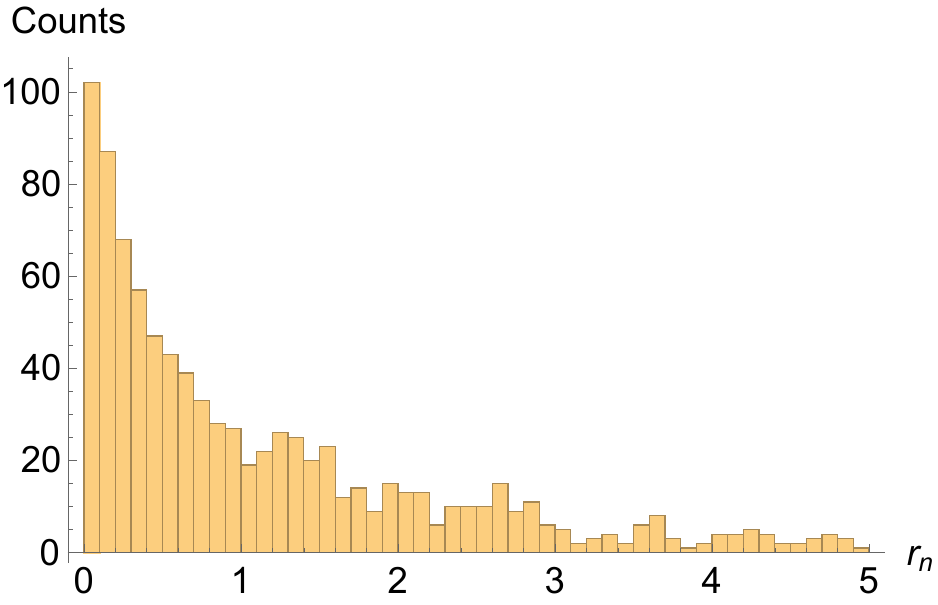}
		\caption{Ratio $r_n=s_n/s_{n-1}$ of adjacent spacings}
		\label{fig:c=1_ratio}
	\end{subfigure}
	\caption{Histograms for the spacings and ratios of the $N_\text{eig}=1077$ eigenvalues of the $\cvec_1$ Lindbladian on $N=11$ sites after the degeneracies have been removed. The chosen symmetry sector is defined by $k=2\pi/11$ ($e^{ik}$ is the eigenvalue under translation by one site) and $(n_1,n_2,n_3,n_4)=(1,2,3,5)$ where $n_i$ is the number of times the local state $i=1,2,3,4$ appears in the tensor-product-state (the $\cvec_1$ sector has local dimension $4$ on each site). This corresponds to the Cartan charges $(J_1^z, J_2^z, \frac{C_1-C_2}{2})=(-2,-3,\frac{3}{2})$. The average ratio of consecutive spacings is $\langle \tilde r \rangle =0.3826$.}
	\label{fig:c=1}
\end{figure}
However, once all these charges have been fixed, there is still a degeneracy in the eigenvalues of the Lindbladian left, which would lead to an artificial large peak at zero in the level-spacing statistic. This is because the Cartan subalgebra for $\mathfrak{sl}(4)$ consists of more than one element, hence there are more than one ``lowering-operator'' and therefore weight-spaces in an irreducible $\mathfrak{sl}(4)$ representation can be more than one-dimensional. But the $\mathfrak{sl}(4)$ symmetry of the Lindbladian ensures that all states in an irreducible $\mathfrak{sl}(4)$ representation have the same eigenvalue and hence, selecting a weight-space (i.e. fixing the Cartan charges) will not lift all degeneracies (as it would do for the $\mathfrak{sl}(2)$ spin chain). We therefore manually deleted all the degenerate copies of eigenvalues from the complete set of eigenvalues in a given symmetry sector and analysed the level-spacing and ratio statistics for the remaining eigenvalues. The procedure is not entirely correct, because there can also be degeneracies in the spectrum solely due to integrability, which would be neglected in our procedure. But it turns out that for the overall statistics this only plays a minor role.

\begin{figure}[t]
	\centering
	\begin{subfigure}[b]{0.32\textwidth}
		\centering
		\includegraphics[width=\textwidth]{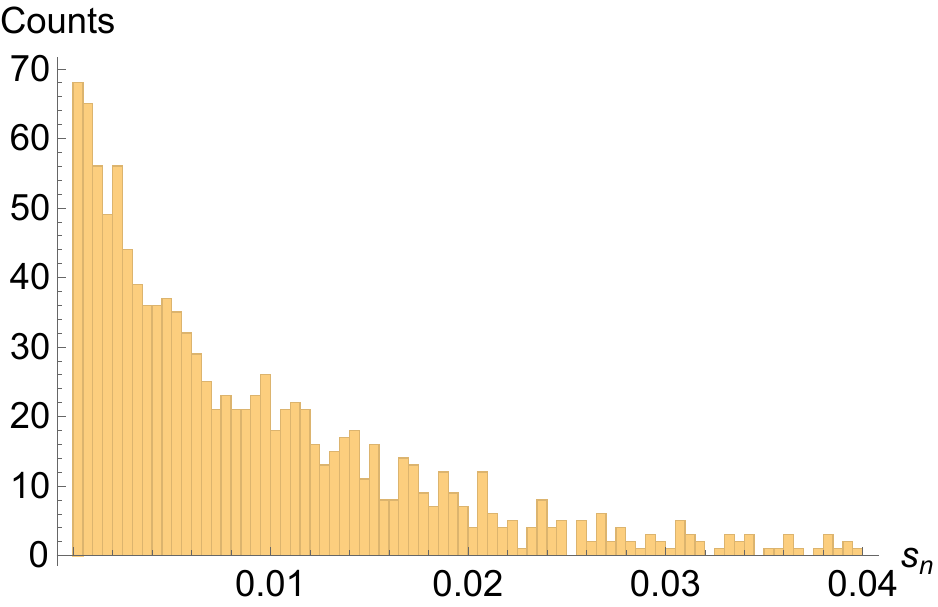}
	\end{subfigure}
	\hfill
	\begin{subfigure}[b]{0.32\textwidth}
		\centering
		\includegraphics[width=\textwidth]{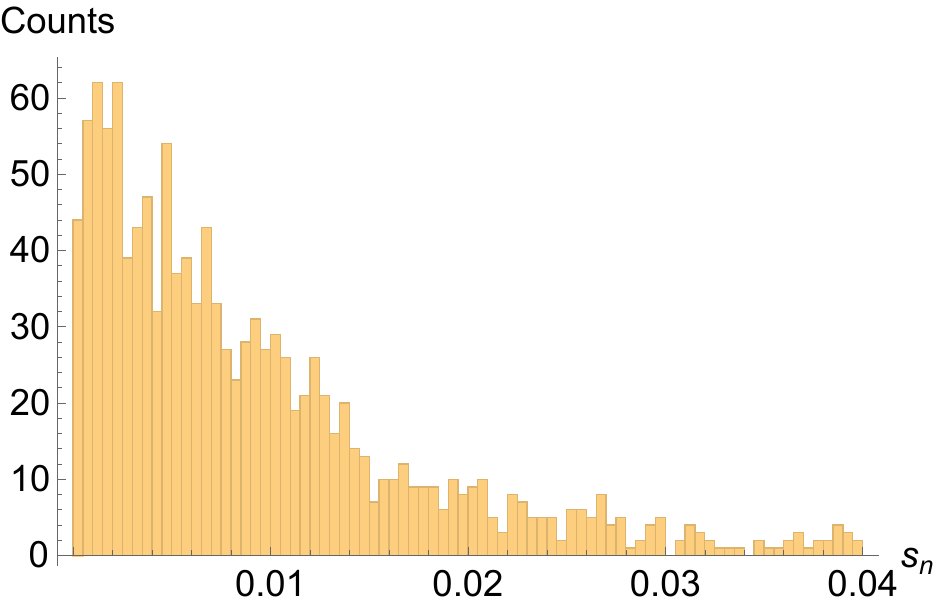}
	\end{subfigure}
	\hfill
	\begin{subfigure}[b]{0.32\textwidth}
		\centering
		\includegraphics[width=\textwidth]{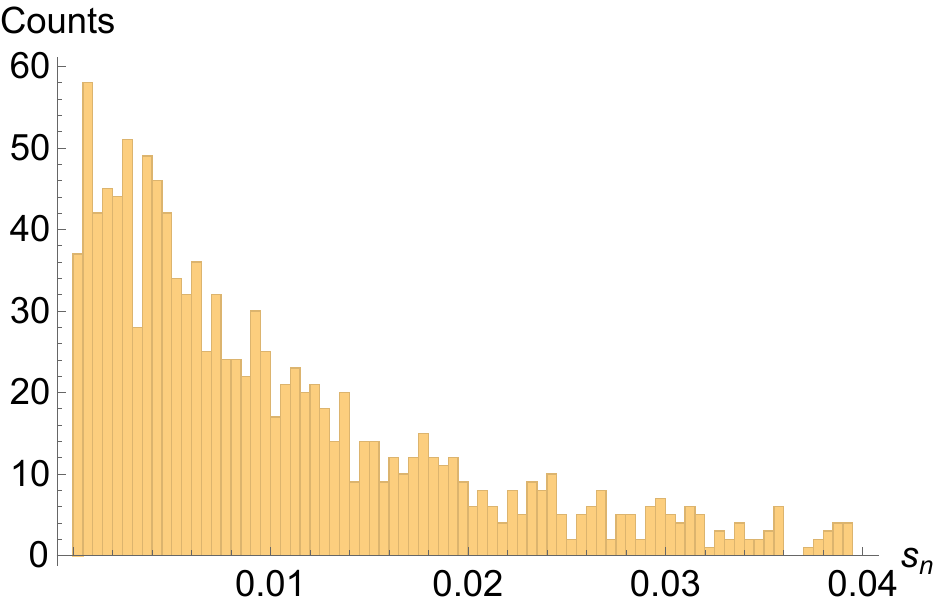}
	\end{subfigure}
	
	\begin{subfigure}[b]{0.32\textwidth}
		\centering
		\includegraphics[width=\textwidth]{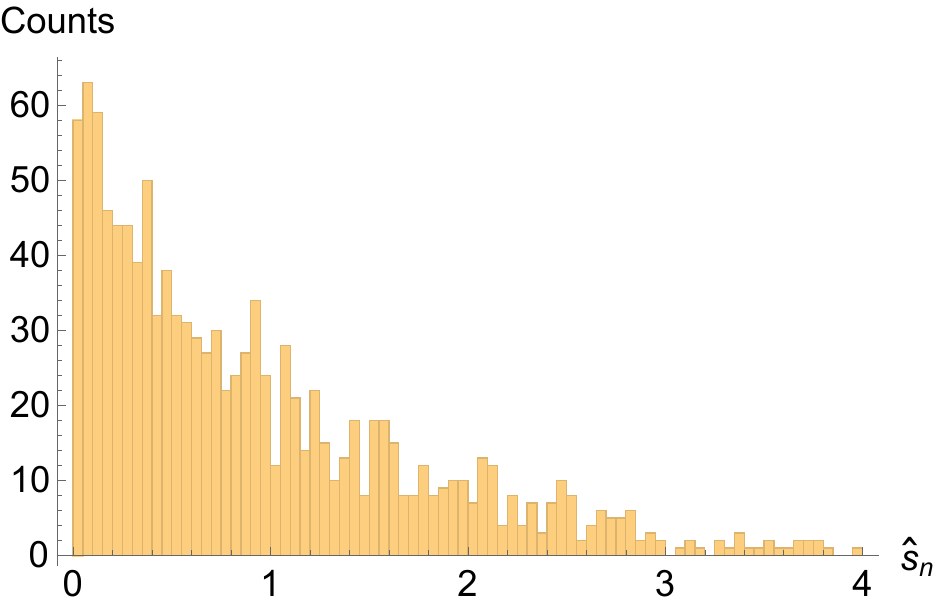}
		\centering\caption{Integrable ($g=1/2$), $N_\text{eig}=1205$, $\langle \tilde r \rangle =0.3817$}
	\end{subfigure}
	\hfill
	\begin{subfigure}[b]{0.32\textwidth}
		\centering
		\includegraphics[width=\textwidth]{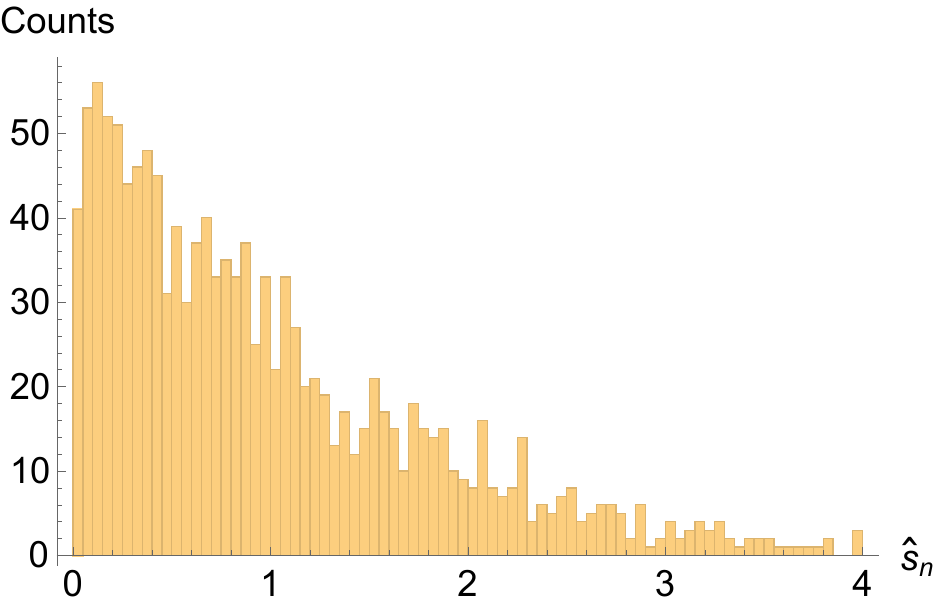}
		\centering\caption{Q-SSEP ($g=1$), $N_\text{eig}=1335$, $\langle \tilde r \rangle =0.3980$}
	\end{subfigure}
	\hfill
	\begin{subfigure}[b]{0.32\textwidth}
		\centering
		\includegraphics[width=\textwidth]{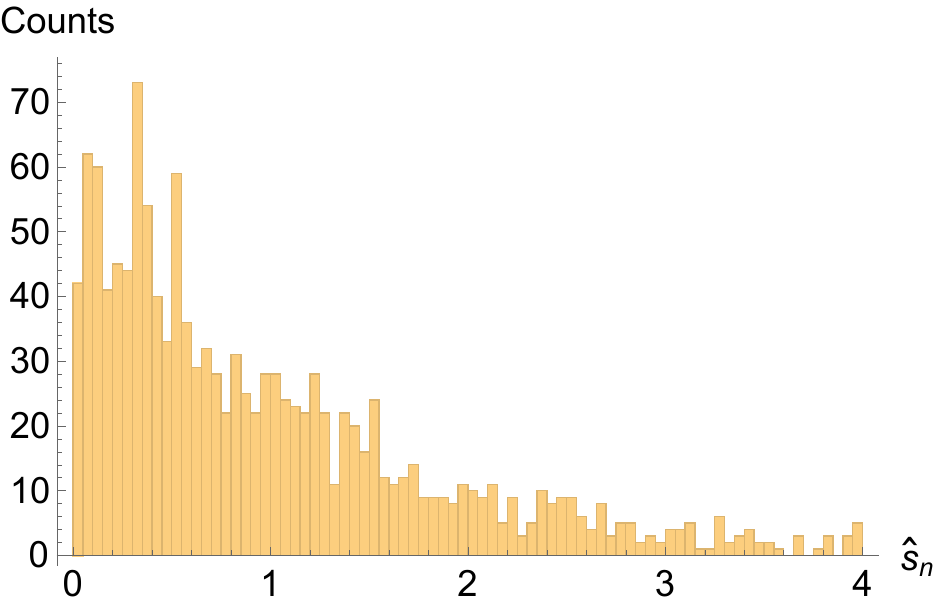}
		\centering\caption{Perturbation with $g=2$, $N_\text{eig}=1336$, $\langle \tilde r \rangle =0.3950$}
	\end{subfigure}
	\hfill
	\caption{Histograms of the spacings between consecutive raw eigenvalues (first row) and unfolded eigenvalues (second row) for the QSSEP two-replica Lindbladian and deformations of it, on $N=10$ sites in the $\cvec_0$ sector. The deformations are parametrised by $g$ and defined as $\mathcal{L}^{(2)}(g)= \sum_j (P_{j;j+1}-gQ_{j;j+1}-1)$ such that $g=1/2$ corresponds to the integrable isotropic $\mathfrak{so}(6)$ spin chain, $g=1$ to QSSEP and $g=2$ to a unknown model that is probably non-integrable. The chosen symmetry sector is defined by $k=2\pi/10$ and Cartan charges $(J_1^z,J_2^z,\frac{C_1-C_2}{2})=(5,3,3)$. The value $N_\text{eig}$ provides the number of eigenvalues left after removing the degeneracies.}
	\label{fig:c=0_L=10}
\end{figure}

\subparagraph{Numerical results.} Fig.~\ref{fig:c=1} shows the results for the $\cvec_1$ sector and both the shape of the distribution and the value for $\langle\tilde r \rangle$ suggest that this model is integrable. The results for the $\cvec_0$ sectors are less clear. In some of the sectors with fixed Cartan elements there is no sign of integrability breaking. In others, such as in Fig.~\ref{fig:c=0_L=10} there are weak signs. There we consider a deformed version of the QSSEP two-replica Lindbladian
\begin{equation}
	\mathcal{L}^{(2)}(g)= \sum_j (P_{j;j+1}-gQ_{j;j+1}-1)
\end{equation} 
such that $g=1/2$ is integrable and $g=1$ is QSSEP. For $g=1/2$ the exponential distribution is correctly reproduced. Increasing the perturbation $g$ one observes a gradual deviation from this distribution. However, a proper Wigner-Dyson distribution is not visible, even for higher values of $g$. We suspect this to be due to the small system size (here $N=10$) we are able to achieve in practice. Also note, that the value for $\langle\tilde r\rangle$ seems to suggest integrability breaking for QSSEP. However, in our example, it does not consistently increase with $g$ as one would expect: The value for $g=2$ is lower than that for $g=1$. Again we think this is due to the limited system size. Finally, it might be interesting to look at the value of $N_\text{eig}$, which describes the number of eigenvalues left after removing the degeneracies. For the known integrable case (a), $N_\text{eig}=1205$ is lower than for the other two cases (b) and (c) where the number almost coincides, $N_\text{eig}=1335$ and $N_\text{eig}=1336$, respectively\footnote{The slight difference could arise due to the numerical imprecision.} This hints, that in the integrable case (a), there were more degenerate eigenvalues than the $\mathfrak{sl}(4)$ symmetry with its higher dimensional weight spaces could explain. We think that these additional degeneracies are probably a result of integrability - or reversely, their absence (as in (b) and (c)) is a sign of integrability breaking. To sum up, all these observations suggest that the quadratic fluctuations in QSSEP are not integrable in the $\cvec_0$ sector.

\chapter{Interplay with free probability}\label{chpt:free_proba}

\section{Introduction to free probability}\label{sec:intro_free_proba}
The section is to a large extend taken from Sec.\ 3 in \cite{Hruza2023Coherent}: Free probability theory is a concept for non-commuting random variables that is in many aspects analogous to what is classical probability theory for commuting random variables. The definition of freeness was proposed by Dan Voiculescu in 1985, who founded the field of free probability theory while working on problems in operator algebras. Details can be found in his book \cite{Voiculescu1997Free} and a good introduction to the subject provide the lecture notes by Roland Speicher \cite{Speicher2019Lecture} as well as the book by Mingo and Speicher \cite{Mingo2017Free}.

In the 1990's, Speicher proposed a complementary combinatorial approach to free probability by introducing what he called \textit{free cumulants}. While a classical cumulant at order $n$ can be written as a sum over all partitions of $n$ elements, free cumulants are defined as sums over non-crossing partitions, as we will see in this section. For us they are important, because they allow us to draw a link between fluctuations of coherences and free probability theory.

\subsection*{Freeness} In classical probability, two variables are independent if (and only if) their
moments factorise at all orders, $\mathbb{E}[X^{n}Y^{m}]=\mathbb{E}[X^{n}]\mathbb{E}[Y^{m}]$
for all $n,m\in\mathbb{N}$. One can therefore determine joint moments of any product of independent variables from the moments of the individual independent variables alone. If instead $X$ and $Y$ are random non-commuting matrices with independent entries, then it is less clear how to factorize e.g.\ $\mathbb E[XYXY]$ into the moments of the independent variables (that is $\mathbb E[X^2],\mathbb E[Y^2], \E[X]$ or $\E[Y]$) on the level of matrices. As a first step, one could replace the expectation value by a linear map from $N\times N$ matrices to $\mathbb C$, for instance $\varphi(\bullet)=\frac{1}{N}\E\tr(\bullet)$. But even then, joint moments with respect to $\varphi$ cannot be expressed in terms of individual moments.

Free probability theory solves this issue by proposing an extension of the notion of independence for non-commutative random variables, called \textit{freeness}. \textit{Free} variables are not only required to be independent in the probabilistic sense, but also to be algebraically independent, in the sense that there are no algebraic relations between the variables. This is similarly to generators in a free group, hence the name "freeness".

\begin{definition}
	Given two non-commuting random variables $a$ and $b$ in some algebra $\mathcal{M}$ (e.g.\ algebra of large random matrices) and a linear functional $\varphi:\mathcal{M}\to\mathbb{C}$ with $\varphi(1)=1$ (that plays the role of the expectation value), then $a$ and $b$ are called \textit{free} if for all polynomials $P_1,\cdots,P_l$ and $Q_1,\cdots,Q_l$ with $\varphi(P_i(a))=0$ and $\varphi(Q_i(b))=0$ we have $\forall l\in\mathbb N$
	\begin{equation}
		\varphi(P_1(a)Q_1(b)\cdots P_l(a) Q_l(b))=0.
	\end{equation}
\end{definition}
\begin{remark}
	The reason to evoke all possible polynomials in the definition is that any element in the subalgebras $\mathcal A$ and $\mathcal B$ generated by $a$ and $b$, respectively, can be written as $P_i(a)$ and $Q_i(b)$. Hence freeness can also be understood as a statement about the subalgebras $\mathcal A$ and $\mathcal B$.
\end{remark}
\begin{example}
	Consider $a$ and $b$ to be free and choose polynomials $P(a)=a-\varphi(a)$ and $Q(b)=b-\varphi(b)$. Then, according to the definition, $\varphi((a-\varphi(a)(b-\varphi(b)))=0$. Simplifying, one finds $\varphi(ab)=\varphi(a)\varphi(b)$. Replacing $a\to a^m$ and $b\to b^n$ one immediately has $\varphi(a^{m}b^{n})=\varphi(a^{m})\varphi(b^{n})$, as for classical independent variable.
	However, additional structure occurs if one interchanges the order such that free variables are no longer grouped together. For example, using the same strategy, one can show that $\varphi(abab)=\varphi(a^2)\varphi(b)^2+\varphi(a)^2\varphi(b^2)-\varphi(a)^2\varphi(b)^2$.
\end{example}
\begin{remark}
	One might be tempted to think that freeness is a generalization of classical independence which appear as a special case for commuting variables with $\phi=\E$. But this is actually not the true: Two commuting independent variables $X$ and $Y$ are not free. According to the definition, freeness would require $\E[P(X)Q(Y)P(X)Q(Y)]=0$. With the polynomials $P,Q$ as above this evaluates to $\E[(X-\E[X])^2]\E[(Y-\E[Y])^2]$ which is in general not zero, but equal to the variance of $X$ and $Y$. So rather as a generalization, one should think about free probability as a new concept for non-commuting variables that is analogous to the concept of independence for commuting variables.
\end{remark}

\subsection*{Classical cumulants\label{subsec:Standard-cumulants}}

Let $\{X_{1},\cdots,X_{N}\}$ be a family of classical random variables
with moment-generating-function
\begin{align*}
	Z[a,u]  &:=\mathbb{E}[e^{u\sum_{i}a_{i}X_{i}}]\\
	&=\sum_{n\ge0}\frac{u^{n}}{n!}\sum_{i_{1}\cdots i_{n}}a_{i_{1}}\cdots a_{i_{n}}\mathbb{E}[X_{i_{1}}\cdots X_{i_{n}}],
\end{align*}
where the power of $u$ provides the order of the joint moment $\mathbb{E}[X_{i_{1}}\cdots X_{i_{n}}]$. The joint cumulant $\mathbb{E}[X_{i_{1}}\cdots X_{i_{n}}]^{c}$ is defined as the term proportional to $a_{i_{1}}\cdots a_{i_{n}}$ in
the expansion of the cumulant generating function $W[a,u]:=\log Z[a,u]$,
\begin{equation}\label{eq:classical_cumulant_genfunc}
	W[a,u]=\sum_{n\ge1}\frac{u^{n}}{n!}\sum_{i_{1}\cdots i_{n}}a_{i_{1}}\cdots a_{i_{n}}\mathbb{E}[X_{i_{1}},\cdots, X_{i_{n}}]^{c}.
\end{equation}
In fact, cumulants and moments are related by a combinatorial formula.
Expanding $Z[a,u]$ in terms of the cumulants and grouping together
terms with the same power of $u$ one can derive that a moment $\mathbb{E}[X_{{1}}\cdots X_{{n}}]$
can be expressed as a sum over partitions $\pi\in P(n)$ of the set $\{1,\cdots,n\}$,
\begin{equation}\label{eq:moments_as_cumulants}
	\mathbb{E}[X_{{1}}\cdots X_{{n}}]=\sum_{\pi\in P(n)}\E_\pi[X_{1},\cdots,X_{n}]^c,
\end{equation}
where 
\begin{equation}
	\E_\pi[X_{1},\cdots,X_{n}]^c=\prod_{p\in\pi}\mathbb{E}[X_{{p(1)}}X_{{p(2)}}\cdots]^{c}
\end{equation}
is a product of cumulants associated to $\pi$ and $p=\{p(1),p(2),\cdots\}$ are the elements of a part of the partition $\pi$. 
\begin{remark}
	The number of partitions of a set of $n$ elements is called the Bell number $B_n$, with recursion relation $B_{n+1}=\sum_{k=0}^n \big({}^n_k\big)B_k$ and $B_1 = 1$, $B_2 = 2$, $B_3 = 5$, $B_4 = 15$ and $B_5 = 52$, etc.
\end{remark}

\begin{example}
	For $n=4$, we can represent the partition $\pi=\{\{1,2\},\{3,4\}\}$ by the following diagram:
	\begin{equation}
		\raisebox{-0.5\height}{\includegraphics{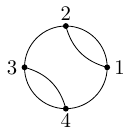}}
	\end{equation}
	The expansion of the moment $\mathbb{E}[X_{1}X_{2}X_{3}X_{4}]$ into the terms $\E_\pi[X_1,\cdots,X_4]^{c}$ summed over all non-crossing partitions $\pi$ becomes
	\begin{align}\label{eq:moment_cumulant_4}
		&\raisebox{-0.5\height}{\includegraphics{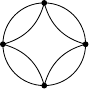}}
		+
		\raisebox{-0.5\height}{\includegraphics{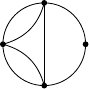}}_{\circlearrowleft4} \hskip -0.2 truecm
		+
		\raisebox{-0.5\height}{\includegraphics{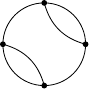}}_{\circlearrowleft2} \hskip -0.2 truecm
		+
		\raisebox{-0.5\height}{\includegraphics{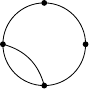}}_{\circlearrowleft4}\nonumber \\
		&+
		\raisebox{-0.5\height}{\includegraphics{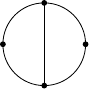}}_{\circlearrowleft2} \hskip -0.2 truecm
		+
		\raisebox{-0.5\height}{\includegraphics{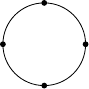}}
		+
		\raisebox{-0.5\height}{\includegraphics{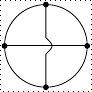}}
	\end{align}
	where ${\circlearrowleft\!k}$ denotes the sum over all $k$ cyclic
	permutation of the diagram. Note the fact, that
	the last diagram (in a dotted box) corresponds to a crossing partition
	$\pi=\{\{1,3\},\{2,4\}\}$. In free probability theory these diagrams
	do not appear as we will see below.
\end{example}

\subsubsection{Classical cumulants of a single variable}
The moment-cumulant relation \eqref{eq:moments_as_cumulants} allows
us to express the cumulants recursively through the moments. In the
case of a single variable $X=X_{1}=\cdots=X_{N}$ we illustrate how this
can be done up to order four. Let us denote by $m_{n}=\mathbb{E}[X^{n}]$
and $c_{n}=\mathbb{E}[X^{n}]^{c}$ the moments and cumulants of this
variable, then
\begin{align}
	m_{1} & =c_{1},\label{eq:moment_cumulants_single_variable}\\
	m_{2} & =c_{2}+c_{1}^{2},\nonumber \\
	m_{3} & =c_{3}+3c_{2}c_{1}+c_{1}^{3},\nonumber \\
	m_{4} & =c_{4}+4c_{1}c_{3}+3c_{2}^{2}+6c_{1}^{2}c_{2}+c_{1}^{4}.\nonumber 
\end{align}
Note that the coefficients correspond exactly to the cyclic multiplicities of the diagrams. This can be solved recursively for $c_{k}$, 
\begin{align}
	c_{1} & =m_{1},\label{eq:free_cumulants_moments_single_variable}\\
	c_{2} & =m_{2}-m_{1}^{2},\nonumber \\
	c_{3} & =m_{3}-3m_{1}m_{2}+2m_{1}^{3},\nonumber \\
	c_{4} & =m_{4}-4m_{1}m_{3}+12 m_{1}^{2}m_{2}-3m_{2}^{2}-6 m_{1}^{4}.\nonumber 
\end{align}
The generating function of cumulants is the logarithm of the moment generating function, as explained above. Below we will see how this formula differs for free cumulants, where it becomes the so-called R-transform.

\subsection*{Free cumulants\label{subsec:free_cumulants}}
\begin{definition}
	Given non-commuting random variables $a_1,\cdots,a_N\in\mathcal M$ and a linear functional $\varphi\colon\mathcal M\to\mathbb C$, the \textit{free cumulants} $\kappa_n$ are multilinear forms, implicitly defined through moments as
	\begin{equation}\label{eq:def_free_cumulants}
		\varphi(a_{1}\cdots a_{n})=:\sum_{\pi\in NC(n)}\kappa_\pi(a_1,\cdots,a_n),
	\end{equation}
	where $NC(n)$ is the set of non-crossing partition of $n$ elements and
	\begin{equation}
		\kappa_\pi(a_1,\cdots,a_n)=\prod_{p\in\pi}\kappa_{|p|}(a_{{p(1)}},a_{{p(2)}},\cdots)
	\end{equation}
	with $|p|$ the number of elements in a part $p=\{p(1),p(2),\cdots\}$ of $\pi$. This means that the family of $\kappa_\pi$'s is \textit{multiplicative} in the sense that $\kappa_\pi \kappa_\sigma=\kappa_{\pi\cup\sigma}$ with $\pi\cup\sigma$ the union of parts of $\pi$ and $\sigma$.
\end{definition} 
\begin{remark}
If we would expand $\varphi(a_{1}a_{2}a_{3}a_{4})$ into free cumulants in analogy to Eq. ~\eqref{eq:moment_cumulant_4}, we would get all diagrams, except the last one which is crossing. Therefore the order of the arguments of $\kappa_n$ becomes important -- even if the $a_i$ were to commute. Hence the separation by the comma.
\end{remark}
\begin{remark}
	The implicit definition of free cumulants in Eq.~\eqref{eq:def_free_cumulants} has a triangular structure and can be inverted by means of the \textit{Möbius function} $\mu(\pi,\sigma)$,
	\begin{align}
		\kappa_n(a_1,\cdots,a_n)&=\sum_{\pi\in NC(n)}\mu(\pi,1_n) \prod_{p\in\pi}\varphi_{|p|}(a_{p(1)},a_{p(2)},\cdots).
	\end{align}
	On the lattice of non-crossing partitions the Möbius function is defined by
	\begin{align}
		\mu(\pi,\pi)&:=1 & \mu(\pi,\sigma):=-\sum_{\substack{\pi\le\tau<\sigma}} \mu(\pi,\tau).
	\end{align}
	where $\tau\in NC(n)$ is finer (consists of more parts) than $\sigma$ and is equal or coarser (consists of less parts) than $\pi$. We denote $0_n$ the partition with $n$ parts, and $1_n$ the partition with of a single part with $n$ elements. The Möbius function $\mu(\pi,\sigma)$ can be efficiently computed by noting that it is multiplicative on the product of intervals that is isomorphic to $[\pi,\sigma]$ on the partially ordered lattice of non-crossing partitions. In particular one can show that, 
	\begin{align}
		[\pi,1_n]&\cong \prod_{d\in\pi^*}NC(|d|) & \mu(\pi,1_n)=\prod_{d\in\pi^*}\mu(0_{|d|},1_{|d|}),
	\end{align}
	where $\pi^*$ is the Kreweras complements (see around Eq.~\eqref{eq:G_moments_etappe} for a definition) and $|d|$ is the number of elements in the subset or part $d$. Note that we used that $NC(|d|) \cong [0_{|d|},1_{|d|}]$.
	Furthermore, it is known that
	\begin{equation}
		\mu(0_n,1_n)=(-1)^{n-1}\text{Cat}_{n-1}.
	\end{equation}
	Therefore, the Möbius function can be efficiently expressed as
	\begin{equation}
		\mu(\pi,1_n)=\prod_{d\in\pi^*}(-1)^{|d|-1}\text{Cat}_{|d|-1}.
	\end{equation}
\end{remark}
\begin{remark}
	Free cumulants satisfy a number of properties that are analogous to properties of classical cumulants:
	\begin{itemize}
		\item Freeness is equivalent to the vanishing of mixed cumulants: $\kappa_n(a_1,\cdots,a_n)=0$ iff there exists among $a_1,\cdots,a_n$ a pair $(a_i,a_j)$ of free variables \cite[thrm.~3.23]{Speicher2019Lecture}.
		\item As a result of multiliniarity and the last bullet point, free cumulants of free variables $a$ and $b$ are additive, $\kappa_n(a+b,\cdots,a+b)=\kappa_n(a,\cdots,a)+\kappa_n(b,\cdots,b)$, see \cite{Novak2011AMS} for a nice discussion.
		\item Any variable $a$ whose free cumulants $\kappa_n(a,\cdots,a)$ vanish for $n\ge3$ is distributed according to Wigner's semi-circle law of random matrix theory for the Gaussian unitary ensemble (GUE). Therefore GUE random matrices are the analogous of Gaussian variables in free probability theory.
	\end{itemize}
\end{remark}
\begin{remark}
	The number of non-crossing partitions of a set of $n$ elements is the Catalan number $C_n=\frac{1}{n+1}\big({}^{2n}_n\big)$, with $C_1=1$, $C_2=2$, $C_3=5$, $C_4=14$ and $C_5=42$, etc.
\end{remark}

\subsubsection{Free cumulants of a single variable.}
In the case of a single variable $a=a_{1}=...=a_{N}$, we denote by $\kappa_{n}:=\kappa_n(a,\cdots,a)$ the $n$-th free cumulants and by $m_{n}:=\varphi(a^{n})$ the $n$-th moment of this variable. Then we have 
\begin{align}
	m_{1} & =\kappa_{1},\label{eq:moments_free_cumulants_4}\\
	m_{2} & =\kappa_{2}+\kappa_{1}^{2},\nonumber \\
	m_{3} & =\kappa_{3}+3\kappa_{2}\kappa_{1}+\kappa_{1}^{3},\nonumber \\
	m_{4} & =\kappa_{4}+4\kappa_{1}\kappa_{3}+2\kappa_{2}^{2}+6\kappa_{1}^{2}\kappa_{2}+\kappa_{1}^{4}.\nonumber 
\end{align}
The equations can be solved for $\kappa_{n}$ recursively, 
\begin{align}
	\kappa_{1} & =m_{1},\label{eq:free_cumulants_moments_4}\\
	\kappa_{2} & =m_{2}-m_{1}^{2},\nonumber \\
	\kappa_{3} & =m_{3}-3m_{1}m_{2}+2m_{1}^{3},\nonumber \\
	\kappa_{4} & =m_{4}-4m_{1}m_{3}+10m_{1}^{2}m_{2}-2m_{2}^{2}-5m_{1}^{4}.\nonumber 
\end{align}
Note, that the difference between standard and free cumulants only
shows up at order $4$ since here a crossing-partition become possible
for the first time. 

For a single variable, the relation between the moments and the free cumulants is phrased in terms of the so-called \textit{R-transform},
\begin{equation}
	R(z) := \sum_{n\geq 1} \kappa_p\, z^{n-1} = \kappa_1  + \kappa_2 z + \kappa_3 z^2 +\cdots.
\end{equation}
It is related to the resolvent
\begin{equation}
	G(z):=\varphi\Big(\frac{1}{z-a}\Big)=\sum_{n\geq 0} m_n z^{-n-1}
\end{equation}
by the relations
\begin{align}\label{eq:r_transform-resolvent}
	G(z)^{-1}+R(G(z))&=z,& G(z^{-1}+R(z))&=z.
\end{align}
In other words, the functions $K(z):=z^{-1}+R(z)$ and $G(z)$ are inverses of each other.

\subsection*{Free probability and Random Matrix Theory}

A relation between free probability theory and random matrices was first observed by Voiculescu in 1991 \cite{Voiculescu1991Limit}. For example, he realised that GUE matrices with independent entries become free in the limit of large matrix size. Ever since, many more connection between other random matrix ensembles and free probability have been found.

Here we will make one of these results more explicit, which applies to matrices that are rotated by Haar random unitaries. Consider $N\times N$ random matrices of the form $X_A=U_{N}A_{N}U_{N}^{\dagger}$ and $Y_N=V_{N}B_{N}V_{N}^{\dagger}$,
where $U_{N}$ and $V_n$ are independently choosen according to the Haar distribution over the unitary group and $A_{N}$ and $B_N$ are deterministic matrices with spectral densities $\mu_{A}$ and $\mu_B$. That is, the moments
\begin{equation}
	m_{k}:=\lim_{N\to\infty}\frac{1}{N}\mathrm{Tr}(A_{N}^{k})=\int\lambda^{k}\mu_{A}(\lambda)d\lambda
\end{equation}
are all finite, and similarly for $B_N$. Then, in the limit $N\to\infty$ and with respect to $\varphi:=\frac 1 N \mathbb E\, \Tr$ where $\E$ is the expectation value of the entries of $X,Y$, the matrices $X_N$ and $Y_N$ become free variables $a$ and $b$ (in some non-commutative probability space) with distributions $\mu_A$ and $\mu_B$. For a proof see \cite[thrm.~7.5]{Speicher2019Lecture}.

It is furthermore known from the HCIZ-integral, cf.\ \cite[thrm.\ 4.5]{Collins2003Moments}, that the classical cumulants of such matrices $X_{N}$ can be expressed as the free cumulants $\kappa_n\equiv\kappa_n(a,\cdots,a)$ of the spectral density $\mu_{A}$. That is,
\begin{equation} \label{eq:large-HCIZ}
	\mathbb{E}[e^{N\mathrm{tr}(X_{N}Q_{N})}]\stackrel{N\to\infty}{\asymp} e^{N\sum_{n=1}\frac{1}{n}\kappa_{n}\mathrm{tr}(Q_N^{n})}, 
\end{equation}
where $Q_{N}$ is a sequence of matrices with fixed rank (such that
$\mathrm{tr}(Q_{N}^k)$ does not scale with $N$), for instance a rank one projector.

\section{Moment-cumulant expansion for a new class of random matrices}\label{sec:moment-cumulant-expansion}
We first recall the three properties which seem to be responsible for the fact that QSSEP satisfies the moment-cumulant formula Eq.~\eqref{eq:QSSEP_moment_cumulants} and we view them as properties shared by a much larger class or random matrices -- than only by QSSEP. Then we provide a derivation of Eq.~\eqref{eq:QSSEP_moment_cumulants}.

Consider a class of random matrices $G$ with measure $\E$ satisfying:
\begin{enumerate}
	\item[(i)] Local $U(1)$-invariance, meaning that in distribution, $G_{ij}\stackrel{d}{=}e^{-i\theta_i}G_{ij}e^{i\theta_j}$ for any angles $\theta_i$ and $\theta_j$;	
	\item[(ii)] Expectation values of loops of order $n$ without repeated indices scale as $N^{1-n}$, meaning that $\mathbb E [G_{i_1 i_2}G_{i_2i_3}\cdots G_{i_n i_1}]=\mathcal O(N^{1-n})$ for all indices $i_k$ distinct;
	\item[(iii)] Factorization of the expectations value of products of loops at leading order,\\$\mathbb E[G_{i_1 i_2}\cdots G_{i_{m} i_1}\,G_{j_{1} j_{2}}\cdots G_{j_n j_{1}}]=\mathbb E [G_{i_1 i_2}\cdots G_{i_{m} i_1}]\,\mathbb E [G_{j_{1} j_{2}}\cdots G_{j_{n} j_{1}}](1+\Ord(\frac{1}{N}))$, even if $i_1=j_1$.
\end{enumerate}
%
Examples of well-known random matrix ensemble satisfying these properties are the Gaussian Unitary Ensemble or matrices rotated by Haar random  unitaries (see Appendix \ref{app:wigner_haar}). But the three properties are more general, since they also apply to so-called \textit{structured} random matrices, which are not invariant, in law, under a permutation of its elements. Indeed, the out-of-equilibrium QSSEP is such an example, since the matrix of coherences $G$ has to respect the different boundary conditions with the reservoirs. 

One can also check that if $G$ satisfies (i)-(iii), then any matrix polynomial in $G$ (or of several independent copies of $G$) also satisfies the properties\footnote{Our attention was drawn to this thanks to Roland Speicher}. In this sense the three properties are closed under polynomial composition \cite{Bernard2024Structured}.

Before moving to the proof, note that these properties ensure that the connected expectation value (classical cumulant) of loops satisfies,
\begin{equation}
	\mathbb E[G_{i_1 i_2}G_{i_2 i_3}\dots G_{i_{n} i_1}]^c=\Ord(N^{1-n}),
\end{equation}
even if two indices are equal: Imagine that  $i_1=i_k$ for some $k\in\{1,\cdots,n\}$ and no other indices are equal. Then, $\mathbb E[G_{i_1 i_2}\dots G_{i_{n} i_1}]$ factorises according to property (iii) into $\mathbb E[G_{i_1 i_2}\dots G_{i_{k-1} i_1}]\mathbb E[G_{i_1 i_{k+1}}\dots G_{i_{n} i_1}]$ at leading order. This product scales according to $\Ord(N^{1-(k-1)})\Ord(N^{1-(n-k+1)})=\Ord(N^{2-n})$. To construct the connected expectation value, one needs to subtract this product of two loops from $\mathbb E[G_{i_1 i_2}\dots G_{i_{n} i_1}]$. What remains, must scale at least one order of magnitude lower in $N$, so it is at most $\Ord(N^{1-n})$.

\paragraph{Proof of moment-cumulant expansion.}
Here we give the proof of Eq.~\eqref{eq:QSSEP_moment_cumulants} as it appears in our original article \cite{Hruza2023Coherent}. Consider the $n$-th moment of $G$ wrt.\ $\varphi:=\E \,\ntr$ where $\ntr=\tr/N$ is the normalized trace,
\begin{equation}\label{eq:moments_G}
	\varphi(G^n):=\frac 1 N\sum_{i_1,\cdots,i_n} \mathbb E[G_{i_1i_2}\cdots G_{i_ni_1}].
\end{equation}
Whenever two indices $i_k$ are equal, the expectation value factorises according to condition (iii). We therefore split the sum into several sums, in each of which all indices are distinct -- except for a given set of indices that are equal,
\begin{equation}
	\begin{aligned}
		\sum_{i_1,\cdots,i_n}=\sum_{\substack{i_1,\cdots,i_n \\ \text{all distinct}}}+\sum_{\substack{i_1=i_2\text{ and }\\ i_3,\cdots,i_n \text{ distinct}}}+\cdots+\sum_{i_1=i_2=\cdots=i_n}
	\end{aligned}.
\end{equation}
Such a splitting can be understood as a sum over partitions $\pi\in P(n)$ into parts $p\in\pi$ that group together all the indices $i_k$ that are equal. For example, the partitions corresponding to the three sums above are $\pi=\{\{1\},\cdots,\{n\}\}$, $\pi=\{\{1,2\},\{3\},\cdots,\{n\}\}$ and $\pi=\{\{1,2,\cdots,n\}\}$. The total sum becomes
\begin{equation}
	\sum_{i_1,...,i_n}=\sum_{\pi\in P(n)} \sum_{\substack{i_1,...,i_n \text{ distinct,} \\ \text{except }i_k=i_l \text{ whenever } k,l \\ \text{are in the same part of }\pi}}.
\end{equation}
For example, for $n=4$ two possible partitions are
\begin{align}\label{eq:pi1}
	\pi_1=\{\{1,3\},\{2\},\{4\}\}=
	\raisebox{-0.5\height}{\includegraphics{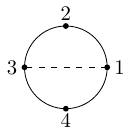}}
\end{align}
and 
\begin{align}\label{eq:pi2}
	\pi_2=\{\{1,3\},\{2,4\}\}=
	\raisebox{-0.5\height}{\includegraphics{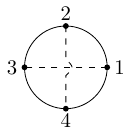}}
\end{align}
This shows very intuitively that $\pi_2$ is a crossing partition while $\pi_1$ is non-crossing. 

It turns out that terms corresponding to crossing partitions in the sum (\ref{eq:moments_G}) are $\Ord(1/N)$ and therefore vanish for a large $N$. One the other hand, all non-crossing partitions are $\Ord(1)$ and survive. Instead of giving a complete derivation, here we illustrate this fact on the two examples above. 

The term corresponding to $\pi_1$ factorises and becomes
\begin{align}
	\frac 1 N \sum_{\substack{i_1=i_3,i_2,i_4 \\ \text{distinct}}} \mathbb E[G_{i_1i_2}G_{i_2i_1}]\mathbb E[G_{i_3i_4} G_{i_4i_3}] = \mathcal O(1).
\end{align}
This is because the expectation value of order-$2$ loops is $\Ord(1/N)$, such that each term in the sum is $\Ord(N^{-3})$. Since the sum carries over three indices, running from $1$ to $N$, the factors of $N$ cancel and the resulting term is $\Ord(1)$. 

In contrast to this, the term corresponding to $\pi_2$ factorises in two different ways and becomes
\begin{align}
	\frac 1 N \sum_{\substack{i_1,i_2 \\ \text{distinct}}} 2\, \mathbb E[G_{i_1i_2}G_{i_2i_1}]^2= \mathcal O(1/N).
\end{align}
The difference to $\pi_1$ is that now there are only two indices to sum over and hence the scaling is of order $1/N$.

For non-crossing partitions, one can ask how a given $\pi$ is related to the resulting product of matrix elements $G_{ij}$? For example, how can we understand that $\pi_1$ leads to $ \mathbb E[G_{i_1i_2}G_{i_2i_1}]\mathbb E[G_{i_3i_4} G_{i_4i_3}]$? A graphical solution is that we connect as many edges of the circle in Eq.~(\ref{eq:pi1}) by solid lines as possible without crossing a dashed line
\begin{equation}
	\raisebox{-0.5\height}{\includegraphics{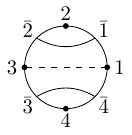}}
\end{equation}
and by associating a product of $G_{ij}$'s to each independent solid line, such that the indices $i,j$ take the value of the nodes that are adjacent to the edges connected by this solid line (in anti-clockwise direction), and by equating nodes connected by a dashed line. For example, the upper solid line in the above diagram corresponds to $\E[G_{i_1i_2}G_{i_2i_1}]$. For the complete contribution associated to $\pi_1$ one takes the product over all solid lines.

\begin{definition}
	This procedure defines a dual non-crossing partition $\pi^*$ on the edges $\{\bar 1, \cdots,\bar n\}$ of a circle with $n$ nodes, which is called the \textit{Kreweras complement}. 
\end{definition}

For $\pi_1$ (dashed lines), the Kreweras complement is $\pi_1^*=\{\{\bar 1,\bar 2\},\{\bar 3,\bar 4\}\}$ (solid lines). In this terminology,
\begin{equation}\label{eq:G_moments_etappe}
	\varphi(G^n)=\frac 1 N \sum_{\pi\in NC(n)} \sum_{\substack{i_1,...,i_n \text{ distinct,} \\ \text{except }i_k=i_l \text{ whenever } k,l \\ \text{are in the same part of }\pi}} \prod_{p\in\pi^*} \mathbb E[G_{i_{p(1)}i_{p(2)}}...G_{i_{p(|p|)}i_{p(1)}}],
\end{equation}
with $NC(n)$ the set of non-crossing partitions of $n$ elements.

Instead of using this slightly akward sum over indices $i_k$ we introduce a modified Kronecker-delta 
\begin{equation}\label{eq:kronecker-delta_pi}
	\delta_\pi\equiv\delta_\pi(i_1,\cdots,i_n)=\prod_{p\in\pi} \delta_{i_{p(1)},\cdots,i_{p(|p|)}}
\end{equation}
that makes indices equal if they belong to the same part. 

Next, we replace $\mathbb E[\cdots]$ by its connected part $\mathbb E[\cdots]^c$, i.e.\ by its classical cumulant, in Eq.~\eqref{eq:G_moments_etappe}. In this way, we can cancel the restriction that all indices must be distinct if not stated otherwise, because $\E[G_{i_1,i_n}\cdots G_{i_n,i_1}]^c\sim\Ord(N^{1-n})$, even if some indices become equal. As a result of this replacement, the leading order of each term stays the same.

Putting all together, we have
\begin{equation}\label{eq:moments_G_expansion}
	\varphi(G^n)=\frac 1 N \sum_{\pi\in NC(n)} \sum_{i_1,...,i_n} \delta_\pi \prod_{p\in\pi^*} \mathbb E[G_{i_{p(1)}i_{p(2)}}...G_{i_{p(|p|)}i_{p(1)}}]^c.
\end{equation}
Replacing the sum by an integral over continuous variables, this becomes Eq.~\eqref{eq:QSSEP_moments_G_cumulants}. 

Note that even though this is close to the the definition of free cumulants, the terms on the right hand side  
\begin{equation}\label{eq:kappa_tilde}
	\tilde \kappa_\pi:= \frac 1 N \sum_{i_1,...,i_n} \delta_\pi \prod_{p\in\pi^*} \mathbb E[G_{i_{p(1)}i_{p(2)}}...G_{i_{p(|p|)}i_{p(1)}}]^c
\end{equation}
are actually not the free cumulants of $G$ since they are not multiplicative, $\tilde\kappa_\pi\tilde \kappa_\sigma\neq\tilde \kappa_{\pi\cup\sigma}$. The reason for this is the contraction with the Kronecker Delta\footnote{Formally, the free cumulants of $G$ are defined as a multiplicative family $\kappa_\pi=\prod_{b\in\pi} \kappa_{|b|}$ with $\kappa_n:=\kappa_{1_n}$ satisfying $\varphi(G^n) = \sum_{\pi\in NC(n)}\kappa_\pi$ and they can be related to $\tilde \kappa_\pi$ by Moebius inversion, $\kappa_n = \sum_{\pi\in NC(n)} \mu(\pi,1_n) \prod_{b\in\pi}\sum_{\sigma\in NC(|b|)}\tilde \kappa_\sigma$.}. We will see in section \ref{sec:spectrum_of_subblocks} that we can give them a meaning as free cumulants in the framework of operator valued probability.

To conclude, we could have inserted an arbitrary test function $h_{i_1,\cdots,i_n}$ into each term on the left hand side of Eq.~\eqref{eq:moments_G}, which would get carried through Eq.~\eqref{eq:moments_G_expansion}. Therefore we can take away the sum over indices,
\begin{equation}\label{eq:moment-cumulant_Q-SSEP}
	\begin{aligned}
		&\mathbb E[G_{i_1i_2}...G_{i_ni_1}]
		\\
		&=\sum_{\pi\in NC(n)}\mathbb \delta_{\pi^*}(i_1,...,i_n)\prod_{p\in \pi}\mathbb E[G_{i_{p(1)}i_{p(2)}}...G_{i_{p(|p|)}i_{p(1)}}]^c,
	\end{aligned}
\end{equation}
We interchanged the role of $\pi$ and $\pi^*$ here, which is possible because they are in one-to-one correspondence. The formula corresponds to what we have claimed in Eq.~\eqref{eq:QSSEP_moment_cumulants}. A full proof of this formula for any $n$ is given in \cite[app.~E.1]{Hruza2023Coherent}.

\section{Spectrum of subblocks of structured random matrices}\label{sec:spectrum_of_subblocks}
This section presents a mathematical result on its own which we obtained in \cite{Bernard2024Structured}. We will derive the spectrum of a random matrix (and of its subblocks) satisfying properties (i)-(iii), from only the knowledge of the joint cumulants of its entries when arranged in a loop, such as in Eq.~\eqref{eq:def_g}. We present two proofs: A direct one that makes use of the tree structure associated with non-crossing partitions, and a proof based on operator-valued free probability. The main ingredient to these derivations is Eq.~\eqref{eq:moments_G_expansion} or equivalently Eq.~\eqref{eq:QSSEP_moments_G_cumulants}, the expansion of moments of the random matrix into non-crossing partitions. We will also find a satisfying answer to the question that came up in the last section, how the terms $\tilde\kappa$ in Eq.~\eqref{eq:kappa_tilde} can be correctly identified with free cumulants. As we will see in the second proof, they can be understood as operator-valued free cumulants (with amalgamation over diagonal matrices).

For the more physically interested reader, the whole section can be skipped, but note that the results presented here have been used to compute the QSSEP entanglement entropy.

\subsection*{Result}
Large parts of the this section are taken from \cite{Bernard2024Structured}: In order to stress, that the result presented here is more general than QSSEP, we denote an element of a random matrix ensemble satisfying properties (i)-(iii) by $M$. The only additional information we require are the joint cumulants of its entries when arranged in a loop, which we recall here (with $x_k=i_k/N\in[0,1]$):
\begin{equation}
	g_n(x_{1},\cdots,x_{n}):=\lim_{N\to\infty}N^{n-1}\mathbb{E}[M_{i_{1}i_{2}}M_{i_{2}i_{3}}\cdots M_{i_{n}i_{1}}]^c.
\end{equation}
For a reason explained in the discussion we will call these functions \textit{local free cumulants}.

To handle the case of an arbitrary number of subblocks, we consider the slightly more general aim of finding the spectrum of \begin{equation}
	M_h:=h^{1/2}Mh^{1/2}
\end{equation}
with $h$ a diagonal matrix. Choosing $h(x)=1_{x\in I}$ (here $h_{ii}=h(i/N)$) to be the indicator function on some interval $I\subset[0,1]$, one recovers the case of subblocks $M_I\subset M$. All the spectral information about $M_h$ is contained in the
generating function,
\begin{align}\label{eq:def_F}
	F[h](z):=\mathbb{E}\,\ntr\log(z-M_h),
\end{align}
where $\ntr=\tr/N$ is the normalized $N$-dimensional trace. 
This function can be seen as a (formal) power series in $1/z$, whose coefficients are the moments of $M_h$. Statements about the domain of convergence of this series can be made if extra global information about the spectrum is available, say about its compactness. The theorem below is formulated with $F[h](z)$ viewed as  power series in $1/z$ (and we use extra analytic inputs in the illustrative examples).

Our main result is:
\begin{theorem}\label{thrm:F}
	$F[h](z)$ is determined by the variational principle
	\begin{equation}\label{eq:action}
		F[h](z)=\underset{a,b}{\mathrm{extremum}}\left[\int_0^1\left[\log(z-h(x)b(x))+a_z(x)b(x)\right]dx-F_{0}[a]\right]
	\end{equation}
	where the information about local free cumulants $g_n$, specific to the random matrix ensemble, is contained in (with $\vec x=(x_1,\cdots,x_n)$) 
	\begin{equation}\label{eq:F_0}
		F_0[p]:=\sum_{n\ge1}\frac{1}{n}\int_0^1 (\prod_{k=1}^ndx_kp(x_{k}))\, g_n(\vec x).
	\end{equation}
\end{theorem}
From there one finds the extremization conditions for $a$ and $b$ to be,
\begin{equation} \label{eq:a-b}
	a_z(x)=\frac{h(x)}{z-h(x)b_z(x)},\quad b_z(x)= R_0[a_z](x) ,
\end{equation}
where
\begin{equation}\label{eq:R_0}
	R_0[a_z](x):= \frac{\delta F_0[a_z]}{\delta a_z(x)}.
\end{equation} 

Note that $F_0[p]$ contains less information than the local free cumulants, since it depends only on a symmetrized version of the family $\{g_n\}_n$. Nevertheless, in the large $N$ limit, it represents the minimal amount of information about the measure $\mathbb{E}$ that is necessary for the spectrum.

\paragraph{Resolvent and Spectrum.}To obtain the spectrum of $M_h$ one takes the derivative $\partial_z F[h](z)=: G[h](z)$ which is the resolvent
\begin{equation}  
	G[h](z) = \mathbb E\,\ntr {(z-M_h)^{-1}} . 
\end{equation}
From Eq.\eqref{eq:action}, we get
\begin{equation}\label{eq:resolvent}
	G[h](z)=\int_0^1 \!\frac{dx}{z-h(x)b_z(x)} ,
\end{equation}
with $b_z$ solution of the extremization conditions.

In the special case where $h(x)=1_{x\in I}$ is the indicator function on an interval $I$ (or on unions of intervals) of length $\ell_I$, we recover the spectral density $\sigma_I$ of the subblock $M_I$ from its resolvent
\begin{equation}\label{eq:resolvent_I}
	G_I(z):=\int_I \frac{d x}{z-b_z(x)}=\int_0^1 \frac{d\sigma_I(\lambda)}{z-\lambda}
\end{equation}
as $G_I(\lambda-i\epsilon) -  G_I(\lambda+i\epsilon)= 2i\pi\sigma_I(\lambda)$.
Writing the total resolvent (including the pole at the origin)
\begin{equation}
	G^\mathrm{tot}_I(z):=G[1_{x\in I}](z)= \frac{1-\ell_I}{z} + \ell_I \int \frac{d\sigma_I(\lambda)}{z-\lambda},
\end{equation}
we can relate the total spectral measure of $M_h$ (including the zero-eigenvalues) to that of a subblock $M_I\subset M$ by \begin{equation}\label{eq:total_spectrum}
	d\sigma^\mathrm{tot}_I(\lambda)=(1-\ell_I)\delta(\lambda)d\lambda+ \ell_I\, d\sigma_I(\lambda).
\end{equation}

\subsection*{Discussion}
Some well-known random matrix ensembles that satisfy properties (i)-(iii) are Wigner matrices or Haar randomly rotated matrices (see Appendix \ref{app:wigner_haar} for how our result applies here). For these ensembles it turns out that the functions $g_n$ are all constant, implying that these ensembles are \textit{structureless}:  In law, these matrices are invariant under permutations of its entries. But our result is more general, it also applies to \textit{structured} matrices \cite{vanHandel2017Structured} where the functions $g_n$ are not constant.

For matrices $M = UDU^\dagger$ rotated by Haar random unitaries $U$ with $D$ diagonal, it is furthermore known (see appendix) that $g_n(x_1,\cdots,x_n)=\kappa_n$ are the free cumulants of the spectral measure of $D$. For structured matrices with non-constant $g_n$, this observation suggests to call $g_n$ the \textit{local free cumulants} of $M$. The name choice is further supported by the fact that the extremization conditions \eqref{eq:a-b} can be rewritten as
\begin{equation}
	zh(x)^{-1}=a_z(x)^{-1}+R_0[a_z](x).
\end{equation} Here $R_0$, the generating function of local free cumulants, resembles a local version of the so-called R-transform of free probability theory and $a_z(x)$ can be seen as a local version of the resolvent of $M$, not in $z$ but in the variable $zh(x)^{-1}$. R-transform and resolvent are related by Eq.~\eqref{eq:r_transform-resolvent} and the above equation can be seen as a local version of this relation.

From the proof via operator-valued free probability theory we will learn that the terms $\tilde \kappa_\pi:=\int g_{\pi^*}(\vec x)\,\delta_\pi(\vec x)d\vec x$ from Eq.~\eqref{eq:kappa_tilde} are in fact the trace of operator-valued free cumulants of $M$ (with amalgamation over diagonal matrices $\mathcal D$),
\begin{equation}
	\tilde \kappa_\pi=\ntr(\kappa_\pi^\mathcal{D}(M,\cdots,M)).
\end{equation}
This shows that formally the correct framework for matrices satisfying (i)-(iii) is rather operator-valued free probability (with amalgamation over diagonal matrices), than scalar free probability.

To conclude, let us invert the variational principle~: Given a generating function $F[h](z)$ that satisfies Eq.\eqref{eq:action}, we can retrieve the initial data $F_0$ as the extremum of
\begin{equation}\label{eq:inverted_action}
	F_0[a]=\underset{h,b_z}{\mathrm{extremum}}\left[\int\left[\log(z-h(x)b_z(x))+a_z(x)b_z(x)\right]dx-F[h](z)\right].
\end{equation}
This is very similar to the Legendre Transformation where the initial function can be retrieved by applying the transformation twice.
Here the inversion works because in extremizing Eq.\eqref{eq:action} we obtain $a=a(h,z)$ and $b=b(h,z)$ as functions of $h$ (and $z$), while  in extremizing Eq.\eqref{eq:inverted_action} we obtain $h=h(a,z)$ and $b=b(a,z)$ as functions of $a$ (and $z$). Through a formal power series, the triple $(a,b,h)$ can be inverted which ensures the variational principle for $F_0$ above.

\subsection*{Proof using a tree structure}\label{subsec:proof_trees}
Eq.~\eqref{eq:moments_G_expansion} (or equivalently Eq.~\eqref{eq:QSSEP_moments_G_cumulants}) can be rewritten in terms of the local free cumulants as
\begin{equation}\label{eq:moment_free_cumulants}
	\varphi_n[h]:=\mathbb E\, \ntr(M_h^n)=\sum_{\pi\in NC(n)}\int \! g_{\pi^*}(\vec x)\,\delta_\pi(\vec x)\, h(x_1)\cdots h(x_n)d\vec x
\end{equation}
where $g_\pi(\vec x):=\prod_{p\in\pi}g_{|p|}(\vec x_p)$ with $\vec x_p=(x_i)_{i\in p}$ the collection of variables $x_i$ belonging to the part $p$ of the partition $\pi$, and $|p|$ the number of elements in this part. By $\delta_\pi(\vec x)$ we denote a product of delta functions $\delta(x_i-x_j)$ that equate all $x_i,x_j$ with $i$ and $j$ in the same part $p\in \pi$. And $\pi^*$ is the Kreweras complement of $\pi$.

Expanding the generating function \eqref{eq:def_F} in terms of the moments $\varphi_n[h]$ one has
\begin{equation}\label{eq:F_expansion}
	F[h](z)=\log(z)-\sum_{n\ge1} \frac{z^{-n}}{n}\varphi_n[h].
\end{equation}
The difficulty here in organising the sum over non-crossing partitions hidden in $\varphi_n$ for any possible integer $n$. To better understand this structure, we note that non-crossing partitions $\pi\in NC(n)$ are in one-to-one correspondence with planar bipartite rooted trees $T_\bullet$ with $n$ edges, if one labels its black and white vertices by the parts of $\pi$ and $\pi^*$. Here is an example for $\pi=\{\{1,3\},\{2\},\{4,5\},\{6\}\}$ (dotted lines) whose Kreweras complement is $\pi^*=\{\{\bar 1, \bar 2\},\{\bar3,\bar5,\bar6\},\{\bar 4\}\}$ (solid lines).
\begin{center}
	\includegraphics[width=0.5 \linewidth]{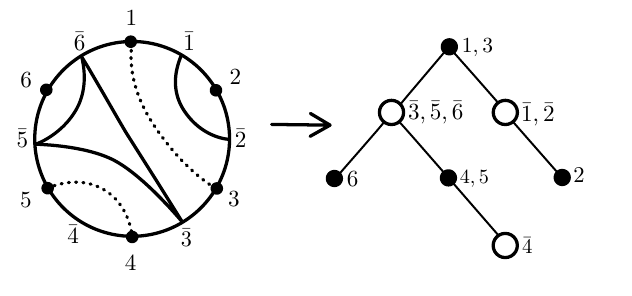}
\end{center}
The parts of $\pi$ are associated with black vertices and parts of $\pi^*$ with white vertices. Two vertices are connected if the corresponding parts of $\pi$ and $\pi^*$ have an element in common (identifying numbers with and without bar, $k\sim\bar k$). The root is (by convention) chosen to be the part $p$ containing $1$.

However, applying this correspondence to Eq.\eqref{eq:moment_free_cumulants} is not directly straightforward, because two partitions $\pi$ and $\pi'$ that are related by a rotation of its elements (in the circle representation) have the same contribution in the sum and thereby complicate the counting of terms. This is due to the integration over $x_1,\cdots, x_n$. If instead, we don't integrate over one of these variables, call it $x$, then $\pi$ and $\pi'$ will give rise to different contributions, because they now depend on $x$. 

This motivates us to define 
\begin{equation}
	\varphi_n[h](x):=\mathbb{E}\langle x|(M_h)^n|x\rangle .
\end{equation}

Note that $\varphi_n[h]\!\!=\!\!\int\! \varphi_n[h](x)dx$. When expanded into non-crossing partitions via Eq.~\eqref{eq:moment_free_cumulants}, we associate the variable $x$ to the part of the partition $\pi$ containing $1$ -- and therefore to the root of the corresponding tree $T_\bullet$. Denoting such a tree by $T_\bullet^x$ we have
\begin{equation}\label{eq:correspondence_trees}
	z^{-n}\varphi_n[h](x)=\sum_{T_\bullet \text{ with $n$ edges}} W(T_\bullet^x).
\end{equation}
where the weight $W(T_\bullet^x)$ of a tree must be defined in accordance with Eq.~\eqref{eq:moment_free_cumulants}: Assign an integration variables $x_i$ to each black vertex, and assign $x$ to the black vertex that constitutes the root. Then assign the value $z^{-k}h(x_1)\cdots h(x_k)g_k(x_1,\cdots,x_k)$ to each white vertex whose neighbouring black vertices carry the variables $x_1,\cdots,x_k$ (one can also think of $z^{-1}h(x_i)$ to live on the edges of the tree). Finally, take the product over all vertices and integrates over all $x_i$ (except for the root $x$). By definition we set the tree consisting of a root without legs to one. Graphically the rules for the weights $W(T_\bullet ^x)$ are
\begin{equation}
	\centering\includegraphics[width=0.6 \linewidth]{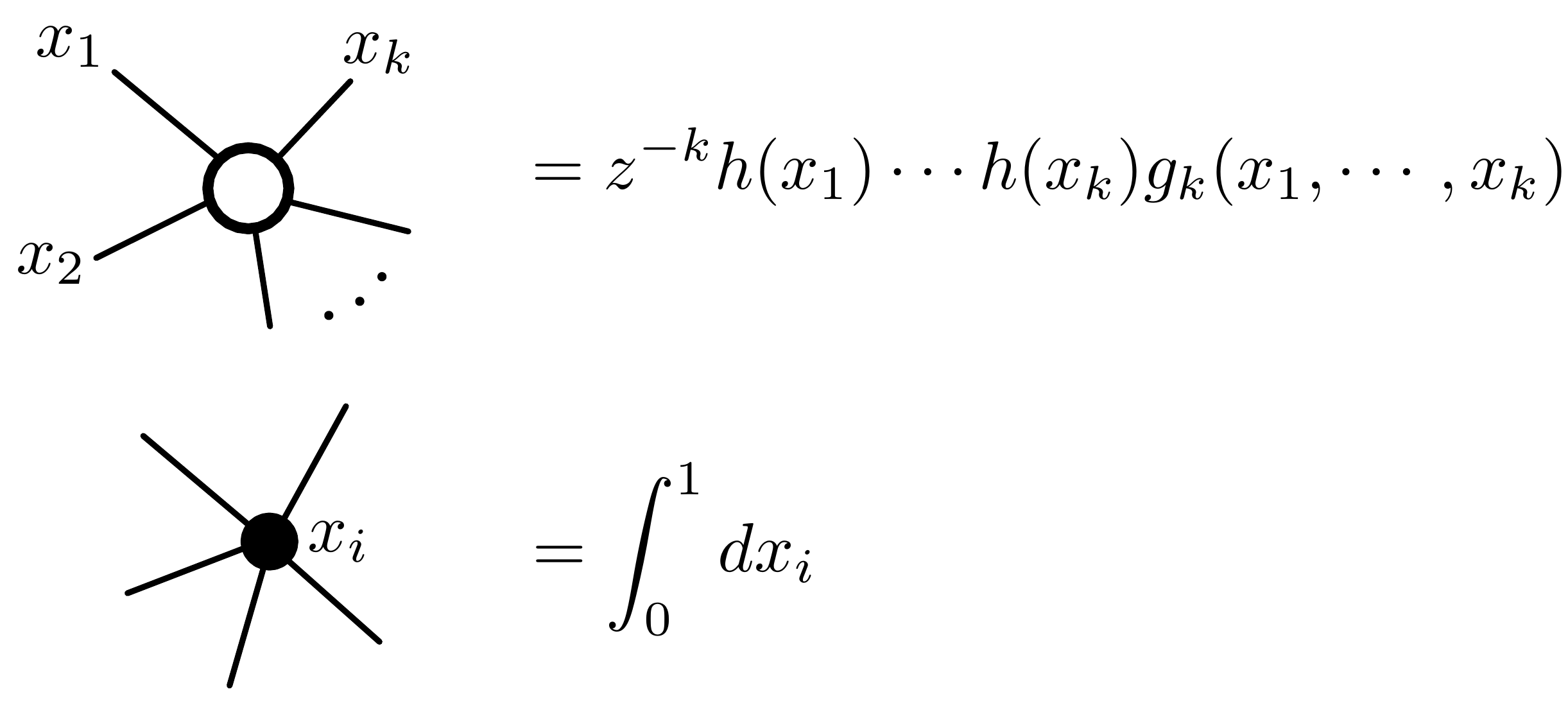}
\end{equation}

Doing the sum over all $n$ is now easy~: We just relax the condition on the sum over trees with $n$ edges to trees of arbitrary size. We consider a generating function involving a sum over $\varphi_n[h](x)$, 
\[
a_z(x):=\mathbb{E}\langle x|\frac{h}{z-M_h}|x\rangle=\frac{h(x)}{z}\sum_{n\ge 0} \frac{\varphi_n[h](x)}{z^n}\overset{!}{=}\frac{h(x)}{z}\sum_{T_\bullet} W(T_\bullet^x),
\]
where the last equality is due to the correspondence with trees in Eq.~\eqref{eq:correspondence_trees}.

In order to establish the relation \eqref{eq:a-b} satisfied by $a_z(x)$ we consider the subset of trees $T_\circ$ whose root (still a black vertex) has a single leg only. This defines
\begin{equation}\label{eq:def_b}
	b_z(x):=\frac{z}{h(x)}\sum_{T_\circ} W(T_\circ^x) .
\end{equation}
Note that the weight $W(T_\bullet^x)$ of a tree whose root has $l$ legs is equal to the product of weights $W(T_{\circ,1}^x)\cdots W(T_{\circ,l}^x)$ of trees with a single leg on their root that arise by cutting the $l$ legs of $T_\bullet^x$. This implies 
\begin{align}
	\sum_{T_\bullet}W(T_\bullet^x)
	&=\sum_{l\ge 0}\sum_{\substack{T_\bullet \text{ with }l \\ \text{ legs on root }}} W(T_\bullet^x)
	=\sum_{l\ge 0} \Big(\sum_{T_\circ} W(T_\circ^x)\Big)^l\nn
	&=	\Big(1-\sum_{T_\circ} W(T_\circ^x)\Big)^{-1} ,
\end{align}
which yields the first relation in Eq.\eqref{eq:a-b}. 

For the second relation, we start with $T_\circ^x$ and cut the $l$ outgoing legs of the first white vertex. This generates a product of $l$ trees $T_{\bullet,i}^{x_i}$ whose weights satisfy 
\begin{equation}
	W(T_\circ^x)=\frac{h(x)}{z}\int 
	\!\left(\prod_{i=1}^ldx_i  \frac{h(x_i)}{z}W(T_\bullet^{x_i})\right)
	g_{l+1}(x,x_1,\cdots,x_l).
\end{equation}
Therefore, taking the sum over all trees $\sum_{T_\circ}=\sum_{l\ge 0} \sum_{T_\circ \text{ with } l \text{ legs}}$, one has
\begin{equation}
	b_z(x)=\sum_{l\ge0}\!\int \!\left(\prod_{i=1}^ldx_i  \frac{h(x_i)}{z}\sum_{T_\bullet}W(T_\bullet^{x_i})\right)g_{l+1}(x,x_1,\cdots,x_l).
\end{equation}
One recognizes the definition of $a_z(x_i)$ in this expression, which then implies the second relation in Eq.\eqref{eq:a-b}.

Both relations in Eq.~\eqref{eq:a-b} are the extremization conditions of the variational principle \eqref{eq:action}. As a last step we should therefore verify that $F[h]$ as defined in Eq.~\eqref{eq:F_expansion} coincides with the solution of the extremization problem from Eq.~\eqref{eq:action}. Here we show that their first derivates with respect to $h$ coincide for any $h$, as well as their value at $h=0$. 

Since $h(x)\,\delta \varphi_n[h]/\delta h(x)=\varphi_n[h](x)$, one calculates from  Eq.\eqref{eq:F_expansion} that
\begin{equation}
	-h(x)\frac{\delta F[h](z)}{\delta h(x)}=\sum_{n\ge 1}\frac{\varphi_n[h](x)}{z^{-n}}=\sum_{T_\bullet} W(T_\bullet^x)-1.
\end{equation}
The $(-1)$ is because the sum over $n$ starts at one and not at zero.
Furthermore, one has $a_z(x)b_z(x)=\sum_{T_\circ} W(T_\circ^x) \sum_{T_\bullet} W(T_\bullet^x)=\sum_{T_\bullet} W(T_\bullet^x)-1$. This leads to 
\begin{equation}
	-h(x)\frac{\delta F[h](z)}{\delta h(x)}=a_z(x)b_z(x).
\end{equation}

On the other hand, starting from Eq.\eqref{eq:action}, one has
\begin{equation}
	h(x)\frac{\delta F[h](z)}{\delta h(x)}=-\frac{h(x)b_z(x)}{z-h(x)b_z(x)}=-a_z(x)b_z(x),
\end{equation}
where we used Eq.~\eqref{eq:a-b} in the last line. Since $F[h=0](z)=\log(z)$ for both definitions \eqref{eq:action} and \eqref{eq:a-b}, the two expressions for $F[h](z)$ coincide.

\subsection*{Proof using operator valued free probability}\label{subsec:proof_op-val}
This section recalls some basic definitions of operator-valued free probability theory and shows how the relation between the R- and the Cauchy-transform (Theorem \ref{thrm:R-Cauchy}) can be used to deduce our main result (Theorem \ref{thrm:F}). Of course, the relation between $R$- and Cauchy transform also uses implicitly the tree structure of non-crossing partitions. We closely follow \cite[chpt. 9]{Mingo2017Free} and \cite{Shlyakhtenko1998Gaussian} and start with the definition of the operator-valued moments for a general unital algebra $\mathcal A$ which later becomes the matrix algebra formed by the matrices $M$.
\vspace{5pt}
\begin{definition}
	Let $\mathcal A$ be a unital algebra and consider a unital subalgebra $\mathcal D \subset \mathcal A$. Then $E^\mathcal{D}:\mathcal A \to \mathcal D$ is called a \emph{conditional expectation value} (with amalgamation over $\mathcal D$) if for all $a\in \mathcal A$ and $d, d'\in \mathcal D$ one has $E^\mathcal{D}[d]\in\mathcal D$ and $E^\mathcal{D}[d a d']=dE^\mathcal{D}[a]d'$. 
	
	For any choice of $d_1,\cdots,d_{n-1}\in\mathcal D$, the \emph{operator-valued} (or \emph{$\mathcal{D}$-valued}) \emph{moments} of $a$ are defined as $E^\mathcal{D}[a d_1a\cdots ad_{n-1}a]\in\mathcal D$ and the collection of all operator-valued moments define the operator-valued distribution of $a$.
\end{definition}

We will now consider the special case where the elements $M\equiv a\in\mathcal A$ are random matrices of size $N$ satisfying properties (i)-(iii), and the elements $\Delta\equiv d\in\mathcal D$ are diagonal matrices of size $N$. Note that $\mathcal{D}$ is indeed a subalgebra of $\mathcal A$ and that in the large $N$ limit we have $\mathcal D\to L^\infty[0,1]$. We also define explicitly a conditional expectation value adapted to our choice of $\mathcal D\subset\mathcal A$. For $M\in \mathcal{A}$,
\begin{equation}\label{eq:cond_expectation}
	E^\mathcal{D}[M]:=\mathrm{diag}(\mathbb E[M_{11}],\cdots,\mathbb E[M_{NN}]).
\end{equation}

As in scalar free probability, one can define operator-valued free cumulants as follows.
\begin{definition}
	The \emph{$\mathcal D$-valued free cumulants} $\kappa^\mathcal{D}_n:\mathcal A^n\to\mathcal{D}$ are implicitly defined by 
	\begin{equation}\label{eq:op-val-free-cumulant}
		E^\mathcal{D}[M_1\cdots M_n]=:\sum_{\pi\in NC(n)} \kappa_\pi^\mathcal{D}(M_1,\cdots,M_n)
	\end{equation}
	where $\kappa_\pi^\mathcal{D}$ is obtained from the family of linear functions $\kappa_n^\mathcal{D}:=\kappa_{1_n}^\mathcal{D}$ by respecting the nested structure of the parts appearing in $\pi$ as explained in the following example.
\end{definition}

\begin{example}\label{ex:nested}
	For $\pi=\{\{1,3\},\{2\},\{4,5\},\{6\}\}$, which corresponds to the dotted lines in the following figure, $\kappa_\pi^\mathcal{D}$ is defined as
	\begin{center}
		\includegraphics[width=0.2 \linewidth]{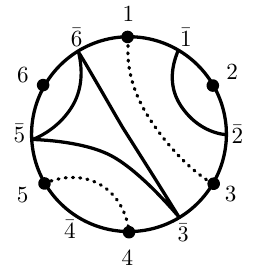}
	\end{center}
	\begin{equation*}
		\kappa_\pi^\mathcal{D}(M_1,M_2,M_3,M_4,M_5,M_6):=\kappa_2^\mathcal{D}(M_1\cdot \kappa_1^\mathcal{D}(M_2),M_3)\cdot \kappa^\mathcal{D}_2(M_4,M_5)\cdot \kappa^\mathcal{D}_1(M_6).
	\end{equation*}
	Note that one deals with matrix products, specifically emphasized by the dot $\cdot$ in this example, which is omitted elsewhere.
\end{example}
\medskip
Next we would like to relate $\kappa_n^\mathcal{D}$ to the local free cumulants $g_n$. In the large $N$ limit with $x=i/N$, we introduce the notation $E^\mathcal{D}[M](x):=E^\mathcal{D}[M]_{ii}\in \mathbb R$ to denote a diagonal elements of $\mathcal D$. By Eq.\eqref{eq:moment_free_cumulants}, we can express the $\mathcal D$-valued moments as
\begin{equation*}
	E^\mathcal{D}[M\Delta_1 M\cdots M\Delta_n M](x)=\sum_{\pi\in NC(n+1)}\int \mathrm d \vec x^{(n)}\Delta_1(x_1)\cdots \Delta_n(x_n)g_\pi(\vec x^{(n)},x) \delta_{\pi^*}(\vec x^{(n)},x)
\end{equation*}
Here we interchanged the roles of $\pi$ and $\pi^*$ which does not change the sum. Comparing to the definition of operator-valued free cumulants, this suggest the following identification.
\begin{proposition} Let $\pi\in NC(n+1)$, we have
	\begin{equation}\label{eq:op-val-cum_pi}
		\kappa_\pi^\mathcal{D}(\underbrace{M,\cdots,M}_{n},M)(x)=\int \mathrm d \vec x^{(n)} g_\pi(\vec x^{(n)},x) \delta_{\pi^*}(\vec x^{(n)},x)
	\end{equation}
\end{proposition}
\begin{proof}
	We must check that this identification can be consistently obtained from the case $\pi=1_{n+1}$ by respecting the nested structure appearing in $\kappa_\pi^\mathcal{D}$. That is, we define
	\begin{equation}\label{eq:op-val-cum_n}
		\kappa_{n+1}^\mathcal{D}(M\Delta_1,\cdots,M\Delta_n,M)(x):=\int \mathrm d \vec x^{(n)} \Delta_1(x_1)\cdots \Delta_n(x_n) g_\pi(\vec x^{(n)},x),
	\end{equation}
	and show that this implies the proposition. It is important to have included the diagonal $\Delta_i$'s in this definition, since this allows us to resolve nested terms such as $\kappa_2^\mathcal{D}(M\, \kappa_1^\mathcal{D}(M),M)$. In fact, one soon notices that Eq.\eqref{eq:op-val-cum_pi} is precisely the definition of the nested structure of $\kappa_\pi^\mathcal{D}$. 
	
	We illustrate this using the above example with $\pi=\{\{1,3\},\{2\},\{4,5\},\{6\}\}$ and Kreweras complement $\pi^*=\{\{\bar 1,\bar 2\},\{\bar 3,\bar 5,\bar 6\},\{\bar 4\}\}$.
	The definition of $\kappa_{n+1}^\mathcal{D}$ implies that the l.h.s of Eq.\eqref{eq:op-val-cum_pi} becomes
	\begin{align*}
	\kappa_2^\mathcal{D}(M \kappa_1^\mathcal{D}(M),M)(x)\, \kappa^\mathcal{D}_2(M,M)(x)\, \kappa^\mathcal{D}_1(M)(x)
	= \int \mathrm d x_1\,g(x_1) g(x_1,x) \int \mathrm d x_2\, g_2(x_2,x) g_1(x)
	\end{align*}
	This corresponds indeed to the r.h.s. where $\delta_{\pi^*}(x_1,\cdots,x_6)=\delta(x_1-x_2) \delta(x_3-x_5)\delta(x_5-x)$ and we identified $x_6\equiv x$. An arbitrary $\pi\in NC(n+1)$ can be tackled in the same way identifying $x_{n+1}\equiv x$.
\end{proof}

This result also explains how the structure of $g_\pi \delta_{\pi^*}$ which we encountered in Eq.\eqref{eq:QSSEP_moments_G_cumulants} fits into the free probability picture. Earlier, we could only ascertain that the family $\tilde \kappa_\pi:=\int g_{\pi^*}(\vec x)\,\delta_\pi(\vec x)d\vec x$ are not the (scalar) free cumulants of $M$ because they where not multiplicative ($\tilde\kappa_\pi\tilde \kappa_\sigma\neq\tilde \kappa_{\pi\cup\sigma}$). Now we understand that they are nonetheless free cumulants, but in the operator valued setting with amalgamation over diagonal matrices. More precisely $\tilde \kappa_\pi =\ntr\left( \kappa_\pi^\mathcal{D}(M,\cdots,M)\right)$. This also suggests that calling the family of functions $g_n$ "local free cumulants" seems to be a good name choice.
\smallskip
\begin{definition} \label{def:op-val-R-and-Cauchy}
	The \emph{$\mathcal{D}$-valued $R$-transform}, $R_M:\mathcal D\to\mathcal D$ of an element $M\in\mathcal A$ is defined by
	\begin{equation}
		R_M(\Delta)=\sum_{n\ge0} \kappa_{n+1}^\mathcal{D}(M\Delta,\cdots,M\Delta,M)
	\end{equation}
	and the \emph{$\mathcal{D}$-valued Cauchy transform} (or \emph{resolvent}) $G_M:\mathcal{D}\to\mathcal D$ is defined by
	\begin{equation}
		G_M(\Delta)=E^\mathcal{D}[\frac 1 {\Delta-M}]=\sum_{n \ge 0} E^\mathcal{D}[\Delta^{-1}(M\Delta^{-1})^n]
	\end{equation}
\end{definition}
\begin{theorem}[see Thrm. 11 in Chpt. 9 of \cite{Mingo2017Free}]\label{thrm:R-Cauchy}	
	Similarly to the scalar-valued case in Eq.~\eqref{eq:r_transform-resolvent}, here the $R$- and Cauchy transforms are related by
	\begin{equation}
		G_M(\Delta)=\frac{1}{\Delta-R_M(G_M(\Delta))}.
	\end{equation}
\end{theorem}
The $\mathcal{D}$-valued Cauchy transform can be related to its scalar analogue $G(z)$ (denoted without the subscript) by 
\begin{equation*}
	G(z):=\ntr(G_M(z\mathbb I))=\E\big[\ntr\big(\frac{1}{z\mathbb I-M}\big)\big].
\end{equation*}
Let us now consider $M_h=h^{1/2}Mh^{1/2}\in\mathcal A$ and define (with $x=i/N$)
\begin{equation}
	\tilde a_z(x):=\lim_{N\to\infty}G_{M_h}(z\mathbb I)_{ii}.
\end{equation}
The scalar Cauchy transform of $M_h$ is then $G(z)=\int_0^1 \mathrm d x \, \tilde a(x)$.
Furthermore, from Eq.\eqref{eq:op-val-cum_n} one sees that
\begin{equation}\label{eq:op-val-R}
	R_{M_h}(\Delta)(x)=\sum_{n\ge 0} \int \mathrm d\vec x^{(n)} \, \Delta(x_1)h(x_1)\cdots \Delta(x_n)h(x_n)h(x)g_{n+1}(\vec x^{(n)},x)
\end{equation}
Together with Theorem \ref{thrm:R-Cauchy} we therefore obtain,
\begin{equation}
	\tilde{a_z}(x)=\frac 1 {z-R_{M_h}(\tilde a_z)(x)}=\frac{1}{z-h(x)R_0[h\tilde a_z](x)}.
\end{equation}
In the last equality we used the definition of $R_0$ from Eq.\eqref{eq:R_0}. Redefining $a_z(x)=h(x)\tilde a_z(x)$ we obtain the extremization conditions in Eq.\eqref{eq:a-b}, which are equivalent to the variational principle in Theorem \ref{thrm:F}.

\chapter{Conclusion and perspectives}
Having started off with the aim to find a quantum generalization of the macroscopic fluctuation theory (MFT) which includes coherent effects, we have to admit that even after three years of work we are still far from this goal. However, we did make important steps into this direction, notably by identifying possible elements of the mathematical structure underlying such a theory, inspired by exact results from the toy model QSSEP. These elements are the three properties of the measure of coherences, which we identified to be responsible for the relation with free probability and which we rederived from a heuristic consideration of more generic mesoscopic systems.

Independent of our guiding question about a quantum coherent mesoscopic fluctuation theory, one should stress that the many mathematical results we obtained about QSSEP bear a value in its own. Being a \textit{minimal model} with just enough structure to observe fluctuating off-diagonal elements of the density matrix, i.e.\ coherences, QSSEP might well be important in physical contexts we are not yet aware of. At the same time, this can be seen as a heavy criticism on the approach we have chosen: We have studied a toy model without having a concrete physical application in mind. Here I would oppose that a concrete physical application has -- at least so far -- not been our aim. Rather, we wanted to explore the space of possible statistical theories for coherent non-equilibrium phenomena which quantum mechanics and a bit of intuition from the macroscopic fluctuation theory would allow.

\paragraph{Summary of results.} We first summarize the mathematical results we have obtained about QSSEP, then comment on its (less rigorous) relation to mesoscopic systems and conclude by recalling the purely mathematical results on the spectrum of random matrices. 

Mathematical results about QSSEP: (1) We showed how to obtain a meaningful description of QSSEP in the scaling limit of large space and times. This is a first necessary step towards a hydrodynamic theory for coherences. (2) We found the surprising relation between moments and cumulants of coherences as a sum over non-crossing partitions, instead of all partitions. This has allowed us to incorporate techniques from free probability into our mathematical toolbox, which otherwise consists mainly of It\={o} calculus for stochastic Hamiltonians. (3) In particular, the structure of non-crossing partitions has allowed us to find a closed expression for connected loop expectation values of coherences in the steady state. (4) We have also been able to obtain a large deviation principle and the generating function for these cumulants, but the rate function for the probability of coherences in the scaling limit is still unknown. (5) Free probability techniques such as the R-transform have allowed us to address the long standing problem of entanglement in the (open) QSSEP. Notably, we found that the quantum mutual information between two adjacent subsystems scales as the volume (in 1D that is the length) of the subsystems. This makes the situation drastically differ from equilibrium, as well as from fully interacting driven random unitary circuits, where the mutual information generically scales as the area. (6) Last, but not least, we have investigated if QSSEP is Yang-Baxter integrable, using both an algebraic and a numerical approach based on the level spacing statistics. While algebraically, QSSEP does not map to any known integrable model, the numerical test shows small deviations from the spectrum of integrable systems which suggests that QSSEP is non-integrable. However, we stress that the numerical results are not decisive enough to completely rule out the possibility that QSSEP might still be integrable.

Next, on the less rigorous side, we have tried to develop the picture, that QSSEP is an effective stochastic description of mesoscopic diffusive systems. To that end we have introduced the notion of small ballistic cells of size comparable to the mean free path, inside which the dynamics is fast and ballistic, but outside of which diffusion emerges. A single site of QSSEP is thought to be the effective noisy description of such a ballistic cell. The picture has allowed us to rederive the three important properties which relate QSSEP to free probability for a generic mesoscopic diffusive system that satisfies MFT. To remind the reader, the measure of coherences satisfies: (i) $U(1)$-invariance, (ii) large deviation scaling of loop expectation values with distinct indices, (iii) factorization of the expectation value of pinched or disconnected loops. We are of course far from "proving" that our picture is correct, which would require to carry out this renormalization or coarse-graining procedure over ballistic cells for an explicit physical microscopic model, but it provides insights into the mathematical structure which a hydrodynamical theory for fluctuating coherences could obey.

Finally, one the purely mathematical side, we have found a variational principle to characterize the spectrum of subblocks of any random matrix satisfying the three properties. This applies in particular for structured random matrices, which are, in law, not invariant under a permutation of their elements. However, except for the QSSEP random matrix class, at the moment we do not know of other random matrix ensemble where this method could lead to new results.

\paragraph{Perspectives.}
It would be interesting to compute the fluctuations of charge such as $\E[(\<Q_t^k\>^c)^2]^c$ within QSSEP and to compare this to the results for mesoscopic diffusive conductors in Eq.~\eqref{eq:disorder_fluctuations}. However, inspection of the protocol for the measurement of charge transfer from Eq.~\eqref{eq:charge_fluctuations_protocol} shows that this would require the knowledge of coherences between different times
\begin{equation}
	G_{ij}(t,t')=\Tr(\rho \,c_i^\dagger(t)c_j(t')),
\end{equation}
here in the Heisenberg picture. We would thus need to carry out the procedure for the scaling limit again and find a stationary solution\footnote{With a big disclaimer: A first guess is that in this case only loop expectation values of the form $\E[G_{i_1,i_2}(t_1,t_2) G_{i_2,i_3}(t_2,t_3)\cdots G_{i_n,i_1}(t_n,t_1)]$ contribute in the scaling limit}.

Another, equally important question is to find a stochastic process for coherences in QSSEP directly in the scaling limit. These equations are a first version of what we would call a quantum coherent extension of MFT. For now the matrix-valued process for $I_{x,t}$ in Eq.~\eqref{eq:dI} provides us with a preliminary answer, but the physical interpretation is completely missing. We do not known how to make the connection between the matrices $I_{x,t}$ defined at a single point and coherences $G_{ij}$ defined at two points.

Furthermore, we would like to have a large deviation rate function for the probability to observe a given fluctuation of coherences. We already have the cumulant generating function, so the rate function obtained as the Legendre transform should not be too hard to find.

Finally, from the mathematical point of view it could be interesting to further explore the consequences of three properties (i)-(iii). For example, what are the constraints on arbitrarily chosen functions $g_n$ such that they can be view as the local free cumulants of some random matrix ensemble, that would then be defined through $g_n$. Even in classical probability theory, not every sequence of numbers defines cumulants of a probability measure. Furthermore, one can ask if the properties (i)-(iii) are stable under non-linear operations on the matrix entries (which has recently been answered in our updated version of \cite{Bernard2024Structured}).

\subparagraph{Comparison to the 3D Anderson model.} 
Finally it would be important to acquire a better understanding of the physical applications of QSSEP. Are there more physical models that can be studied numerically and be compared to the exact results of QSSEP \cite{HruzaJin2024Fluctuating}? A very suitable model for this purpose might be the 3D Anderson model which has been very successful in describing mesoscopic diffusive transport in disorder media. For example, it reproduces the weak localization correction to the mean conductance, or the universal conductance fluctuations \cite{Markos2006Numerical}. And it has been shown in Ref.~\cite{Gullans2019Entanglement} that its mutual information satisfies a volume law (if all sides of the 3D sample are of same length). At the same time, it naturally comes with a notion of classical noise -- the static disorder -- such that the fluctuation of coherences with changing disorder $\E[G_{ij}G_{ji}]^c$ has a meaning and can be compared to the prediction of QSSEP. 

To be more precise, one would have to reduce the 3D Anderson model to one dimension by a spatial average over transverse slices. In addition to this, if one wants to take seriously the picture about ballistic cells, it might be necessary to also average over the mean free path $\ell$ in the longitudinal direction. The combined disorder and spatial average leaves several possibilities for the definition of cumulants. The most natural would be to first take the cumulant with respect to disorder $\E_\text{dis}$ and then perform the spacial average. Denoting $i=(i_x,i_y,i_z)$ and $i_\perp=(i_x,i_y)$ this would mean
\begin{equation}
	\E_\text{QSSEP}[G_{{i_z}{j_z}} G_{{j_z}{i_z}}]^c=\sum_{i_\perp, j_\perp}\Big(\E_\text{dis}[G_{ij}G_{ji}] -\delta_{ij} \E_\text{dis}[G_{ii}]^2\Big).
\end{equation}
But there are more possibilities, notably when including an average over the mean free path in the longitudinal direction, which needs to be explored. On the technical side, the disordered 3D Anderson model allows a very efficient numerical solution in terms of the transfer matrix method. Up to about 20 sites in each direction can be simulated without too much efforts \cite{Markos2006Numerical}.

\subparagraph{Experiments.}
In case the comparison to the Anderson model is successful, it would be interesting to work closer with experimentalists to understand to what extend such results could be verified. While it is possible in general to directly measure coherences in an experiment on well controlled ballistic conductors, this seems still to be a big challenge for diffusive disordered conductors. In this regard, it might be beneficial to find other quantities within QSSEP that depend on coherences, but could be measured by simpler experiments, such as the sample-to-sample variation of cumulants of the transported charge.
\appendix
\chapter{Appendices}

\section{Large deviation theory}\label{app:large_deviations}
A review of large deviation theory is provided in \cite{Touchette2009}. Here we derive the large deviation principle for a particular example, the density fluctuations in a system at equilibrium.
\paragraph{Large deviations in equilibrium.}
Consider a small volume $v$ inside a big volume $V$ with $N$ particles at inverse temperature $T$.
\begin{center}
	\includegraphics[width=0.4\textwidth]{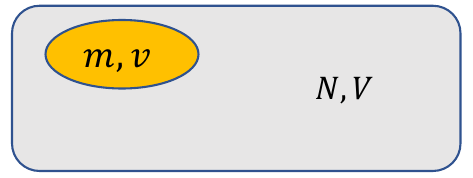}
\end{center}
What is the probability $P_{v}(m)$ that there are $m$ particle in the small volume? As we will show now, this question leads to a large deviation principle. We start by counting all configuration that correspond to this situations weighed by their probability. This is given by the canonical partition function $Z(N,V)=\sum_{\mathcal C=\mathcal C(N,V)} e^{-E(\mathcal C)/kT}$. Neglecting interactions on the boundary of the small volume, one has
\begin{equation}
	 P_{v}(m)=\frac{Z(m,v)Z(N-m,V-v)}{Z(N,V)}
\end{equation}
Expressing this via the system's free energy density at particle density $n$
\begin{equation}
	f(n)=-kT\lim_{V\to\infty}\frac{\log Z(n V,V)}{V}
\end{equation}
one has
\begin{align}
	- kT\,\log P_{v}(m)=v f\left(\frac m v\right) + (V-v) f\left(\frac{N-m}{V-v}\right) - V f\left(\frac N V\right).
\end{align}
Expanding 
\begin{equation}
	f\left(\frac{N-m}{V-v}\right)\approx f\left(\frac N V\right)-\frac{m-v\frac{N}{V}}{V-v}f'\left(\frac N V \right)
\end{equation}
and setting $n^*=N/V$ to be the mean density, one finds that the probability for a large deviation from the mean density is,
\begin{equation}
	P_v(m)=e^{-vI(m/v)}
\end{equation}
with rate function
\begin{equation}
	I(n=m/v)=\frac{1}{kT}\left(f(n)-f(n^{*})-(n-n^{*})f'(n^{*})\right).
\end{equation}

If one partitions the big volume $V$ into $l$ small volumes $v=\frac V l$ and one asks for the probability that they are filled with particle densities $n_1,\cdots,n_l$, one applies the same reasoning and has
\begin{equation}
	P(n_1,\cdots,n_l)=\frac{Z(n_1 v,v)\cdots, Z(n_l v,v)}{Z(N,V)}.
\end{equation}
From there one continues
\begin{equation}
	-kT \,\log P(n_1,\cdots,n_l) = \sum_{i=1}^l v f(n_i)- Vf(n^*)
\end{equation}
and therefore 
\begin{align}
	P(n_1,\cdots,n_l)=e^{-VI(n_1,\cdots,n_l)} \qquad \text{with }I(n_1,\cdots,n_l)= \frac{1}{kT}\frac{1}{l} \sum_{i=1}^l [f(n_i)-f(n^*)]
\end{align}
This is the discrete version of Eq.~\eqref{eq:density_large_deviation}.

\section{Gaussian fermionic states}\label{app:fermionic_gaussian}
We derive a few identities of fermionic Gaussian states. They are defined as 
\begin{equation}
	\rho=\frac{1}{Z(M)}\exp(\vec c^{\:\dagger} M \vec c), \qquad Z(M):=\Tr[\exp(\vec c^{\:\dagger} M \vec c)]
\end{equation}
with $\vec c=(c_1,\cdots,c_N)$ and $M$ a Hermitian matrix of size $N$. We denote $\Tr(\cdots)$ the trace over the fermionic Fock space of dimension $2^N$ and $\tr(\cdots)$ the trace over the one-particle Hilbert space of dimension $N$. To evaluate $Z(M)$, first note that $\Tr(e^{\mu c^\dagger c})=1+e^\mu$. Then, we diagonalize $M=V^{-1}DV$ with $D=\text{diag}(\mu_1,\cdots,\mu_N)$ and introduce $\vec d=V\vec c$ such that $\vec c^{\:\dagger} M c^\dagger = \sum_i \mu_i\, d^\dagger_i\, d_i$. Then
\begin{equation}
	Z(M)=\prod_i(1+e^\mu_i)=\det(1+e^M).
\end{equation}
Furthermore, we use that $\tau_*(E_{ij}):=c^\dagger_i c_j$ defines a Lie algebra representation of $\mathfrak{gl}(N)$, with $E_{ij}$ the matrix with a one at position $(ij)$. This is because it satisfies the $\mathfrak{gl}(N)$ commutation relations, cf.\ Eq.~\eqref{eq:gl_commutator}, $[c_i^\dagger c_j,c_k^\dagger c_l]=\delta_{jk} c_i^\dagger c_l-\delta_{li} c_k^\dagger c_j$. Then $\tau(e^M):=\exp(c^\dagger M c)$ is a group representation, and in particular we have that $\tau(e^M e^A)=\exp(c^\dagger M c)\exp(c^\dagger A c)$. Since $Z[M]=\Tr[\tau(e^M)]$, we use this to evaluate
\begin{equation}
	\frac{\Tr[e^{\vec c^{\:\dagger} M \vec c}e^{\vec c^{\:\dagger} A \vec c}]}{Z(M)}
	= \frac{\Tr[\tau(e^M eA)]}{Z(M)}
	=\det\left(\frac{1+e^M e^A}{1+e^M}\right).
\end{equation}
The term proportional to $A_{ij}$ on the left hand side is $G_{ij}:=\Tr(\rho c_i^\dagger c_j)$. The right hand side is equal to $\det(1+\frac{e^M}{1+e^M}(e^A-1))\approx 1+\tr(\frac{e^M}{1+e^M} A)$ to first order. Therefore, to first order in $A$
\begin{equation}
	G=\frac{e^M}{1+e^M}.
\end{equation}
\paragraph{Wick's theorem.}
Another useful identity is Wick's theorem. Let $\rho$ be a fermionic Gaussian state as above and $\eta_i$ denotes either $c_i^\dagger$ or $c_i$, then
\begin{align}
	\Tr(\rho \,\eta_1 \eta_2 \eta_3 \eta_4\cdots)=&\Tr(\rho\,\eta_1 \eta_2)\Tr(\rho \,\eta_3 \eta_4 \cdots)-\Tr(\rho\,\eta_1 \eta_3)\Tr(\rho\, \eta_2 \eta_4 \cdots) \nn
	&+\Tr(\rho\,\eta_1 \eta_4)\Tr(\rho \,\eta_2 \eta_3 \cdots)-\cdots
\end{align}

\section{Free probability glossary}\label{app:glossary}
Taken from \cite{Bernard2024Structured}: In the following, let $\sigma$ be a (classical) measure for a random variable $X$ with expectation value $\E_\sigma$. The notation is choose in accordance with Speicher \cite{Speicher2019Lecture}.
\begin{itemize}
	\item 
	The \textit{resolvent} $G_\sigma(z):=\mathbb{E}_\sigma[\frac{1}{z-X}]$ generates the moments $m_n:=\mathbb{E}_\sigma[X^n]$
	\begin{align*}
		G_\sigma(z) &:= \sum_{n\geq 0} m_n\, z^{-n-1} = z^{-1} + m_1 z^{-2} + m_2 z^{-3} +\cdots\\
		M_\sigma(z)&:= z^{-1}G_\sigma(z^{-1}) = \sum_{n\geq 0} m_n\, z^{n} =  1 + m_1 z + m_2 z^{2} +\cdots
	\end{align*}
	
	\item 
	The \textit{$R$-transform} $R_\sigma$ is a generating function of the free cumulants $\kappa_p:=\kappa_p(\sigma)$
	\begin{align*}
		R_\sigma(z) &:= \sum_{p\geq 1} \kappa_p\, z^{p-1} = \kappa_1  + \kappa_2 z + \kappa_3 z^2 +\cdots \\
		K_\sigma(z) &:=z^{-1}+R_\sigma(z) = \sum_{p\geq 0} \kappa_p\, z^{p-1} = z^{-1}+ \kappa_1  + \kappa_2 z + \kappa_3 z^2 +\cdots \\
		C_\sigma(z) &:=zR_\sigma(z)= \sum_{p\geq 1} \kappa_p\, z^{p} = \kappa_1 z  + \kappa_2 z^2 + \kappa_3 z^3 +\cdots
	\end{align*} 
	(Our definition of $C_\sigma$ differs from Speicher's by starting the sum at $p=0$.)
	
	\item
	The function $G_\sigma$ and $K_\sigma$ are \textit{inverses} of each other, thus
	\begin{equation*}
		K_\sigma(G_\sigma(z))=z,\quad G_\sigma(K_\sigma(z))=z
	\end{equation*}
	The previous relation then reads
	\begin{equation*}
		zG_\sigma(z) = 1 + C_\sigma(G_\sigma(z))\ ,\ 
		\hat M_\sigma(z) = 1 + C_\sigma(z\hat M_\sigma(z)) .
	\end{equation*}
	\item 
	The \textit{$S$-transform} can be defined by
	\begin{equation*}
		C_\sigma(zS_\sigma(z))=z,\quad C_\sigma(z)\,S_\sigma(C_\sigma(z))=z
	\end{equation*}
	The function $S_\sigma$ exists, as a formal power in $z$, whenever $\kappa_1\not= 0$ : $S_\sigma(z) = \frac{1}{\kappa_1} - \frac{\kappa_2}{\kappa_1^3}z + \cdots$.
	Using $G_\sigma(K_\sigma(z))=z$, this relation can alternatively be written as
	\[
	G_\sigma\left(\frac{1+z}{zS_\sigma(z)}\right)= zS_\sigma(z),\quad S_\sigma(zG_\sigma(z)-1)=\frac{G_\sigma(z)}{zG_\sigma(z)-1} .
	\]
	Setting $w=zG_\sigma(z)-1$, the above formula can be written as $S_\sigma(w)=\frac{w+1}{zw}$ with $z(w)$ determined by solving $zG_\sigma(z)=w+1$.
	
	\item 
	For two measures $\sigma$ and $\nu$, the \textit{additive free convolution} is defined 
	\[ R_{\sigma \boxplus\nu}(z) = R_{\sigma}(z)  + R_{\nu}(z) ,\]
	that is, we add the free cumulants. Thus if $a$ and $b$ are (relatively) free then $R_{a+b}(z)=R_a(z)+R_b(z)$.
	
	\item 
	For two measures $\sigma$ and $\nu$, the \textit{free multiplicative convolution} $\sigma \boxtimes\nu$ is defined via their $S$-transform
	\[ S_{\sigma \boxtimes\nu}(z) = S_{\sigma}(z)  S_{\nu}(z) ,\]
	that is, we multiply the $S$-transforms. Thus, if $a$ and $b$ are (relatively) free, then $S_{ab}(z)=S_a(z)S_b(z)$ (instead of $ab$ we could have considered $a^{1/2}ba^{1/2}$).
\end{itemize}


%
%

\section{Wigner and Haar rotated matrices}\label{app:wigner_haar}
The paragraphs are taken from \cite{Bernard2024Structured}: Here we show how the result on the spectrum of subblocks applies to two well-known matrix ensembles.
\paragraph{Wigner matrices}
Wigner matrices are characterized by the vanishing of its associated free cumulants of order strictly bigger than two. Thus, for Wigner matrices only $g_1$ and $g_2$ are non vanishing and both are $x$-independent. All $g_n$, $n\geq 3$, are zero. Without loss of generality we can choose $g_1=0$ and we set $g_2=s^2$. Then $F_0[p]=\frac{s^2}{2}\int \!dx dy\, p(x)p(y)$ and $R_0[p]=s^2 \int\! dx\, p(x)$. For the whole interval $h(x)=1$ (considering a subset will be equivalent), the extremization equations \eqref{eq:a-b} become
\[
a = \frac{1}{z- b},\ b = s^2\, A, 
\]
with $A=\int \!dx\, a(x)$. This yields a second order equation for $A$, i.e. $A^{-1}=z-s^2 A$. Solving it, with the boundary condition $A\sim \frac{1}{z}+\cdots$ at $z$ large, gives
\[ 
A = \frac{1}{2s^2}\left( z - \sqrt{z^2-4s^2}\right)
\]
Thus the cut is on the interval $[-2s,+2s]$ and the spectral density is
\begin{equation}
d\sigma(\lambda) = \frac{d\lambda}{2\pi s^2}\, \sqrt{4s^2-\lambda^2}\ 1_{\lambda\in[-2s,+2s]}
\end{equation}
Of course, that's Wigner's semi-circle law.

\paragraph{Inhomogeneous Wigner.}
Next we consider $N\times N$ Wigner matrices with zero mean and variance
\[
\mathbb{E}[M_{ij}M_{kl}] = \frac{1}{N} \delta_{jk}\delta_{il}\, g_2(\frac{i}{N},\frac{j}{N}) ,
\]
with $g_2(x,y)$ a (smooth) function. It is clear that the three fundamental properties (i)-(iii) are satisfied. We restrict to diagonal covariances $g_2(x,y)=s^2(x)\delta(x-y)$, because otherwise we cannot find closed expressions for the spectrum. The saddle point equation is then a quadratic equation for $a_z(x)$ which, in the case $h(x)=1$, reads $a_z(x)(z-s^2(x)a_z(x))=1$ so that
\[
a_z(x) = \frac{1}{2s^2(x)}(z-\sqrt{z^2-4s^2(x)}) .
\]
The resolvent is $G(z)=\int\! dx\, a_z(x)$. Its discontinuity at the cut is the sum of the discontinuities for each value of $x$. This yields for the spectral density
\begin{equation}
d\sigma(\lambda) = \frac{d\lambda}{2\pi}\int\! \frac{dx}{s^2(x)}\, \sqrt{4s^2(x)-\lambda^2}.
\end{equation}

\paragraph{Haar-randomly rotated matrices.} \label{subsec:Haar}
We consider matrices of the form $M=UDU^\dag$, with $U$ Haar distributed over the unitary group and $D$ a diagonal matrix with spectral density $\sigma$ in the large $N$ limit. For such matrices, it is known that the local free cumulants are constant and equal to the free cumulants of $\sigma$, that is
\begin{align} \label{eq:Haar-loop}
	g_n(\vec x):=\lim_{N\to\infty}N^{n-1}\mathbb{E}[M_{i_{1}i_{2}}M_{i_{2}i_{3}}\cdots M_{i_{n}i_{1}}]^c=\kappa_n(\sigma) .
\end{align}
\begin{proof}
	From the HCIZ integral we had Eq.~\eqref{eq:large-HCIZ}, which we state again,
\[
\mathbb{E}[e^{z N \tr(QM)}] \asymp_{N\to\infty} \exp\Big( N \sum_{k\geq1} \frac{z^k}{k}\tr(Q^k)\,\kappa_k(\sigma) \Big)
\]
where $\kappa_n(\sigma)$ are the free cumulants of the density $\sigma$ and $Q$ is any finite rank matrix.

Let us prove that this implies that the local free cumulants are $g_n=\kappa_n(\sigma)$, that is
\begin{align} \label{eq:Haar-loop}
	\mathbb{E}[M_{12}M_{23}\cdots M_{n1}]=N^{1-n}\, \kappa_n(\sigma)\, (1+O(N^{-1}))
\end{align}
Note that due to $U(N)$ invariance (which in particular includes permutations), all sets of distinct indices $i_1,i_2,\cdots,i_n$ are equivalent.

Choose $Q=P_n$ the cyclic permutation $(12\cdots n)$, so that $\tr(P_nM)=M_{12}+M_{23}+\cdots +M_{n1}$. It is easy to see (using $U(1)^N\subset U(N)$ invariance), that the first non-vanishing term in $\mathbb{E}[e^{zN \tr(P_nM)}]$ is of order $z^n$ and given by $z^n N^n \mathbb{E}[(\tr(P_nM))^n]$. Furthermore, (this can be proved say by recurrence)
\[
\mathbb{E}[(\tr(P_nM))^n] = \mathbb{E}[(M_{12}+M_{23}+\cdots +M_{n1})^n]{=} n!\, \mathbb{E}[M_{12}M_{23}\cdots M_{n1}]
\]
Thus 
\[
\mathbb{E}[e^{z N \tr(P_nM)}] = z^nN^n\, \mathbb{E}[M_{12}M_{23}\cdots M_{n1}] + O(z^{n+1})
\]
Since $\tr(P_n^k)=0$ for $k<n$ and $\tr(P_n^n)=n$, we have 
\[
e^{N \sum_{k\geq0} \frac{z^k}{k}\tr(P_n^k)\,\kappa_k(\sigma)}= N z^n \kappa_n(\sigma)+ O(z^{n+1})
\]
Comparing the two last equations proves Eq.\eqref{eq:Haar-loop}.
\end{proof}

Of course the spectrum of the whole matrix $M$ is that of $D$ with spectral density $\sigma$. Let us check this within our variational principle. With  $g_n=\kappa_n(\sigma)$ and $h(x)=1$ (we consider the whole matrix $M$), Eqs.\eqref{eq:a-b} become 
\[
A =\frac{1}{z- b_z(A)},\  b_z(A)= \sum_{k\geq 1} A^{k-1}\kappa_k(\sigma), 
\]
with $A=\int\!dx\, a_z(x)$. Let us recall some basics definition from free probability. For any measure $\sigma$ of some random variable $X$, let $G_\sigma(z)=\mathbb{E}[\frac{1}{z-X}]= \sum_{n\geq 0}z^{-n-1}m_n(\sigma)$  and $K_\sigma(z)=\sum_{n\geq 0}z^{n-1}\kappa_n(\sigma)$, with $m_n$ and $\kappa_n$ the $n$-th moments and free cumulants, respectively. As well known from free probability, $G_\sigma$ and $K_\sigma$ are inverse functions, i.e. $K_\sigma(G_\sigma(z))=z$. Comparing with the previous equation, we see that $b_z(A)=K_{\sigma}(A)-A^{-1}$. The equation $A=1/(z- b_z(A))$ can thus be written as $z= b_z(A)+A^{-1}=K_{\sigma}(A)$, and hence 
\[
A=G_{\sigma}(z)
\]
As a consequence, the resolvent of $M$ is equal to $G_{\sigma}(z)$ and the spectral density of $M$ is indeed that of $D$, as it should be.

\section{Explicit $\mathfrak{gl}(4)$ generators} \label{app:gl4_generators}
Taken from \cite{Bernard2022DynamicsClosed}: We chose an explicit basis of the local fermionic two-replica Hilbert space $\Hil_j$:
\begin{center}
	\begin{tabular}{|c|c|c|c|c|c|c|c|}
		\hline
		$|0\rangle$ &
		$|1\rangle$ &
		$|2\rangle$ &
		$|3\rangle$&
		$|4\rangle$ &
		$|5\rangle$ &
		$|6\rangle$ &
		$|7\rangle$ \\
		\hline
		$c^\dagger_1c_2$
		&$n_1n_2$
		&$n_1(1-n_2)$
		&$(1-n_1)n_2$
		&$(1-n_1)(1-n_2)$
		&$c_1c^\dagger_2$
		&$n_1c_2$
		&$c_1n_2$\\
		\hline
	\end{tabular}
\end{center}
\begin{center}
	\begin{tabular}{|c|c|c|c|c|c|c|c|}
		\hline
		$|8\rangle$ &
		$|9\rangle$ &
		$|10\rangle$ &
		$|11\rangle$ &
		$|12\rangle$ &
		$|13\rangle$ &
		$|14\rangle$ &
		$|15\rangle$ \\
		\hline
		$(1-n_1)c_2$
		&$ c_1(1-n_2)$
		&$n_1c_2^\dagger$
		&$c_1^\dagger n_2$
		&$(1-n_1)c_2^\dagger$
		&$ c_1^\dagger(1-n_2)$
		&$c_1c_2$
		&$c_1^\dagger c_2^\dagger$\\
		\hline
	\end{tabular}
\end{center}
The $\mathfrak{gl}(4)$ generators $G^{AB}$ can now be written out as $16$-dimensional matrices. Denote by ${\cal E}_{a,b}=|a\rangle \langle b|$ a matrix with a one at position $(a,b)$ and zero otherwise. Then we have
\begin{align}
	G^{12}&={\cal E}_{1,3}+{\cal E}_{2,4}+{\cal E}_{6,8}+{\cal
		E}_{10,12}\ ,\nn
	G^{13}&={\cal E}_{0,4}+{\cal E}_{1,5}-{\cal E}_{6,9}-{\cal E}_{11,12}\ ,\nn
	G^{14}&={\cal E}_{0,3}-{\cal E}_{2,5}-{\cal E}_{6,7}+{\cal E}_{13,12}\ ,\nn
	G^{23}&=-{\cal E}_{0,2}+{\cal E}_{3,5}-{\cal E}_{8,9}+{\cal E}_{11,10}\ ,\nn
	G^{24}&=-{\cal E}_{0,1}-{\cal E}_{4,5}-{\cal E}_{8,7}-{\cal E}_{13,10}\ ,\nn
	G^{34}&={\cal E}_{2,1}+{\cal E}_{4,3}+{\cal E}_{9,7}+{\cal E}_{13,11}\ ,\nn
	G^{11}&={\cal E}_{0,0}+{\cal E}_{1,1}+{\cal E}_{2,2}+{\cal E}_{6,6}+{\cal  E}_{10,10}+{\cal  E}_{11,11}+{\cal  E}_{13,13}
	+{\cal  E}_{15,15}\ ,\nn
	G^{22}&={\cal E}_{0,0}+{\cal E}_{3,3}+{\cal E}_{4,4}+{\cal  E}_{8,8}+{\cal E}_{11,11}+{\cal  E}_{12,12}+{\cal  E}_{13,13}+{\cal  E}_{15,15}\ ,\nn
	G^{33}&={\cal E}_{2,2}+{\cal E}_{4,4}+{\cal E}_{5,5}+{\cal  E}_{7,7}+{\cal  E}_{10,10}+{\cal  E}_{11,11}+{\cal  E}_{12,12}+{\cal  E}_{15,15}\ ,\nn
	G^{44}&={\cal E}_{1,1}+{\cal E}_{3,3}+{\cal E}_{5,5}+{\cal E}_{9,9}+{\cal E}_{10,10}+{\cal  E}_{12,12}+{\cal E}_{13,13}+{\cal E}_{15,15}\ .
\end{align}
and the remaining ones are related by $(G^{ab})^\dagger=G^{ba}$

\section{Unfolding and level-spacing for a Poisson process.}\label{app:unfolding}
Taken from \cite{Bernard2022DynamicsClosed}: Given $N$ i.i.d random variables $X_i$ (the eigenvalues of an integrable Hamiltonian) taking values in $\mathbb{R}$ with density $p_X(x)$, we perform a local change of variables such that the new variable $\hat x(x)$ describes the average number of old variables below $x$,
\begin{equation}
	x\to \hat x(x)=N \int_{-\infty}^x p_X(x')dx'.
\end{equation}
This procedure is called "unfolding the spectrum" and it ensures that the density of the new variables $\hat p_X(\hat x) d \hat x = p_X(x) dx$ is indeed uniform, $\hat p_X(\hat x)=1/N$. 

The probability to find $N_s=k$ of the new random variables in the interval $[0,s]$ is now is given by the Poisson distribution
\begin{equation}
	\mathbb{P}[N_s=k]= \frac{(\lambda s)^k}{k!}e^{-\lambda s}, 
\end{equation}
with average "even rate" $\lambda=1$. We can define the probability $p_S(s)$ to observe a spacing $s$ between adjacent eigenvalues by 
\begin{equation}
	\mathbb{P}[N_s=0]=\int_s^\infty p_S(s')ds'.
\end{equation}
Deriving w.r.t $s$ provides us with the exponential distribution $p_S(s)=e^{-s}$.

%
%
%

\printbibliography[heading=bibintoc]

\clearpage
\begin{figure}
	\centering\includegraphics[width=\textwidth]{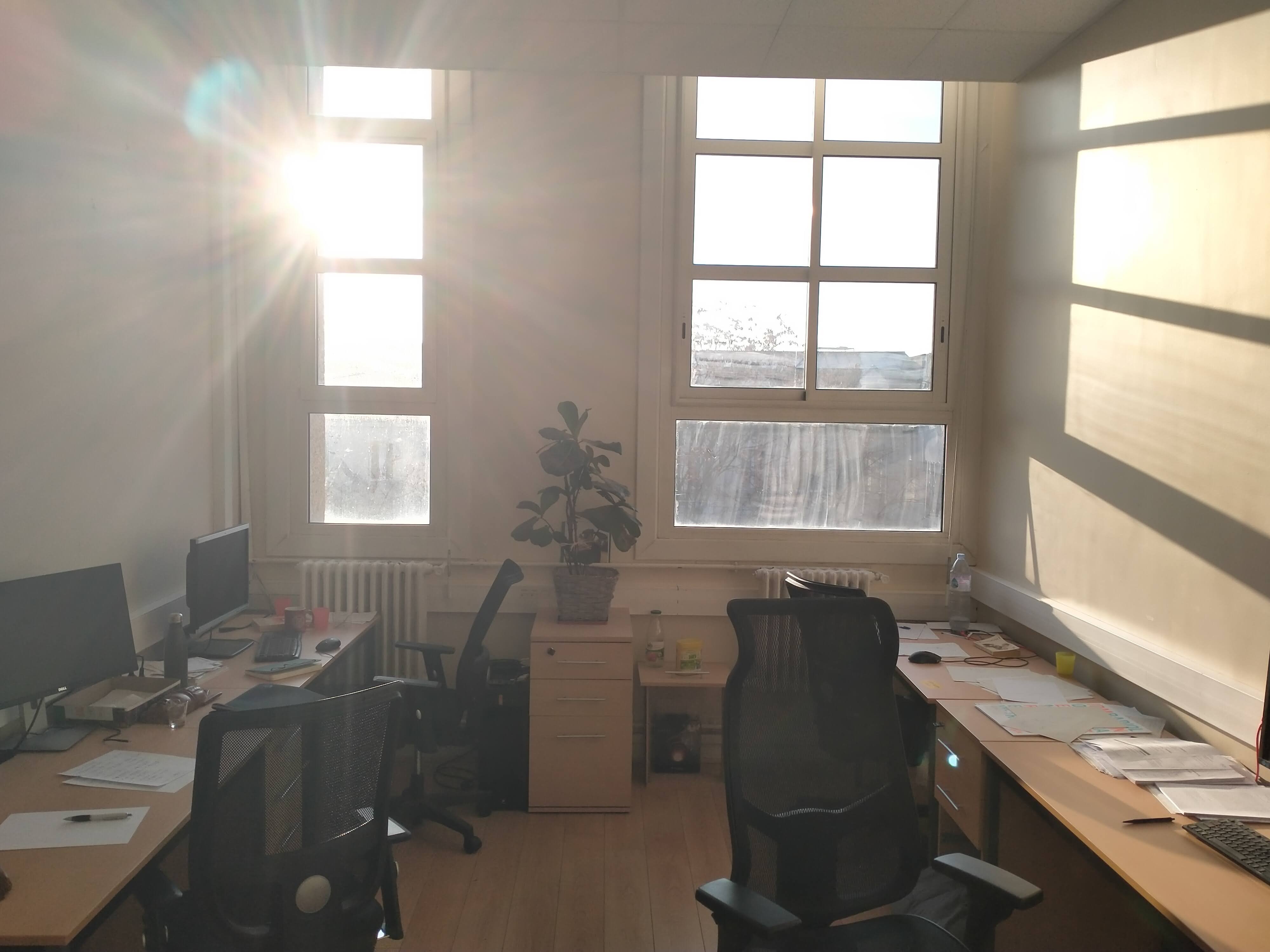}
	\caption*{People leave, our office stays -- and the sun comes back, every day anew.}
\end{figure}
\clearpage

\end{document}